\documentclass[twoside,11pt,aasms,eqsecnum]{article}
\usepackage{epsfig}

\usepackage{hyperref}
\usepackage{url}
\usepackage{amssymb}
\usepackage{amsmath}

\usepackage{mathabx}
\usepackage{bm}

\usepackage{xcolor}
\newcommand{\B}{\color{blue}}

\newcommand{\G}{\color{gray}}

\definecolor{green2}{rgb}{0,0.5,0}
\definecolor{mangeta2}{rgb}{0.57,0.36,0.51}
\definecolor{mangeta3}{rgb}{0.5,0.,0.}

\newcommand{\ms}{\scriptscriptstyle}

\def\Unit{1\kern -4pt 1}

\def\sp{\kern +3pt}
\def\sm{\kern -3pt}
\def\spQ{\kern +6pt}

\def\sfrac#1#2{{\textstyle \frac{#1}{#2}}}

\usepackage{subcaption}

\newcommand{\be}{\begin{equation}}
\newcommand{\ee}{\end{equation}}
\newcommand{\ba}{\begin{eqnarray}}
\newcommand{\ea}{\end{eqnarray}}

\begin{document}

\title{ \vspace{1cm} Electromagnetic Transition Form Factors of Baryon Resonances}
\author{G.\ Ramalho$^{1}$, 
  M.T.\ Pe\~na$^{2,3}$,\\
$^1$Department of Physics and OMEG Institute, Soongsil University, \\
Seoul 06978, Republic of Korea \\
$^2$LIP, Laborat\'orio de Instrumenta\c{c}\~ao e F\'{i}sica 
Experimental de Part\'{i}culas, \\
Avenida Professor Gama Pinto, 1649-003 Lisboa, Portugal \\
$^3$Departamento de F\'{i}sica e Departamento de Engenharia e Ci\^encias Nucleares, \\
Instituto Superior T\'ecnico (IST),
Universidade de Lisboa, \\
Avenida Rovisco Pais, 1049-001 Lisboa, Portugal}
\maketitle

\begin{abstract}
Recent experimental and theoretical advancements have led to significant progress
in our understanding of the electromagnetic structure of nucleons ($N$),
nucleon excitations ($N^\ast$), and other baryons.
These breakthroughs have been made possible by the capabilities of modern facilities,
enabling the induction of photo- and electro-excitation of nucleon resonances.
These experiments have specifically probed the evolution of their
electromagnetic structure across
a range of squared momentum transfer scales,
from $Q^2= 0-0.01$ GeV$^2$ up to $Q^2=5$ or $8$ GeV$^2$.
These experimental advances have sparked notable developments in theoretical approaches.
New theoretical methods have been tested and proven to be robust,
marking the beginning of a new era in our understanding on baryons.
This includes the study of newly discovered exotic hadrons with various multiquark components. 
We present a comprehensive review of progress in experimental data on
$\gamma^\ast N \to N^\ast$  reactions. Additionally,
we discuss various analyses and theoretical results, such as quark models
in combination (or not) with meson cloud excitations of the baryon quark cores, lattice QCD,
Dyson-Schwinger equations, chiral effective field theory, the large $N_c$ limit,
and AdS/CFT correspondence, among others.
Some of these methods have matured in their predictive power,
offering new perspectives on exotic hadrons with multiquark components.
We place special emphasis on both the low-$Q^2$ and large-$Q^2$
regions to reinforce crucial physical constraints on observables that
hold in these limits. Furthermore, we illustrate that the combination of lattice QCD
with chiral effective field theory and quark models,
respectively, proves beneficial in interpreting data and applying constraints
within those different regimes.
As a practical contribution and for future reference,
we review the formulas for helicity amplitudes,
multipole form factors and the relations between these two sets of functions
for transitions to resonances with general spin $J \geq \frac{1}{2}$.
These formulas are ubiquitous and play a pivotal role in experimental
and theoretical studies on baryon structure.
Notably, the multipole transition form factors for $J \ge  \frac{3}{2}$ resonances momentum
serve as valuable tools to test perturbative QCD results in the large-$Q^2$ region,
thanks to the correlations between electric and magnetic transition form factors.
\end{abstract}

\tableofcontents


\renewcommand{\theequation}{1.\arabic{equation}}
\setcounter{equation}{0}

\renewcommand{\thefigure}{1.\arabic{figure}}
\setcounter{figure}{0}

\section{Introduction}

Hadrons, consisting of two families, baryons and mesons,
constitute most part of the visible universe, and they are manifest
in a wide range of energy scales.
At low energies, baryons can become bound into stable
or unstable nuclei of atoms in normal matter,
where mesons are also created and annihilated. In addition,
together with mesons, they can also be free with short or long lifetimes,
either in high energy cosmic rays showers from violent cosmic events,
or in debris from high energy accelerator driven collisions.

The composite nature of hadrons is revealed 
by the pole positions of scattering amplitudes, as well as by decay 
channel couplings and their branching ratios.
Along the years these properties have been extracted from a variety of scattering experiments.
Hadrons and their decays were at the root of the development of the early quark models in 1964.

These models opened the way to the fundamental theory of the interaction between the elemental 
constituents of hadrons,  quarks and gluons --
Quantum ChromoDynamics (QCD)~\cite{PDG2022,HalzenMartin,Griffiths}. However, 60 years after the quark model was postulated,
hadrons are still
quizzical objects and have recently generated intriguing questions, because
only in  the last decade or so  a broad range
of experiments could be technically materialized
in forefront energy and detection conditions.
They covered electro- and photo-production experiments,
as well as $e^+e^-$ annihilation and heavy-ion collisions.
The recent achievements include the discovery
of some long predicted but missing baryon states~\cite{Crede13a,Kamano16a,Anisovich17a,Mokeev22a},
the high-precision determination of the evolution
of their electromagnetic couplings with
squared 4-momentum transfer ($q^2$)
supported by independent sophisticated analyses of single
and double pion production channels~\cite{Mokeev22a,Aznauryan12a,Burkert04,NSTAR,Carman23a},
and the revelation of exotic tetra-~\cite{LHCb22a}
and penta-quark~\cite{LHCb19a,LHCb21a,LHCb23a} hadrons.

This wealth of recent experimental findings is changing
the way we understand hadrons
and their structure. In parallel to experimental methods,
theoretical methods motivated by the new data have also
achieved remarkable success, not only in consistency and uncertainty control, but also in scope.
For instance, only recently, many years after the origin of the quark model and QCD,
lattice QCD calculations and also QCD functional methods in the continuum,
known as Dyson-Schwinger methods,
provided a unified picture and the observed relative energy level ordering
of the four
lightest baryon isospin-doublets
($I= \frac{1}{2}$) with angular momentum  $J= \frac{1}{2}$ and 
positive and negative parity.
In particular, the many years theoretical puzzle of
the low mass of the positive parity $N(1440)$
nucleon excitation relatively to the negative parity
$N(1535)$ resonance state
was deciphered, and novel QCD based results demonstrated the importance of the
combination of different diquark correlations
in baryon structure~\cite{Burkert19a,Eichmann19a}.
The combination of theory and experiment in
Hadron Physics is thus entering an era of unified knowledge of hadron structure.
However, the interpretation of some observed baryon properties
is still under discussion.

Our main focus here is on the lowest lying nucleon excitations $N^\ast$
(made of $u$ and $d$ quarks).
The reason for this choice is that in recent times they constitute the group
of the baryon family that has been studied in more detail and with increasing precision,
due to unique instrumentation opportunities in  luminosity, energy and detector
acceptance frontiers in worldwide laboratories.
Several experimental programs (as Jefferson Lab, ELSA, MAMI, GRAAL, SPring-8)
accumulated a vast body of detailed experimental information on
nucleon resonances,
by combining phenomenology and data analyses.
In particular, the very precise electron-scattering data from CLAS at
Jefferson Lab (JLab) on light nucleon resonances had a leading role in the current progress.
Therefore the lowest lying nucleon excitations $N^\ast$ are ideal
to scrutinize theoretical ideas and methods.
Nevertheless, we also present briefly the more sparse but recent results 
for baryons with strange and heavy quarks. And although the bulk of the contents
of this review paper
is mostly concerned with electromagnetic excitation reactions of low-lying mass baryons,
it nevertheless already unveils that baryons are very diverse
in their inner structure.

An additional  interest on light baryons and their couplings comes from the role they may have
in astrophysical phenomena, such as the formation of galaxies and the evolution of stars.
In the present era of gravitational wave data, the understanding of the properties of light baryons
may help shaping our knowledge on compact and dense stars,
for instance. Since the light $\Delta$ (isospin $\frac{3}{2}$, spin $\frac{3}{2}$) baryons
are about 30\% heavier than the nucleon, but lighter
than the heaviest baryons of the ($uds$) spin $\frac{1}{2}$ baryon octet,
they can mix together with hyperons (baryons with strange quarks)
in nuclear medium densities about 2--3 times the nuclear saturation
density. Therefore, $\Delta$ baryons may contribute to
increase the maximum value of neutron star mass, compensating for the effects of
hyperon degrees of freedom that decrease
it~\cite{Marquez22a}.
   
Electromagnetic transition form factors have a central role in this review,
since they are a fundamental quantity encapsulating  composition and dynamical properties of hadrons.
For instance, in the spacelike kinematic regime
(where one has for 4-momentum transfer $q^2 = -Q^2 \le 0$), the elastic form factors
give the charge and magnetic distributions of the hadron interior.
For example, dipole shaped form factors mean that those distributions
decrease as exponential functions from the center, and form factors
with zeros and oscillations indicate that the distributions
vanish more abruptly.
Although relativistic effects make this interpretation
slightly more complex,
they enable us to build an even richer
picture, e.g.~tomographic images of transverse densities
in the impact parameter transverse plane. 
In the timelike regime ($q^2 = -Q^2>0$) form factors carry the signatures
of particle creation and are tied to spectroscopy,
with a revealing structure of bumps for negative $Q^2$ closer to the photon point $Q^2= 0$. 
In the other extreme $Q^2 \to -\infty$ limit,
because of general physical and mathematical principles as unitarity,
in the timelike regime form factors become exactly identical
to the spacelike form factors in the symmetric limit $Q^2 \to +\infty$,
which provides useful constraints for theory~\cite{Pacetti14a}.

For experimentalists transition form factors or helicity amplitudes 
are extracted from cross sections:
in spacelike kinematics, from the cross sections of electron
induced excitation on the proton followed by hadron decay;
in timelike kinematics from the cross sections of electron-positron collisions
inducing hadron production~\cite{BESIII21a}.
For theorists they enter into the definition and then construction 
of the electromagnetic current that describes the photon interaction with the hadron.

Through the $Q^2$-evolution of electromagnetic form factors one can chart
the running of QCD strong coupling, quark clusterization effects, and find signatures
of quark mass dynamical generation~\cite{Carman23a}.
We will see here that for large $Q^2$ photon virtualities multipole form factors
for nucleon resonances $N^\ast$ with $J \ge \frac{3}{2}$
are good probes of the outbreak of the perturbative QCD regime,
due to simple correlations that can be extracted between electric
and magnetic transition form factors in the large-$Q^2$ region. 
In contrast, for low momenta they exhibit signs of the baryon {\it enlargement}
via the so-called meson cloud screening of the baryons dressed-quark core,
or other possible multiquark configurations: collected data
on several nucleon excitations indeed show that form factor calculations
based exclusively on valence quark degrees of freedom fail in that region.

Another main lesson learnt from this exercise is that hadron
physics structure and spectroscopy quests have combined, and still need to combine,
phenomenology and data analyses.
We are living impressive technical advances of QCD simulations in a discrete
spacetime lattice (lattice QCD)~\cite{Briceno18a} and of applications
of functional methods in QCD for quantitative predictions of spectroscopic
and structure properties of hadrons.

However, it is expected that the interpretation of experiments
will keep combining phenomenology and sophisticated data analysis,
for instance based on machine learning algorithms, and  will guide the demand
on uncertainty control in the results from theoretical methods based in first principles. 
Progress is then a result of  both {\it bottom-up} approaches
that start only with scale setting and current quark masses as input,
and more {\it top-down} approaches with different weights of
phenomenological assumptions, on degrees of freedom and modeling
by adjusting the calculation through fits to the observables.

The subject of this review, electromagnetic
 induced nucleon excitations is very vast. 
Therefore this review is anchored, updates or complements
past and valuable reviews, as the ones
on Refs.~\cite{Crede13a,Aznauryan12a,Burkert04,NSTAR,Capstick00,Eichmann16,Thiel22a,Pascalutsa07}.
Along this work we use the
Particle Data Group (PDG) edition of 2022~\cite{PDG2022}
as information reference source of experimental results,
unless mentioned otherwise.
Given the central role of the quark model in the studies of hadron structure, 
and because it has established a general picture of hadrons, we present 
a short summary of its main pillars in Section~\ref{secQM}.

As we will expand on detailed results on the several nucleon excitations,
we found pertinent to provide short summaries
(labeled {\it Short notes} in the text)
with intermediate conclusions on the most relevant points specific to each resonance,
to allow for quick access to concise information.
One of our objectives is that this review is used
as a first reading guide to students and younger scientists in the field,
providing them with basic information and references, upon which depth can be build
following their interests and motivation.
     
This work is organized as follows: In Section~\ref{secQM}
we briefly revisit the origin and
essential foundations of the quark model and give the recent spectroscopic data on
the lower-lying nucleon resonances;
in Section~\ref{sec-TransitionFF} we obtain and compile the formulas
for the helicity amplitudes and multipole form factors of the electromagnetic reactions,
for nucleon excitations of  general spin $J \geq \frac{1}{2}$.
The formulas are written in a general and compact form,
and update notation and results from several sources;
in Section~\ref{sec-Expermental} we describe the experimental
facilities with specially dedicated contributions to this subject
and different prevailing methods of data analyses;
Section~\ref{sec-Models} highlights some representative theoretical approaches;
in Section~\ref{secData} we organize some of the results of those approaches in comparison
to diverse experimental data for the nucleon resonances with masses up to 2 GeV.
Results for baryons with heavy quarks are discussed in Section~\ref{secBaryons}.
Section~\ref{sec-conclusions} summarizes main conclusions.
Notation and calculations details can be found in Appendices~\ref{appNotation} to~\ref{appLargeQ2}. 
Appendix~\ref{appMultipole} complements the results of Section~\ref{sec-TransitionFF},
while Appendix~\ref{appLargeQ2} assists the analyses
in Section~\ref{secData} on the large-$Q^2$ behavior of the form factors --
which is of high general interest as evidence of the outset of the perturbative QCD regime.


\renewcommand{\theequation}{2.\arabic{equation}}
\setcounter{equation}{0}

\renewcommand{\thefigure}{2.\arabic{figure}}
\setcounter{figure}{0}

\section{Baryons and the Quark Model \label{secQM}}

The quarks  that make up baryons and mesons are of six distinct species referred to as flavors:
$u$ and $d$  ({\it up} and {\it down}), that define the  light quark sector,
$s$ ({\it strange}) with intermediate mass, and
$c$, $b$,  and $t$
(respectively {\it charm}, {\it bottom} and {\it top}), in the heavy quark sector.
Besides the mass, the quarks are characterized by flavor quantum numbers
such as, charge, isospin, strangeness and others~\cite{PDG2022}.

Most of the composite quark states are unstable resonances, experimentally manifest
in decays to final states with two or more particles.
Light nucleon resonances $N^\ast$, the main subject of this review,
are made of light quarks $u$ and $d$,
with fleeting contributions of $s$ quarks through quark-antiquark excitations,
resulting for example in $\eta$ meson production channels.
The $N^\ast$ states with  isospin $I= \frac{1}{2}$ ($N$ states) 
are
mixed symmetric flavor states, and the ones with $I= \frac{3}{2}$ ($\Delta$ states)
are symmetric flavor states~\cite{PDG2022,HalzenMartin,Griffiths}.

Since the quark model was the influential phenomenological precursor 
of QCD, we sum up in this section its key features and limitations.
In general, the expression
{\it quark model}  refers to the description of hadrons 
assuming the confinement property (since no free quarks and gluons
are directly observed) 
with a minimum content of {\it constituent quarks}~\cite{Giannini90}.
It defines a framework that classifies hadrons into two classes:
mesons, valence quark-antiquark ($q \bar q$) 
systems, and baryons, three valence quarks ($qqq$) systems.

\subsection{\it The early quark model}

In the very early history of the quark model, Fermi and Yang~\cite{Fermi49a}
introduced the (wrong) hypothesis that the pion was formed by a nucleon 
and an anti-nucleon, under a force different from ordinary nuclear forces, 
and of such short range character that it could not be seen in scattering.
This idea was around for some time leading to the so-called Sakata model 
(picturing meson and baryon resonances as made up of {\it elementary} protons, neutrons and 
$\Lambda$s and their antiparticles) but it led to an unsatisfactory asymmetric 
treatment of the strange baryons  $\Lambda$, $\Sigma$ and $\Xi$.  
The compositeness of hadrons ceased being an issue in 1962~\cite{GellMann62a}
 with Gell-Mann's  field theoretic based paper where equal-time relations were 
abstracted, and a mathematical classification scheme of baryons emerged from the condition 
of {\it unitary symmetry} (meaning the conservation of
strangeness-changing vector currents).
In such classification scheme, stable and unstable baryons
($N$, $\Lambda$, $\Sigma$ and $\Xi$) 
form a degenerate baryon octet in the limit of {\it unitary symmetry},
suggesting that this symmetry could be broken. 
Gell-Mann pointed out that of all the groups generated by the vector weak currents, 
$SU(3)$ is the smallest giving rise in a natural way to the rules 
$|\Delta I | =\frac{1}{2}$, $|\Delta S|=1$ and $|\Delta Q| =0, 1$
($I$, $S$ and $Q$ represent isospin, strangeness quark number and charge).
And similarly to local $U(1)$ gauge invariance which necessary generates the existence 
of a massless photon, local gauge invariance for $SU_c(3)$ color generates eight 
massless gluons, the carriers of the (strong) color force.
Thus QCD was launched
(a very recent comprehensive review on QCD, from its beginnings
up to its future explored in many facilities,
can be found in Ref.~\cite{Gross22a}).

By then one still hoped that
the discovered meson resonances were composite states of
one another related by unitarity (mechanism called bootstrap).
In a parallel route to Gell-Mann's theory,
George Zweig came up with the concept of
{\it constituent quarks}
({\it aces}, in his original language) to explain the puzzling $\phi$ meson decay into the 
$l=1$ $K \bar{K}$ channel and the suppression of the $l=0$ $\pi \rho$ decay channel
($l$ is the relative orbital angular momentum).
There was no symmetry to impose this suppression.
While Gell-Mann focused on symmetries and elementary quarks, Zweig tinkered with dynamics and effective, 
dressed or {\it constituent quarks}:  he assumed that when a meson $a \bar{a}$
decays into two other mesons, as the separation between the 
elements  of the $a \bar{a}$ pair increases, two new {\it aces} $b$ and $\bar{b}$ 
pop out separately of the vacuum and each one recombines with a quark from the
previous two {\it aces}, completing the decay into the mesons $a \bar{b}$ and $\bar{a}b$.

QCD establishes that a quark carries one of the 3 charges (colors) that form the basis of an 
$SU(3)$ group. Baryons do not have this crucial property.
This means that all baryons 
are color singlets. At quark level the color $SU(3)$ symmetry is the only 
internal symmetry that is exact (different from the flavor
$SU_F(3)$ symmetry which is an approximate symmetry). 
How does QCD lead to colorless baryons was not yet directly established from the QCD Lagrangian. 
But meson radii are about million times smaller than atom radii and  the excitation energies 
of quarkonia levels of the order of hundreds of MeV, with hyperfine and spin-orbit splittings 
of order of tens to hundreds of MeV. 
Therefore the strong color forces acting between quarks are much stronger than electromagnetism. 
Following the conceptual picture of  Zweig, in an initial droplet of quark matter 
the attractions by the strong or color force make the droplet evolve 
to grains of separate clusters of 3 quarks. Each cluster is colorless, hiding color in its interior.
In other words, the baryon wave function is an antisymmetric state of the 3 $SU(3)$ 
colors, and therefore a symmetric state in the other configurations altogether
(spin, flavor and orbital parts).

This symmetry constraint alone has decisive consequences.
The strong interaction, through the Pauli principle, imposes constraints on quark clusters that 
are opposite of those on nucleon clusters, affecting deeply the electromagnetic structure of baryons. 
This was already well illustrated by F.~Close~\cite{Close82}
by a simple example,
the comparison of magnetic moments of the proton ($uud$ quark content)
and neutron ($ddu$ quark content) 
with the magnetic moments of the 
$^3$He ($ppn$) and $^3$H ($nnp$) nuclei:
because of the antisymmetry in color in the proton and neutron wave functions, 
the flavor symmetric pairs ($uu$ or $dd$) inside the proton or the neutron
have spin $1$ and do contribute 
to the nucleon's magnetic moments, while their nuclear analogue pairs ($pp$ or $nn$), 
being colored symmetric have spin $0$ and therefore have no contribution
to the net nuclear magnetic moments.

Since its early days in the 60's of last century, the quark model evolved into a diversity 
of quark models with different levels of formulations, from the almost dynamics-free picture 
to a microscopic dynamics with direct input from QCD. The so-called traditional quark potential models 
described the baryon spectrum by calculating 
the quantum-mechanical three quark system with 
a confining potential~\cite{Capstick00,Isgur82,Isgur78a,Koniuk80}, as presented
in Sections~\ref{secPotential} and \ref{secWF}.

\subsubsection{\it General features of constituent quark potential models \label{secPotential}}

The Isgur-Karl model~\cite{Isgur78a,Isgur79a,Isgur79b} was the first 
successful quark potential model. It was developed in the late 70's 
and different potential versions followed~\cite{Isgur82,Koniuk80,Beg64,Giannini15,Giannini89,Isgur80}. 
Good reviews on the quark model are in Refs.~\cite{Capstick00,Giannini89,Pascalutsa07a,Richard92a}.
The Isgur-Karl model~\cite{Isgur78a,Isgur79a} assumes  an harmonic oscillator confining potential 
to effectively represent the gluon fields where the quarks move. 
The picture is valid only for processes that probe the structure with soft 
or low energy and momentum transfers compared to the confinement scale, 
such that the light $u$ and $d$ constituent quarks are seen not as the partons 
in deep inelastic scattering,
but rather as extended objects with masses of about  200 to 350 MeV,  
and strange quarks 150-200 MeV heavier. 
Moreover, the model is limited to low-mass baryons
where gluonic excitations  are very unlikely
and with small mass shifts due to their couplings to decay channels. 
The Hamiltonian used as input into the  three quark Schr\"odinger 
equation includes also an hyperfine interaction with spin-spin and tensors terms,
with a strength fixed by the nucleon $\Delta(1232)$ mass splitting.
The model is defined by 5 parameters: 
2 related to the quark masses, 1 to the harmonic oscillator level spacing, 
1 to the strength of the hyperfine interaction, and 1 defining the unperturbed level
of the non strange states. The wave functions and  their
corresponding energies are found by diagonalizing the Hamiltonian in
a large basis of harmonic oscillator states.
It describes the main features  of the low-lying baryon resonances
and the hyperfine tensor interaction explains the observed strong decays.
Its successor, the Isgur-Godfrey model~\cite{Isgur79a}
used a {\it relativized} Hamiltonian with a relativistic kinetic energy operator, 
and added a spin-orbit term to the interaction since there was evidence
of spin-orbit splitting, for instance between the
$\Delta (1620)$ and the $\Delta(1700)$ baryons. 
To incorporate the finite extension of the constituent quarks 
the  Isgur-Godfrey model also smeared
the inter-quark coordinates $r_{ij}$ over mass-dependent distances,  
by convoluting the interactions terms with a function 
proportional to $\sigma^3_{ij} \exp(-\sigma^2_{ij} r^2_{ij})$ where 
$\sigma_{ij}$ smears the coordinated over distances depending on the quark masses. 
The quark mass  $m_i$ are also replaced by the energies $E_i$ in the 
kinematic factors entering the non relativistic reduction of the potentials.

The relativized model led to the calculations of Ref.~\cite{Capstick86b} that 
shows some improvements relatively to the original non relativistic one, 
but two problems persisted: there are more states predicted from the model 
than the ones observed 
and the resonance $N(1440)$ ($J=\frac{1}{2}$, positive parity)
is predicted with a mass that is too high with respect
the first negative parity state $N(1535)$ ($J=\frac{1}{2}$, negative parity),
the problem being even worse for the $J= \frac{3}{2}$ case.
To fix the last problem Glozman, Plessas, Varga and 
Wagenbrunn~\cite{Glozman96,Glozman96b,Glozman98a} introduced in their quark model 
a hyperfine interaction originated not from gluon exchange but from pion exchange 
as well as vector and  a scalar meson exchanges.
The rational of such boson-exchange models is that a flavor dependent hyperfine interaction 
term produces in each flavor sector different negative and positive splitting effects which 
are large enough for the inversion of masses of the positive 
and negative first excited parity states of the nucleon and of the 
$\Delta(1232)$ relatively to the unperturbed harmonic or linear confinement levels.
Meanwhile, recent  works~\cite{Burkert19a,Eichmann19a}
solved this spectrum ordering problem from more first-principled QCD based
calculations as Dyson-Schwinger equations
(see Section~\ref{secDS}).

The variety of families of non relativistic and relativistic quark models is too vast
to be addressed here in full completeness. The harmonic oscillator  potential 
$V(r_{ij})=\frac{1}{2}K r_{ij}^2 $ was often used as confining interaction.
But anharmonic linear terms were also considered leading to different spectrum results.
Reference~\cite{Trawinski14a}
clarified that this issue does not mean that one of the two shapes is in reality
favored by Nature: the shape of the confinement potential depends on the form of dynamics chosen.
More precisely,  the same spectrum is obtained either within instant form dynamics
and a linear potential, or front form dynamics and a quadratic potential.
In simple terms, what happens is that  the  eigenvalues of the
instant form Hamiltonian are the masses of the hadrons, while
the eigenvalues of the Hamiltonian in front form dynamics are the hadron masses squared.
In this last case then, the quadratic term of the binding energies contribute
then effectively to the front form potential and this one has to be quadratic
in the non relativistic limit. Moreover, the authors show that the
instant form potentials extracted from lattice calculations are
in agreement with potential forms of models using front form, by comparing the coefficient of the
{\it return distance}, where the kinetic energy vanishes and turns
into potential energy~\cite{Trawinski14a}.

To the potential models above, one should add the algebraic string quark model 
from Bijker and Iachello~\cite{Bijker94,Bijker96a,Bijker97a,Bijker00}, 
the hypercentral quark
models~\cite{Giannini15,Giannini01a,Bijker09a,Sanctis05,Aiello96,Santopinto12,Bijker16a}, 
the large $N_c$ models~\cite{Goity03a,Matagne05a,Brodsky15a,Jenkins94a,Dashen95,Jenkins02},
the relativistic Bethe-Salpeter based equation models~\cite{Loring01a,Ronniger13a} 
and models that include in addition $(qqq)q \bar q$ 
components~\cite{Helminen02a,Li07a,Li06a,An09a,An06a,JDiaz06a,An09c} in the wave functions.
Comprehensive reviews and interesting
references are also in~\cite{Crede13a,Capstick00,Giannini89,Pascalutsa07a,Richard92a}.

\subsubsection{\it Symmetries, baryon spectroscopy and wave functions in quark model calculations
  \label{secWF}}

Although the potential quark model framework does not solve QCD for quark dynamics, 
it applies the symmetries imposed by the QCD Lagrangian 
to establish the form of the wave functions and thus guiding the identification of states.
The total angular momentum $J$ is a good quantum number because of the invariance 
of the QCD Lagrangian under the Poincar\'e group,
which includes rotations, and invariance under parity transformation
makes parity $P$ a good quantum number too ($P=\pm$). 
Isospin is another good quantum number because of the invariance of
$SU(3)$ flavor under isospin rotations.

Our notation here for the baryon states is then the commonly
used in recent  literature: $B(m) J^P$,
where $B$ represent a $N^\ast$ state with 
$B=N$, for $I=\frac{1}{2}$, and $B=\Delta$, for $I=\frac{3}{2}$.
The  approximated mass $m$ in MeV distinguishes different states with common $J^P$.
One writes then $N(1440)\frac{1}{2}^+$ and $N(1710)\frac{1}{2}^+$ for the first
resonances with the quantum number of the nucleon,
and $\Delta(1232) \frac{3}{2}^+$, $\Delta(1600) \frac{3}{2}^+$
for the first $\Delta$ resonances.
The notation extends to baryons with strange and heavier quarks by replacing
$B$ for the corresponding name symbol
($\Lambda, \Sigma, \Xi, \Omega, \Lambda_c$,...)~\cite{PDG2022,Capstick00}.
For the notation associated with baryons with strange quarks
we direct the reader to the chapter ``Quark Model'' from PDG 2022~\cite{PDG2022}.

About 10 years ago the nucleon resonance spectroscopy
were based almost exclusively on the $\pi N$ decays of the $N^\ast$ states. 
Instead of the good quantum numbers $(I,J^P)$, an older notation used  $L_{(2I)(2J)}$,
with $L$ the angular momentum of the pion on the pion-nucleon decay
expressed in terms of the spectroscopic indices ($L=S,P,D,...$).
Since more data independent of the pion production 
became increasingly available, the old notation became obsolete.
In this historically notation of the Particle Data Group before 2012~\cite{PDG2010}, 
the states $\Delta(1232)\frac{3}{2}^+$, $N(1440) \frac{1}{2}^+$,
$N(1520) \frac{3}{2}^-$, $N(1535) \frac{1}{2}^-$, $\Delta(1620) \frac{1}{2}^-$ 
were labeled as $\Delta(1232)P_{33}$, 
$N(1440)P_{11}$, $N(1520)D_{13}$, $N(1535)S_{11}$, 
$\Delta(1620)S_{31}$ etc.~(see chapter
``$N$ and $\Delta$ resonances'' from Ref.~\cite{PDG2022} for details).

We consider now the wave functions associated with the $SU(6)$ spin-flavor group,
which combines the quarks $u$, $d$ and $s$, and not only the light quarks,
as for the nucleon resonances.
Within the non relativistic framework the eigenstates of the Hamiltonian 
in the potential quark model are then represented as sums of components
of a basis of states $\left|B^{2S+1} L_J\right>$,
where $S$ is the sum of the spin of the quarks and 
$L$ the total quark orbital momentum (again $L=S,P,D,...$).
Again, $B$ can be $N$, $\Delta$, or label
a class of hyperons or baryons with heavy quarks ($\Lambda, \Sigma, \Xi$,...).

The general form of the $SU(6)$ spin-flavor baryon wave functions
(see for example chapter "Quark Models" from PDG 2022~\cite{PDG2022}
and Ref.~\cite{Eichmann22a})
is obtained by forming all the 6 combinations of states 
$\left|B^{2S+1} L_J\right>_{ijk}$ of the three quarks $(ijk) \, \,   (i \ne j \ne k; 1,..,3)$
that transform into a given subspace under any of the 6 permutations of the 
$S_3$ permutation group.
For example, the symmetric ($S$) combination
of the $\left|B^{2S+1} L_J\right>_{ijk}$ states is a sum of 6 terms
\ba
S= \sum_{i\ne j \ne k} \left|B^{2S+1} L_J\right>_{ijk}
\ea
and one of the two mixed-antisymmetric ($M_A$) combinations,
under permutation of quarks 1 and 2 but not under all possible permutation of quarks, is
\ba
M_A= \left|B^{2S+1} L_J\right>_{132}-\left|B^{2S+1} L_J\right>_{231}+
\left|B^{2S+1} L_J\right>_{312}-\left|B^{2S+1} L_J\right>_{321},
\ea
In total, one has one symmetric ($S$) and one antisymmetric ($A$) combination of states 
under any permutation of $S_3$,  two mixed-symmetric combinations ($M_S$) and
two mixed-antisymmetric combinations ($M_A$) under exchange of one pair of
three quarks but without definite symmetry under all permutations. 
Specifying then the 27 $SU_F(3)$ combinations of quarks $u$, $d$ and $s$
into the six $S_3$ irreducible states, 
at the end one obtains 8 mixed-symmetric, 8 mixed-antisymmetric,
10 symmetric and 1 antisymmetric $SU_F(3)$ combinations. 
In other words, one decomposes the $SU(3)$ 
group into a baryon decuplet of symmetric states, two baryon octets 
of mixed symmetry states and one antisymmetric baryon singlet state.
This is usually represented as
\ba
3 \otimes 3 \otimes 3 = 10_S \oplus  8_{M_A} \oplus 8_{M_S} \oplus 1_A
\ea

To obtain the baryon states of spin  $J=\frac{1}{2}$ and $J= \frac{3}{2}$ one has in addition to combine 
these quark flavor states with their spin states.
The resulting total wave functions are then organized into $SU(6)$ multiplets. 
The baryons of each multiplet  do not have exactly the same masses,
since flavor symmetry is broken which lifts mass degeneracy.  
Finally, using the fact that the color part of the wave function
is totally antisymmetric, the orbital parts  $\left|B^{2S+1} L_J\right>$ of the states 
are constrained to have a matching symmetry with the symmetry of the flavor-spin component, 
such that the $\Psi_{\rm orbital}  \otimes \Psi_{\rm flavor} \otimes \Psi_{\rm spin}$ 
is symmetric under the interchange of equal mass quarks. 
Taking then all  possible flavor multiplets (octet, decuplet and singlet)
and the possible spin wave functions one concludes from the combinatorial analysis that 
\begin{enumerate}
\item
  the orbital wave functions are necessarily symmetric when the flavor and spin states
  are both either symmetric or of mixed symmetry; for $S= \frac{1}{2}$ this happens
  for the flavor octet case and for
  $S=\frac{3}{2}$ for the flavor decuplet case; this originates respectively
  $8 \times 2=16$ and $10 \times 4=40$ states. They form the $[56]$-plet of $SU(6)$
  where the nucleon and the $\Delta (1232)$ belong to.
\item
The mixed symmetric orbital wave function can combine with the 3 flavor multiplets  
(octet, decuplet and singlet) for $S=\frac{1}{2}$, and only with flavor octet states for $S=\frac{3}{2}$;  
this originates respectively additional 
$2\times 8+2\times 10+2\times 1=38$ and $8\times 4=32$ states. They form the $[70]$-plet.
\item 
The antisymmetric orbital wave functions can combine with the flavor octet states 
for $S= \frac{1}{2}$ and with the flavor singlet for $S= \frac{3}{2}$; this originates 
$2 \times 8 + 4 \times 1=20$ states. They form the $[20]$-plet.
\end{enumerate}

Cases 2) and 3) necessarily require some non zero orbital excitation in the quark pairs.
In total one has then 216 states decomposed as 
\ba
6 \otimes 6 \otimes 6 = 56_S \oplus  70_{M_A} \oplus 70_{M_S} \oplus 20_A
\ea

All ground states of the baryon multiplets are experimentally known.  
The identification of the spectrum is usually based on 
quark models that combine the $SU(6) \otimes O(3)$ symmetry group 
with some confinement mechanism and, as explained in the
previous section by solving the
Schr\"odinger equation.
Then, the constituent quark model the baryon wave functions
may then be written in terms of the traditional 
Jacobi variables for the two internal relative momenta
of three quark system. In momentum space they are:
\ba
k_\rho = \frac{1}{\sqrt{2}}(k_1 - k_2), \hspace{1cm}
k_\lambda = \sqrt{\frac{2}{3}}(k_1 + k_2) - \frac{1}{\sqrt{6}} P,
\label{eqJacobi}
\ea
where $P= k_1 + k_2 + k_3$ is total 3-body momentum, $k_\rho$  
is the relative momentum of quarks 1 and 2 and $k_\lambda$
the momentum of the (spectator) quark 3 relatively to that pair.
In the configuration space we use
the conjugated coordinates ${\bf \rho}$ and ${\bf \lambda}$.
The first Jacobi variable is anti-symmetric under the change of quarks (12) and
the second is symmetric under the same change. 
In the baryon rest frame the dependence of $P$ can be omitted,
${\bf P}={\bf 0}$ and use 
$k_\rho \to r= \frac{1}{2}(k_1-k_2)$ and $k_\lambda \to k \equiv k_1 + k_2$~\cite{Capstick00,Isgur77a}.

It turns out that the comparison of experimental level state results
with potential quark model calculations has established an identification of the
3D-harmonic oscillator excitation $N=2n+l$ band levels
for $N=0,1$ with baryon energy states.
Here $n= n_{\rho}+n_{\lambda}$, $l= l_{\rho}+l_{\lambda}$,
where ($n_{\rho}$, $n_{\lambda}$) specifies the two possible radial excitations and
($l_ {\rho}$,  $l_{\lambda}$) the two orbital internal excitations, with $\rho$ associated to the
quark-quark pair and $\lambda$ to the quark-diquark system.
Not all $N=2$ and $N=3$ states have been observed.
This is possibly due to limitations of the experimental analysis
that is dominantly restricted to $\pi N$ scattering.
It was also proposed, and is still questionable, whether degrees of freedom
as diquarks or other multiquark structures involving $q \bar q$ pairs
are causing this failure (see chapter ``Quark Model'' of PDG 2022~\cite{PDG2022}),
as well as intrinsically relativistic quantum
field theory orbital excitations~\cite{Mokeev22a,Santopinto15a,Galata12a,Barabanov21}.

\begin{table}[t]
\begin{center}
\begin{tabular}{c   c  l l l l l l l l l l l l}
\hline
\hline 
\vspace{.1cm}
&        & &    & &  \\
Particle & $J^P$ & Status &  $\gamma N$ & $\pi N$ & 
$\pi \Delta$  & $\sigma N$ & $\eta N$ & $K \Lambda$  & $K \Sigma $ &
$\rho N$ & $\omega N$ & $\eta^\prime N$ & $SU(6)$\\[.3cm]
\hline
           &                   &      &      &     &       &       &  \\[-.15cm]
$N(1440)$  &  $\frac{1}{2}^+$  & **** & **** & **** & **** &  ***   &  &&&&&& $[56,0_2^+]$\\
$N(1520)$  &  $\frac{3}{2}^-$  & **** & **** & **** & **** &  **  & ****  & &&&&&  $[70,1_1^-]$  \\
$N(1535)$  &  $\frac{1}{2}^-$  & **** & **** & **** & ***  &  *   & ****  & &&&&&  $[70,1_1^-]$  \\
$N(1650)$  &  $\frac{1}{2}^-$  & **** & **** & **** & ***  &  *   & **** & *  & &&&&  $[70,1_1^-]$ \\
$N(1675)$  &  $\frac{5}{2}^-$  & **** & **** & **** & **** &  *** & *    & *   & *  &&&&  $[70,1_1^-]$\\
$N(1680)$  &  $\frac{5}{2}^+$  & **** & **** & **** & **** &  *** & *    & *   & *  &&&&  $[56,2_2^+]$\\
$N(1700)$  &  $\frac{3}{2}^-$  & ***  & **   & ***  & ***  &  *   & *    &     &    & * & &  & $[70,1_1^-]$\\
$N(1710)$  &  $\frac{1}{2}^+$  & **** & **** & **** & *    &      & ***  & **  & *  & * & * & &  $[70,0_2^+]$\\
$N(1720)$  &  $\frac{3}{2}^+$  & **** & **** & **** & ***  &  *   & *    & ****& *  & * & * & &  $[56,2_2^+]$\\
$N(1875)$  &  $\frac{3}{2}^-$  & ***  & **   & **   & *    &  **  & *    & *   & *  & * & * \\
$N(1880)$  &  $\frac{1}{2}^+$  & ***  & **   & *    & **   &  *   & *    & **  & ** &   & ** \\
$N(1895)$  &  $\frac{1}{2}^-$  & **** & **** & *    & *    &  *   & **** & **  & ** & * & * & **** \\
$N(1900)$  &  $\frac{3}{2}^+$  & **** & **** & **   & **   &  *   & *    & **  & ** &   & * & ** \\[.1cm]
\hline
\hline
\end{tabular}
\end{center}
\caption{\footnotesize
$N$ resonances (isospin $\frac{1}{2}$) with mass below 2 GeV with status *** and ****,
 and dominant decays. 
\label{tableN-star}}
\end{table}

\begin{table}[t]
\begin{center}
\begin{tabular}{c   c  l l l l l l l l l l l l}
\hline
\hline 
\vspace{.1cm}
&        & &    & &  \\
Particle & $J^P$ & Status &  $\gamma N$ & $\pi N$ & 
$\pi \Delta$  & $K \Sigma $ &
$\rho N$ & $\eta \Delta$ &  $SU(6)$\\[.3cm]
\hline
                &                   &      &      &     &    \\[-.15cm]
$\Delta(1232)$  &  $\frac{3}{2}^+$  & **** & **** & **** &      &  &&& $[56,0_0^+]$  \\
$\Delta(1600)$  &  $\frac{3}{2}^+$  & **** & **** & ***  & **** &  &&& $[56,0_2^+]$ \\
$\Delta(1620)$  &  $\frac{1}{2}^-$  & **** & **** & **** & **** &   &&&  $[70,1_1^+]$\\
$\Delta(1700)$  &  $\frac{3}{2}^-$  & **** & **** & **** & **** &  *  & * &&  $[70,1_1^+]$ \\
$\Delta(1905)$  &  $\frac{5}{2}^+$  & **** & **** & **** & **   &  *   & *    & ** & $[56,2_2^+]$ \\
$\Delta(1910)$  &  $\frac{1}{2}^+$  & **** & ***  & **** & **   &  **  &      & *  &  $[56,2_2^+]$ \\
$\Delta(1950)$  &  $\frac{7}{2}^+$  & **** & **** & **** & **   &  *** &  & & $[56,2_2^+]$\\[.1cm]
\hline
\hline
\end{tabular}
\end{center}
\caption{\footnotesize
$\Delta$ resonances (isospin $\frac{3}{2}$) with mass below 2 GeV with status ****,
and dominant decays. 
\label{tableDelta-star}}
\end{table}

In Tables \ref{tableN-star} and \ref{tableDelta-star} 
we present a list of the best experimentally known baryons up to 2 GeV
and the information about the decay modes
according with the data of the PDG edition of the year 2022~\cite{PDG2022}.
By convention PDG assigns the rating **** or *** only
to those resonances which are confirmed by independent analyses 
of data sets that include precision differential cross sections and
polarization observables.  All other signals that do not fulfill these conditions are
given ** or * status. The existence of  a **** resonance is certain and 
very likely for *** resonances. The label ** indicates that the evidence of existence is fair, while
* indicates that evidence of existence is poor at the moment.
Notice that the overall status and the status associated 
with the decay channels changes with the time (and PDG edition).
However, the tendency is for an upgrade
of the ratings with time and cumulative experiments.

The last column of Tables~\ref{tableN-star} and \ref{tableDelta-star} gives 
the multiplet of $SU(6)\otimes O(3)$ to where the state belongs to.
The multiplets are labeled by $[{\cal D},L_N^P]$ where ${\cal D}$ stands for the number of members
and $L^P$ for the quark total orbital angular momentum and parity $P$ of the state.
$N$ is the harmonic oscillator excitation index mentioned above ($N=0,1,2,...$).
The classifications with $N=2$ are partly tentative~\cite{PDG2022}.
This index $N$ is important to distinguish the ground states, 
$N(940)$ and $\Delta(1232)$, members of the multiplet $[56,0_0^+]$
from the radial excitations $N(1440)$ and $\Delta(1600)$,
members of the multiplet $[56,0_2^+]$.

The quark model based identification of states has two model dependent limitations.
First, the spectroscopic identification of the $J^P$ quantum numbers is not sufficient to 
discriminate the structure of the baryon wave function. States with same
$J^P$ can mix different $L$, $S$ combinations,
and the extent of this mixing is model dependent.
Although this may affect more decay patterns than mass positions in the spectrum in general,
it makes the spectroscopic assignment of states less clear, specially for excited states. 
Example: the states $N(1440)\frac{1}{2}^+$, $N(1710)\frac{1}{2}^+$
and $N(1880)\frac{1}{2}^+$ share the $J^P$ quantum number 
but they do not all correspond to pure radial excitations of the nucleon.
Second, the ordering in the harmonic oscillator bands is only approximate 
since it comes from non relativistic calculations.
In relativistic calculations one obtains extra components of
the baryon's wave functions with a different parity which are completely suppressed
in a non relativistic framework, for e.g.~$P$-waves in the relativistic 
$J^P=\frac{1}{2}^+$ nucleon state.
The full implementation of relativity amplifies 
the number of structures included in the baryon wave function
and expands the complexity of the spectrum~\cite{Capstick00,Eichmann16}.
For example, in Dyson-Schwinger functional methods, 
the Schr\"odinger equation for the wave function is replaced by quark 
$n$-point Bethe-Salpeter equations that include
intrinsically relativistic components. We will come to this later and especially
in Section~\ref{secDS}.

\begin{figure}[t]
\centerline{
    \includegraphics[width=12cm]{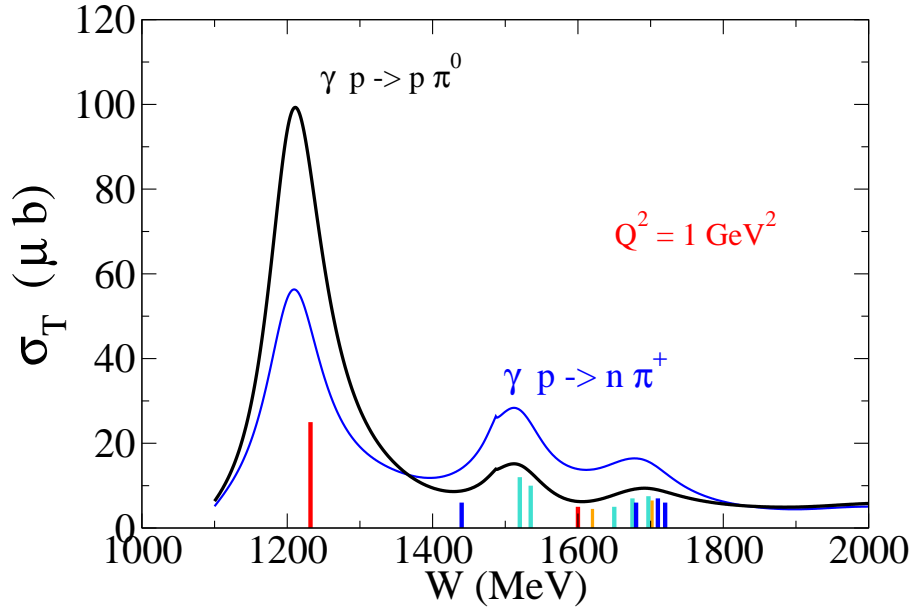}}
\caption{\footnotesize Representation of the transverse  $\gamma p \to \pi^0 p$ and 
$\gamma p \to \pi^+ n$ cross sections with $Q^2=1$ GeV$^2$.
Notice the bumps associated to the first, second and 
third resonance region.
Calculation using the MAID 2007 parametrization~\cite{Drechsel07,SAID-website}.
The vertical lines indicate the $N^\ast$ states discussed in the present work.
It includes $N(J^+)$ (blue), $N(J^-)$ (cyon),
$\Delta(J^+)$ (red) and $\Delta(J^-)$ (orange) states.
\label{fig-resonance-regions}}
\end{figure}

Finally, Fig.~\ref{fig-resonance-regions} is the classic figure
that shows the consistency between the spectroscopic assignments
of the lowest lying resonances and the profile of the
pion photo-production transverse cross section, that clearly identifies
the states in the three resonance energy regions that dominate that cross section.

\subsection{\it Outlook: beyond the constituent quark model \label{secOutlookQM}}

Although successful, there are three problems with the quark model:
the first is the old puzzle, known as the
{\it missing resonances problem}~\cite{Mokeev22a,Burkert04,Capstick00,Isgur78a,Bijker16a}, already mentioned
in the previous sections: 
up to $2.4$ GeV about half of the states predicted 
by quark models was not observed or firmly established experimentally~\cite{PDG2022},
even though a few new states have meanwhile been identified~\cite{Anisovich17a,Mokeev22a}. This  hints for
other degrees of freedom not theoretically considered, or non dominant decays not experimentally sought
(the searches were mostly restricted to $\pi N$ scattering). 
Second, in the other direction, since 2015~\cite{LHCb15a} and even more recently~\cite{LHCb19a}
experimental evidence from the LHCb collaboration exist
for non expected states or exotic states, as charmed pentaquark states which have
a $(qqq)q\bar q$ composition and masses 4.60 - 4.74  times the proton mass.
The energies of these states are very close to the energy production thresholds of meson-baryon pairs.
Finally, the third problem is that non relativistic constituent
quark models do not address how the constituent quark mass emerges from the bare quark mass, i.e.~the
QCD feature of dynamical mass generation -- which accounts for  about  
$99\%$ of the light baryon masses.

All these three problems lead to the revision
of the traditional quark model. Quantum Field Theory approaches become inevitable
by adding particle creation and annihilation mechanisms and subsequent meson {\it dressing}
extensions of the {\it bag} baryon picture.
Quantum Field Theory-based functional methods of QCD as Dyson-Schwinger methods,
that solve $n$-point Bethe-Salpeter equations, have the advantage over quark models of also
 treating consistently both the mechanisms for
quark clusterization components of the baryon wave functions, and for mass acquisition
through dynamical chiral symmetry breaking.

The second problem above, on newly found multiquark compositeness
(with about 30 new baryons detected by the LHC collaboration~\cite{Chen23a}) adds to the vast experimentally 
known complexity of meson-baryon decay channels
and of electromagnetic excitations of light nucleon resonances
on which we focus here.
Understanding these recently found exotic tetraquark and pentaquark
with significant {\it molecular-like} and/or compact structure components tests
computational techniques that
explain a variety of quark clusterization,
and may decide on which structure 
(compact or a molecular-like state of five quark state?) is predominating.
QCD based Dyson-Schwinger equations determine the weight in the
baryon wave function of molecular-like
versus compact state configurations, and the
relative importance of different diquark structures~\cite{Barabanov21},
absent in constituent quarks traditional models.
This disclosure of diquarks
may shed light also on the role of collective degrees of freedom
other than constituent quarks, for the solution of the first open problem above.
Dyson-Schwinger equations can in principle also compare
baryon components with three-quark and five-quark
components.

In lattice QCD calculations it is also possible to evaluate
the weight of these components, through overlaps between 
local three-quark interpolating fields and non local
five-quark meson-baryon states.
But this is less direct, and has to recur to  methods as the L\"uscher method
for finite volume effects 
and comparison of effective field theory with coupled-channels
results~\cite{Wu18b} (more in Section \ref{sec-LatticeQCD}).

As a final remark: even with these advances in QCD based dynamics,
a comprehensive and quantitative understanding of all observed
baryon properties is not yet closed: 
how elementary degrees of freedom, quarks and gluons, 
are not observed due to color confinement remains a theoretical challenge.


\renewcommand{\theequation}{3.\arabic{equation}}
\setcounter{equation}{0}

\renewcommand{\thefigure}{3.\arabic{figure}}
\setcounter{figure}{0}

\section{Transition amplitudes and transition form factors 
\label{sec-TransitionFF}}

The data associated with the electroexcitation of the nucleon 
($\gamma^\ast N \to N^\ast$) can be translated into structure functions 
with an evolution with the invariant squared momentum transfer $q^2$.
The properties of these structure functions 
depend on the nucleon resonance $N^\ast$, 
particularly on the spin-parity state $J^P$.
In this section we discuss the structure 
of the current associated with a $\gamma^\ast N \to N^\ast$ 
transition, where the final state can be $J^P= \frac{1}{2}^\pm$,   
$\frac{3}{2}^\pm$, $\frac{5}{2}^\pm, ...$, and the initial state
($N$) is a $J^P = \frac{1}{2}^+$ state.

We start introducing some notation necessary for clarity of the discussion.
The helicity amplitudes which parametrize the transitions 
in terms of the photon and nucleon polarizations are defined in Section~\ref{sec-amplitudes}.
The transition form factors are discussed in 
the Sections~\ref{sec-spin12} (for $J= \frac{1}{2}$) and \ref{sec-spin32} (for $J \ge \frac{3}{2}$).
In the last case, we also define the 
magnetic dipole ($G_M$), the electric ($G_E$) and the Coulomb ($G_C$) 
quadrupole transition form factors, also known as multipole form factors.
The properties of the transition amplitudes  
at large $Q^2$ are discussed in Section~\ref{sec-largeQ2}.
In Section~\ref{sec-Siegert}, we discuss the constraints
that one needs to take into account at low $Q^2$, 
when we use either helicity amplitudes or multipole form factors.

\subsection{\it Notation} 

A covariant representation is necessary throughout this review.
Here we define our notation.
A 4-vector $a^\mu$ is written as $a^\mu =(a_0, {\bf a})$,
where $a_0$ is the time component and ${\bf a}$ is the space component.
In general, we omit the covariant index ($\mu$), 
unless we refer to the contravariant component $a_\mu = (a_0, - {\bf a})$.
More information about the covariant notation can be found in 
Appendix~\ref{appNotation}.

Throughout this paper, we use natural units, $\hbar = c =1$, 
with masses, 3-momenta and energies given in GeV.
To convert energies in distances, we use the relation 
$\hbar c \simeq 0.197327$ GeV fm $=1$, 
meaning that \mbox{1 fm $\simeq 5.07$} GeV$^{-1}$.

\begin{figure}[t]
\begin{center}
\epsfig{file=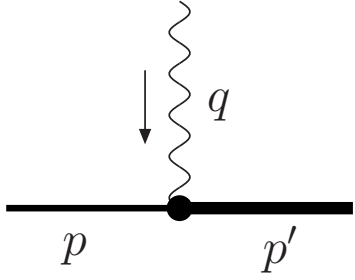,scale=1.3}
\end{center}
\caption{\footnotesize
 Kinematic associated to the $\gamma^\ast N \to N^\ast$ transition.
\label{fig-VertexNR}}
\end{figure}

To describe the kinematics associated with a general electromagnetic
$\gamma^\ast N \to N^\ast$ transition, we use $p$ for 
the initial nucleon momentum and $p^\prime$ 
for the resonance momentum, as described in Fig.~\ref{fig-VertexNR}.
We label by $q$ the reaction 4-momentum transfer and by $P$ the average of the nucleon 
and resonance 4-momenta 
\ba
q= p^\prime - p, 
\hspace{2cm}
P= \frac{1}{2}(p^\prime + p).
\label{eqQP}
\ea
In this notation $p= P -\sfrac{1}{2}q $ and $p^\prime = P + \sfrac{1}{2} q$.
We denote the mass of the nucleon ($N$) by $M$ and the nucleon resonance ($N^\ast$) by $M_R$.

The invariant squared momentum transfer $q^2$
is non positive and therefore defined in the spacelike kinematic region
($q^2 \le 0$)\footnote{The sign of $q^2$ can be determined 
by the change of momenta in the electrons 
on the transition $e^- N \to e^- N^\ast$.
A simple analysis of the kinematics at the Breit frame
shows that $q^2 = - 4{\bf k}^2$, where ${\bf k}$ 
is the 3-momentum of the final electron. Thus  $q^2 \le 0$.}. 
As it is usual in the literature, we define $Q^2=-q^2$, and use the variable $Q^2>0$
to parametrize the structure functions associated 
with the electroexcitations of the nucleon resonances.
(Photoexcitations or radiative decays correspond to $Q^2=0$.)

Along this work it is useful to define two variables that will often enter 
into the formulas describing the
helicity amplitudes, discussed next:
\ba
Q_\pm^2 = (M_R \pm M)^2 + Q^2.
\label{eqQpm}
\ea

\subsection{\it Helicity transition amplitudes \label{sec-amplitudes}}

A general nucleon electromagnetic excitation $\gamma^\ast N \to N^\ast$ 
is determined by the transition current $J^\mu$.
By definition this current encapsulates the full complexity of 
the $\gamma^\ast N \to N^\ast$ vertex.

The helicity amplitudes are transition matrix elements 
of the projections of this current on the three possible photon polarization states.
The helicity amplitudes of the transition 
$N  \to N^\ast \left(J^P\right)$ are
defined at the resonance rest frame as\footnote{The 
amplitude $A_{3/2}$ is equivalent to an amplitude defined 
by the transition $S_z= -\frac{1}{2} \to S_z^\ast = -\frac{3}{2}$ 
and photon polarization $\epsilon_\mu^{(-)}$.}
\ba
A_{1/2} & =& \sqrt{\frac{2 \pi \alpha}{K}}
\left< 
N^\ast, S_z^\ast= + \frac{1}{2} \right|   
\epsilon_\mu^{(+)} J^\mu  \left| 
N, S_z= - \frac{1}{2} \right>,    
\label{eqA12}
\\
A_{3/2} & =& \sqrt{\frac{2 \pi \alpha}{K}}
\left< 
N^\ast, S_z^\ast= + \frac{3}{2} \right|   
\epsilon_\mu^{(+)} J^\mu  \left| 
N, S_z=  + \frac{1}{2} \right>, \label{eqA32} \\
S_{1/2} & =& \sqrt{\frac{2 \pi \alpha}{K}}
\left< 
N^\ast, S_z^\ast=  + \frac{1}{2} \right|   
\epsilon_\mu^{(0)} J^\mu  \left| 
N, S_z= + \frac{1}{2} \right> \frac{|{\bf q}|}{Q},
 \label{eqS12}
\ea
where $\alpha = \frac{e^2}{4\pi^2}$ is the fine structure constant,
$K = \frac{M_R^2 - M^2}{2 M_R}$, $\epsilon^{({\ms \lambda_\gamma})}_\mu$ 
is the photon polarization vector ($\lambda_\gamma= 0,\pm$),
$Q= \sqrt{Q^2}$ 
and $|{\bf q}|$ is the magnitude of the photon 3-momentum.
The current $J^\mu$ is in units of the elementary charge ($e$).
The nucleon and resonance asymptotic states have no dimensions and the 
helicity amplitudes have units of GeV$^{-1/2}$.

The above definitions of helicity amplitudes are valid for any $N^\ast$ final state.
However, the amplitude $A_{3/2}$ is defined only for states $N^\ast$ with $J \ge \frac{3}{2}$. 
Otherwise, there are only the other two amplitudes.
The scalar amplitude $S_{1/2}$ cannot be determined 
at the photon point ($Q^2=0$) since there are no real photons 
with longitudinal polarization, 
but it can be measured for values arbitrarily close to $Q^2=0$.

The kinematic variables in the $N^\ast$ rest frame,
with ${\bf q}$ along the $\hat z$ direction, read
\ba
p= (E, 0, 0, -|{\bf q}|), 
\hspace{1.2cm}
p^\prime= (M_R, 0, 0, 0), 
\hspace{1.2cm}
q= (\omega, 0,0, |{\bf q}|),
\ea
where $E= \sqrt{M^2 +|{\bf q}|^2 }$ is the nucleon energy and 
$\omega = M_R- E$ is the photon energy.
These variables can be written in a covariant form as
\ba
|{\bf q}|= \frac{\sqrt{Q_+^2 Q_-^2}}{2 M_R}, 
\hspace{1.2cm}
E= \frac{M_R^2 + M^2 + Q^2}{2 M_R},
\hspace{1.2cm}
\omega=
\frac{M_R^2 - M^2 - Q^2}{2 M_R},
\label{eqQEw}
\ea
where for $|{\bf q}|$ the variables defined in Eqs.~(\ref{eqQpm}) were used.
The photon polarization vectors are
\ba
\epsilon^{(0)}_\mu = \frac{1}{Q} ( |{\bf q}|, 0, 0, - \omega), 
\hspace{1.4cm}
\epsilon^{(\pm)}_\mu = \pm \frac{1}{\sqrt{2}}(0, 1, \mp i,0).  
\label{eqEpsilon}
\ea

Differently than the longitudinal amplitudes $A_{3/2}$ and $A_{1/2}$ 
corresponding to photons with transverse polarization, the scalar amplitude $S_{1/2}$, 
corresponding to photons with zero polarization, includes an extra 
factor $|{\bf q}|/Q$ to the projection $\epsilon^{(0)} \cdot J$.
This factor is not included in some definitions of the helicity amplitudes,
as for instance the helicity amplitudes in the Breit frame.
Applying the current conservation condition  $q \cdot J =0$ 
to the product $(|{\bf q}|/Q) (\epsilon^{(0)} \cdot J)$,  
we conclude that $(|{\bf q}|/Q) (\epsilon^{(0)} \cdot J) = J^0$, i.e., 
the definition (\ref{eqS12}) is equivalent to the matrix element of 
the charge component of the current.

The helicity amplitudes in the limit $Q^2=0$ 
are important for the determination 
of the radiative decay width of the resonances.
The $N^\ast \to \gamma N$ decay width of a resonance with spin $J$
takes the form~\cite{Aznauryan12a,PDG2018}
\ba
\Gamma_{N^\ast \to \gamma N} =
\frac{K^2}{\pi} \frac{2 M}{(2J +1) M_R} 
\left[ |A_{1/2}(0)|^2 + |A_{3/2}(0)|^2 \right],
\label{eqGammaPDG}
\ea
where $A_{3/2}$ and $A_{1/2}$ are the helicity amplitudes 
from Eqs.~(\ref{eqA12}) and  (\ref{eqA32}). 
In the case $J=\frac{1}{2}$, we use $A_{3/2} = 0$.

The electron scattering on nucleons ($\gamma^\ast N \to N^\ast$ transitions) 
probe the region $Q^2 >0$.
Near the photon point ($Q^2=0$), the helicity amplitudes 
are physically defined only in the region $|{\bf q}| \ge 0$.
The minimum of $|{\bf q}|$ is zero when the photon 3-momentum vanishes ($|{\bf q}|=0$).
This point is called pseudothreshold.
At the pseudothreshold the photon energy component is $\omega = M_R -M$ 
and $Q^2=- (M_R -M)^2$.
This means that, although the helicity amplitudes measured 
in the $\gamma^\ast N \to N^\ast$ transitions
are defined in the region $Q^2 \ge 0$, 
they can be extended to the region $-(M_R-M)^2 \le Q^2 < 0$. 
The region $Q^2 < 0$ can be probed by $N^\ast \to \gamma^\ast N$ transitions, 
the source of the $N^\ast$ Dalitz decay ($N^\ast \to e^+ e^- N$).
As discussed later, the helicity amplitudes near the pseudothreshold 
are subject to some conditions that constraint the $Q^2$-dependence 
of those functions.

\begin{figure}[t]
\begin{center}
\epsfig{file=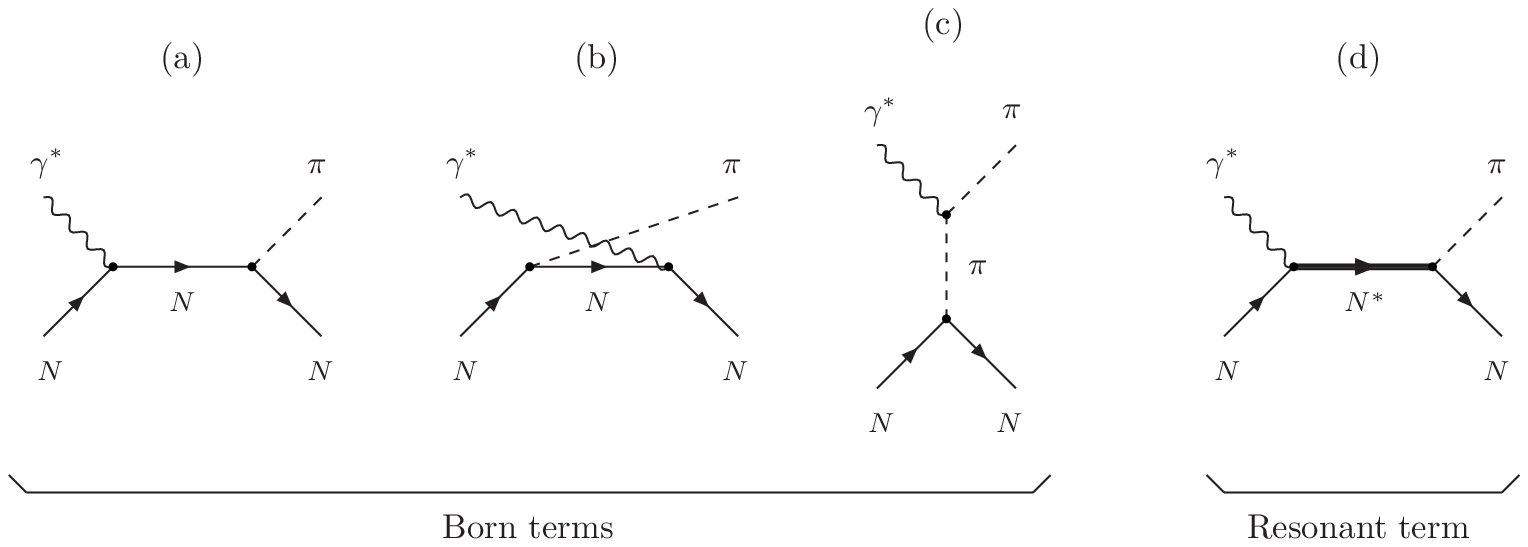,scale=1.0}
\end{center}
\caption{\footnotesize
 Born terms and resonance terms in the $\gamma^\ast N \to \pi N$ reaction.
\label{fig-Total}}
\end{figure}

The basic ingredient of the amplitudes (\ref{eqA12}), (\ref{eqA32}) and (\ref{eqS12}) 
is the transition current which is a theoretical object.
But how are helicity amplitudes extracted from the data? Since the $N^*$ resonances 
decay into baryon and meson final states, the transition current or $\gamma^\ast N \to N^\ast$ 
is necessarily part of the pion photoproduction 
and electroproduction off nucleons.
The experimentally extracted amplitudes, labeled $A_{1/2}^{\rm exp}$, 
$A_{3/2}^{\rm exp}$ and $S_{1/2}^{\rm exp}$, are obtained from data analyses of 
pion electroproduction (or meson electroproduction in general) processes
represented in Fig.~\ref{fig-Total}, where the contribution 
of diagram (d) has to be separated.
Pion electroproduction combines a
electromagnetic component ($\gamma^\ast N \to N^\ast$), 
which define the helicity amplitudes, and the component of
the resonance strong decay ($N^\ast \to \pi N$) proportional 
to the pion-nucleon-resonance coupling constant $g_{\pi N N^\ast}$~\cite{Devenish76}.
The extraction of the helicity amplitudes from the 
pion production amplitudes depends on relative signs 
of the $N \to \pi N$ (present in the Born term) 
and $N^\ast \to \pi N$ couplings.

In Refs.~\cite{Aznauryan08a,Aznauryan07} it is shown that 
the theoretical calculations of the helicity amplitudes 
by Eqs.~(\ref{eqA12}), (\ref{eqA32}) and (\ref{eqS12})
with the experimentally measured $A_{1/2}^{\rm exp}$, $A_{3/2}^{\rm exp}$ and $S_{1/2}^{\rm exp}$
are related through a global sign $\zeta$ 
\ba
A_{1/2}^{\rm exp} = \zeta A_{1/2},
\hspace{1.5cm} 
A_{3/2}^{\rm exp} = \zeta A_{3/2},
\hspace{1.5cm} 
S_{1/2}^{\rm exp} = \zeta S_{1/2},
\label{eqAmpsEXP}
\ea
where 
\ba
\zeta = - {\rm sign}(g_{\pi NN^\ast}/g_{\pi NN}).
\label{eqPhaseZeta}
\ea 
It is understood that $g_{\pi NN^\ast}$  and $g_{\pi NN}$ 
are determined by the same framework, and
as a consequence the r.h.s.~is independent of 
relative signs of the nucleon and resonance states~\cite{Aznauryan12a,Aznauryan08a}.  
Detailed discussions on the subject can be found  
in Refs.~\cite{Aznauryan12a,Aznauryan08a,Bauer14a}.

In quark model calculations
the phase between the calculated amplitudes and the measured amplitudes,
which depend on the relative sign between 
of the $N^\ast$ resonance and nucleon states\footnote{In the elastic
transitions $\gamma^\ast N \to N$ or 
$\gamma^\ast N^\ast \to N^\ast$ 
the normalization is fixed by the charge of the baryon ($N$ and $N^\ast$)
and the result is independent of the sign of the normalization constant.
In inelastic transition the form factors 
depend on the relative phase of the two states.}
can be determined directly by the data in the following way:
one adjusts the sign of the resonance wave function 
by the comparison of the model calculations 
with the helicity amplitude data for large $Q^2$, where the 
valence quark degrees of freedom dominate 
and the meson cloud excitations are negligible~\cite{Aznauryan08a,Aznauryan07,N1440-1}.

In the previous discussion 
we gave the $\pi N$ decay channel as an example, 
since it is the dominant source 
of information about the $\gamma^\ast N \to N^\ast$ transition.
But the analysis can include other decay modes 
as the $\eta N$ channel in the $N(1535)$ state 
or the $\pi \Delta(1232)$ channel in the  
$N(1440)$ state~\cite{Aznauryan12a,Burkert04,CLAS09,CLAS12,Denizli07,Dalton09}.
The implementation of the method becomes more complicated 
for resonances with a weak coupling to the $\pi N$ channel 
or with dominance of the $\pi  \pi N$ channels~\cite{Aznauryan12a,Mokeev22a,CLAS12,CLAS16a}.

In the next sections, we present the expressions for 
the helicity amplitudes 
based on the relations (\ref{eqA12})--(\ref{eqS12}), 
and multipole transition form factors for 
resonances with $J^P= \frac{1}{2}^\pm$, $\frac{3}{2}^\pm$, $\frac{5}{2}^\pm, ...$.
The conventions associated to the gamma matrices 
and the normalizations of the states 
are presented in Appendix~\ref{appNotation}.
We start writing the $\gamma^\ast N \to N^\ast$ transition current 
for a resonance characterized by the spin-parity $J^P$,  
using Lorentz structures built from the momentum variables $q$ and $P$, 
compatible with gauge invariance 
and free of kinematic singularities~\cite{Devenish76,JonesScadron}.
The transition current is then expressed   
in terms of independent kinematic-singularity-free 
form factors $G_i$, which are used to write down the expressions 
for the helicity amplitudes and multipole form factors.
We follow the formulation from Devenish et al.~\cite{Devenish76}
and Aznauryan and Burkert~\cite{Aznauryan12a}, and adopt 
the normalization of Aznauryan-Burkert.
Early formulations of transition currents can be found 
in Refs.~\cite{JonesScadron,BjorkenWalecka}.
The results for the case $J^P=\frac{1}{2}^\pm$ and 
$J^P= \frac{3}{2}^\pm, \frac{5}{2}^\pm, ...$, are presented 
in Sections~\ref{sec-spin12} and \ref{sec-spin32}, respectively.

It is worth noticing that the transition form factors  $G_i$
used in the parametrization of the  $\gamma^\ast N \to N^\ast$
transition current  are kinematic-singularity-free functions
and form an independent basis, meaning that they
are uncorrelated and free of constraints.
As for the helicity amplitudes and multipole form factors 
expressed in terms of the kinematic-singularity-free form factors $G_i$,
they are not independent in the kinematic limits $Q_\pm^2=0$,
and are therefore subject to some constraints.
The constraints associated to the helicity amplitudes and 
multipole form factors near the pseudothreshold, $Q^2=-(M_R-M)^2$, 
below the $Q^2$-range reveled by electron scattering  ($Q^2 > 0$), 
are discussed in Section~\ref{sec-Siegert}.

\subsection{\it Form factors $\frac{1}{2}^\pm$ \label{sec-spin12}}

With the notation for the momenta variables of Eq.~(\ref{eqQP}), 
the most general structure for the transition current associated 
with the $\gamma^\ast N  
\to N^\ast \left(\frac{1}{2}^\pm\right)$ vertex can be 
represented by the following combination of the two possible vector 
Lorentz structures 
\ba
J^\mu = 
G_1 ( q^\mu {\not \! q} - q^2 \gamma^\mu) 
\left( 
\begin{array}{c}  \Unit \cr \gamma_5 \cr \end{array}
\right) + 
G_2  [{\not \! q} P^\mu - (P \cdot q) \gamma^\mu]
\left( 
\begin{array}{c}  \Unit \cr \gamma_5 \cr \end{array}
\right).
\label{eqcurrentgeral}
\ea
The previous equation describes the current 
associated to the $J^P= \frac{1}{2}^\pm$ states
in the following way:
the upper operator stands for the upper parity index ($P=+$) 
and the lower operator stands for the lower parity index ($P=-$).
This notation will be used along the article.
The transition form factors $G_1$ and $G_2$
have dimensions of inverse of square masses ($G_1, G_2 \sim 1/M_R^2$).

Another common representation of the current used in the 
literature~\cite{N1440-1,N1535-1,Wilson12,Lin08} is
\ba
J^\mu = 
F_1 \left( \gamma^\mu  - \frac{{\not \! q} q^\mu }{q^2}\right) 
\left( 
\begin{array}{c}  \Unit \cr \gamma_5 \cr \end{array}
\right) + 
F_2 \frac{i \sigma^{\mu \nu} q_\nu}{M_R + M}
\left( 
\begin{array}{c}  \Unit \cr \gamma_5 \cr \end{array}
\right), 
\label{eqJ12b}
\ea
where $F_1$ and $F_2$ are dimensionless and given by 
$F_1= Q^2 G_1$ and $F_2= - \frac{1}{2}(M_R \mp M) (M_R + M) G_2$.
The form factors $F_1$ and $F_2$ are known as
Dirac and Pauli form factors, respectively, 
as in the case of the nucleon elastic form factors.
Comparing to Eq.~(\ref{eqcurrentgeral}) we notice that
the Lorentz operator of the first term has a singularity at $q^2=0$.
This singularity is canceled by the factor $Q^2=-q^2$ in $F_1= Q^2 G_1$, 
leading to the constraint $F_1(0)=0$.
The representation (\ref{eqcurrentgeral}) has the advantage of 
avoiding singularities on the Lorentz structure operators.
The representation (\ref{eqJ12b}) is  advantageous for the comparison 
with the elastic limit for $J^P= \frac{1}{2}^+$.

From the definitions above, we obtain the relations between the 
kinematic-singularity-free form factors $G_i$
and the helicity amplitudes 
introduced in the previous section for $J^P= \frac{1}{2}^\pm$
\ba
& & A_{1/2} (Q^2)= \frac{e}{2} {\cal B}_\mp
\left[2 Q^2 G_1  (Q^2)- (M_R^2 -M^2) G_2 (Q^2)  
\right], 
\label{eqA12-s12}\\
& & S_{1/2} (Q^2)= \pm \frac{e}{2\sqrt{2}} {\cal B}_\mp |{\bf q}|
\left[ 2 (M_R \pm M) G_1 (Q^2) + (M_R \mp M) G_2 (Q^2)
\right],
\label{eqS12-s12}
\ea
with
\ba
{\cal B}_\mp = \sqrt{\frac{Q_\mp^2}{4 M M_R K}}.
\label{eqB-s12}
\ea

In these and in the following formulas, we use a compact notation 
for the obtained signs: the upper and lower sign
specifies respectively the parity index, 
corresponding respectively to the 
upper and lower operators in Eq.~(\ref{eqcurrentgeral}).
The upper signs holds then for $J^P= \frac{1}{2}^+$ 
and the lower signs holds for $J^P= \frac{1}{2}^-$.

The helicity amplitudes in terms of the Dirac and Pauli form factors are
\ba
& & A_{1/2} (Q^2)= e {\cal B}_\mp
\left[ F_1  (Q^2) + \frac{M_R \pm M}{M_R + M} F_2 (Q^2)  \right], 
\label{eqA12-b} \\
& & S_{1/2} (Q^2) = \pm \frac{e}{\sqrt{2}} {\cal B}_\mp 
\frac{(M_R + M) |{\bf q}| }{Q^2}
\left[ \frac{M_R \pm M}{M_R + M} F_1 (Q^2) - \tau F_2 (Q^2)
\right],
\label{eqS12-b}
\ea
with $\tau = \frac{Q^2}{(M_R + M)^2}$.
In this case, we can calculate the inverse relations
\ba
& & F_1 (Q^2)= \frac{1}{e B_\mp} \frac{Q^2}{Q_\pm^2} 
\left[  A_{1/2} (Q^2) \pm \sqrt{2} \frac{M_R \pm M}{|{\bf q}|} S_{1/2} (Q^2)
\right],  \label{eqF1-b}\\
& & F_2 (Q^2)= \frac{1}{e B_\mp}
\frac{(M_R + M)(M_R \pm M)}{Q_\pm^2} 
\left[ A_{1/2} (Q^2) \mp \sqrt{2} \frac{Q^2}{(M_R \pm M)|{\bf q}|} S_{1/2} (Q^2)
\right].  \label{eqF2-b}
\ea
Importantly, the determination of $F_1$ and $F_2$ data 
is only possible when there are data for  $A_{1/2}$ and $S_{1/2}$ 
at the same value of $Q^2$.
The exception is the point $Q^2=0$, where $S_{1/2}$ 
cannot be measured and $F_1(0)=0$, by construction. 
From Eq.~(\ref{eqF2-b}), one has $F_2(0) \; \propto \; A_{1/2} (0)$.

In the representation by the  $F_1$ and $F_2$ form factors, 
there are no singularities in the amplitude $S_{1/2}$, because in (\ref{eqS12-b}) 
the  prefactor $1/Q^2$ combines with $F_1(Q^2)$ such that in the limit $Q^2 \to 0$ the ratio
$F_1/Q^2$ converges to 
a constant, $F_1^\prime(0)$, and $\tau /Q^2 \to 1/(M_R + M)^2$.

In the case $J^P= \frac{1}{2}^+$ the amplitudes 
take the form  $S_{1/2} \; \propto \; (F_1 - \tau F_2)$
and  $A_{1/2} \; \propto \; (F_1 + F_2)$, similar to 
the relations $G_E \; \propto  \; F_1 - \tau F_2$ and 
$G_M \;  \propto  \; F_1 + F_2$
for the nucleon elastic form factors~\cite{Hyde04a,Perdrisat07,Arrington07a}.
Notice, however, that the functions $F_1$ and $F_2$ have different meanings 
in the case of elastic and inelastic transitions.
In the inelastic transitions, one has always $F_1 \; \propto \; Q^2$.
In the elastic transitions this happens only for neutral baryons,
as in the case of the neutron.

Some authors define the Coulomb ($G_C$) and magnetic/electric ($G_M/G_E$) form factors 
for the states $J^P=\frac{1}{2}^\pm$ based on the 
direct relations with the amplitudes $A_{1/2}$ 
and the magnetic/electric amplitudes~\cite{Drechsel07,Krivoruchenko02,N1535-TL}.
One obtains then for $J^P = \frac{1}{2}^+$
\ba
G_M =  F_1 + F_2 = \frac{1}{e {\cal B}_-} A_{1/2}, \hspace{1cm}
G_C =  F_1 - \tau F_2 =
\frac{\sqrt{2}}{e {\cal B}_-} \frac{Q^2}{M_R + M}
\frac{S_{1/2}}{|{\bf q}|}.
\label{eqFFs12p}
\ea 
The combination $F_1 - \tau F_2$ is relabeled as Coulomb form factor~\cite{Drechsel07,Krivoruchenko02}.
As for $J^P = \frac{1}{2}^-$, one has  
\ba
G_E =  F_1 +\eta F_2 = \frac{1}{e {\cal B_+}} A_{1/2}, \hspace{1cm}
G_C = - \frac{M_R}{2}\frac{M_R+ M}{Q^2}
\left[\eta  F_1 - \tau F_2 \right]= 
\frac{M_R}{\sqrt{2} e {\cal B}_+}
\frac{ S_{1/2}}{|{\bf q}|},
\label{eqFFs12m}
\ea 
with $\eta = \frac{M_R - M}{M_R + M}$. 

The magnetic/electric 
and Coulomb form factors  for $J^P= \frac{1}{2}^\pm$ 
are useful for the study of the 
Dalitz decay ($N^\ast \to e^+ e^- N$)~\cite{Krivoruchenko02,N1535-TL}. 
For completeness, we point out that alternative 
representations of the $\gamma^\ast N \to N^\ast \left( \frac{1}{2}^\pm \right)$
transition currents  
can be found in Refs.~\cite{Pace99a,Jido08a,Braun09}.

\subsection{\it Form factors $\frac{3}{2}^\pm, \frac{5}{2}^\pm, ...$ \label{sec-spin32}}

In this section, we present the formulas for the helicity amplitudes 
and the multipole transition form factors of the high spin cases, 
$J \ge \frac{3}{2}$.
These functions are also expressed in terms of 
independent kinematic-singularity-free form factors.
For simplicity, we introduce auxiliary functions $h_i$ labeled as 
helicity form factors~\cite{Devenish76}.
The explicit forms depend on the quantum number
\ba
l= J -\frac{1}{2},
\ea
and on the parity $P=\pm$~\cite{Aznauryan12a,Devenish76}.

There are similarities between the case $J^P= \frac{3}{2}^\pm$ 
and the cases $J^P= \frac{5}{2}^\mp,  \frac{7}{2}^\pm, ...$.
Notice the alternation of parity, when we increase $J$.
Because of the similarities, it is convenient 
to separate the case $J^P=\frac{3}{2}^\pm$ from the high order spins.
Thus, in Section~\ref{sec32}, we discuss the results for $J^P= \frac{3}{2}^\pm$,
and in Section~\ref{sec52}, we discuss  
the results for  $J^P= \frac{5}{2}^\mp,  \frac{7}{2}^\pm, ...$.

We can anticipate that for a fixed $J$, the transverse helicity amplitudes and 
the electric and magnetic form factors for negative parity states can be obtained 
from the formulas for positive parity through simple transformations.

\subsubsection{\it Case $J^P= \frac{3}{2}^\pm$ \label{sec32}}

Let us start with the case $J^P=\frac{3}{2}^\pm$ for clarity.
The projection of the transition current $J^\mu$
on the asymptotic states $\left| N^\ast \right>$ and  $\left| N \right>$ 
can be expressed as
\ba
\left< N^\ast \right|  J^\mu  \left| N  \right>=
\bar u_\alpha (p') \;
\Gamma^{\alpha \mu} (q) 
\left( 
\begin{array}{c}  \gamma_5 \cr \Unit \cr \end{array}
\right) \; 
u(p), 
\label{eqGamma0}
\ea
where $u_\alpha$ and $u$ are the resonance 
Rarita-Schwinger and nucleon Dirac spinors, respectively.
The operator $ \Gamma^{\alpha \mu} $ is in units of the elementary charge ($e$).
According to the notation adopted, the upper operator is 
associated with the upper parity index (state $J^P= \frac{3}{2}^+$) 
and the lower operator is associated with the lower parity index
(state $J^P= \frac{3}{2}^-$). 
The spin projectors of the $N^\ast$ state and of the nucleon are omitted for simplicity.

The operator $\Gamma^{\alpha \mu} $ can be expressed 
in the gauge-invariant form~\cite{Aznauryan12a}
\ba
\Gamma^{\alpha \mu}  (q)=
\left[q^\alpha \gamma^\mu - {\not \! q} g^{\alpha \mu} \right] G_1 (Q^2) + 
\left[q^\alpha p^{\prime  \mu} -  (p^\prime \cdot q)  g^{\alpha \mu} \right] G_2 (Q^2) +
\left[q^\alpha q^\mu -  q^2  g^{\alpha \mu} \right] G_3 (Q^2), 
\label{eqGamma1}
\ea
where $G_i$ are independent kinematic-singularity-free form factors.
The dimensions of the form factors are 
$G_1 \sim 1/M_R$, $G_2 \sim 1/M_R^2$ and  $G_3 \sim 1/M_R^2$.

Using the form factors $G_i$, we can define the helicity form factors~\cite{Aznauryan12a,Devenish76}
\ba
& &
h_1(Q^2) = 4 M_R G_1(Q^2) + 4 M_R^2 G_2 (Q^2) + 2 (M_R^2 - M^2 - Q^2) G_3(Q^2)\label{eqH1}, \\ 
& &
h_2(Q^2) = - 2(M_R \pm M) G_1(Q^2) - (M_R^2 -M^2 -Q^2) G_2(Q^2)   + 2 Q^2 G_3(Q^2), \label{eqH2}\\
& &
h_3(Q^2) = 
- \frac{2}{M_R} \left[Q^2 +  M (M \pm M_R) \right] G_1(Q^2) 
+ (M_R^2 -M^2 -Q^2) G_2(Q^2)   - 2 Q^2 G_3(Q^2).
\label{eqH3}
\ea
For $l=1$ the helicity form factors $h_i$ are dimensionless.

The helicity amplitudes can now be calculated using 
\ba
& &
A_{1/2}(Q^2) =  {\cal A}_{1 \mp} \, h_3(Q^2), \hspace{1.4cm}
S_{1/2}(Q^2) =   \pm \sqrt{2} {\cal A}_{1 \mp} \, \frac{|{\bf q}|}{2 M_R} h_1(Q^2),
\nonumber \\
& &
A_{3/2}(Q^2) =  \pm \sqrt{3} {\cal A}_{1 \mp} \, h_2(Q^2), \hspace{.6cm}
\label{eqAmps32}
\ea
where 
\ba
{\cal A}_{1 \mp} = 
\frac{1}{2 \sqrt{3}} \sqrt{\frac{2 \pi \alpha}{K}}
\sqrt{\frac{Q_\mp^2}{4 M M_R}}.
\label{eqA1}
\ea
Again, the previous equations are valid for the $J^P= \frac{3}{2}^\pm$ states,
where the upper signs hold for the upper parity index ($J^P= \frac{3}{2}^+$)
and the lower signs hold for the lower parity index ($J^P= \frac{3}{2}^-$).
We use here the Aznauryan-Burkert representation~\cite{Aznauryan12a} which
is equivalent to the Devenish representation~\cite{Devenish76}
when we redefine $G_i \to \sqrt{\frac{2}{3}} G_i$.

The multipole form factors can now be expressed in 
terms of the helicity amplitudes.
For $J^P = \frac{3}{2}^+$, one has~\cite{Aznauryan12a,Drechsel07,Capstick95}
\ba
& &
G_M (Q^2)= -F_{1 +} \left( \sqrt{3} A_{3/2} (Q^2)+ A_{1/2} (Q^2)  \right), 
\label{eqGM1p}
\\
& &
G_E (Q^2)= -F_{1 +} \left( \frac{1}{\sqrt{3}} 
A_{3/2} (Q^2) - A_{1/2} (Q^2) \right),\label{eqGE1p} \\
& &
G_C (Q^2) = \sqrt{2}  
F_{1+} \frac{2 M_R}{|{\bf q}|}   S_{1/2} (Q^2), 
\label{eqGC1p}
\ea 
and for $J^P = \frac{3}{2}^-$
\ba
& &
G_M (Q^2)= -F_{1 -} \left( \frac{1}{\sqrt{3}} 
A_{3/2} (Q^2) - A_{1/2} (Q^2) \right), 
\label{eqGM1m}
\\
& &
G_E (Q^2)= -F_{1 -} \left( \sqrt{3} A_{3/2} (Q^2)+ A_{1/2} (Q^2)  \right), 
\label{eqGE1m} \\
& &
G_C (Q^2) =
- \sqrt{2}  F_{1-}   \frac{2 M_R}{|{\bf q}|}  S_{1/2} (Q^2),
\label{eqGC1m}
\ea 
where 
\ba
F_{1 \pm} &= &
 \frac{M}{|{\bf q}|} \frac{2 M}{(M_R \pm M)} 
\sqrt{\frac{K}{4 \pi \alpha}} \sqrt{\frac{Q^2_\pm}{4 M M_R}} \nonumber \\
& =& 
\sqrt{\frac{3}{2}}   \frac{M}{6(M_R \pm M)} \frac{1}{{\cal A}_{1 \mp}}. 
\label{eqFl1a}
\ea 
The explicit expressions for $G_M$, $G_E$ and $G_C$ 
in terms of $G_i$ are presented in Appendix~\ref{appMultipole}.
The previous formulas show that we can obtain the 
electric/magnetic form factors for  $J^P= \frac{3}{2}^-$ 
from the formulas for $J^P= \frac{3}{2}^+$ 
interchanging $G_M/F_{1\pm} \leftrightarrow G_E/F_{1\mp}$ 
and  $G_C/F_{1\pm} \leftrightarrow - G_C/F_{1\mp}$. 

For convenience, we also compile here the inverse relations 
for the helicity amplitudes in terms 
of the multipole form factors, for $J^+= \frac{3}{2}^+$:
\ba
& &
A_{1/2} (Q^2) = - \frac{1}{4 F_{1+}} \left[ G_M(Q^2) - 3 G_E (Q^2)  \right], 
\hspace{1.cm} 
S_{1/2} (Q^2) = \frac{1}{\sqrt{2} F_{1+}} \frac{|{\bf q}|}{2 M_R} G_C(Q^2),
\nonumber \\
&& 
A_{3/2} (Q^2) = -\frac{\sqrt{3}}{4 F_{1+}} \left[G_M (Q^2) + G_E(Q^2)  \right].
\label{eqAmp32p}
\ea
and for  $J^+= \frac{3}{2}^-$:
\ba
& &
A_{1/2} (Q^2) = - \frac{1}{4 F_{1-}} \left[G_E(Q^2) - 3 G_M (Q^2)\right],
\hspace{1.cm} 
S_{1/2} (Q^2) = 
 - \frac{1}{\sqrt{2} F_{1-}} \frac{|{\bf q}|}{2 M_R} G_C(Q^2),
\nonumber \\
&& 
A_{3/2} (Q^2) = -\frac{\sqrt{3}}{4 F_{1-}} \left[G_E (Q^2) + G_M(Q^2)  \right].
\label{eqAmp32m}
\ea
As a consequence, the relations 
between the $J^P= \frac{3}{2}^+$ and  $J^P= \frac{3}{2}^-$ 
multipole form factors, we can obtain the amplitudes for $J^P= \frac{3}{2}^-$ 
from the formulas for $J^P=\frac{3}{2}^+$ 
replacing $G_{E/M} \leftrightarrow G_{M/E}$, $G_C \to - G_C$ 
and $F_{1+} \rightarrow F_{1-}$.

For the comparison with the literature we add that 
the formulation from Devenish~\cite{Devenish76} 
and Jones and Scadron~\cite{JonesScadron} includes 
an isospin factor $\sqrt{\frac{2}{3}}$ 
associated to the $\gamma^\ast p \to \Delta^+ $ transition 
on the vertex (\ref{eqGamma0}) which leads to different 
transformations between $G_i$ and multipole form factors.
These transformations are discussed on Appendix~\ref{appMultipole}.
The present formulation and the Devenish/Jones and Scadron formulation are, 
however, equivalent, and give the same 
multipole form factors and helicity amplitudes (see Appendix~\ref{appMultipole}).

To prepare the discussion of the asymptotic behavior of the helicity amplitudes
in the following sections, we present a relation between the 
functions $h_2$ and $h_3$.
From Eqs.~(\ref{eqH2}) and (\ref{eqH3}), one obtains 
\ba
h_2 + h_3 = - 2 \frac{Q_\pm^2}{M_R} G_1,
\ea
and we conclude that
\ba
G_1 \; \propto \; \frac{G_M -G_E}{Q_+^2}
\; \; \mbox{for} \; J^P= \frac{3}{2}^+,
\hspace{1.2cm}
G_1 \; \propto \; \frac{G_M}{Q_-^2}
\; \mbox{for}\; J^P= \frac{3}{2}^-.
\label{eqG1-32}
\ea

There are alternative representations for $J= \frac{3}{2}$ currents 
that can be generalized for  $J > \frac{3}{2}$~\cite{Devenish76,Eichmann18,Vereshkov10a,Vereshkov07a}.
Some of the representations are presented 
in Devenish et al.~\cite{Devenish76} and 
Jones and Scadron~\cite{JonesScadron}.
Several authors~\cite{Kondratyuk01a,Pascalutsa03a,Lorenz15a} propose
another representation which also defines three kinematic-singularity-free 
form factors\footnote{In the original form~\cite{Pascalutsa07,Pascalutsa05b}, 
the form factors have a kinematic zero at the threshold 
$Q^2= -(M_R+M)^2$, but they can be redefined 
without the kinematic zero~\cite{Eichmann18}.} usually labeled as $g_M$, $g_E$ and $g_C$.
The explicit forms and the conversion 
into the Jones and Scadron multipole form factors $G_E$, $G_M$ and $G_C$ 
can be found in Refs.~\cite{Pascalutsa07,Eichmann18,Pascalutsa05b}.
A difference to the form factors $G_1$, $G_2$ and $G_3$, defined above, 
is that $g_M$, $g_E$ and $g_C$ are correlated at large $Q^2$~\cite{Eichmann18}.
Pascalutsa and Vanderhaeghen have used extensively this representation
within the  
chiral effective field theory and large $N_c$ limit frameworks
in calculations of transition 
form factors~\cite{Pascalutsa07a,Pascalutsa05b,Pascalutsa06a,Pascalutsa08a}, 
two-photon contributions to transition form factors~\cite{Pascalutsa06b} 
and in the Compton scattering~\cite{Pascalutsa07,Pascalutsa03a}. 
The modified parametrization from Ref.~\cite{Eichmann18} 
has shown to be particularly appropriate
for the study the falloff of the transition form factors with $Q^2$ and for the
calculation of the nucleon resonance contributions to the nucleon Compton scattering.
The transition form factors have monotonous falloffs with $Q^2$, 
except for radial excitations,  where we expect some zero crossing~\cite{Eichmann18}.
Another representation for transition to states $J \ge \frac{3}{2}$ 
proposed by Vereshkov and collaborators~\cite{Vereshkov10a,Vereshkov07a},
uses simple linear combinations of the 
kinematic-singularity-free form factors $G_i$.
This representation has some advantages in the study of the 
asymptotic behavior of the multipole form factors~\cite{Vereshkov10a,Vereshkov10b,Vereshkov13a}.

\subsubsection{\it Cases $J^P= \frac{5}{2}^\mp$, 
$\frac{7}{2}^\pm, ...$ \label{sec52}}

For the states $J^P= \frac{5}{2}^\mp$, 
$\frac{7}{2}^\pm, ...$ ($l=2,3,...$), 
the transition current reads~\cite{Aznauryan12a}

\ba
\left< N^\ast \right|  J^\mu  \left| N  \right>=
\bar u_{\alpha_1 \alpha_2 ... \alpha_{\; \, {\sm l-1}} \alpha }(p') \; q^{\alpha_1}  
\; q^{\alpha_1} \cdot \cdot \cdot 
 q^{\alpha_{{\ms l-1}}} 
\Gamma^{\alpha \mu} (q) 
\left( 
\begin{array}{c} \gamma_5  \cr \Unit \end{array}
\right) \; 
u(p),
\label{eqVertexL}      
\ea 
where $\Gamma^{\alpha \mu} (q)$ takes the form (\ref{eqGamma1})
and is in units of the elementary charge ($e$).
The states $u_{\alpha_1 \alpha_2 ... \alpha_{\; \, {\sm l-1}} \alpha }$ 
are discussed in Appendix~\ref{appSpinJ}.

Equation (\ref{eqVertexL}) is valid within the adopted notation:
the upper operator holds for the states  $J^P= \frac{5}{2}^-$, $\frac{7}{2}^+, ...$
and the lower operator holds for the states $J^P=\frac{5}{2}^+$, $\frac{7}{2}^-, ...$.
It is important to realize that in Eq.~(\ref{eqVertexL}) the effective vertex 
for $l > 1$ includes, in addition to $\Gamma^{\alpha \mu} (q)$, 
the momenta $q^{\alpha_{\ms i}}$, and differ from the case $l=1$ 
in the dimensions of $G_i$.

The expressions for the helicity amplitudes 
associated with the states $J^P= \frac{5}{2}^\mp$, 
$\frac{7}{2}^\pm, ...$, can be written using Eqs.~(\ref{eqH1})--(\ref{eqH3}) as
\ba
& &
A_{1/2}(Q^2) = (-1)^{l+1} 
{\cal A}_{l \mp} \, h_3(Q^2), \hspace{1.4cm}
S_{1/2}(Q^2) =  \pm (-1)^{l+1} \sqrt{2}
 {\cal A}_{l \mp} \, \frac{|{\bf q}|}{2 M_R} h_1(Q^2),
\nonumber \\
& &
A_{3/2}(Q^2) =  \pm (-1)^{l+1} 
\frac{C_l}{l} {\cal A}_{l \mp} \, h_2(Q^2),   
\label{eqAmps52}
\ea
where 
\ba
{\cal A}_{l \mp} = 
\frac{1}{2\sqrt{2}} \frac{1}{\sqrt{z_l}}
\sqrt{\frac{2 \pi \alpha}{K}} 
\sqrt{\frac{Q_\mp^2}{4 M M_R}}
|{\bf q}|^{l-1},
\hspace{1cm}
z_{\ms l} = \frac{(2 l +2 )!}{ 2^{l+1} [(l+1)!]^2}.
\label{eqAaZ}
\ea
The upper signs hold for the states $J^P= \frac{5}{2}^-$, $\frac{7}{2}^+,...$,
and the lower signs hold for the states $J^P= \frac{5}{2}^+$, $\frac{7}{2}^-, ...$
($l=2,3,...$).
The same sign convention is used in the functions $h_i$ 
defined by Eqs.~(\ref{eqH1})--(\ref{eqH3}).
The expressions presented above are calculated directly 
from the definitions (\ref{eqA12})--(\ref{eqS12}) 
using the spin $J$ spinors associated to the resonances. 
The factor $C_l$ entering in the amplitude $A_{3/2}$ 
is a product of Clebsch-Gordan 
coefficients for the $J= l + \frac{1}{2} $ states\footnote{Given $J= l +\frac{1}{2} > \frac{3}{2}$,
 we can write $C_l = l \sqrt{2 z_l} Z$,
where $Z$ is the product of $(l-1)$ Clebsch-Gordan coefficients
$\left< j_1 j_2 ; m_1 m_2 | j m \right>$:
\ba
Z = \left< 1\; J-1; 0 +\sfrac{3}{2} \right| \left. J  +\sfrac{3}{2}\right> 
\left< 1\; J-2; 0 +\sfrac{3}{2} \right| \left. J-1  +\sfrac{3}{2}\right> 
.....
\left< 1\; \sfrac{3}{2}; 0 +\sfrac{3}{2} \right| \left.
 \sfrac{5}{2}  +\sfrac{3}{2}\right>. 
\nonumber
\ea}
and can be written in the form 
\ba
C_l = \sqrt{l(l+2)}.
\ea 

From the previous relations, we can conclude 
that the  $J > \frac{3}{2}$ helicity amplitudes can also be written 
in terms of the helicity form factors $h_i$, defined previously for $J=\frac{3}{2}$,
with the conversion factors depending on $l$.
This dependence appears on the factor ${\cal A}_{l \pm}$ 
for all the amplitudes and on the factor $1/l$ for $A_{3/2}$.

Differently than for $l=1$, the helicity form factors have now dimensions $h_i \sim 1/M_R^{l-1}$,
since ${\cal A}_{l \pm}$ has dimensions $M_R^{l+ 1/2}$, 
and the helicity amplitudes have dimension $1/M_R^{1/2}$.
The kinematic-singularity-free form factors 
carry the dimensions $G_1 \sim 1/M_R^{l+1}$, 
$G_2 \sim 1/M_R^{l+2}$ and  $G_3 \sim 1/M_R^{l+2}$,
and the multipole form factors $G_M$, $G_E$ and $G_C$ discussed next,
have also dimensions $1/M_R^{l-1}$.

For a compact and simple representation of the multipole form factors, 
we define
\ba
F_{l \pm} &= & \sqrt{\frac{2}{3} z_l} 
\frac{M}{|{\bf q}|^{l}} \frac{2M}{M_R \pm M}
\sqrt{\frac{K}{4 \pi \alpha}} \sqrt{\frac{Q_\pm^2}{4 M M_R}}  
\nonumber \\
&=& 
\sqrt{\frac{3}{2}}
\frac{M}{6(M_R \pm M)} \frac{1}{{\cal A}_{l \mp}}.
\ea

For $J^P= \frac{5}{2}^-, \frac{7}{2}^+, ...$,  we obtain
\ba
& &
G_M (Q^2)= -F_{l +} \left( \frac{l +2}{C_l} A_{3/2} (Q^2) 
+ A_{1/2} (Q^2) \right) \frac{2l}{l+1}  (-1)^{l+1}, 
\label{eqGM2m}
\\
& &
G_E (Q^2)= -F_{l +} \left( 
\frac{l}{C_l} A_{3/2} (Q^2)-  A_{1/2} (Q^2)  \right)  \frac{2}{l+1} (-1)^{l+1}, 
\label{eqGE2m}\\
& &
G_C (Q^2) = \sqrt{2}  F_{l +} \frac{2 M_R}{|{\bf q}|} S_{1/2} (Q^2) (-1)^{l+1}, 
\label{eqGC2m}
\ea
\ba
& &
A_{1/2} (Q^2) = - \frac{1}{4 F_{l + }} 
\left[ G_M(Q^2) - (l+2) G_E(Q^2) \right] (-1)^{l+1}, 
\hspace{0.8cm}
S_{1/2} (Q^2)= \frac{1}{\sqrt{2}  F_{l +}} \frac{|{\bf q}|} {2 M_R} 
G_C (Q^2) (-1)^{l+1},
\nonumber\\
& &
A_{3/2} (Q^2)= - \frac{C_l}{4 F_{l +}} 
\left[ \frac{1}{l}G_M(Q^2) + G_E(Q^2) \right] (-1)^{l+1}.
\label{eqAmpsM}
\ea

For $J^P= \frac{5}{2}^+, \frac{7}{2}^-, ...$, one has
\ba
& &
G_M (Q^2)= -F_{l -} \left( 
\frac{l}{C_l} A_{3/2} (Q^2) - A_{1/2} (Q^2)  \right)  \frac{2}{l+1} (-1)^{l+1}, 
\label{eqGM2p}\\
& &
G_E (Q^2)= -F_{l -} \left( \frac{l +2}{C_l} A_{3/2} (Q^2) 
+  A_{1/2} (Q^2) \right) \frac{2l}{l+1} (-1)^{l+1}, 
\label{eqGE2p} \\
& &
G_C (Q^2) = 
- \sqrt{2}  F_{l-} \frac{2 M_R}{|{\bf q}|} S_{1/2} (Q^2) (-1)^{l+1}, 
\label{eqGC2p}
\ea 
\ba
& &
A_{1/2} (Q^2) = - \frac{1}{4 F_{l-}} 
\left[G_E (Q^2) - (l+2) G_M(Q^2) \right] (-1)^{l+1}, 
\hspace{0.8cm}
S_{1/2} (Q^2)= 
- \frac{1}{\sqrt{2}  
F_{l -}} \frac{|{\bf q}|} {2 M_R} G_C (Q^2) (-1)^{l+1}, \nonumber \\
& &
A_{3/2}(Q^2) = - 
\frac{C_l}{4 F_{l-}} \left[ \frac{1}{l}G_E(Q^2) 
+ G_M (Q^2)\right] (-1)^{l+1}.
\label{eqAmpsP}
\ea

Again, negative parity 
amplitudes can be obtained from positive parity amplitudes by replacing $G_E \leftrightarrow G_M$,
$G_C \leftrightarrow - G_C$, and $F_{l \pm} \leftrightarrow F_{l \mp}$.
The explicit expressions for $G_M$, $G_E$ and $G_C$ 
in terms of the kinematic-singularity-free form factors $G_i$ 
are presented in Appendix~\ref{appMultipole}.

The relations presented here for the helicity amplitudes,
multipole form factors and the conversion relations 
are also valid for the particular case $l=1$ ($J=\frac{3}{2}$).
The separation in the cases $J=\frac{3}{2}$ and  $J >\frac{3}{2}$
is justified by the need to emphasize the differences 
associated with the use of spinors with $l$ indices (see Appendix~\ref{appSpinJ}).
The more relevant differences come from the 
inclusion of factors depending on $l$ in the 
expressions for the multipole form factors and the phase $(-1)^{l+1}$.

The relations (\ref{eqGM2m})--(\ref{eqGE2m})
and (\ref{eqGM2p})--(\ref{eqGE2p}) for the multipole form factors are in agreement with the 
radiative decay widths from Devenish et al.~\cite{Devenish76,Krivoruchenko02}, 
for a generic resonance of spin $J$, 
and with Eq.~(\ref{eqGammaPDG}).
The results for the helicity amplitudes 
and for the conversion helicity amplitudes/multipole form factors 
differ from Aznauryan-Burkert~\cite{Aznauryan12a} for $J > \frac{3}{2}$ ($l > 1$). 
The differences are in the phase $(-1)^{l+1}$, absent in Aznauryan-Burkert,
and in the factor $C_l$ for $l > 1$. Notice, however, that the radiative decay widths are insensitive 
to the signs of the helicity amplitudes.

From the theoretical point of view, the multipole form factors 
for $J \ge \frac{3}{2}$
are very convenient for the study of the large-$Q^2$ region 
due to the correlations between electric and magnetic transition form factors.
These correlations can be deduced from the helicity amplitudes 
(\ref{eqAmp32p}), (\ref{eqAmp32m}), (\ref{eqAmpsM}) and (\ref{eqAmpsP}), 
and is discussed in the next section.

\subsection{\it Large $Q^2$ \label{sec-largeQ2}}

The behavior of the $\gamma^\ast N \to N^\ast$ helicity amplitudes 
and multipole form factors can be estimated in the
perturbative QCD (pQCD) regime when $Q^2$ is very 
large~\cite{Brodsky73,Brodsky75,Lepage80a,Lepage79b,Carlson86a,Stoler93a}.
These estimates use hadron helicity conservation 
and dimensional counting rules related to the effective 
number of quarks or antiquarks~\cite{Krivoruchenko02,Lepage80a}.
The calculations assume that at very large $Q^2$ 
the photon momentum is redistributed by the three quarks 
(the dominant Fock state) through two gluon exchanges. 
The explicit calculations require the use of distribution 
amplitudes derived from pQCD which include the three quark component 
of the baryon wave functions 
at short distances~\cite{Lepage80a,Lepage79b,Stoler93a}.
Estimates of the helicity amplitudes for the $N(1535)$ and 
$\Delta(1232)$ systems at large $Q^2$~\cite{Carlson98a,Carlson88a,Carlson99a} 
are discussed in Sections~\ref{sec-N1535} and \ref{sec-D1232}, respectively.

\subsubsection{\it Breit frame amplitudes}

The analysis of the asymptotic or pQCD results is simplified
when one uses the Breit frame (BF), 
where ${\bf p}^\prime = - {\bf p} =  \frac{1}{2}{\bf q}_{\ms B}$,
and the subscripts $B$ stands for Breit frame.

Choosing the 3-momentum of the resonance along 
the $\hat z$ axis, 
one can write $p^\prime = (E_R, 0, 0, \sfrac{1}{2}|{\bf q}_{\ms B}|)$,
$p= (E, 0, 0, -\sfrac{1}{2}|{\bf q}_{\ms B}|)$, 
where $E_R = \frac{3 M_R^2 + M^2 + Q^2}{2 P_{\ms 0B}}$, 
$E= \frac{M_R^2 + 3 M^2 + Q^2}{2 P_{\ms 0B}}$,  
$2 P_{\ms 0B}= \sqrt{2 (M_R^2 + M^2) + Q^2}$ 
and $|{\bf q}_{\ms B}| = \frac{\sqrt{Q_+^2 Q_-^2}}{2 P_{\ms 0B}}$. 
The momentum transfer is then 
$q_{\ms B}= (\omega_{\ms B}, 0, 0, |{\bf q}_{\ms B}|)$, 
where $ \omega_{\ms B} = \frac{M_R^2 -M^2}{2 P_{\ms 0B}}$.

The Breit frame is useful to study the large-$Q^2$ limit 
because it maximizes the 3-momentum transfer: for large $Q^2$
the energy component $\omega_{\ms B}$ is suppressed 
($\propto \; 1/Q$), while $|{\bf q}_{\ms B}| \; \propto \; Q$.
For comparison with the rest frame, 
the relation between the magnitude of the 
photon 3-momentum in the two frames is 
$|{\bf q}_{\ms B}|= \frac{M_R}{P_{\ms 0B}} |{\bf q}|$.

In the Breit frame, we can define the three amplitudes~\cite{Carlson98a}
\ba
G_m = \left< N^\ast, \lambda_R = m - \frac{1}{2}\right| \epsilon_\mu^{(m)} J^\mu
\left| N, \lambda_N=  + \frac{1}{2} \right>_{\rm BF} ,
\ea
where $m$ labels the photon polarization ($m=0,\pm$),  
$\lambda_R = S_z^\ast$ and $\lambda_N = -S_z$ represent the 
final and initial helicities, respectively 
(recall that the initial state has momentum $- \frac{1}{2}{\bf q}_{\ms B}$, 
consequently the spin projection in the direction $\hat z$ is $-\lambda_N$).
In comparison to the formulas by C.~Carlson et al.~\cite{Carlson98a}, 
in these definitions we have omitted the factor $\sqrt{\frac{M_R}{M}}$.

The explicit form for the three possible amplitudes
take the form~\cite{Carlson86a,Carlson98a,Carlson98b}
\ba
& & 
G_+ = \left< N^\ast, \lambda_R = + \frac{1}{2}\right| \epsilon_\mu^{(+)} J^\mu
\left| N, \lambda_N= + \frac{1}{2} \right>_{\rm BF}, 
\label{eqGp}
\\
& & 
G_0 = \left< N^\ast, \lambda_R = - \frac{1}{2}\right| \epsilon_\mu^{(0)} J^\mu
\left| N, \lambda_N= + \frac{1}{2} \right>_{\rm BF}, 
\label{eqG0} \\
& & 
G_- = \left< N^\ast, \lambda_R = - \frac{3}{2}\right| \epsilon_\mu^{(-)} J^\mu
\left| N, \lambda_N= + \frac{1}{2} \right>_{\rm BF}.
\label{eqGm}
\ea
Note that the previous relations 
resemble the amplitudes (\ref{eqA12}), (\ref{eqS12}) and (\ref{eqA32})
when we use the spin projections $S_z^\ast$ and $S_z$.
The main difference is that $G_0$ is not modified by the factor $|{\bf q}|/Q$, 
like $S_{1/2}$.

Since the amplitudes at the Breit frame (\ref{eqGp}), (\ref{eqG0})
and (\ref{eqGm}) and the amplitudes at the rest frame 
(\ref{eqA12}), (\ref{eqS12}) and (\ref{eqA32}) 
can both be expressed in terms of the 
kinematic-singularity-free transition form factors  
(Sections~\ref{sec-spin12} and \ref{sec-spin32}),
one can relate the two sets 
for the cases $J^P= \frac{1}{2}^\mp$, $\frac{3}{2}^\pm$,
 $\frac{5}{2}^\mp, ...$, by
\ba
A_{1/2} = \sqrt{\frac{2 \pi \alpha}{K}} G_+, \hspace{.5cm}
S_{1/2}  = \mp  \sqrt{\frac{2 \pi \alpha}{K}}  \frac{|{\bf q}|}{Q}  G_0, 
\hspace{.5cm}
A_{3/2} = \mp \sqrt{\frac{2 \pi \alpha}{K}}  G_-. 
\label{eqAmpLQ2}
\ea 
The amplitudes $A_{3/2}$ and $G_-$ are not defined when $J=\frac{1}{2}$. 
As before, we use the upper symbols for the states
 $\frac{1}{2}^-$, $\frac{3}{2}^+$, $\frac{5}{2}^-, ...$ 
and the lower symbols for the states
$\frac{1}{2}^+$, $\frac{3}{2}^-$, $\frac{5}{2}^+, ...$.
The sign in the amplitudes 
$S_{1/2}$ and $A_{3/2}$ follows from
the properties of the space reflection 
of the states in rest frame and in the Breit frame~\cite{Blin21a}.

For completeness, we mention that some authors~\cite{Stoler93a}
define also helicity amplitudes at the Breit frame 
$\bar A_{1/2}$, $\bar A_{3/2}$ and $\bar S_{1/2}$ 
using
$\tilde A_{|m -1/2|} =  {\cal Z} \, G_m$ ($m=\pm$)
and $\tilde S_{1/2} = {\cal Z} \, G_0$,
where ${\cal Z} =   \sqrt{\frac{2 \pi \alpha}{K}}$.
In this case, we obtain the correspondence
$A_{|m- 1/2|} = \tilde A_{|m - 1/2|}$ ($m=\pm$) and 
$S_{1/2} = \frac{|{\bf q}|}{Q} \,  \tilde S_{1/2}$, 
where $\frac{|{\bf q}|}{Q} = \sqrt{\frac{Q_+^2 Q_-^2}{4 M_R^2 Q^2}}$.

\subsubsection{\it Asymptotic behavior \label{secPQCD}}

The calculation of the leading order dependence on $Q^2$ of the Breit 
frame amplitudes by C.~Carlson et al.~\cite{Carlson86a,Carlson88a,Carlson99a} 
shows that the dominant amplitude at large $Q^2$ is the amplitude $G_+$, 
which preserves the helicity.
For large $Q^2$, one has $G_+ \; \propto \; 1/Q^3$.
The calculations show also that each flip of the helicity of a quark 
introduces a suppression of the amplitudes by a factor $1/Q$.
The expressions for the amplitudes $G_m$ 
for large $Q^2$ in leading order 
can then by written as~\cite{Carlson98a,Carlson98b}
\ba
G_+ \; \propto \; \frac{1}{Q^3}, \hspace{1.3cm}
G_0 \; \propto  \; \frac{1}{Q^4}, \hspace{1.3cm}
G_-  \; \propto \; \frac{1}{Q^5}.
\label{eqGLQ2}
\ea 
Combining the relations (\ref{eqAmpLQ2}) and  (\ref{eqGLQ2}),
we conclude that
\ba
A_{1/2} \; \propto \; \frac{1}{Q^3}, \hspace{1.3cm}
S_{1/2} \; \propto \; \frac{1}{Q^3}, \hspace{1.3cm}
A_{3/2} \; \propto \; \frac{1}{Q^5}, 
\label{eqPQCD}
\ea 
using the relation $\frac{|{\bf q}|}{Q} \; \propto \; Q$ for large $Q^2$.

In the derivation of the relations (\ref{eqGLQ2}) and (\ref{eqPQCD})
only the leading order dependence on $Q^2$ has explicitly been taken into account.
Smother contributions related to  logarithmic corrections
from different sources are omitted in the previous relations.
As a consequence, the relations (\ref{eqGLQ2}) and (\ref{eqPQCD})
are valid only apart log corrections
(modulo logarithms)~\cite{Carlson98a,Carlson98b}.

The logarithmic contributions to the leading order power laws 
(\ref{eqGLQ2}) and (\ref{eqPQCD}) have different sources,
and can be expressed in terms of powers 
of $\log \frac{Q^2}{\Lambda^2}$, where  $\Lambda$ 
is a soft pQCD scale~\cite{Lepage80a,Stoler93a,Belitsky03a}.
The scale $\Lambda$ is a low-energy regulator associated with 
the size of the baryons, and has typical values 
of 0.2--0.3 GeV~\cite{Stoler93a,Belitsky03a,Idilbi04}.
The logarithmic corrections are due to the strong 
running coupling constant $\alpha_s (Q^2)$ 
in the form $\alpha_s^2 \; \propto \; 1/(\log \frac{Q^2}{\Lambda^2})^2$~\cite{Lepage80a,Lepage79b,Carlson86a,Stoler93a,Belitsky03a,Stoler91b},
to soft contributions associated to the 
light cone distribution amplitudes of the baryons~\cite{Lepage80a,Lepage79b,Stoler93a,Carlson87a},
and also due to the regularization 
of the divergent integrals with the introduction 
of a low-energy cutoff~\cite{Belitsky03a,Idilbi04}.
The global logarithmic calculations can be determined 
analytically for the nucleon 
elastic form factors~\cite{Lepage80a,Lepage79b,Belitsky03a},
but are very sensitive to the light-cone 
distribution amplitudes for the transition form factors~\cite{Carlson87a}.
The study of leading order dependence of the 
form factors (\ref{eqGLQ2}) or helicity amplitudes (\ref{eqPQCD}),
can be improved when we scale the data by $\alpha_s^2 (Q^2)$
for a given scale $\Lambda$, or when we consider ratios 
between form factors or helicity amplitudes.
This way, we eliminate the global logarithmic effect 
of the strong coupling constant~\cite{Vereshkov10b,Vereshkov13a,Lepage79b,Belitsky03a,Idilbi04}.
In the following, we assume that apart the global factor $\alpha_s^2$,
the logarithmic corrections are similar for 
all Breit frame amplitudes, $G_+$, $G_0$ and $G_-$, 
and correspond to smooth functions that can be regarded as constants
in a first approximation.

The expressions for the helicity amplitudes (\ref{eqPQCD})
can be transposed for the form factors, modulo logarithm.
In the case $J=\frac{1}{2}$, we can use (\ref{eqF1-b}) and (\ref{eqF2-b}) 
to conclude that 
\ba
F_1 = Q^2 G_1  \; \propto \; \frac{1}{Q^4}, \hspace{1.5cm}
F_2, G_2 \; \propto \; \frac{1}{Q^6}.
\ea 

In the case $J \ge \frac{3}{2}$ the asymptotic behavior 
of the form factors $G_i$ and $h_i$, can be derived from  
the relations between helicity amplitudes and form factors 
(presented in Appendix~\ref{appMultipole}).
The results for $J \ge \frac{3}{2}$ are
\ba
G_1 \; \propto \; \frac{1}{Q^{2l+4}}, \hspace{1.3cm}
G_2 \; \propto \; \frac{1}{Q^{2l+6}},  \hspace{1.3cm}
G_3 \; \propto \; \frac{1}{Q^{2l+6}},  \\
h_1 \; \propto \; \frac{1}{Q^{2l+4}}, \hspace{1.35cm}
h_2 \; \propto \; \frac{1}{Q^{2l+4}}, \hspace{1.35cm}
h_3 \; \propto \; \frac{1}{Q^{2l+2}}, \\
G_E \; \propto \; \frac{1}{Q^{2l+2}}, \hspace{1.2cm}
G_M \; \propto \; \frac{1}{Q^{2l+2}}, \hspace{1.2cm}
G_C \; \propto \; \frac{1}{Q^{2l+4}}. 
\label{eqGEMC-LQ2}
\ea
The details of the calculations are in Appendix~\ref{appLargeQ2}. 
As for the helicity amplitudes, 
the above relations are valid apart logarithmic corrections.

\begin{figure}[t]
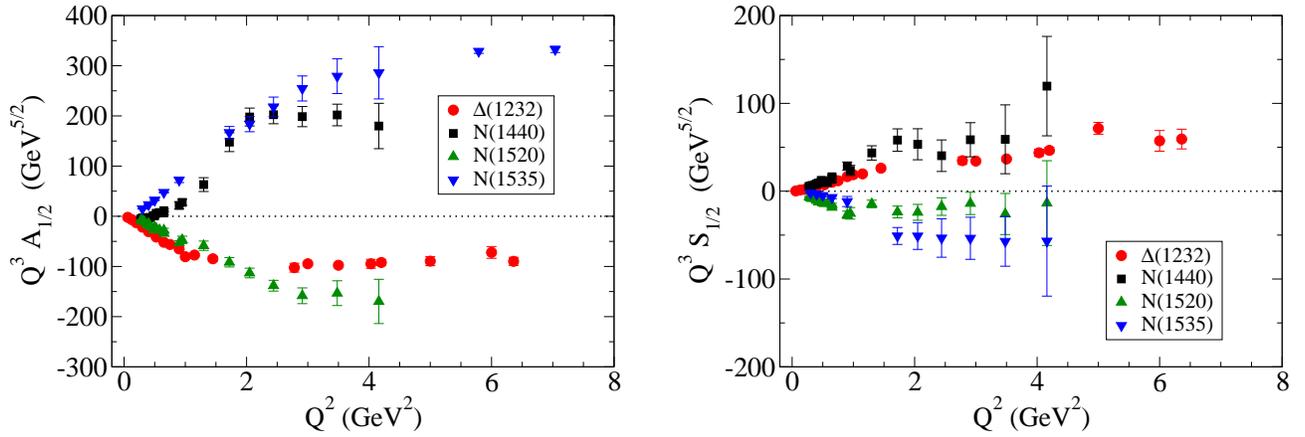

\begin{center}
\mbox{
\includegraphics[width=3.2in]{Q3-A12}} \hspace{.5cm}
\includegraphics[width=3.2in]{Q3-S12} 
\end{center}
\caption{\footnotesize
The experimental helicity amplitudes $A_{1/2}$ and $S_{1/2}$ 
multiplied by $Q^3$ 
for the states $\Delta(1232)$, 
$N(1440)$, $N(1520)$ and $N(1535)$.
Most of the data are from JLab/CLAS~\cite{CLAS09}.
For $N(1440)$ and $N(1520)$, we include the 
results from JLab/CLAS for two pion production~\cite{CLAS12,CLAS16a}.
We include also data from JLab/Hall C  for larger values of $Q^2$
for the $\Delta (1232)$~\cite{Villano09,Frolov98} 
and $N(1535)$~\cite{Dalton09}.
The results, combine statistical and systematic errors. 
\label{fig-Amps-LQ2}}
\end{figure}

The relations (\ref{eqPQCD})  have been tested at the maximum values of 
$Q^2$ available for the resonances 
$\Delta(1232)$, $N(1440)$, $N(1520)$, and $N(1535)$ 
up to $Q^2 \simeq 4$, 6 or 7 GeV$^2$, depending on the case.
In Fig.~\ref{fig-Amps-LQ2}, we present the experimental data 
for the amplitudes $A_{1/2}$ and  $S_{1/2}$ for the mentioned resonances,
multiplied by the factor $Q^3$. 
The graphs show a trend compatible 
to a constant in the range $Q^2=3$--4 GeV$^2$,
as suggested by Eqs.~(\ref{eqPQCD}),
although the error bars are large 
for the amplitude $S_{1/2}$~\cite{Aznauryan12a}.
Future data from JLab-12 GeV or future JLab updates~\cite{Mokeev22a}
may help to confirm this trend.
The large $Q^2$ behavior of the resonances $N(1535)$, 
$\Delta(1232)$ and 
$N(1520)$ are discussed in detail in Sections~\ref{sec-N1535}, \ref{sec-D1232} 
and \ref{sec-N1520}, in that order.

We do not present the test for the amplitude $A_{3/2}$,
expected to scale with $1/Q^5$,
since the results for the $J=\frac{3}{2}$, for $N(1520)$ and $\Delta(1232)$ 
are less conclusive for different reasons.
In the case of the $N(1520)$, the uncertainties are large.
In the case of the $\Delta(1232)$, the present data suggests 
that the convergence may happen only for very large $Q^2$, 
as discussed in Section~\ref{sec-D1232}.

The study of the asymptotic behavior can be very interesting when we use the 
multipole form factors (\ref{eqGEMC-LQ2}).
According to Eqs.~(\ref{eqAmp32p}) and (\ref{eqAmp32m}) 
the results $A_{1/2} \; \propto \; 1/Q^3$ and 
$A_{3/2} \; \propto \; 1/Q^5$ are equivalent to
\ba
G_M  = - G_E + {\cal O} \left(\frac{1}{Q^6} \right)
\hspace{.1cm} \mbox{for} \hspace{.1cm} J^P= \frac{3}{2}^\pm,
\label{eqGEGM-J3/2}
\ea
This large-$Q^2$ relation is valid in particular for the $\Delta(1232)\frac{3}{2}^+$ 
and $N(1520)\frac{3}{2}^-$ states.

The correlations between electric and magnetic 
form factors can also be observed for higher spin resonances 
$J \ge \frac{5}{2}$.
We use again $l = J - \frac{1}{2}$.
The explicit form can be derived from the relations 
(\ref{eqGM2m})--(\ref{eqGC2m}) and (\ref{eqGM2p})--(\ref{eqGC2p})
and the asymptotic forms $A_{1/2} \; \propto \; 1/Q^3$ and
 $A_{3/2} \; \propto \; 1/Q^5$.
One has then for $J^P = \frac{5}{2}^-$, $\frac{7}{2}^+, ...$
\ba 
G_M = - l G_E + {\cal O} \left(\frac{1}{Q^{2l +4}} \right),
\label{eqGM-LL1}
\ea
and for the cases  $J^P = \frac{5}{2}^+$, $\frac{7}{2}^-, ...$
\ba 
G_E = - l G_M + {\cal O} \left(\frac{1}{Q^{2l +4}} \right).
\label{eqGM-LL2}
\ea

It is important to mention, however, that the relations 
(\ref{eqGLQ2}) and (\ref{eqGEMC-LQ2}) 
are derived under very general assumptions.
One important assumption is that the 
kinematic-singularity-free form factors $G_i$ are non zero.
A simple illustration of a deviation from 
the expected leading order behavior can be 
obtained for instance for the $J^P=\frac{3}{2}^\pm$ resonances if $G_1 \equiv 0$.
According to the relations (\ref{eqG1-32}),
if $G_1$ vanish, we would have $G_M \equiv G_E$ for $J^P = \frac{3}{2}^+$,
and $G_M \equiv 0$ for $J^P = \frac{3}{2}^-$.
As a consequence, from Eqs.~(\ref{eqAmp32p}) one obtains the relation 
$A_{3/2} = \mp \sqrt{3} A_{1/2} \; \propto \; 1/Q^5$, 
in contradiction with the  relations (\ref{eqPQCD}) 
for $J^P = \frac{3}{2}^\pm$.
This illustrates that deviations of the general 
power laws in some particular cases can be expected.

The study of the falloff tail of transition form factors 
and of the helicity amplitudes at large $Q^2$ is important because
it enables us to deduce the contribution from higher order 
Fock states, namely the ones associated 
to quark-antiquark excitations.
At very large $Q^2$, one expects the leading order 
form factors $F_1= Q^2 G_1$ for $J=\frac{1}{2}$ 
and $G_1$ for $J=\frac{3}{2}$ to be dominated by terms on $1/Q^{2(n-1)}$,
where $n$ is the number of effective constituents~\cite{Carlson98a,Carlson99a}.
The power of the asymptotic falloff is associated
with the minimum number of gluons exchanged between the constituents.
When the change from $n=3$ (falloff $1/Q^4$) to a system of three quarks and a quark-antiquark pair,
one has 5 constituents ($n=5$), and expect than 
that the asymptotic behavior follows $1/Q^8$.
The price of the change from $(qqq)$ to $(qqq)q \bar q$ 
is then the extra $1/Q^4$ suppression.
The previous estimate can be generalized for next leading order form factors 
($F_2$, $G_2$ and $G_3$) accounting for the faster falloffs 
($1/Q^6$, $1/Q^8$, etc.).
The test of the asymptotic behavior of the amplitudes 
or of the multipole form factors can then be useful 
to establish the range where the valence quark degrees of freedom start to dominate,
since the relations are based on the 
properties of the valence quarks.

Concerning logarithmic contributions to the leading order from Eqs.~(\ref{eqPQCD}),
there are attempts to parametrize these corrections by Vereshkov 
and collaborators~\cite{Vereshkov10a,Vereshkov07a}. 
We notice, however, that although the corrections may be relevant for some amplitudes, 
to find their explicit experimental signals may require much larger values for $Q^2$ 
and broader distributions of data.

\subsection{\it Empirical parametrizations of the helicity amplitudes 
\label{sec-Siegert}}

In theoretical and phenomenological calculations, it is convenient to 
use parametrizations of the data obtained for the helicity amplitudes 
or for the transition form factors ($G_1$, and $G_2$ for $J=\frac{1}{2}$, and 
$G_E$, $G_M$ or $G_C$ for   $J \ge \frac{3}{2}$) based 
on simple analytic functions of $Q^2$.
Examples of successful parametrizations of the data 
are the MAID parametrizations~\cite{Drechsel07,MAID2009,MAID2011}, 
rational parametrizations~\cite{Eichmann18}, 
parametrizations based on the JLab/CLAS data for the resonances $N(1535)$ and $N(1520)$
presented in Refs.~\cite{Aznauryan12a,CLAS09}, 
and more recently a complete set of parametrizations 
of 12 $N^\ast$ states below 1.8 GeV 
associated with states $J=\frac{1}{2}^\pm$,  
$\frac{3}{2}^\pm$ and $\frac{5}{2}^\pm$ based on JLab/CLAS data~\cite{JLab-website}.
A collection of the JLab/CLAS data and relevant data from 
other groups can be found in Ref.~\cite{JLab-database}.

In this section, we discuss the constraints on 
the helicity amplitudes and transition form factors that should 
be taken into account near the pseudothreshold,
where the photon 3-momentum vanishes $|{\bf q}| \to 0$ and  $Q^2 \to - (M_R-M)^2$.
As discussed in Sections~\ref{sec-spin12} and \ref{sec-spin32}, 
the helicity amplitudes and the multipole form factors 
are linear combinations of the unconstrained 
kinematic-singularity-free form factors $G_i$.
Near the pseudothreshold, however, there are constraints 
on the helicity amplitudes and the multipole form factors,
which cannot be ignored when we consider empirical parametrizations 
of these structure functions.
However, some phenomenological parametrizations of the data
ignore those constraints at low $Q^2$, and end up with 
$Q^2$-dependencies incompatible with the pseudothreshold constraints.
The impact of the constraints on phenomenological 
parametrizations of the data may be especially significant 
for light nucleon resonances because the pseudothreshold $Q^2=-(M_R-M)^2$
is very close to the photon point $Q^2=0$.
When the parametrizations are implemented on the kinematic-singularity-free 
form factors $G_i$ or an equivalent  kinematic-singularity-free basis,
the pseudothreshold constraints are automatically 
satisfied~\cite{Eichmann18,LowQ2param}.
An example of such phenomenological parametrization is in Ref.~\cite{Eichmann18}.

The constraints on the helicity amplitudes and 
multipole form factors are a consequence of the structure of the transition current 
and its behavior near the pseudothreshold~\cite{Devenish76,JonesScadron,BjorkenWalecka}.
The gauge-invariant structure of the transition current $J^\mu$, based on
the independent kinematic-singularity-free form factors $G_i$, 
and the kinematics in the $N^\ast$ rest frame  
imply that the helicity amplitudes near the pseudothreshold
must have an explicit dependence on $|{\bf q}|$, and also that some amplitudes 
are correlated~\cite{Buchmann98a,Drechsel92,Amaldi79,DeForest66a,Atti78a}.

The leading order dependence of the helicity amplitudes in terms of $|{\bf q}|$
can be derived from Eqs.~(\ref{eqA12-s12}), (\ref{eqS12-s12}) and 
(\ref{eqB-s12}) for $J^P= \frac{1}{2}^\pm$,  
and Eqs.~(\ref{eqAmps32}), (\ref{eqA1}), (\ref{eqAmps52}) and (\ref{eqAaZ}) 
for $J \ge \frac{3}{2}$, combined with the 
fact that the form factors $G_i$ are free of singularities 
(finite and well defined functions in the limit $|{\bf q}|=0$).
Noticing that for small $|{\bf q}|$, based on Eq.~(\ref{eqB-s12}),
we can write ${\cal B}_- \; \propto \; |{\bf q}|$ and ${\cal B}_+ \; \propto \; 1$,
we conclude that the leading order dependence of the amplitudes 
$A_{1/2}$ and $S_{1/2}$ corresponds to the expressions displayed 
in the first and second columns of Table~\ref{table-Siegert1} 
for $J^P= \frac{1}{2}^\pm$.
From Eqs.~(\ref{eqA1}) and (\ref{eqAaZ}), we can also 
conclude for $l= 1,2,...$ that ${\cal A}_{l-} \; \propto \; |{\bf q}|^l$ 
and  ${\cal A}_{l+} \; \propto \;|{\bf q}|^{l-1}$ for small  $|{\bf q}|$.
Using these relations for ${\cal A}_{l\pm}$,
we conclude that for $J^P = \frac{3}{2}^+$, $\frac{5}{2}^-$, $\frac{7}{2}^+, ...$,
one has
\ba
A_{1/2} \; \propto \; |{\bf q}|^l, \hspace{1.4cm}
A_{3/2} \; \propto \; |{\bf q}|^l, \hspace{1.3cm}
S_{1/2} \; \propto \; |{\bf q}|^{l+1},
\ea
and for $J^P = \frac{3}{2}^-, \frac{5}{2}^+, \frac{7}{2}^-, ...$, we obtain
\ba
A_{1/2} \; \propto \; |{\bf q}|^{l-1}, \hspace{1cm}
A_{3/2} \; \propto \; |{\bf q}|^{l-1}, \hspace{1cm}
S_{1/2} \; \propto \; |{\bf q}|^{l}.
\ea
The leading order dependencies for $J \ge \frac{3}{2}$ 
are also presented in the first and second columns of Table~\ref{table-Siegert1}.

The most well known effect related to the constraints at the 
pseudothreshold is known as Siegert's theorem or 
long wavelength theorem~\cite{Buchmann98a,Drechsel92,Amaldi79,DeForest66a,Atti78a,Siegert37}.
The theorem states that the scalar amplitude $S_{1/2}$ 
and the electric amplitude $E$ (combination of the transverse amplitudes $A_{1/2}$ and $A_{3/2}$) 
are related by $S_{1/2} \; \propto \;  E \; |{\bf q}|$.
A simple consequence of this relation is that the scalar amplitude 
$S_{1/2} \; \propto \; \left< J^0 \right> \; \propto \; |{\bf q}|^n$ ($n \ge 1$),
where $\left< J^0 \right> $ represents the projection of $J^0$ 
between equal spin projections states ($S_z= S_z^\ast =\pm \frac{1}{2}$),
vanishes when $|{\bf q}| \to 0$.
The condition $\left< J^0 \right>=0$, when both states are at rest ($|{\bf q}| =0$),  
is equivalent to the orthogonality of the $N$ and $N^\ast$ states~\cite{Siegert3,Siegert-N1535}.
The explicit relations between the scalar amplitude 
and the electric amplitude, expressed in terms of $A_{1/2}$ and $A_{3/2}$, 
are presented in the third column of Table~\ref{table-Siegert1}.
For the state $J^P = \frac{1}{2}^+$ there is no relation 
associated to Siegert's theorem, because the electric amplitude 
is not defined~\cite{Drechsel07,Devenish76,Krivoruchenko02}.
At this respect, recall the discussion at the end of Section~\ref{sec-spin12}.
In addition to the correlation between electric and scalar amplitudes,
in the cases $J^P= \frac{3}{2}^-, \frac{5}{2}^+, \frac{7}{2}^-, ...$,
there is also a correlation between the transverse amplitudes $A_{1/2}$ and $A_{3/2}$,
equivalent to the condition that the magnetic amplitude vanishes at the 
pseudothreshold.
This condition is included in the last line of 
the third column of Table~\ref{table-Siegert1}, for $J^P= \frac{3}{2}^-$
and for $J^P= \frac{5}{2}^+, \frac{7}{2}^-, ...$

\begin{table}[t]
\begin{center}
{\small  
\begin{tabular}{c | c   c   c  c}
\hline
\hline 
\vspace{.1cm}
$\frac{1}{2}^+$ & $A_{1/2} \; \propto \; |{\bf q}|$, 
&  $S_{1/2} \; \propto \; |{\bf q}|^2 $  &  \\[.1cm]
$\frac{1}{2}^-$ & $A_{1/2} \; \propto \; 1 $, &  $S_{1/2} \; \propto \; |{\bf q}| $  &   
$A_{1/2} = \sqrt{2}(M_R -M) \frac{S_{1/2}}{|{\bf q}|}$  &
$G_E = 2 \frac{M_R -M}{M_R} G_C$
\\[.2cm]
\hline
$\frac{3}{2}^+$ & $A_{1/2} \; \propto \; |{\bf q}|$, 
&  $S_{1/2} \; \propto \; |{\bf q}|^2 $  & \spQ \spQ 
$\left( A_{1/2} - \frac{1}{\sqrt{3}} A_{3/2}\right)\frac{1}{|{\bf q}|}
= \sqrt{2} (M_R -M) \frac{S_{1/2} }{|{\bf q}|^2}$  &
$G_E = \frac{M_R -M}{2M_R} G_C$  \\[.1cm]
                & $A_{3/2} \; \propto \; |{\bf q}|$  &                         &  
\\[.1cm]
$\frac{3}{2}^-$ & $A_{1/2} \; \propto \; 1$, &  $S_{1/2} \; \propto \; |{\bf q}| $  
& 
$A_{1/2} + \sqrt{3} A_{3/2}  = 
 - 2 \sqrt{2} (M_R -M)   \frac{S_{1/2} }{|{\bf q}|} $  & 
$G_E = -\frac{M_R -M}{M_R} G_C$
\\[.1cm]
                & $A_{3/2} \; \propto \; 1$ &                         &   
$ A_{1/2} = \sfrac{1}{\sqrt{3}} A_{3/2} $  &  $G_M \; \propto \; |{\bf q}|^2$\\[.2cm]
 \hline
$\frac{5}{2}^-, \frac{7}{2}^+, ...$ & $A_{1/2} \; \propto \; |{\bf q}|^{l}$, 
&  $S_{1/2} \; \propto \; |{\bf q}|^{l+1} $  & \spQ \spQ 
$\left(A_{1/2} - \frac{l}{C_l} A_{3/2} \right)\frac{1}{|{\bf q}|^l}
=  \frac{l+1}{2} \sqrt{2}  (M_R -M)  \frac{S_{1/2} }{|{\bf q}|^{l+1}}$  &
$G_E = \frac{M_R -M}{2M_R} G_C$  \\[.1cm]
                & $A_{3/2} \; \propto \; |{\bf q}|^{l}$ &                         &  
\\[.2cm]
$\frac{5}{2}^+, \frac{7}{2}^-, ...$
& $A_{1/2} \; \propto \; |{\bf q}|^{l-1}$, 
&  $S_{1/2} \; \propto \; |{\bf q}|^{l}$  
& 
$\left( A_{1/2} + \frac{l+2}{C_l} A_{3/2} \right)  \frac{1}{|{\bf q}|^{l-1}}
=  - \frac{l+1}{l} \sqrt{2}(M_R -M) \frac{S_{1/2} }{|{\bf q}|^l} $  & 
$G_E = -\frac{M_R -M}{M_R} G_C$
\\[.2cm]
                & $A_{3/2} \; \propto \; |{\bf q}|^{l-1}$  &                         &   
$ A_{1/2} = \sfrac{l}{C_l} A_{3/2} $  &  $G_M \; \propto \; |{\bf q}|^2$\\[.1cm]  
\hline
\hline
\end{tabular} }
\end{center}
\caption{\footnotesize 
Constraints on the $\gamma^\ast N \to N^\ast$ helicity amplitudes 
at the pseudothreshold.
The first two columns presents the leading order dependence 
of the transverse and scalar amplitudes.
The third column shows the correlations between amplitudes at the pseudothreshold. 
The fourth column translates the correlations between 
amplitudes to relations between transition form factors.
The relations are similar for the cases $J=\frac{3}{2}^+$ and
$J=\frac{5}{2}^-$,  $\frac{7}{2}^+, ...$.
There is also a similarity between the cases $J=\frac{3}{2}^-$
and $J=\frac{5}{2}^+$,  $\frac{7}{2}^-, ...$.}
\label{table-Siegert1}
\end{table}

The correlations between amplitudes can be converted 
into correlations between multipole form factors,
using the conversion formulas of the previous sections.
The condition associated with Siegert's theorem leads
to relations between the electric and 
Coulomb quadrupole form factors~\cite{Devenish76,JonesScadron,BjorkenWalecka}.
The explicit expressions are displayed in 
the fourth column of Table~\ref{table-Siegert1}.
The condition for $J^P= \frac{1}{2}^-$
is also the consequence of the definitions (\ref{eqFFs12m})~\cite{N1535-TL}.
The relations for $J^P= \frac{3}{2}^+$ are equivalent for 
the states $J^P= \frac{5}{2}^-,\frac{7}{2}^+, ...$.
Similarly, the relations for $J^P= \frac{3}{2}^-$ are equivalent for 
$J^P= \frac{5}{2}^+,\frac{7}{2}^-,...$.
The relation for the magnetic form factor  $G_M \; \propto  \;  |{\bf q}|^2$
for $J^P= \frac{3}{2}^-$
is also valid for the states $J^P= \frac{5}{2}^+,\frac{7}{2}^-,...$.
In this case the relation between the transverse amplitudes
can be written in the general form $A_{1/2} -\sfrac{l}{C_l}A_{3/2} \; \propto  \; |{\bf q}|^{l+1}$.
The relations for the multipole form factors 
can be also derived from expressions for the multipole form factors 
in terms of the form factors $G_i$, presented in Appendix~\ref{appMultipole},
considering expansions in powers of $|{\bf q}|$~\cite{Siegert2}.
In the following, and along the article,
we focus on the parametrizations 
for $J=\frac{1}{2}, \frac{3}{2}$.
In the states with $J > \frac{3}{2}$ the pseudothreshold $Q^2= -(M_R-M)^2$ 
is more distant from the photon point, and 
there is also a lack of data below $Q^2=0.5$ GeV$^2$~\cite{JLab-database}.

A note on the possible analytical form
of the helicity amplitudes is in order.
Parameterizations based on smooth analytical forms, like rational functions,
may not be appropriate for the transverse amplitudes 
of the excitations of $N(1440)\frac{1}{2}^+$ and 
$\Delta(1232) \frac{3}{2}^+$, even when compatible 
with the pseudothreshold constraints~\cite{LowQ2param}.
This happens because of the proportionality to $|{\bf q}|$,
due to the difficulty of approximating the factor $\sqrt{Q_-^2}$
by a rational function, 
for light resonances (small $M_R-M$).
Recall that near the pseudothreshold $|{\bf q}| \; \propto \; \sqrt{Q_-^2}$.
This effect is illustrated in Section~\ref{sec-Delta-lQ2} 
for the $\Delta(1232)$ helicity amplitudes.

As mentioned, the constraints described above are not always 
considered in parametrizations of the data,  
that erroneously considered the amplitudes 
as independent functions in the whole region of $Q^2$.
Even the parametrizations developed by 
the MAID group~\cite{Drechsel07,MAID2009,MAID2011}, 
that take into account some features 
of the helicity amplitudes~\cite{Drechsel99a,Tiator07,Tiator16},
with particular emphasis on the relation 
$S_{1/2} \; \propto \; E |{\bf q}|$ (Siegert's theorem), 
fail in obtaining parametrizations fully consistent 
with the pseudothreshold constraints, and some of the relations displayed 
in columns one, two, three and  four of Table~\ref{table-Siegert1} are violated.
The MAID parametrizations are in general
compared with the MAID data analysis from different groups,
as described in Refs.~\cite{Drechsel07,MAID2009,MAID2011}.
We refer to this MAID data analysis, 
collected in the website~\cite{MAID-database}, as MAID data.
Some of these issues are discussed by Tiator in Ref.~\cite{Tiator16} within 
the context of the MAID parametrizations.

\begin{figure}[t]
\begin{center}
\epsfig{file=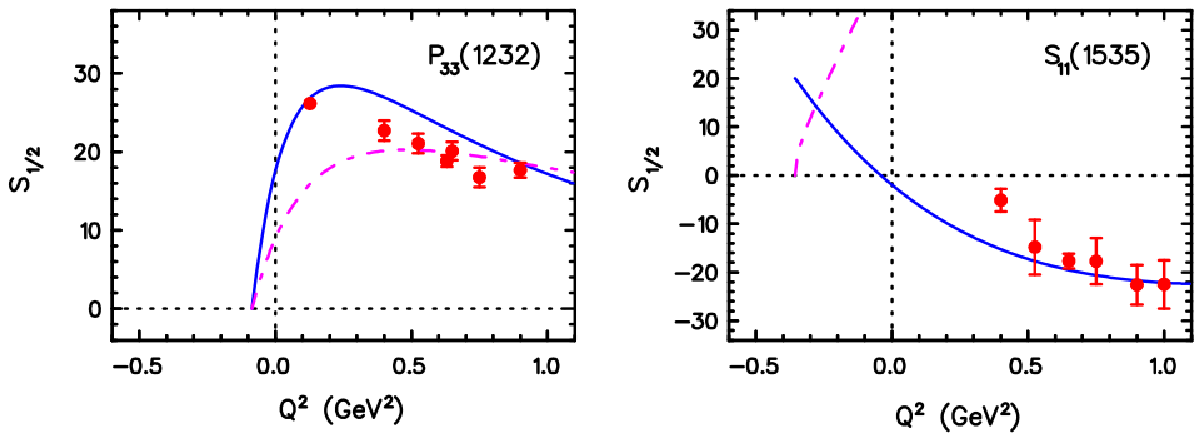,scale=1.4}
\mbox{
\includegraphics[width=3.in]{D1232-Siegert2} \hspace{.75cm}
\includegraphics[width=3.in]{N1535-Siegert2} }
\end{center}
\caption{\footnotesize 
Test of consistency of amplitudes with Siegert's theorem 
for $\Delta(1232)$ (at the left) and $N(1535)$ (at the right).
The amplitudes $S_{1/2}$ (solid lines) are compared with $\lambda_R \, E \, |{\bf q}|$ (dot-dashed lines),
where $\lambda_R = 1/(\sqrt{2}(M_R-M))$ for both cases 
(see Table~\ref{table-Siegert1}).
The electric amplitude is defined by  
$E= A_{1/2} - A_{3/2}/\sqrt{3}$ for the $\Delta(1232)$
and $E= A_{1/2}$ the for $N(1535)$. 
Helicity amplitudes in 10$^{-3}$ GeV$^{-1/2}$.
{\bf Upper panel:}
Results of the MAID2007 parametrization~\cite{Tiator16}.
Data from MAID~\cite{Drechsel07,MAID2009,MAID2011,MAID-database}.
{\bf Lower panel:} 
On the left: calculations of 
a quark model with pion cloud dressing 
for the $\Delta(1232)$~\cite{Siegert3}.
Data from Refs.~\cite{JLab-database,Siegert3}. 
On the right: MAID type parametrization 
compatible with Siegert's theorem~\cite{Siegert-N1535}.
Data from MAID~\cite{Drechsel07,MAID2009,MAID2011,MAID-database}.
The figure of the upper panel is a courtesy of Lothar Tiator. \\
Reprinted with permission from
\href{https://link.springer.com/article/10.1007/s00601-016-1158-1}{L.~Tiator,
  Few Body Syst. 57, 1087 (2016).}
Copyright (2016) by Springer. 
\label{fig-Tiator1}}
\end{figure}

The results of the MAID parametrization for the amplitude $S_{1/2}$ 
for the states $\Delta(1232)$ and  $N(1535)$ are presented 
in the upper panel of Fig.~\ref{fig-Tiator1} (solid lines).
The results are compared with the estimate given by $E\, |{\bf q}|$
(dot-dashed lines)
according with the results from Table~\ref{table-Siegert1}.
At the first glance, the figure suggests that 
Siegert's theorem is verified in the case of the $\Delta(1232)$,
since the two lines converge at the pseudothreshold,
and that there is a clear violation of the theorem for $N(1535)$, 
since the parametrization fails to reproduce $S_{1/2} \; \propto \; |{\bf q}|$.
In the case of the $\Delta(1232)$, a closer look
reveals that the two calculations have different derivatives.  
This means that, although the amplitudes $S_{1/2}$ and 
$E= A_{1/2} - A_{3/2}/\sqrt{3}$ have the correct shape at 
pseudothreshold ($S_{1/2} \; \propto \; |{\bf q}|^2 $ and $E \; \propto \; |{\bf q}|$)
the relation between the amplitudes is not consistent with 
the condition 
$E/|{\bf q}| = \sqrt{2}(M_R -M) S_{1/2}/|{\bf q}|^2$,
displayed on Table~\ref{table-Siegert1}.
Notice the importance of including the factors $1/|{\bf q}|$ and $1/|{\bf q}|^2$.
The condition for the form factors (fourth column) is equivalent 
to the condition for the amplitudes, only when the amplitudes scale 
with the correct power of $|{\bf q}|$~\cite{Siegert-N1535,Siegert2}.

In the lower panel of Fig.~\ref{fig-Tiator1}, 
we present calculations of the 
$\Delta(1232)$ and $N(1535)$ amplitudes compatible with Siegert's theorem.
In the case of the $\Delta(1232)$ (left panel), we include calculations 
of a quark model which take into account 
pion cloud contributions for the form factors $G_E$ and $G_C$
valid in the low-$Q^2$ region~\cite{Siegert3,LatticeD,NDeltaD}.
The results are consistent with the relation $G_E= \frac{M_R-M}{2 M_R} G_C$
per construction~\cite{Siegert3}, and satisfy consequently 
the equivalent relation for the amplitudes. 
The results for the $\Delta(1232)$ at low $Q^2$ are discussed in more detail 
in Section~\ref{sec-D1232}.
The results for the $N(1535)$ presented on the right panel 
are obtained using a parametrization based on a MAID form 
that take into account the expected shape for $S_{1/2}$ and 
the constraint $A_{1/2} = \sqrt{2}(M_R -M) S_{1/2}/|{\bf q}|$
associated with Siegert's theorem    
(see Table~\ref{table-Siegert1}).  
The free parameters are fixed by the MAID data~\cite{Siegert-N1535}.
The new parametrization reveals a turning point in the function $S_{1/2}$ 
near the pseudothreshold~\cite{Siegert-N1535}, 
as a consequence of the positive values for the function $E \equiv A_{1/2}$ 
and Siegert's theorem condition. 
The properties of the $N(1535)$ helicity amplitudes at low $Q^2$ 
are discussed in detail in Section~\ref{sec-N1535}.
The impact of the pseudothreshold constraints 
on other resonances are discussed in Section~\ref{secData}.

The discussion of this section answers the question about how to 
test the consistence of the parametrizations with the pseudothreshold constraints.
Another possible question is, given a set of parametrizations of the 
helicity amplitudes for a $N^\ast$ resonance in a certain range of $Q^2$, 
but incompatible with the pseudothreshold constraints,
if it possible to make it compatible below a given value of $Q^2$.
This subject was discussed in Ref.~\cite{LowQ2param},
where a method is proposed to extrapolate
parametrizations defined for $Q^2 \ge 0$, below a point $Q_P^2$ 
to the pseudothreshold, using smooth analytic extensions 
of the original parametrizations, while ensuring them to be consistent 
with the pseudothreshold conditions.
The method was applied to the JLab parametrization~\cite{JLab-website,Blin19a},
but it can be applied in any analytic parametrizations of the nucleon resonance helicity amplitudes.
Examples of the method are discussed in Sections~\ref{sec-N1535} 
and \ref{sec-D1232} for the cases of the $N(1535)$ and $\Delta(1232)$.
The state $N(1535)$ turns out to be a special case 
where low-$Q^2$ data are crucial to 
establish the sign of the amplitudes $A_{1/2}$ and $S_{1/2}$ near $Q^2=0$,
as discussed on Section~\ref{sec-N1535-lQ2}.


\begin{table}[h]
\begin{center}
\begin{tabular}{c c  c  c c }
\hline
\hline 
Reaction & $N^\ast$ &  $Q^2$ (GeV$^2$) & Functions &  Lab/Experiment\\
\hline
\hline 
$e p \to e  p \eta$ & $N(1535)\frac{1}{2}^-$ & 2.4, 3.6 &  $A_{1/2}$ &  JLab/Hall C~\cite{Armstrong99} (1999) \\
$e p \to e p \eta$ & $N(1535)\frac{1}{2}^-$ & 1.8--4.0 &  $A_{1/2}$ &   JLab/CLAS~\cite{Thompson01} (2001) \\
$e p \to e p \eta$ & $N(1535)\frac{1}{2}^-$ & 0.17--3.1 &  $A_{1/2}$ &  JLab/CLAS~\cite{Denizli07} (2007)  \\
$e p \to e p \eta$ & $N(1535)\frac{1}{2}^-$ & 5.7, 7.0 &  $A_{1/2}$ &   JLab/Hall C~\cite{Dalton09} (2009)  \\
$e p \to e p \eta$ & $N(1535)\frac{1}{2}^-$ & 0.38, 0.75 &  $A_{1/2}$, $S_{1/2}$ &  JLab/CLAS~\cite{Aznauryan05b} (2005)  \\[.2cm]
$e p \to e N \pi$ & $N(1440)\frac{1}{2}^+$  & 0.40, 0.65 &  $A_{1/2}$, $S_{1/2}$ &  JLab/CLAS~\cite{Aznauryan05b} (2005)   \\
                  & $N(1520)\frac{3}{2}^-$  & 0.40, 0.65 &  $A_{1/2}$, $A_{3/2}$, $S_{1/2}$ &  \\    
                  & $N(1535)\frac{1}{2}^-$  & 0.40, 0.65 &  $A_{1/2}$, $S_{1/2}$ &  \\     [.2cm]         
$e p \to e N \pi$ & $\Delta(1232)\frac{3}{2}^+$ & 0.3--6.0 & $A_{1/2}$, $A_{3/2}$, $S_{1/2}$ & JLab/CLAS~\cite{CLAS09} (2009)  \\
                &  $N(1440)\frac{1}{2}^+$     & 0.30--4.16 & $A_{1/2}$, $S_{1/2}$ & \\
                &  $N(1520)\frac{3}{2}^-$     & 0.30--4.16 & $A_{1/2}$, $A_{3/2}$, $S_{1/2}$ & \\
                &  $N(1535)\frac{1}{2}^-$     & 0.30--4.16 & $A_{1/2}$, $S_{1/2}$ & \\  [.2cm]
$e p \to e n \pi^+$ & $N(1675)\frac{5}{2}^-$ & 1.8--4.0 &  $A_{1/2}$, $A_{3/2}$, $S_{1/2}$ &  JLab/CLAS~\cite{CLAS15} (2015) \\
                    & $N(1680)\frac{5}{2}^+$ & 1.8--4.0 &  $A_{1/2}$, $A_{3/2}$, $S_{1/2}$ &  \\  
                    & $N(1710)\frac{1}{2}^+$ & 1.8--4.0 &  $A_{1/2}$, $S_{1/2}$ &  \\  [.2cm]
$e p \to e p \pi^0$ & $N(1440)\frac{1}{2}^+$  &  0.1     &  $S_{1/2}$ &  MAMI-A1~\cite{Stajner17} (2017) \\ [.2cm]
$e p \to e p \pi^+ \pi^-$ &  $N^\ast$  & 0.65 & $A_{1/2}$, $S_{1/2}$ & JLab/CLAS~\cite{Aznauryan05a} (2005)  \\
                         &  $N^\ast$  & 0.65 & $A_{1/2}$, $A_{3/2}$, $S_{1/2}$ & \\[.2cm]
$e p \to e p \pi^+ \pi^-$ & $N(1440)\frac{1}{2}^+$     & 0.28--0.58 & $A_{1/2}$, $S_{1/2}$ & JLab/CLAS~\cite{CLAS12} (2012) \\
                          & $N(1520)\frac{3}{2}^-$     &  0.28--0.58 & $A_{1/2}$, $S_{1/2}$ & \\[.2cm] 
$e p \to e p \pi^+ \pi^-$ & $N(1650)\frac{1}{2}^-$      & 0.65, 0.95, 1.30 & $A_{1/2}$, $S_{1/2}$ & JLab/CLAS~\cite{Mokeev14a} (2014) \\
                          & $N(1680)\frac{5}{2}^+$      & 0.65, 0.95, 1.30 & $A_{1/2}$,  $A_{3/2}$, $S_{1/2}$ & \\
                          & $\Delta(1700)\frac{3}{2}^-$      & 0.65, 0.95, 1.30 & $A_{1/2}$,  $A_{3/2}$, $S_{1/2}$ & \\[.2cm]
$e p \to e p \pi^+ \pi^-$  & $N(1440)\frac{1}{2}^+$     & 0.65, 0.95, 1.30 & $A_{1/2}$, $S_{1/2}$ & JLab/CLAS~\cite{CLAS16a} (2016)\\
                          & $N(1520)\frac{3}{2}^-$      &  0.65, 0.95, 1.30 & $A_{1/2}$, $A_{3/2}$, $S_{1/2}$ & \\
                          & $\Delta(1620)\frac{1}{2}^-$  & 0.65, 0.95, 1.30 & $A_{1/2}$, $S_{1/2}$ & \\
$e p \to e p \pi^+ \pi^-$  & $N(1720)\frac{3}{2}^+$    & 0.65, 0.95, 1.30 & $A_{1/2}$, $S_{1/2}$ & JLab/CLAS~\cite{Mokeev20a} (2020) \\
\hline
\hline
\end{tabular}
\end{center}
\caption{\footnotesize $N^\ast$ data.
The label $N^\ast$ in  JLab/CLAS~\cite{Aznauryan05a} (2005) 
corresponds to $N^\ast =$ $N(1440)\frac{1}{2}^+$,  $\Delta(1620)\frac{1}{2}^-$, $N(1650)\frac{1}{2}^-$ 
or $N^\ast =$  $N(1520)\frac{3}{2}^-$, $N(1680)\frac{5}{2}^+$, $\Delta(1700)\frac{3}{2}^-$, $N(1720)\frac{3}{2}^+$.
\label{table-Data1} }
\end{table}

\begin{table}[t]
\begin{center}
\begin{tabular}{c   c  c c }
\hline
\hline 
Reaction & $Q^2$ (GeV$^2$) & Functions &  Lab/Experiment\\
\hline
\hline
$e p \to e p \pi^0, e n \pi^+$  &   0.0     &   $R_{EM}$ & MAMI~\cite{Beck00a} (2000) \\
$e p \to e p \pi^0, e n \pi^+$  &   0.0     &   $R_{EM}$ & LEGS~\cite{LEGS97,LEGS01} (2001) \\
$e p \to e p \pi^0$ &         0.127           &    $R_{SM}$ & ELSA~\cite{ELSA97} (1997) $\dagger$ \\
$e p \to e p \pi^0$ &  2.8, 4.0  & $G_M$, $R_{EM}$, $R_{SM}$  & JLab/Hall C~\cite{Frolov98} (1999) \\
$e p \to e p \pi^0$ &         0.121         &   $R_{SM}$ & MAMI~\cite{Pospischil01} (2001) $\ddagger$\\
$e p \to e p \pi^0$ &  0.4--1.45, 1.8  & $R_{EM}$, $R_{SM}$  & JLab/CLAS~\cite{CLAS02} (2002) \\
$e p \to e p \pi^0$ &         0.126, 127       &  $R_{EM}$, $R_{SM}$ & MIT-Bates~\cite{MIT-Bates01,MIT-Bates03} (2003) $\ddagger$\\
$e p \to e p \pi^0$ &   0.127 &   $G_M$, $R_{EM}$, $R_{SM}$ & MIT-Bates~\cite{Sparveris05a} (2005) $\ddagger$\\
$e p \to e p \pi^0$ &   0.06, 0.200  & $G_M$, $R_{EM}$, $R_{SM}$ & 
MAMI-A1~\cite{Stave06,Sparveris07,Stave08} (2008)  $\ddagger$\\
$e p \to e p \pi^0$ &    1.0     &  $R_{EM}$, $R_{SM}$      &  JLab/Hall A~\cite{Kelly05b,Kelly07a} (2007) \\
$e p \to e p \pi^0$ &      0.127   &     $R_{SM}$  &    MAMI-A1~\cite{Sparveris13} (2013) \\ 
$e p \to e p \pi^0$ &  0.09, 0.13  &  $R_{EM}$     &  JLab/Hall A~\cite{Blomberg16} (2016) \\
                    &  0.04, 0.09, 0.13  &  $R_{SM}$  & JLab/Hall A~\cite{Blomberg16} (2016)    \\
$e p \to e p \pi^0$ &  3.0--6.0  &   $G_M$, $R_{EM}$, $R_{SM}$  &  JLab/CLAS~\cite{CLAS06b} (2006)  \\  
$e p \to e p \pi^0$ &  0.16--0.28  & $R_{EM}$, $R_{SM}$  & JLab/CLAS~\cite{CLAS09} (2009)$\star$ \\  
                    &  0.30--1.45, 3.0--6.0  & $G_M$, $R_{EM}$, $R_{SM}$  &   JLab/CLAS~\cite{CLAS09} (2009)$\star$  \\  
$e p \to e p \pi^0$ &  6.4       &  $G_M$, $R_{EM}$, $R_{SM}$  & JLab/Hall C~\cite{Villano09} (2009) \\
                    &  7.7       &      $G_M$               & JLab/Hall C~\cite{Villano09} (2009) \\
\hline
\hline
\end{tabular}
\end{center}
\caption{\footnotesize $\Delta(1232)$ single pion production.
For $Q^2=0$ we take as reference the average from PDG~\cite{PDG2022}.
$\dagger$ Early measurement, very large magnitude 
(excluded in most comparisons). 
$\ddagger$ Problems in the analysis, much larger than 
more recent analysis~\cite{Blomberg16}.
\mbox{$\star$ Include} analysis of 
previous CLAS experiments~\cite{CLAS02,CLAS06b}.
\label{table-Data2}}
\end{table}

\renewcommand{\theequation}{4.\arabic{equation}}
\setcounter{equation}{0}

\renewcommand{\thefigure}{4.\arabic{figure}}
\setcounter{figure}{0}

\section{Experimental facilities and methods of analysis of the data  
\label{sec-Expermental}}

In this section we discuss the experimental facilities 
and methods used in the analysis of the data.
We start with the list of current experimental facilities dedicated to 
study of the electromagnetic structure of the nucleon resonances 
in the spacelike region, and with large contribution
to the knowledge of the properties of the $N^\ast$ resonances 
and $\gamma^\ast N \to N^\ast$ helicity amplitudes.
Next, we discuss the methods used in the analysis of the data 
for the determination of the properties of the baryon resonances.
The properties, discussed here include pole positions, resonance masses, 
decay widths, and branching ratios.
To close the section we discuss the class of models 
for the meson-baryon interaction known as dynamical reaction models,
which are used in the analysis of the meson electroproduction data.

\subsection{\it Experimental facilities}

The Thomas Jefferson National Accelerator Facility (JLab for short), 
located in Newport News (USA),
and known before 1996 as the
Continuous Electron Beam Accelerator Facility (CEBAF),
operates a continuous wave electron accelerator in the luminosity frontier,
that already for about two and a half decades
produces the highest intensity
electron beam~\cite{Mokeev22a,Aznauryan12a,Burkert04,NSTAR,Arrington22a}. 
Till 2014 the maximum energy attained by the electron beam was 6 GeV.
In 2014 the energy was extended to 10.5 GeV. 
The extension to a beam of 12 GeV started in 2018.
The first results from the JLab-12 GeV upgrade were published
in 2022~\cite{Mokeev22a,Carman23a}.
Under study is the extension of the JLab-12 GeV program to JLab-22 GeV~\cite{Carman23a,Arrington22a}.

In parallel to the $\gamma^\ast N \to N^\ast$ program, we notice that JLab conducted as well a
variety of important experiments, as the precise measurements of the proton form factor
at low $Q^2$, of neutron radii in nuclei, and of quark distribution functions
from $^3$He and $^3$H targets.
For a complete review on past achievements
and future possibilities enabled by large luminosity and acceptance power
consult Ref.~\cite{Arrington22a}.
At JLab the beam produced in the main accelerator is distributed by
four main experimental Halls 
where the collision with the targets and the interaction with a multitude
of detectors happens.
The first Halls are labeled as Hall A, B and C. Hall D was built on 2014.
The properties of the different Halls are described below. 

Other  laboratories that are also mentioned here are ELSA and MAMI (Germany) 
and MIT-Bates (USA), which produced relevant studies of nucleon resonances 
at low energy and low $Q^2$.

\subsubsection{\it JLab - CLAS (Hall B)}

The Hall B of JLab houses the 
CEBAF Large Acceptance Spectrometer (CLAS) detector, and was key for the development of 
the nucleon resonance program focused on 
the measurement of transition form factors 
in the first three energy resonance regions~\cite{Aznauryan12a,NSTAR}.

This detector can operate with electron beams 
or with energy tagged photon beams for meson electro- and photo-production,
and was designed specifically for the detection 
of multiparticle final states~\cite{Burkert04}.
The protons and pions can be separated according to their momentum 
within certain limits.
The identification of  particles over a wide range 
of resonance energies ($W$) allows the study of 
a complete set of reactions important to the $N^\ast$ program.
JLab/CLAS contributed with about 95\% of the data 
associated with different $N^\ast$ resonances 
as can be inferred from the analysis 
of Tables~\ref{table-Data1} and \ref{table-Data2}.
JLab/CLAS, now JLab/CLAS 12 is the most complete and advanced facility
in the field~\cite{Burkert04}.
A small caveat of the CLAS detector is the impossibility 
or measuring observables below $Q^2=0.3$ GeV$^2$. 

The results from CLAS are dominated by the analysis 
of the $\gamma^\ast N \to \pi N$ reaction.
More recently, since 2014, the $\pi \pi N$ channel 
was included in the analysis and used in the study of 
resonances from the second and third resonance region~\cite{CLAS12,CLAS16a}.

\subsubsection{\it JLab - Hall A}     

The JLab/Hall A includes a high resolution spectrometer 
which can be used to measure the reaction 
$\vec{e}p\to e \vec{p} \pi^0$ in the $\Delta(1232)$ region~\cite{Burkert04}.
Of particular relevance are the measurements of the 
$\gamma^\ast N \to \Delta(1232)$ quadrupole form factor ratios 
at momentum transfer as low as $Q^2=$0.04, 0.09 GeV$^2$~\cite{Blomberg16}.

\subsubsection{\it JLab - Hall C}      

The JLab/Hall C has a high momentum spectrometer.
One of the applications of this detector was the measurement of  
$\gamma^\ast N \to \Delta(1232)$  and $\gamma^\ast N \to N(1535)$ transitions  
at large $Q^2$, up to 5 or 8 GeV$^2$ depending on the resonance.
At such large values of $Q^2$ it is not possible 
to separate the effects of the different amplitudes 
but it is possible to extract the leading order amplitude 
or form factor~\cite{Dalton09,Villano09}.

\subsubsection{\it ELSA, MAMI and MIT-Bates}

Examples of electron accelerators different than CEBAF are MAMI (Mainz, Germany), ELSA (Bonn, Germany) 
and MIT-Bates (USA)~\cite{Burkert04}.

The MIT-Bates is a linear accelerator which has been used for the study
of the $\Delta^+(1232) \to \pi^0 p$ decay.
The facility is no longer in use~\cite{Aznauryan12a} 
but was decisive in the study of the $\gamma^\ast N \to \Delta(1232)$ 
quadrupole form factors at very low $Q^2$~\cite{Stave06,Bernstein07a}.

MAMI at Mainz is a microtron accelerator which has been used 
to probe the structure of the $\Delta(1232)$ and $N(1440)$ resonances also 
at very low $Q^2$ (near 0.1 GeV$^2$)~\cite{Stajner17,Stave06,Stave08}.

ELSA at Bonn is a electron synchrotron that has studied meson electroproduction.
It includes the Elan apparatus which has been used 
in the measurements of the $\Delta^+(1232) \to \pi^0 p$ 
and  $\Delta^+(1232) \to \pi^+ n$ decays~\cite{Burkert04,ELSA97}.

\subsubsection{\it LEGS (Brookhaven National Laboratory)}

Of interest are also LEGS experiments at  Brookhaven National Laboratory (USA) 
in the $\Delta(1232)$ region at the photon point ($\gamma N \to \Delta(1232)$ transition)
which provided one of the first accurate measurement of the ratio $R_{EM} (0)$.

For a more detailed discussion of these and other experimental facilities 
check Ref.~\cite{Burkert04}.
The facilities mentioned above were focused on the study 
of $\gamma^\ast N \to N^\ast$ transitions, with emphasis 
to the reactions $\gamma^\ast N \to \pi N$,  $\eta N$, $\pi \pi N$.

\subsection{\it Methods of analysis of the data \label{sec-analysis}}

We discuss now methods for the analyses of the different transitions 
$\gamma^\ast N \to N^\ast \to M B$, where 
$M$ and $B$ are the final meson and baryon states, respectively.
Originally the methods have been developed for the $\pi N$ final state,
but can be generalized for other combinations, such as $\eta N$, 
$\rho N$, $\omega N$, etc.~\cite{Aznauryan12a,Burkert04}.
The combined analysis of different channels is important 
to determine the properties of the states which couple weakly 
to $\pi N$~\cite{Ronchen13a}.
When we increase the energy $W$ of the $\gamma^\ast N$ system
we open the possibility of creation of particles 
with strange quarks, starting with the channels 
$K \Lambda$, $K \Sigma$, $\phi N$, etc.~\cite{Mokeev22a}.
The generalization of the method for three-body final states
(two mesons and a baryon) is straightforward~\cite{Burkert04}.
The analysis from JLab/CLAS of the $\gamma^\ast N \to \pi \pi N$ transitions 
is an example of the generalization of the method~\cite{CLAS12,CLAS16a}.
At the present the analyses of MAID, ANL-Osaka
and J\"ulich-Bonn groups
include states with three-body final states.
In ANL-Osaka, ANL stands for Argonne National Laboratory.

The analysis of  nucleon resonances was initiated
by $\pi N$ scattering studies of the 
$\pi N \to \pi N$ transition alone, and most of the methods 
of analysis were based on resonance decays 
into the $\pi N$ and $\gamma N$ channels.
This is why till very recently the $N^\ast$ states were labeled by
the properties of the $\pi N$ decay channel using the 
old spectroscopic notation described in Section~\ref{secWF}.
The complexity of the analysis of the data beyond the 
$\pi N$ channel comes from the $\Delta(1232)$ resonance 
not being the only resonance to be taken into account, as
Fig.~\ref{fig-resonance-regions} illustrates.
Although for $W < 1.4$ GeV the $\Delta(1232)$ resonance is clearly isolated 
in energy, spin and parity, 
and the interference of the resonance with the background is then residual,
the resonances from the second and third resonance region 
are very close to each other and effects from states with similar
mass have to be disentangled~\cite{Aznauryan12a,Burkert04,Thiel22a}.
Going beyond the $\pi N$ channel made then necessary the simultaneous
analysis of different resonance production reactions, 
and instead of the old spectroscopic notation, the good quantum numbers $(I,J^P)$
are used to classify the resonant states
$N(J^P)$ and $\Delta(J^P)$~\cite{PDG2022}. Theoretically, these good quantum numbers
of the three quark states come from the invariance of the QCD Lagrangian
under the Poincar\'e group which consists of translations,
rotations, boosts and parity.
Nevertheless, since each $N(J^P)$ and $\Delta(J^P)$ state,
can be decomposed into multipole amplitudes for a reliable identification of the properties 
of the different  $N(J^P)$ and $\Delta(J^P)$ states,
it is necessary to make an analysis based on a partial wave decomposition, 
which extracts the partial wave amplitudes (PWA)  
associated to the possible meson-baryon channels~\cite{Burkert04,Thiel22a,Cutkosky79a}.

In addition, a significant information about the nucleon resonances comes 
also from the analysis of meson production on nucleons induced by photons.
Meson photoproduction gives information beyond the 
$\pi N \to \pi N $ scattering, in particular on $N^\ast$ states
that couple weakly to the
$\pi N$ channel.
A high quality data set is already available
(for a review check Ref.~\cite{Ireland20a}).
Recently a large variety of  $\gamma N \to M B$ reactions,
has been investigated for increasing values of $W$~\cite{Thiel22a}.
The analysis of those experiments includes not only
specific masses of the nucleon resonances but also 
photo and meson decay widths.
Although the experiments do not probe the nucleon resonances 
for non zero photon virtualities, the PWA 
of those experiments provide constraints on the position 
of the resonance poles which can then be analyzed in more detail 
in electroproduction experiments.

The study of the $\gamma^\ast N \to M B$ tied to $M B \to M' B'$ transitions,
where $M'$ and $B'$ are generic meson and baryon states,
can be done in general through the scattering $T$-matrix formalism.
The $T$-matrix framework is derived from the scattering matrix ($S$-matrix) 
under the requirements of Unitary, Analyticity and Crossing Symmetry 
(symmetry between $s$- and $u$-channel)~\cite{Burkert04,Thiel22a,Drechsel07}.
For a detailed discussion of the $S$- and $T$-matrix formalism 
we recommend Ref.~\cite{Thiel22a}.
The $T$-matrix elements are projected into the possible $\gamma^\ast N$, 
$MB$ and $M'B'$ initial and final states, generating a system of coupled-channel states.
The explicit calculation of the $\left< M' B' \right| T \left| M B \right>$
and $\left<  M B  \right| T \left| \gamma^\ast N \right>$ matrix elements 
can be done directly with the resolution of coupled-channel integral equations.
In alternative, there are methods based on the $K$-matrix, discussed below,
which avoid the use of integral equations.

The $T$-matrix formalism can be implemented using  either
Hamiltonian or Lagrangian
formulations~\cite{Aznauryan12a,Burkert04,Pascalutsa07,Matsuyama07a,Kamano13a,Capstick08a},
and data analysis models discussed below apply effective Lagrangians
to describe the meson-baryon interactions.
The interactions between baryons and mesons, 
defined by a Lagrangian at tree-level, 
are used to calculate contributions from the background 
and explicit contributions from the resonances $N^\ast$ 
in terms of the asymptotic states $MB$.
Details can be found in Section~\ref{secDM}.
The model analysis provides a partial wave decomposition of the data with defined total angular
momentum and parity ($J^P$).
The resonant states  $N(J^P)$ or $\Delta(J^P)$ 
are identified by the poles of the $T$-matrix elements
in the complex plan of the variable $W$.
These poles determine the resonance mass $M_R$ and total decay width $\Gamma_R$
in the formula $W = M_R - \frac{i}{2} \Gamma_R$.

Furthermore the PWA of a given $N(J^P)$ or $\Delta(J^P)$ state generates
an explicit multipole decomposition, and thus identifies 
transverse amplitudes $M_{\ell \pm} (W,Q^2)$ and $E_{\ell \pm}(W,Q^2)$,
which can be associated to magnetic and electric amplitudes, 
and scalar amplitude $S_{\ell \pm} (W,Q^2)$, 
which are correlated to the longitudinal amplitude $L_{\ell \pm} (W,Q^2)$ 
through gauge invariance\footnote{The current conservation condition 
implies that $|{\bf q}|L_{\ell \pm} (W,Q^2) = \omega S_{\ell \pm} (W,Q^2)$, 
using the $\gamma^\ast N$ rest frame representation for $q$: 
$q=(\omega, 0, 0, |{\bf q}|)$.}. 
The labels $\ell \pm$ are related with the spin $J$ of the final state,
according to $J= \ell \pm \frac{1}{2}$. 
The multipoles $M_{\ell \pm} (W,Q^2)$, $E_{\ell \pm}(W,Q^2)$ and  $S_{\ell \pm} (W,Q^2)$
are complex functions and are traditionally defined at the resonance rest frame 
where $P_R= (W, 0,0,0)$.  
The resonance rest frame is defined by the center of mass frame of the $MB$ system.
The transition form factors are determined by the imaginary part 
of the amplitudes 
$M_{\ell \pm} (W,Q^2)$, $E_{\ell \pm}(W,Q^2)$ and  $S_{\ell \pm} (W,Q^2)$
at the resonant mass $W=M_R$~\cite{Burkert04,Drechsel07}.  
In more simplified models the poles of the multipole amplitudes are introduced
apriori through a Breit-Wigner form,
but they can also be generated naturally by amplitudes determined 
by some dynamical reaction model, as discussed in the next sections.

The first analyses of the photo- and electro-production of pions on nucleon 
were based on dispersion 
relations~\cite{Aznauryan12a,Burkert04,Amaldi79,Cutkosky79a,Aznauryan03b,Aznauryan03a,Chew57a,Hanstein97a}.
More recently, the analyses of the data have been done 
preferentially with $K$-matrix
methods~\cite{Drechsel99a,Davidson91a,Shklyar07a},  
and dynamical reaction models.
The $K$-matrix formalism relation  
with the  $T$-matrix: $T=K (1 -i K)^{-1}$, where $K$ is a hermitian operator 
($K= K^\dagger$). The matrix elements of $K$ are necessarily real
in order to preserve the unitary of the $T$-matrix.
In the practical applications, however, additional restrictions 
are imposed to the $K$-matrix in order to simplify the calculations
and preserve the necessary properties of the $K$- and $T$-matrices.

In a particular model known as
the Unitary Isobar Model (UIM)~\cite{Aznauryan12a,Burkert04,Drechsel99a},
the background is unitarized for each multipole amplitude 
using the $K$-matrix representation, with
the resonance contributions  written 
in Breit-Wigner energy dependent forms.
The SAID, MAID, and JLab/CLAS groups have applied different forms of UIMs .

Representative groups that developed for the
analysis of the reactions $\gamma N \to MB$,
$\gamma^\ast N \to M B$ and $M B \to M' B'$ are discussed next.

\subsubsection{\it SAID analysis \label{secSAID}}

SAID (Scattering Analysis Interactive
Dial-in)~\cite{SAID-website,Arndt02,Kamalov02,Arndt06a,Anisovich16a,Pavan02,Workman12b} 
is the name 
of the analysis performed at George Washington University (GWU) group (USA).
It is characterized as  a $K$-matrix approach multi-channel analysis
and is focused on the pion nucleon scattering and 
pion photoproduction data~\cite{Aznauryan12a,Burkert04,Thiel22a}.
It uses polynomial parametrizations of the $K$-matrix elements with
masses, widths and hadronic couplings
from pion induced production reactions $\pi N$ and $\eta N$~\cite{Arndt02,Anisovich16a,Workman12b}.
Photocouplings are obtained from  photoproduction reactions.

Although the SAID analyses are not applied directly to the $\gamma^\ast N \to N^\ast$ transitions,
their results for  $\gamma N \to \pi N$ and $\pi N \to \pi N$ 
are used as reference of several groups, 
including the MAID, ANL-Osaka, J\"ulich-Bonn and Kent University State analysis~\cite{Thiel22a},
discussed in the following sections.

\subsubsection{\it MAID analysis \label{secMAID}}

MAID (MAinz unitary Isobar moDel) is the analysis developed 
by the Mainz group (Germany)~\cite{Drechsel07,MAID2009,MAID2011,Kamalov02,MAID-website}.
As the name suggests it is based on an UIM 
as defined in Ref.~\cite{Drechsel99a}.

The MAID analysis is applied to the 
$\gamma^\ast N \to \pi N$ and $\pi N \to \pi N$ reactions.
Thus, the MAID model is not a coupled-channel analysis 
in the sense that does not go above the $\pi N$ channel.
The inclusion of other channels has been considered 
in specific applications of MAID, 
including the Eta-MAID, Kaon-MAID and 
EtaPrime-MAID~\cite{MAID-website,Tiator18a}.

The $T$-matrix for the $\gamma^\ast N \to \pi N$ transition 
takes the form~\cite{Pascalutsa07,Drechsel07,MAID2009,MAID2011}
\ba
t_{\ms \gamma N, \pi N} (W) = 
v_{\ms \gamma N, \pi N} (W) + v_{\ms \gamma N, \pi N} (W) g_0(W) t_{\ms \pi N, \pi N} (W),
\label{eqMAID1}
\ea
where $t_{\ms \pi N, \pi N}$ is the $\pi N$ scattering matrix 
($\pi N \to \pi N$ transition),
$v_{\ms \gamma N, \pi N}$ is the $\gamma^\ast N \to \pi N$ transition potential,  
and $g_0$ is the free $\pi N$ propagator, 
as defined in the Lippmann-Schwinger formalism.
The integration in the intermediate momenta 
is implicit in Eq.~(\ref{eqMAID1}).
In the present notation $\gamma N$ stands generically for   
the $\gamma^\ast N$ channel for both real and virtual photons.

The MAID analysis decomposes the interaction term
into background (bg) and resonant (R) contributions 
\ba
v_{\ms \gamma N, \pi N} (W) = 
v^{\rm bg}_{\ms \gamma N, \pi N} (W) + v^{\rm R}_{\ms \gamma N, \pi N} (W).
\label{eqMAID2}
\ea
The potential $v^{\rm bg}_{\ms \gamma N, \pi N}$ is described by the Born terms 
including a mixture pseudoscalar-pseudovector $\pi NN$ coupling 
and $t$-channel vector meson exchanges. 
The term $v^{\rm R}_{\ms \gamma N, \pi N} $ includes 
the bare contributions of the resonance.

From Eqs.~(\ref{eqMAID1}) and (\ref{eqMAID2}), 
one concludes that~\cite{Drechsel07,Kamalov01a}
\ba
t_{\ms \gamma N, \pi N} (W) = 
t^{\rm bg}_{\ms \gamma N, \pi N} (W)  + t^{\rm R}_{\ms \gamma N, \pi N} (W),
\label{eqMAID3}
\ea 
where
\ba
&& 
t^{\rm bg}_{\ms \gamma N, \pi N} (W) =
v^{\rm bg}_{\ms \gamma N, \pi N} (W) + 
v^{\rm bg}_{\ms \gamma N, \pi N} (W) g_0 (W) t_{\ms \pi N, \pi N} (W),
\label{eqMAID4} \\
&& 
t^{\rm R}_{\ms \gamma N, \pi N} (W) =
v^{\rm R}_{\ms \gamma N, \pi N} (W) + 
v^{\rm R}_{\ms \gamma N, \pi N} (W) g_0 (W) t_{\ms \pi N, \pi N} (W).
\label{eqMAID5} 
\ea
In this representation $t^{\rm bg}$ and $t^{\rm R}$ are both unitary.
From Eqs.~(\ref{eqMAID4}) and (\ref{eqMAID5}), we conclude that 
the background and resonant contributions can be determined 
once the $\pi N$ scattering matrix $t_{\ms \pi N, \pi N}$ is known.
Notice also that $t^{\rm bg}_{\ms \gamma N, \pi N}$ 
comprises contributions from both the background 
and the resonances since the amplitude $t_{\ms \pi N, \pi N}$ 
includes all the mechanisms~\cite{Aznauryan12a,Drechsel07}.

The previous relations define what is known as the
Dubna-Mainz-Taipei (DMT) model~\cite{Kamalov01a,Kamalov99a}.         
The MAID analysis differs from the DMT model 
in the amplitude $t^{\rm bg}_{\ms \gamma N, \pi N}$ defined by Eq.~(\ref{eqMAID4}).
Instead of considering the explicit integration, 
the determination of the background contributions 
as in the original UIM~\cite{MAID2011,Drechsel99a}, replaces the explicit integral in Eq.~(\ref{eqMAID4}) 
by the term of the elastic cut only, neglecting the contribution from the
principal part of the integration~\cite{Burkert04,Drechsel07}.
This procedure gives for the MAID contribution of the
background~\cite{Drechsel07,Drechsel99a}   
\ba
t^{{\rm bg}, \alpha}_{\ms \gamma N, \pi N} (W,Q^2)=  
v^{{\rm bg}, \alpha}_{\ms \gamma N, \pi N} (W,Q^2) 
\left[ 1 + i  t^{\alpha}_{\ms \pi N, \pi N}  (W) \right],
\label{eqMAID6}
\ea
where the amplitude  $t^{\alpha}_{\ms \pi N, \pi N}$
is the on-shell pion-nucleon elastic amplitude, 
as determined by SAID in terms of phase-shifts and 
inelasticity parameters~\cite{Thiel22a,Drechsel07}, and $\alpha= (\xi,\ell, j,I)$ specifies
the orbital angular momentum ($\ell$), 
the total angular momentum ($j$), the isospin ($I$), 
and the multipole label $\xi =M,E,S$.
The dependence on $Q^2$ is included explicitly.

As for the resonant term, MAID considers
an energy dependent Breit-Wigner type amplitude parametrization
\ba
t^{{\rm R}, \alpha}_{\ms \gamma N, \pi N} (W,Q^2)=  
{\cal A}^{{\rm R}}_\alpha (W,Q^2) f_{\ms \gamma N}(W)
\frac{\Gamma_{\rm tot} (W) M_R}{M_R^2 - W^2 - i M_R \Gamma_{\rm tot}(W)} 
f_{\ms \pi N}(W) e^{i \phi_R},
\label{eqMAID7}
\ea
where ${\cal A}^{{\rm R}}_\alpha (W,Q^2)$ are 
resonance R electromagnetic couplings, $\alpha$ labels the multipoles ($M,E,S$),
$M_R$ and $\Gamma_{\rm tot} (W)$ are the mass 
and total decay width of the resonance respectively,
$f_{\ms \gamma N}$ and $f_{\ms \pi N}$ are known parametrizations 
associated to the radiative decay and $\pi N$ decay 
in terms of the energy $W$, and $\phi_R$ 
is a phase necessary for the unitarization 
of the amplitudes~\cite{Thiel22a,Drechsel07,Tiator18a}.
Its explicit form can be found in 
Refs.~\cite{Aznauryan12a,Drechsel07,MAID2011}.
The contribution of all intermediate resonance
is summed over to make the total resonant contribution.

For the determination of the $\gamma^\ast N \to N^\ast$ 
transition form factors it is important to discuss 
the form of the amplitudes ${\cal A}^{{\rm R}}_\alpha (W,Q^2)$.
The MAID parametrization of the amplitudes 
${\cal A}^{{\rm R}}_\alpha (W,Q^2)$ 
takes the general form
\ba
{\cal A}^{{\rm R}}_\alpha (W,Q^2)
= {\cal A}^{{\rm R}}_\alpha (W,0)
\left( 1 + a_1 Q^2 + a_2 Q^4 + ... \right) \exp(-b_1 Q^2), 
\label{eqMAID8}
\ea
where $a_i$ ($i=1,2,...$) and $b_1$ are adjustable parameters.
For most resonances, MAID considers ${\cal A}^{{\rm R}}_\alpha (W,0)$
determined by the value of $W$ at the pole~\cite{Thiel22a}.
This combination of polynomials and exponential functions 
provides a simple and economical way (due to the reduced number of parameters)
to describe functions of $Q^2$, which may have complicated shapes.
In the case of the $\Delta(1232)$ resonance,  
some multiplicative functions are also considered, 
including the dipole function $G_D= 1/\left( 1 + Q^2/\Lambda_D^2\right)^2$,
where $\Lambda_D^2=0.71$ GeV$^2$, 
and a factor $|{\bf q}|/K$, where 
$|{\bf q}|$ is the magnitude of the photon 
3-momentum and $K= \frac{W^2- M^2}{2 W}$~\cite{Drechsel07}.
The inclusion of these factors is important to ensure consistency 
with the constraints at the pseudothreshold, particularly for the case of 
the $\Delta(1232)$, as discussed in Section~\ref{sec-Siegert}.
The inclusion of similar factors for other resonances is under 
consideration~\cite{LowQ2param,Siegert-N1535,Siegert2,Tiator16}.

The MAID parametrizations (\ref{eqMAID8}) deserve some discussion.
Extrapolations for $Q^2 <0$ should be taken with caution
due to the possibility of significant exponential enhancement.
It is assumed that the parametrizations
are limited to a certain range of $Q^2$ determined
by the available data.
The upper range can then be 4.5 or 8 GeV$^2$
depending on the resonance.
One notice, however, that the
expected asymptotic large-$Q^2$ dependence (see Section~\ref{sec-largeQ2})
are not considered in the MAID parametrizations.
As an example, notice that the form factors $G_M$ and $G_E$ 
are in general regulated by different values of $b_1$, thus
implying a dominance of one over 
the another for very large $Q^2$.
In any case, the 
asymptotic behavior of the MAID parametrization 
can be redefined above a certain limit if necessary.
A discussion of the modification of the 
MAID exponential shape of amplitudes for large $Q^2$ can be found in 
Ref.~\cite{Siegert2}.

\subsubsection{\it JLab/CLAS analysis \label{secJLabY}}

The method used by the JLab/CLAS group, 
sometimes mentioned as the JLab/Yeveran model~\cite{Burkert04}
is a model that combines fixed-$t$ dispersion relations
with a UIM in the form proposed by Drechsel et al.~\cite{Drechsel99a}.
The UIM method is extended with the incorporation of Regge poles 
for the large-$W$ region~\cite{Aznauryan12a,Aznauryan03b}.
The approach is described in detail in Refs.~\cite{Aznauryan05b,Aznauryan05a,Aznauryan03b}.

The comparison between the two separate
approaches, dispersion relations and extended UIM, allowed to study the sensitivity
of the analysis of  JLab/CLAS data on the transition amplitudes of the
different low-lying energy resonances covering differential cross sections,
longitudinally polarized beam asymmetries, and longitudinal target and beam-target asymmetries
and established the model uncertainty in the final data~\cite{CLAS09}. 
The method has been used in the analysis of most of the JLab/CLAS 
publications of $\gamma^\ast N \to N^\ast$  transition amplitudes,
in particular in the Refs.~\cite{CLAS09,Aznauryan05b,CLAS15}.
Other results based on the UIM are published in Refs.~\cite{CLAS02,CLAS06b,Aznauryan03b}.

\subsubsection{\it JLab-Moscow analysis}

Another tool for the analysis of the high precision JLab/CLAS
$\gamma^\ast N \to \pi \pi N$ data 
is the phenomenological 
Jefferson Laboratory-Moscow State University (JLab-Moscow) model~\cite{Mokeev09a}.
It includes contributions from all known resonances with
$\pi \Delta$ and $\rho p$ decay channels, needed for the analysis of 
$\gamma^\ast p \to \pi^+ \pi^- p$ reactions, and it applies an unitarized Breit-Wigner
ansatz for the resonant amplitudes consistent with a general unitarity
constraint~\cite{CLAS12,CLAS16a,Mokeev14a}.

With these ingredients it recently allowed the analysis of $\pi \pi  p $ electroproduction
and photoproduction cross sections measured by CLAS for
$W=1.6$--1.8 GeV, and photon virtualities $Q^2 < 1.5$ GeV$^2$~\cite{Mokeev20a}.
The analysis covered the
resonances $N(1440)$, $N(1520)$ and $\Delta(1620)$.
Although the first results based on the data from JLab-6 GeV are restricted to 
a range $Q^2=0.5$--1.5 GeV$^2$,
preliminary analysis of JLab-12 GeV data indicates 
that the method can be extended to the range $Q^2=2$--5 GeV$^2$~\cite{Mokeev22a}.

\subsubsection{\it J\"ulich-Bonn  analysis}

The J\"ulich-Bonn coupled-channel 
model~\cite{Ronchen13a,Ronchen18a,JulichBonn21a,JulichBonn22a,JBW-website,Doring09a}
was developed for the analysis of the $\pi N$ scattering
and $\gamma N \to \pi N$ reactions.
The analysis includes the channels $\pi N$, $\eta N$,   $K \Lambda$ and  $K \Sigma$, and 
in addition to channels $\pi \Delta$, $\sigma N$ and $\rho N$
to simulate the $\pi \pi N$ channel. 
The J\"ulich-Bonn is also a dynamical model.
Some of the properties of the model are discussed
in Section~\ref{secDM3}.

\subsubsection{\it Other analyses}

Besides the analyses described above, there are other groups that gave important contributions to 
the spectroscopy of nucleon excitations and light baryon spectroscopy in general. Examples are the groups of
Giessen, Kent University State and  Bonn-Gatchina.
Detailed discussions about these and others groups work can be found in
Refs.~\cite{Aznauryan12a,Burkert04,Thiel22a,Pascalutsa07}.

The Giessen group~\cite{Shklyar07a,Penner02a,Shklyar16a} 
developed a coupled-channel model based on the $K$-matrix formalism, 
where the matrix $K$ is determined by an interaction kernel $V$.
The interaction kernel is decomposed into background and resonance terms 
evaluated at the tree level~\cite{Burkert04,Penner02a}.
The model has been used mainly for the analysis 
of $\pi N$ and $\gamma N$ reactions.

The Kent University State model~\cite{KSU12a,KSU19b}
is based on the $S$-matrix framework, 
where the resonant part of the $T$-matrix 
has an energy dependent Breit-Wigner form~\cite{Burkert04,Thiel22a,KSU19b}.
The model {\G has been} used in the study of the
$\gamma N \to \pi N$ and 
$\pi N \to \pi N$, $\eta N$, $K \Lambda$, $\pi \pi N$.
More recently the analysis was extended to $\gamma N \to \eta N$ 
and $\gamma p \to K^+ \Lambda$ reactions~\cite{Thiel22a,KSU19b}.

The Bonn-Gatchina 
model~\cite{Anisovich17a,Anisovich16a,Sarantsev08a,Bonn-Gatchina-website,Anisovich05a,Anisovich06a,Anisovich12a}
provides a partial wave analysis of many reactions within a coupled-channel formalism.
The model uses the $K$-matrix formalism 
for a set of channels which include the $\gamma N$ channel,
in addition to the  $\pi N$, $\eta N$, $\pi \Delta$ and $K \Lambda$ channels.
Three-body states $\pi \pi N$, $\pi \eta N$ are also included. 
The $K$-matrix is decomposed into background and resonant (or pole) terms 
with a given analytic form~\cite{Anisovich16a}.

\subsection{{\it Dynamical Models} \label{secDM}}

Dynamical coupled-channel (DCC) reaction models, or dynamical models for short, 
are a family of models used for the analysis of  photo- and
electro-production of
mesons on nucleons where the resonance propagators and decay vertices
are computationally tied to meson-baryon interactions and consistently
calculated from them.  Compared to more phenomenological analyses, as the ones
discussed in the previous section, no special form for the resonance
widths is assumed and no simplifications to the full non perturbative
meson-baryon rescattering series are done.

Examples of dynamical coupled-channel models
applied to the $\gamma^\ast N \to N^\ast$ transitions
are the Sato-Lee model~\cite{SatoLee96,SatoLee01},
the DMT (Dubna-Mainz-Taipei)
model~\cite{Drechsel07,Kamalov01a,Kamalov99a,Chen07a},
the EBAC model~\cite{Matsuyama07a,EBAC-website,JDiaz07b,Kamano09b},
the ANL-Osaka model~\cite{Kamano16a,Kamano13a,Nakamura15a},
the Utrecht-Ohio model~\cite{Pascalutsa07a,Caia04a,Caia05a}, 
the J\"ulich-Bonn model~\cite{Ronchen13a,Ronchen18a,Doring09a} and the
J\"ulich-Bonn-Washington model~\cite{JulichBonn21a,JulichBonn22a,JBW-website}.

DCC reaction models are not formulated at the level of quark degrees of freedom,
but rather at the level of mesons and baryon degrees of freedom.
The underlying assumptions of dynamical models are then
i) internal quark degrees of freedom are manifest
in meson-baryon scattering processes through effective meson-baryon interaction vertices~\cite{SatoLee01,JDiaz07b,JDiaz08a},
valid in the energy domain of their description of the meson photo-
and electro-production data; ii)  these interaction vertices dress the
three-quark baryon propagators (referred to as bare propagators) and the 
bare meson-baryon-resonance vertices for meson-baryon and photo-baryon channels.
The two effects (re-scattering and dressing) are coupled together
through integral equations as explained below.

In the absence of fully consistent microscopic
calculations based on QCD
for the meson-baryon scattering states, this is the best approach possible.

Meson-baryon scattering is determined explicitly 
by the solution of integral equations for the  transition $T$-matrix
of coupled multiple meson-baryon scattering
states~\cite{Burkert04,Aznauryan12a,Pascalutsa07}.
Dynamical models use a Bethe-Salpeter formulation to take 
into account off-shell rescattering effects at the level of meson-baryon systems and
obtain non perturbative solutions of the meson-baryon scattering
coupled-channel system~\cite{Aznauryan12a,JDiaz07a}. The calculations apply some prescription 
for the intermediate energy of the meson-baryon propagators
changing the 4D Bethe-Salpeter formulation to a 3D Lippmann-Schwinger
formulation or similar~\cite{Pascalutsa07,SatoLee01} which is
defined by time-ordered perturbation theory and automatically
generates unitarity constraints.

To illustrate the methodology of dynamical reaction models 
we may start by looking at the equations of the DMT model 
(\ref{eqMAID3})--(\ref{eqMAID5})~\cite{Kamalov01a,Kamalov99a} introduced  in~\ref{secMAID}.
These equations determine the $\gamma^\ast N \to \pi N$ $T$-matrix 
in terms of the  $\pi N \to \pi N$ $T$-matrix. In turn this one is determined by   
the solution of the integral equation
\ba
t_{\ms \pi N, \pi N} (W) = 
v_{\ms \pi N, \pi N} (W) + v_{\ms \pi N, \pi N} (W) g_0(W) t_{\ms \pi N, \pi N} (W),
\label{eqMAID9}
\ea
which takes into account the $\pi N \to \pi N$  scattering interaction kernel $v_{\ms \pi N, \pi N}$ in all orders.
The DMT model~\cite{Kamalov01a,Kamalov99a} is restricted to the $\gamma N$
and $\pi N$ channels,
but the procedure can be generalized for $M B \to M' B'$ reactions,
as discussed below. From the knowledge of the $M B \to M' B'$ $t$-matrices,
and generalizing Eqs.~(\ref{eqMAID3})--(\ref{eqMAID5}) and (\ref{eqMAID9})
we can determine the $\gamma^\ast N \to M B$  $t$-matrices 
for a given set of meson-baryon states $MB$~\cite{Kamalov01a,Kamalov99a}.

\subsubsection{{\it General formalism} \label{secDM2}}

Dynamical models can be formulated 
using an Hamiltonian approach to multichannel 
reactions~\cite{Burkert04,Ronchen13a,JDiaz07b,JDiaz07a,Suzuki10a,JDiaz09a}.
The starting point is the Hamiltonian~\cite{Burkert04,SatoLee96,SatoLee01}
\ba
H = H_0 + V, 
\hspace{1.5cm}
V= v^{\rm bg} + v^R,
\ea
where $H_0$ is the free meson-baryon Hamiltonian
and $V$ is the interaction term (kernel).
The Hamiltonian formalism requires 
assumptions about bare baryon masses  seen as quasi-particles, and bare
meson-baryon-resonance ($M B \to B^\ast$)
and photon-baryon-resonance ($\gamma^\ast B \to B^\ast$)
vertices, where $B$ and $B^\ast$ represent baryon resonances~\cite{Burkert04,Burkert19a}.
These properties are not observed and cannot be tested by experiment, since at the end
one can only measure the properties of the physical baryon 
modified by the meson-baryon interactions~\cite{Burkert04,Burkert19a,JDiaz07b}.
Dynamical models consider these bare properties together with effective
meson-baryon interactions into the interaction term $V$.
Therefore, the interaction kernel contains  Born terms for the
background meson-baryon potential $v^{\rm bg}$ added to
the resonance potential term $v^R$ associated with meson-baryon-resonance
vertex ($M B \to B^\ast$).

The background term corresponds to tree level $t$- and $u$-channel
exchange diagrams
and contact terms for the meson-baryon interaction; the resonant term
comprises bare $s$-channel diagrams.
In the case of the $\gamma N \to \pi N$ transition, 
the Born terms are the diagrams shown in Fig.~\ref{fig-Total}.
Namely, processes with intermediate mesons ($\pi$, $\rho$, $\omega$, $\sigma$, etc.)
as diagram (c) in Fig.~\ref{fig-Total}
are part of the background term.
Intermediate resonant states $B_i^\ast$ which decay 
into $M B$ and $\gamma B$ states as diagram (d)  in Fig.~\ref{fig-Total} are part of
resonant term $v^R$ which is written 
as sums of the product of vertices $\Gamma_{\ms B_i^\ast, MB}$ 
and $\Gamma_{\ms B_i^\ast , \gamma B}$ of
the resonance meson and electromagnetic decays. 
The complete set of interactions includes also the $\gamma B$ channel.

It can be shown that the meson-baryon $T$-matrix obtained from the integral
equation with the full kernel $V=v^{\rm bg} +v^{\rm R}$ can be decomposed as~\cite{Burkert04,JDiaz07b}
\ba
T_{\ms M B,M' B'} (E)= t_{\ms MB, M'B'}^{\rm bg} (E) + t_{\ms M B, M'B'}^R (E),  
\label{eqT}
\ea
where $ t_{\ms MB,M'B'}^{\rm bg}$, $t_{\ms MB,M'B'}^R$ are usually referred as
background (or non resonant) and resonant contributions;
and $E$ is the energy in the $MB$ center of mass frame.
The 3-momenta of initial and final states are omitted for simplicity.

The non resonant term is a solution 
of the integral equation~\cite{Burkert04,Matsuyama07a} 
\ba
t_{\ms MB,M'B'}^{\rm bg} (E) =
v_{\ms MB,M'B'}^{\rm bg} + 
\sum_{\ms M_1 B_1} v_{\ms MB,M_1B_1}^{\rm bg} \;
g_{\ms M_1B_1}(E) \, t^{bg}_{\ms M_1B_1,M'B'} (E),
\label{eqT1} 
\ea  
where 
\ba
g_{\ms M_1 B_1}(E)= \frac{1}{E - H_0 - \Sigma_{\ms M_1 B_1}(E) + i \epsilon}
\label{eqGMB}
\ea
is the dressed propagator of the meson-baryon state $M_1 B_1$ which
includes the self-energy $\Sigma_{\ms M_1 B_1}$ where the 
meson emission and absorption vertices
$\Gamma_{\ms B_i^\ast, M_1B_1}$ originated by $v^{\rm R}$ enter.
The self-energy have real and imaginary components and is written down below;
its real part gives a mass difference due to the co-existence of a resonance with a meson
in flight, its imaginary part indicates that the state has a finite lifetime.
The kernel $v_{\ms MB,M'B'}^{\rm bg}$ does not carry any energy dependence because
at tree level the kernel is independent of the energy $E$.
We use this notation throughout  this section.
Notice that Eq.~(\ref{eqT1}) is not
a single equation but it is in fact a set of coupled integral equations for all possible
$MB \to M'B'$ transitions,
with $t_{\ms MB,M'B'}^{\rm bg}$, appearing on the l.h.s.~and r.h.s.
The set correlates dynamically all $MB$ channels, which  justifies 
its classification as a dynamical coupled-channel reaction model.

The resonant amplitude can be written in the form
\ba
t_{\ms MB,M'B'}^{R}(E) 
= \sum_{\ms B_i^\ast B_j^\ast} 
\bar \Gamma^\dagger_{\ms B_i^\ast, MB} (E) [G (E)]_{i,j} 
\bar \Gamma_{\ms B_j^\ast,M'B'} (E),
\label{eqTBG}
\ea
where $\bar \Gamma^\dagger_{\ms B_i^\ast, MB}$ is the dressed 
vertex associated to the $B_i^\ast \to MB$ decay,
and $[G(E)]_{i,j}$ is the $B_i^\ast$ dressed propagator
\ba
[G^{-1} (E)]_{i,j} =
(E -M_{\ms B_i^\ast}^0)  \delta_{i,j}- \Sigma_{i,j}(E),
\label{eqGbare}
\ea
where $M_{\ms B_i^\ast}^0$ is the bare mass of the resonance 
and the self-energy $\Sigma_{i,j}$ is determined by
\ba
\Sigma_{i,j} (E) = \sum_{\ms MB} \Gamma_{\ms B_i^\ast,MB}^\dagger \; g_{\ms MB}(E) 
\bar  \Gamma_{\ms B_j^\ast,MB}(E). 
\label{eqSigma}
\ea 
The last equation includes the bare vertex $\Gamma_{\ms B_i^\ast MB}$
and the dressed vertex $\bar \Gamma_{\ms B_i^\ast MB}$
The dressed vertex can be calculated from the bare vertex 
using 
\ba
\bar  \Gamma_{\ms B^\ast,MB}(E)
=  \Gamma_{\ms B^\ast,MB}  + \sum_{\ms M_1 B_1}  \Gamma_{\ms B^\ast,M_1 B_1} \;   
g_{\ms M_1 B_1}(E) \, t_{\ms M_1 B_1,MB}^{\rm bg} (E).
\label{eqGammaD}
\ea
This calculation can only be performed after the 
determination of $t_{\ms M_1 B_1,MB}^{\rm bg}$ using Eqs.~(\ref{eqT1}).
There is no dependence of the  bare vertices $\Gamma_{\ms B^\ast,MB}$
in the energy because the energy of $B^\ast$  
is determined by its mass $M_{B^\ast}$.
In  Eqs.~(\ref{eqT1}), (\ref{eqTBG}), (\ref{eqSigma}) and (\ref{eqGammaD})   
the integration on the intermediate momenta
of the $MB$ states is implicit.

Relations (\ref{eqT})--(\ref{eqGammaD}) define the basis of DCC models.
The $T$-matrix is decomposed into the background 
and resonant contributions.
The resonant component (\ref{eqTBG}) is defined in terms of the dressed vertex 
$\bar \Gamma_{\ms B^\ast, MB}$, which include 
meson decays and electromagnetic decays ($MB$ and $\gamma B$ channels).
The background term is determined by the interaction $v^{\rm bg}_{\ms MB, M'B'}$
associated with the bare $MB$ systems, and it is the solution
of the integral equation (\ref{eqT1}), which is also used to
calculate the dressed vertex (\ref{eqGammaD}).
The parametrizations of $v^{\rm bg}_{\ms MB, M'B'}$,  
$\Gamma_{\ms B^\ast, MB}$ and $M_{\ms B_i^\ast}^0$
are the starting point of the calculations.
The functions define the bare couplings and
the mass provides the estimate of the bare resonance mass.

We turn now to the study of the electromagnetic structure of the baryon states.
The equations associated to the electroproduction (channel $\gamma B$)
can be obtained from Eq.~(\ref{eqT1}) 
\ba
t_{\ms \gamma B,M'B'}^{\rm bg} (E) =
v_{\ms \gamma B,M'B'}^{\rm bg} + 
\sum_{\ms M_1 B_1} v_{\ms \gamma B,M_1B_1}^{\rm bg} \;
g_{\ms M_1B_1}(E) \, t^{bg}_{\ms M_1B_1,M'B'} (E),
\label{eqT1g} 
\ea
and it is not an integral equation because it 
calculates $t_{\ms \gamma B,M'B'}^{\rm bg} (E)$ in terms of $t_{\ms M B,M'B'}^{\rm bg} (E)$,
a function already determined by the  $MB \to M'B'$
coupled-channel integral system (\ref{eqT1}).
As for the dressing of the $\Gamma_{\ms B^\ast, \gamma B}$,
we can reorganize Eqs.~(\ref{eqGammaD}) for $\bar \Gamma_{\ms B^\ast, MB}$,
exchanging the initial and final states, obtaining~\cite{JDiaz07b}
\ba
\bar \Gamma_{\ms B^\ast, \gamma B} (E)=
\Gamma_{\ms B^\ast, \gamma B} + \sum_{\ms M_1B_1} 
\bar \Gamma_{\ms B^\ast, M'B'} (E) \; g_{\ms M_1B_1} (E) \, v^{\rm bg}_{\ms M'B', \gamma B},
\label{eqGamma-gB}
\ea
meaning that also the dressed vertex $\bar \Gamma_{\ms B^\ast, \gamma B}$ 
can be evaluated using $\bar \Gamma_{\ms B^\ast, M_1B_1}$ 
determined by Eq.~(\ref{eqGammaD}) for the meson-baryon coupled-channel system.

We can now discuss the process of meson cloud dressing in the 
context of the dynamical models.
The contribution associated with the meson cloud dressing 
can clearly be observed in the Eqs.~(\ref{eqT1}) and (\ref{eqGammaD})
for the background contribution to the $T$-matrix, and the dressed vertex 
$\bar \Gamma_{\ms B^\ast, MB}$.
The integral term at the r.h.s.~includes  
the meson-baryon propagator, $g_{\ms MB}$, which combined with $t_{\ms MB, M'B'}^{\rm bg}$ 
takes into account the multiple interactions 
from the baryon cores with the mesons, 
corresponding to the meson cloud dressing in all orders.

\begin{figure}[t]
\begin{center}
\includegraphics[width=5.5in]{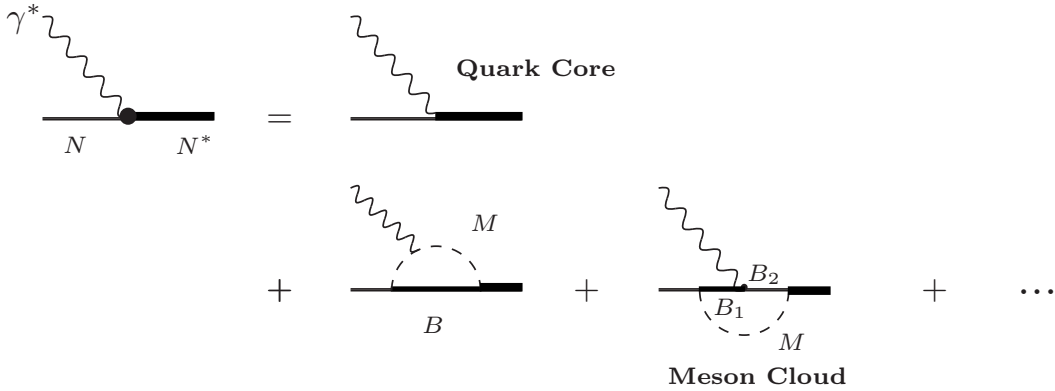} 
\caption{\footnotesize
Bare and meson cloud contributions to 
transition form factors. \label{figPionCloud1}}
\end{center} 
\end{figure}

The processes that contribute to the $\gamma^\ast N \to \Delta(1232)$ 
transition form factors are present in Fig.~\ref{figPionCloud1}.
These diagrams can be interpreted in light of Eq.~(\ref{eqGamma-gB}).   
The first term of the r.h.s.~corresponds to the 
(bare) contribution from $\Gamma_{\Delta, \gamma N}$.
The second term includes one-pion loop corrections
as represented diagrammatically.
Higher order terms are included in the dots.

How does one then evaluate meson cloud dressing effects in the 
context of dynamical models? Their assessment can proceed
by comparing the result of the full calculation with truncated results
obtained with meson-baryon and meson-photon couplings turned off in the input.

Up to now, we have restricted the equations to single meson channels $MB$.
But the formalism can be extended to include also
explicit $\pi \pi N$ states~\cite{Kamano16a,Matsuyama07a,Kamano13a,Kamano09a}.
In the simplest model, at low energy, the $\Delta(1232)$ 
is the only relevant resonance, and
the calculation involves $t_{\ms \gamma N, \pi N}$ 
and $t_{\ms \pi N, \pi N}$~\cite{Pascalutsa07,SatoLee96,SatoLee01,JDiaz07b}.
This procedure was discussed here in the context of the DMT model (Section~\ref{secMAID}).

In summary, in the analysis 
of the baryon transition electromagnetic reactions,
the DCC formalism distinguishes between two separate processes: 
the meson-baryon reactions and the three-quark or bare resonances
associated to the
baryons~\cite{Matsuyama07a,SatoLee01,JDiaz07b,Tiator04}. 
The meson-baryon reaction components include the 
non resonant processes through diagrams with vertices of the background
$v^{\rm bg}$ meson-baryon interactions,
and determine the meson-baryon coupling constants. 
The resonant part, described by a
bare vertex $\Gamma_{\ms B^\ast, MB}$ and a bare mass $M_{B^\ast}^0$,
sub-sums the microscopic quark-quark and quark-gluon 
interactions of the QCD Lagrangian.
Non resonant  and resonant parts,
are intertwined by a set of coupled integral equations.
By turning off  in that set the meson-baryon couplings
one obtains the results from the bare baryon
(also named quark core)~\cite{Capstick08a,SatoLee01,JDiaz07b,Tiator04}.

The DCC formalism can be applied in different ways.
One possibility is to use some simple ansatz 
for the bare contributions as in the first applications 
of the Sato-Lee~\cite{Matsuyama07a,SatoLee96,SatoLee01,JDiaz07b} 
and DMT models~\cite{Kamalov01a,Kamalov99a,Chen07a,Tiator10a}.
Another alternative is to use the input of quark models.
A third possibility is to adjust the unknown 
parameters to the data, including the bare baryon mass, which has no physical reality meaning.
The last option is very insightful since it allows us 
to obtain an estimate of the bare contributions to 
the transition form factors, compare it to
quark model predictions or lattice QCD simulations~\cite{LatticeD,SatoLee01,JDiaz07b,Tiator04},
and finally conclude on the role
of meson-baryon resonance generation.

\subsubsection{\it Summary and Discussion about dynamical models 
\label{secDM3}}

The previous discussion closely followed the 
Sato-Lee model~\cite{Burkert04,Matsuyama07a,SatoLee01,JDiaz07b}.
The connection with other dynamical models
can be made with minor modifications.
The differences between different dynamical models are mainly 
due to different treatments 
of the background contributions as well as differences in the regularizations 
of the $\pi N$ and $\gamma N$ amplitudes~\cite{Burkert04}.
A detailed discussion of differences between the Sato-Lee and 
DMT models can be found in Refs.~\cite{Aznauryan12a,Burkert04}.

An important aspect about dynamical models is how the integral in 
the intermediate momentum states is performed.
More specifically, how the energy component 
of the $MB$ propagator is determined.
The Sato-Lee dynamical model~\cite{SatoLee96,SatoLee01} uses the prescription associated 
to the Lippmann-Schwinger equation.
For a more detailed discussion of 
the subject see Refs.~\cite{Pascalutsa07,Matsuyama07a}.

Dynamical models differ from the model analyses based on the $K$-matrix
discussed in Section~\ref{sec-analysis}
in the treatment of the momentum dependence 
 of the interaction and scattering matrix functions.
By taking the on-shell approximation in the kernel $v^{\rm bg}$ and fixing the energy 
of the $MB$ propagator using the same condition,  
one reduces the equation for the $T$-matrix 
to an $K$-matrix equation, as in Eq.~(\ref{eqMAID6})~\cite{MAID2011,JDiaz08a}.
This approximation has been used in the MAID and SAID partial wave analysis.

The Sato-Lee model was developed along the work of the 
Excited Baryon Analysis Center (EBAC) at JLab~\cite{EBAC-website}.
Its extended coupled-channel analysis was renamed as 
EBAC DCC model~\cite{Matsuyama07a,JDiaz07b,Kamano09b,JDiaz08a,JDiaz07a,JDiaz09a,Kamano09a,Kamano12a}.   
The activity of the EBAC group started in 2006 and ended in 2012.
The most recent developments of the EBAC model 
are known as ANL-Osaka DCC model~\cite{Kamano16a,Kamano13a,Nakamura15a,Kamano19a,Kamano18a,Sato16a}.
A recent extension of the DMT model is
the DMT meson-exchange model~\cite{MAID2011,Chen07a,Tiator10a}.
The J\"ulich-Bonn-Washington
model~\cite{JulichBonn22a,JBW-website} is an extension
of the J\"ulich-Bonn model with the
incorporation of the single meson electroproduction
($\gamma^\ast N \to M B$ reactions)
analysis of the SAID/GWU group~\cite{SAID-website}
in a gauge invariant approach.
Calculations of the PWA for the more relevant  $N^\ast$ states are in progress.
Results for the  $N^\ast$ transition form factors are 
expected in a near future~\cite{JulichBonn21a}.

Before getting into the details of the comparison
between dynamical models, it is worth noticing that
the dynamical models mentioned above
have less freedom in adjusting the parameters to the PWA than 
most of the analyses of Section~\ref{sec-analysis},
because the meson-baryon interactions 
are assumed to have underlying meson-exchange mechanisms
and are not fixed on phenomenological grounds alone~\cite{Ronchen13a,Kamano13a}.

The J\"ulich-Bonn model has the property of generating additional 
resonances (dynamically generated resonances) from the 
non perturbative interactions induced by the scattering
integral equations~\cite{Thiel22a,Ronchen13a,Ronchen18a}.
The generated resonances are independent from 
the resonances assumed to describe the data.

The ANL-Osaka and J\"ulich-Bonn-Washington models 
are similar in their principles and methods to 
determine the scattering $T$-matrix pole positions~\cite{Thiel22a}.
The calculation of the poles is based on Eq.~(\ref{eqGbare}) 
for  $i=j$,
where $M_{B_i^*}^0$ is the bare mass and $\Sigma_{i,i}(E)$ is the resonance self-energy.
The self-energy term includes a real and an imaginary part.
The pole is determined by analytic continuation to the complex plan of
the pole condition 
$z= M_{B_i^*}^0 + {\rm Re}[\Sigma_{i,i}(z)] + i \,{\rm Im}[\Sigma_{i,i}(z)]$
for the resonance labeled by $i$~\cite{Ronchen13a,Suzuki10a}.
The DMT meson-exchange model~\cite{MAID2011,Chen07a,Tiator10a}
also uses an analytic continuation to calculate the pole 
properties of the nucleon resonances. 

However, there is an important conceptual difference between 
the ANL-Osaka and DMT models on one side,
and the J\"ulich-Bonn-Washington model on the another hand-side.
The first ones make an attempt to connect the 
properties of the bare cores and bare couplings with 
the internal content of the baryons, in principle determined by the  
quark-gluon structure~\cite{Kamalov01a,Kamalov99a,SatoLee01,JDiaz07b,JDiaz09a}.
The interpretation of the bare contributions 
as a consequence of valence quark effects is motivated 
by quark model calculations of the $N^\ast$ states and by 
the inference from data of the effect of the meson cloud dressing, 
interpreted as the result of meson-baryon interaction
couplings which are suppressed 
when $Q^2$ increases~\cite{Kamalov01a,JDiaz07b,JDiaz09a}.
In this perspective the bare contributions for the transitions 
form factors are related with quark-gluon substructure, regulated by QCD, 
and may be compared with calculations based on valence 
quark degrees of freedom~\cite{Burkert19a,Capstick08a,Kamalov01a,JDiaz07b}.

The J\"ulich-Bonn-Washington model avoids making any 
interpretation of the bare parameters of the model,
including bare masses and bare couplings,
introduced by an effective Lagrangian.
These quantities are given no physical meaning, 
since they are scheme dependent, and only the dressed quantities 
(renormalized quantities) have physical interpretation~\cite{Ronchen13a,Doring09b}.
The main argument is related with the separation between the resonant term 
of the scattering matrix (the pole term in their language) 
from the non-pole term, which can be regarded as the background term
in other models. This separation does not ensure that the pole term contains 
only the pole contribution, since Eq.~(\ref{eqT1}) unitarizes $v^{\rm bg}$ and
leads to poles in $t_{\ms MB,M'B'}^{\rm bg}$. In turn, $t_{\ms M B, M'B'}^R$
contains terms beyond poles terms, as constant terms and others, as the
Lippmann-Schwinger equation contains all the interaction terms (background and pole terms)
in its kernel. This implies that the decomposition of Eq.~(\ref{eqT})
into background and resonance parts does not correspond
to a separation of the $T$-matrix into singularity free and
non-singularity free parts; $t_{\ms M B, M'B'}^R$ includes also constant 
terms at the pole, meaning that there are 
ambiguities in the separation~\cite{Ronchen13a,Ronchen18a,Doring09b}
and no separate physical meaning can be given to the two terms of
Eq.~(\ref{eqT})~\cite{Ronchen13a}. However, since the ambiguities do not affect the position of the poles,
the J\"ulich-Bonn model decomposes for convenience
the $T$-matrix  into pole and non-pole contributions.
The decomposition has a practical advantage since its computationally efficient
and therefore convenient, allowing the more time consuming calculation of
$t_{\ms MB,M'B'}^{\rm bg}$  to be done once for the fitting procedure of the resonance parameters.
One may then say that independently of interpretations and physical inferences,
in all models the parameters of the pole interaction terms
are tested by the data on $\gamma^\ast N \to MB$ and $MB \to M'B'$ reactions.

As a last remark, the Hamiltonian formalism has also been used in descriptions 
of baryon resonances based on valence quark degrees of freedom.
The first calculations of this type applied the
bag models and the cloudy bag models~\cite{Thomas84,Theberge83a,Kaelbermann83,Bermuth88}.
Recent works relate the coupled-channel formalism 
and the cloudy bag model~\cite{Golli11a,Golli13a,Golli19,Golli18a,Golli08a,Golli09a}.
Other calculations can been found in in Refs.~\cite{Bijker09a,Capstick08a,Santopinto10a}.
In all these calculations the baryon wave functions are based on 
quark properties, and the $T$-matrix 
and the dressed $\Gamma_{\ms B^\ast, MB}$
vertex computed from Eqs.~(\ref{eqMAID9}) and (\ref{eqGammaD}).
They differ from dynamical model calculations in the input for
the contributions from the valence quarks, which are calculated 
instead of fitted to the data as in DCC models.


\renewcommand{\theequation}{5.\arabic{equation}}
\setcounter{equation}{0}

\renewcommand{\thefigure}{5.\arabic{figure}}
\setcounter{figure}{0}

\section{Theoretical models \label{sec-Models}}

\subsection{\it Single Quark Transition Model \label{secSQTM}}

The single quark transition 
model (SQTM)~\cite{Hey74a,Cottingham79a,Burkert-SQTM} 
is a model assuming both $SU(6)$ 
spin-flavor and  $O(3)$ orbital symmetries, and
the $SU(6)\otimes O(3)$ structure of the baryons wave functions
explained on Section~\ref{secWF}.
Moreover, it uses the impulse approximation, by assuming that the interaction 
with the electromagnetic fields affects only the properties 
of the interacting quark (i.e.~proceeds through single quark interaction).
Within SQTM the transverse quark current given by the 
projection of the photon transverse polarization $\epsilon_\mu^{(+)}$
into the quark current $j^\mu$, $j^+ = j \cdot \epsilon^{(+)}$,
can be decomposed into  the general terms~\cite{Hey74a,Cottingham79a}
\ba
j^+ = A \, L^+ + B \, \sigma^+ L_z + C \, \sigma_z L^+ 
+ D \, \sigma^- L^+ L^+,
\label{eqJ-SQTM}
\ea
where $\sigma$ is the Pauli spin operator, 
and $L$ is orbital angular momentum operator.
The operators act on the quark spatial wave function,  
changing the component of the orbital angular momentum
along the direction of motion~\cite{Burkert-SQTM}.
The coefficients $A$, $B$, $C$ and $D$ are functions 
of $Q^2$ that depend on the initial 
and final baryon states.

As discussed in Section~\ref{secWF}, we can classify the
baryon states in $SU(6)\otimes O(3)$ multiplets: 
$[{\cal D},L^P_N]$ where ${\cal D}$ is the number of baryon states
including all spin projections, $L$ is the 
angular momentum quantum number, $P$ is the parity
and $N$ is the harmonic oscillator excitation index.
Since in the SQTM the transverse current take the form (\ref{eqJ-SQTM})
the transverse amplitudes between two $SU(6)\otimes O(3)$ multiplets
can be written
in terms of the same set of $A$, $B$, $C$ and $D$ coefficients~\cite{Burkert-SQTM}.
This is possible because all members of the multiplet have the same 
radial and orbital wave function, for each of the two possible values of 
total quark spin, $S_{3q}=\frac{1}{2}$ and $S_{3q}=\frac{3}{2}$.
Predictive power is then generated by this decomposition, once (only) 4 coefficients of
the same multiplet are known.
The model can be used to determine 
transition amplitudes for proton and neutron targets.
Transition amplitudes between the ground state multiplet $[56,0_0^+]$
and the multiplets $[70,1_1^-]$ and $[56,0_2^+]$  have 
been calculated within the SQTM~\cite{Hey74a,Cottingham79a}.

Notice that according to the assumptions of the SQTM,
the input for the calculations of the coefficients
should be based exclusively on valence quarks,
and not on parametrizations of the data, 
since those are in principle contaminated with meson cloud effects~\cite{SQTM}.
As a consequence the results from the SQTM 
are expected to be valid for intermediate and large $Q^2$ only (say $Q^2 > 1.5$ GeV$^2$).

From all the multiplets, the multiplet $[70,1^-_1]$,  
which contains $N$ and $\Delta$ states with $J^P = \frac{1}{2}^-, \frac{3}{2}^-$,
as the states $N(1535) \frac{1}{2}^-$ and $N(1520) \frac{3}{2}^-$,
is a special case.
For this multiplet only the first three terms 
of Eq.~(\ref{eqJ-SQTM}) contribute to the transition current matrix element
(the contribution of the last term vanishes),
meaning that the transition amplitudes
between the multiplet $[56,0_0^+]$ (which include the nucleon) 
and the multiplet $[70,1_1^-]$ are determined
only by three coefficients~\cite{Hey74a,Cottingham79a,Burkert-SQTM}.
This is what enables us in Section~\ref{secSQTM2} to give predictions for
transitions to the $[70,1_1^-]$ states $N(1650) \frac{1}{2}^-$,  $N(1700) \frac{3}{2}^-$
$\Delta(1620) \frac{1}{2}^-$ and  $\Delta(1700) \frac{3}{2}^-$ from
the coefficients obtained from the transitions to the
$N(1535) \frac{1}{2}^-$ and $N(1520) \frac{3}{2}^-$ states.

For the transition between the multiplet $[56,0_0^+]$ 
and the multiplet $[56,2_2^+]$, which include the $N(1720)\frac{3}{2}^+$
and $N(1680) \frac{5}{2}^+$ states, the transverse amplitudes depend 
on the 4 independent coefficients~\cite{Hey74a}.
Once the four coefficients are known, we can make predictions 
for the states $\Delta(1910)\frac{1}{2}^+$,  $\Delta(1920)\frac{3}{2}^+$,
$\Delta(1905)\frac{5}{2}^+$ and  $\Delta(1950)\frac{7}{2}^+$,
all four stars states~\cite{PDG2022}.
At the moment, however, the application of the method is limited
by the lack of information about the resonances of the multiplet.
The measurements of the amplitudes associated to the multiplet $[56,2_2^+]$
are difficult because most states couple weakly with the $\pi N$ channel,
and more strongly with the $\pi \pi N$ channels~\cite{Burkert-SQTM}.
The data for $N(1720)\frac{3}{2}^+$ are scarce and limited to $Q^2 < 1.5$ GeV$^2$.
The data for $N(1680)\frac{5}{2}^+$ are more abundant, although 
distributed in the region $Q^2 > 1.8$ GeV$^2$~\cite{CLAS15}.
Both states  $N(1720)\frac{3}{2}^+$ and $N(1680) \frac{5}{2}^+$ are discussed
in Sections~\ref{sec-N32p} and \ref{sec-N52p}, respectively. New JLab data, based on two pion production 
and the upcoming data from the JLab-12 GeV upgrade,  
may help to explore the states of this multiplet.

\subsection{\it Covariant Spectator Quark Model \label{secCSQM}}

A particular constituent quark model is the covariant spectator quark model.
The covariant spectator quark model is built within
the covariant spectator theory~\cite{Gross93,Gross69a},
a field-theoretic based framework already applied both to nuclei
and quark systems~\cite{Stadler97a,Gross06a,NSTAR2017,Gross08d,Pinto10a,Biernat14a,Biernat14b,Leitao17a,Leitao17b}.
In this framework the baryons are defined as systems of three quarks,
where two of the quarks are on-shell~\cite{Nucleon,Omega,NDelta,Nucleon2}.
The formalism has advantages in the calculation of
electromagnetic transition between baryon states due to the
separation between off-shell and on-shell quark states.
Under the relativistic impulse approximation, 
the photon couples with the off-shell quark and
one can integrate into the degrees of freedom 
of the two spectator on-shell quarks obtaining a vertex function $\Gamma_B$ 
associated to the quark-diquark system with an off-shell quark
and an effective on-shell diquark with an average mass $m_D$~\cite{Gross06a,Nucleon}.
In this approach confinement is ensured because in
the three quark vertex $\Gamma_B$ vanishes in such a way that
it cancel the singularities of the propagators of the three
on-shell quarks~\cite{Stadler97a,Gross06a,Nucleon}.
The baryon wave function $\Psi_B(P,k)$
where $P$ and $k$ are the baryon and the diquark momenta, respectively,
is then defined from in terms of the vertex function $\Gamma_B$
for the quark-diquark system.

Contrarily to other constituent quark models, 
which are based on wave equations calculated from 
some confining potential, in the calculations on  baryons within the 
covariant spectator quark model the wave functions are built from 
the internal symmetries only, and their shape (its radial part) is determined by 
experimental data, or lattice QCD data 
for some ground state systems~\cite{NSTAR2017,Nucleon,Omega,NDelta}.
For that reason the model is not used to predict the baryon spectrum,
but uses the experimental masses as an input.
The wave functions of the baryons contain 
all possible quark-diquark individual Lorentz structures
states compatible with the internal symmetries of the baryon 
(color, flavor, spin, momentum etc.).
The states are first defined in the baryon rest frame 
and then boosted to an arbitrary frame by appropriated 
Lorentz transformations. The wave functions are then fully covariant.
In the rest frame and taking their non relativistic limit
the wave functions generate the components of the 
non relativistic quark models~\cite{NDeltaD,Nucleon,NDelta,Nucleon2,Fixed-Axis}.

The contribution of the valence quarks for a $\gamma^\ast B \to B'$ transition current 
in relativistic impulse approximation $J_{\rm B}^\mu$
(B standing for {\it bare} constituent quark contribution)
is expressed 
in terms of the quark-diquark wave functions $\Psi_B$ and $\Psi_{B'}$ 
by~\cite{Nucleon,Octet1}
\ba
J_{\rm B}^\mu = 3 \sum_{\Gamma} \int_k 
\overline{\Psi}_{B'}(P_{B'},k) j_q^\mu \Psi_B (P_B,k),
\hspace{1.5cm}
j_q^\mu = j_1(Q^2) \gamma^\mu + j_2(Q^2) \frac{i \sigma^{\mu \nu}}{2 M},
\label{eqJimpulse}
\ea
where $j_q^\mu$ is the quark current operator, discussed below, 
$P_{B'}$, $P_B$ and $k$ the are the final, initial and diquark momenta, respectively, 
and $\Gamma$ labels the scalar diquark and vector diquark polarizations.
The factor 3 takes into account the contributions 
from the different diquark pairs,
and the integration symbol represents the covariant 
integration over the diquark on-shell momentum
$\int_k \equiv \int \frac{d^3{\bf k}}{(2 \pi)^3 2 E_D}$,
where $E_D = \sqrt{m_D^2 + {\bf k}^2}$ is the diquark energy.

The quark current operator $j_q^\mu$ has 
the generic structure for a constituent (extended) quark
in terms of the Dirac ($j_1$) and Pauli ($j_2$) quark operators 
that act on the $SU(3)$ flavor states.
$M$ is the nucleon mass, as before. The Dirac and Pauli components
$j_i$ ($i=1,2$) of the quark electromagnetic current
are decomposed into a sum of operators 
acting on the $SU(3)$ flavor space
\ba
j_i(Q^2)=
\sfrac{1}{6} f_{i+} (Q^2)\lambda_0
+  \sfrac{1}{2}f_{i-} (Q^2) \lambda_3
+ \sfrac{1}{6} f_{i0} (Q^2)\lambda_s,
\label{eqJq}
\ea
where $\lambda_0$,  $\lambda_3$ and $\lambda_s$ 
are the flavor operators based on the Gell-Mann matrices~\cite{Omega}
that act on the flavor component of the quark wave function
$q=  (\begin{array}{c c c} \! u \, d \, s \!\cr
\end{array} )^T$. 
The functions  $f_{i+}$, $f_{i-}$ ($i=1,2$) 
represent the quark isoscalar and isovector 
form factors, respectively, based on 
specific combinations of  quarks $u$ and $d$;
the functions $f_{i0}$ ($i=1,2$) represent 
the structure associated with the strange quark.  
These three quark form factors effectively describe the structure of the 
constituent quarks due to
gluon and quark-antiquark dressing\footnote{In the present formalism 
we can consider the baryons as three-quark systems with vertices 
which include gluon exchanges. This description differs 
from formalisms which consider pointlike quarks 
and require the introduction of higher order Fock states $(qqq)g$, 
$(qqq)q \bar q$, etc.,
where the complexity is on the wave function (not in the vertex).
A more detailed discussion of the subject can be found in Ref.~\cite{Nucleon}.}.
To parametrize this structure, including its evolution in momentum space we use 
a vector meson dominance (VMD) representation where 
the functions associated to the isoscalar, isovector and strange quark components 
are written in terms of vector meson poles such as 
the $\omega$, $\rho$ and $\phi$, and some effective higher mass poles 
to modulate the short range structure.

The explicit expressions for $f_{i\pm}$ and $f_{i0}$ can be found 
in Refs.~\cite{Omega,Octet2,Octet2Decuplet,HyperonFF,Medium}.
The free parameters of the current were calibrated 
by the nucleon elastic form factor data and
the baryon decuplet lattice QCD data~\cite{Nucleon,Omega},
and used further is other calculations.
The advantage of a VMD parametrization is that 
it can be generalized to different regimes.
For instance, VMD parametrizations enables directly an extension of the model 
to the lattice QCD regime (more details below), 
to the timelike
region~\cite{N1535-TL,N1520TL,Timelike,Timelike2,DecupletDecays2,DecupletDecays3},
and to the nuclear medium~\cite{Medium,Medium2}.

For the calculation of transition form factors
between baryon states, it is necessary also to define the
radial wave functions of the baryon states.
Since the baryons are on-shell and the 
intermediate diquark in the covariant spectator quark model 
is taken also to be on-shell,
the radial wave functions $\psi_B$
can be written in terms of the dimensionless variable 
$\chi_{\ms B} = \frac{(M_B-m_D)^2 - (P-k)^2}{M_B m_D}$,
where $M_B$ is the baryon mass~\cite{Nucleon}.
The radial wave functions 
for the nucleon and octet baryon are then given 
the analytic Hulthen form
$\psi_B (P,k) =
\frac{N_B}{m_D (\alpha_i + \chi_{\ms B})(\alpha_j + \chi_{\ms B})}$, 
where $N_B$ is a normalization constant and
$\alpha_i$ and $\alpha_j$ are 
momentum-range parameters in units $M_B m_D$~\cite{Nucleon,Octet2,Medium}.

Other analytic forms can be used with a different number of
factors $1/(\alpha_i + \chi_{\ms B})$, depending on 
the angular momentum states
and on the baryon flavor structure~\cite{NDeltaD,Octet2,Octet2Decuplet,Omega2,Omega3}.
The radial wave functions of the $\Delta(1232)$ $S$-state, for instance
use the form $\psi_\Delta (P,k) =
\frac{N_\Delta}{m_D (\alpha_i + \chi_{\ms B})(\alpha_j + \chi_{\ms B})^2}$~\cite{LatticeD,Omega}.
The falloff of the transition form factors is shaped 
by the parametrizations of these radial wave functions. 
It turns out that although fixed by data still away from that large-$Q^2$ regime,
that falloff
is in general consistent with the power laws 
dictated by quark counting rules discussed in Section~\ref{sec-largeQ2}.

As just described, the covariant spectator
quark model takes into account only valence quark degrees of freedom.
However, some processes, such as  meson exchanges between
the different quarks inside the baryon, cannot
be reduced to processes associated to the dressing of a
single quark and included in the constituent quark picture.
Those processes are regarded as meson exchanges between the different
quarks inside the baryon, and therefore classified as meson
cloud dressing effects of the hadron as a whole,
and are included phenomenologically~\cite{N1520TL,N1520SL,SemiR,DecupletDecays}.

This separation of the transition form factors into
valence quark and meson cloud contributions 
is model dependent~\cite{Hammer04,Meissner07}, and the identification of the bare baryon
states depends on the calibration of the background~\cite{Burkert04,Ronchen13a}.
To mitigate this problem, the option was to reduce
the impact of the model dependence by
matching the valence quark
contributions of the model to lattice QCD simulations for large pion masses,
since quenched and unquenched lattice QCD
simulations include some meson cloud effects, those effects
are reduced for large pion masses~\cite{Pascalutsa07,Alexandrou08a}.

\subsubsection*{\it Extension of the model to the lattice QCD regime}
\label{secLatticeCSQM}

We give now some details on the calibration the valence quark degrees
of freedom component of the covariant spectator quark model with lattice QCD data.
The extension to the lattice QCD simulate the properties of the hadrons
in a regime where the hadron masses and couplings are modified according
with the value of the pion mass of the simulation.

The extension of the covariant spectator quark model to
the lattice QCD regime in then performed as follows:
the radial wave functions $\psi_B$ (in terms of  the variable $\chi_{\ms B}$)
were kept with their shapes determined by the physical regime
but calculated for the baryon mass from lattice simulations;
the quark electromagnetic current operator $j_q^\mu$ which is defined by
a VMD parametrization (and the mass of the nucleon) was calculated 
using the nucleon masses and the vector meson masses associated
to the lattice simulation pion mass~\cite{LatticeD,Octet2,Medium,Lattice}.
By having running baryon masses in the wave functions
together with vector meson masses in the VMD current ansatz
with the pion mass, we took into account the properties
of the lattice QCD implicitly into the calibration
of the valence quark model with the lattice results.
The application of the covariant spectator quark model
to the $\gamma^\ast N \to \Delta(1232)$ transition in the lattice QCD
regime is discussed in Section~\ref{sec-latticeQCD-Delta}.

The generalization to lattice QCD is justified mainly by two arguments:
dominance of valence quark degrees of freedom in the lattice regime;
the expressions for the transition form factors are independent of the diquark mass.
The dominance of the valence quark degrees of freedom
is expected since the effects of the meson cloud are suppressed 
for large pion masses, in particular for
$m_\pi > 400$ MeV~\cite{LatticeD,Detmold01,Ashley04}.
The independence of the calculations on the diquark mass
happens because the diquark mass scales out from the current and consequently 
from the form factors due to the form chosen 
to the radial wave function $\psi_B$ in terms of the variable $\chi_{\ms B}$~\cite{Nucleon,Octet2}.
The presumption is then that main dependence on the pion mass
comes from the hadron masses and the electromagnetic couplings,
and therefore it is expected only a weak dependence on the range 
parameters ($\alpha_i$) close to the physical region.
This was supported by nucleon and $\gamma^\ast N \to \Delta(1232)$
studies of the lattice QCD data for masses $m_\pi= 400$--600 MeV~\cite{Lattice}.

\subsubsection*{\it Semirelativistic approximation}

The covariant spectator quark model can also be used in combination 
with the semirelativistic approximation~\cite{N1535-TL,NSTAR2017,SemiR}.
The description of the $\gamma N \to N^\ast$ transitions can be 
considerable simplified in such approximation
that does not take into account the energy component of the baryon's momenta.
In that case, the orthogonality condition, defined by the 
projections of the zero component of the current $J^0$ 
onto the asymptotic states, i.e.~the condition $\left< J^0\right>=0$
when $|{\bf q}|=0$ (and then both particles are at rest) is automatically insured. 
In a relativistic framework the discussion of orthogonality is more intricate.
One of the difficulties is that the two states cannot be at rest in the same frame,
unless $M_R = M$.
Noting that   $|{\bf q}|= \frac{M_R^2 -M^2}{2 M_R}$ in the limit $Q^2=0$,
we conclude that the condition   $\left< J^0\right>=0$ at  $|{\bf q}|=0$,
is valid only in the limit $M_R =M$~\cite{NSTAR2017,SemiR}.

In the semirelativistic approximation~\cite{SemiR}, we start by considering the 
approximation $M_R =M$,
using the average of the two masses for the calculation of the 
kinematic-singularity-free form factors $F_i$ or $G_i$.
But in the final calculation of the 
helicity amplitudes and the multipole form factors,
which are linear combinations of the kinematic-singularity-free form factors
with coefficients which depend on $M_R$ and $M$, we take the different values for these masses.
Combined with the kinematic approximation,  
we relate the radial wave functions of the state $N^\ast$ 
with the radial wave function of the nucleon: 
$\psi_{N^\ast} (P_R,k)$ defined by $\psi_{N} (P_N,k)$
replacing $M$ by $M_R$ and $P_N$ by $P_R$
(we use here $P_N$ and $P_R$ for the nucleon and resonance momenta, respectively).

With these two procedures, kinematic approximation and definition 
of the $N^\ast$ radial wave function, 
we obtain model for the $\gamma N \to N^\ast$ transition 
form factors that calculate the helicity amplitudes with no adjustable parameters.
The direct comparison with the measured data turns out to be successful
in the large-$Q^2$ region, as will be seen in Section~\ref{secData}.

In conclusion, the semirelativistic approximation enables us to achieve two goals:
we keep the nice analytic properties of form factors obtained for some transitions 
when for $M_R \ne M$ and we impose the orthogonality between states 
in a simple way.
The approximation is particularly useful for 
transitions to negative parity states, where the implementation of the 
orthogonality is more cumbersome~\cite{N1535-TL,SemiR}.
In the case of the $J^P = \frac{1}{2}^+,  \frac{3}{2}^+$  states
the orthogonality comes more naturally~\cite{N1440-1,NDeltaD,NDelta,N1710}.
Finally, the approximation can be expected to work very well when the mass difference 
is not too large and for values of $Q^2$ not too close to the photon point.
Calculations for the $N(1535)\frac{1}{2}^-$ and  $N(1520)\frac{3}{2}^-$ states
are performed in Ref.~\cite{SemiR} and discussed in
Sections~\ref{sec-N1535} and \ref{sec-N1520}, respectively.

\subsubsection*{\it Summary of the Covariant Spectator Quark Model}

The predictive power of the covariant spectator quark model for baryons
is noteworthy, especially considering that those calculations
do not include yet dynamical calculations of the wave functions,
and only start by imposing the internal symmetries.
The model has been applied to the study 
of the nucleon, including nucleon electromagnetic form factors, 
axial form factors and nucleon deep inelastic 
scattering~\cite{Nucleon,Nucleon2,Lattice,Nucleon-DIS,Nucleon-Axial}.
It was also extended to the study of the octet baryon and decuplet baryon form 
factors and other transitions between baryon
states~\cite{Omega,Octet1,Octet2,Octet2Decuplet,HyperonFF,DecupletDecays2,Omega3,DecupletDecays,DeltaFF1,DeltaFF0,DeltaFF2,Octet4,LambdaStar,Delta1600}
and to several $N^\ast$ resonances discussed in Section~\ref{secData}.
A review of the results for several $N^\ast$ states can be found in Ref.~\cite{NSTAR2017}.
The formalism was also used to study of $N^\ast$ resonances
in the timelike region~\cite{N1535-TL,N1520TL,Timelike,Timelike2}.

Because the baryon wave functions are built by imposing internal symmetries only,
in certain cases extra work is necessary to have orthogonality
between different states and gauge invariance of the current.
This condition is crucial for the properties of
the transition form factors near the photon point.
For transitions between resonances with small mass differences
and different parity, the semirelativistic approximation allows us to 
implement the orthogonality condition in a natural way,
allowing also for economy in the number of the model parameters.

\subsection{\it Chiral Effective Field Theory/Chiral Perturbation Theory \label{secEFT}}

Of particular importance at low $Q^2$ are the excitations 
due to the production of pions, the lightest meson state.
Chiral symmetry assigns a special role to 
the pion due to its pseudoscalar character 
and the small mass compared to the scale
of the chiral symmetry breaking ($\Lambda_{\chi }\simeq 1$ GeV),
close to the mass of the nucleon~\cite{Burkert04,Pascalutsa07,NDeltaD,Stave06,Bernstein03}.
Since little energy is necessary to create a physical pion,  
the nucleon resonances can also be regarded 
as baryon cores surrounded by pion or meson clouds.
As a consequence, effects associated with 
the pion cloud are expected to contribute significantly to
the nucleon excitations at low $Q^2$.
Methods based on chiral symmetry replace the degrees of freedom from QCD by the 
effective degrees of freedom at low-$Q^2$,
the nucleon and nucleon resonances and light meson fields,
but maintain consistency with the relevant symmetries 
and scales of QCD~\cite{Pascalutsa06a}.
The starting point of chiral effective field theories (EFT)
is to write down the most general Lagrangian for the observed
asymptotic fields in agreement with the fundamental chiral symmetry of QCD~\cite{Bernard08}. 
There are different types of chiral EFTs depending on the
fundamental degrees of freedom included in the
Lagrangian~\cite{Meissner07,Bernard08,Bernard95,Bernard01a,Jenkins91a,Jenkins93a}.

Because at low energies dynamical breaking of chiral symmetry occurs --
an essential property of  QCD -- an effective field theory equivalent
to QCD at low energy can be formulated.
This is chiral perturbation theory (ChPT)~\cite{Bernard08,Jenkins91a,Jenkins93a},
which below a certain scale uses momentum expansion as well as expansion in the quark mass.
The unknown parameters are low energy constants (LECs) determined by selected 
physical and/or lattice QCD data.
The heavy fields, beyond light mesons as the pion, not included in the theory,
can be seen through virtual effects included in these LECs.
Along this work we use {\it chiral theories} to refer to
chiral perturbation theory and chiral EFTs.

Calculations based on chiral theories evaluate the transition form factors 
from loops contributions to pion-baryon diagrams 
up to a certain order~\cite{Jenkins91a,Jenkins93a,Butler93,Gellas99}.
The $Q^2$-dependence of the form factors are limited to the range of application of the expansion.
Chiral EFTs calculations are important
to relate results from lattice QCD with physical results at low $Q^2$.
Considering on expansions on the pion mass for fixed values of $Q^2$,
we can compare lattice QCD results and physical results
in terms of the pion mass.
This subject is discussed in the next section and in Section~\ref{sec-Delta-lQ2},
for the $\gamma^\ast N \to \Delta(1232)$ transition.

Chiral EFTs have also been used to
study effective meson-baryon interactions in a Hamiltonian formulation
to interpret lattice results in terms of baryon-meson
dynamical states~\cite{Wu18b,Liu17a,Kiratidis17a,Liu16a}.


\subsection{\it  Lattice QCD \label{sec-LatticeQCD}}

Lattice QCD is a non perturbative approach to solve QCD 
in an euclidean discrete spacetime lattice. The  gluon ($\rm{A}$)
and quark ($\psi$) fields are quantized on this discrete lattice,
where the intersections represent a point
in spacetime~\cite{Burkert19a,Wilson74a,Gattringer10}.
The foundation or starting point of the method
is the partition function $\cal {Z}$ defined by
the path integral where the QCD Lagrangian $\cal{L}$ enters,
\ba
{\cal{Z}} = \int {\cal{D}} {\rm{A}} {\cal{D}} \psi {\cal{D}} {\bar{\psi}}
\, e^{-\cal{L}}.
\label{eqpartition}
\ea

A physical property $\rm{Q}$ defined by a general operator $\cal {Q}$
(which may involve fundamental fields, gluons and quarks, or even composite fields)
is then obtained from Green functions or correlation functions.
These functions give the expectation values of the operators $\cal {Q}$,
\ba
\left< Q \right> =
\frac {\int {\cal{D}} {\rm{A}} {\cal{D}}
\psi {\cal{D}} {\bar{\psi}} \, Q \, e^{-\cal{L}} }{\cal{Z}},
\label{eqexe}
\ea
and encompass all dynamical information,
as momentum dependent quark mass functions and running coupling constants.
The partition function enables to treat the fields
as a statistical ensemble that is dealt with Monte-Carlo techniques.
This generates automatically a statistical error bar of the calculation
and control of the results. 
In addition, other sources of systematic errors
come from the dependence of the final results
on the lattice spacing $a$ (and the finite lattice volume)
related to the Compton wavelength of the hadrons, in particular 
with the mass of the pion.

The main limitation of lattice QCD calculations is that
in practice lattice QCD results may not be directly compared
to the physical regime where $m_\pi = 138$ MeV,
or that the lattice volumes are insufficient to simulate the continuous limit.
In the last years, however, a significant number
of lattice QCD calculations have been performed
for several physical quantities at or close to
the physical pion mass~\cite{Lattice-ARehim15a,Lattice-Bhattacharya16a,Lattice-Green17a,Lattice-Alexandrou19a,Lattice-Bali19b,Lattice-Alexandrou20b,Lattice-Lin22a,Lattice-Alexandrou18a,Lattice-Bali19c,Lattice-ETM18,Lattice-Bali19a,Lattice-Hua21a}.
At the moment, however, there are no calculations of transition form factors
between the nucleon and nucleon resonances near the physical pion mass.

The link between the lattice data
and the experimental data in the physical world has been made
with the assistance of chiral EFTs,
which provides analytic expansions 
on the pion mass for fixed values of $Q^2$, at low $Q^2$.
For the correct comparison with lattice QCD
it is necessary to take into account the dependence
of the different variables on $m_\pi$.
The expressions of the baryon masses in terms of $m_\pi$ 
include non-analytic dependencies like $\sqrt{m_q}$ and $\log m_q$,
where $m_\pi^2 = 2 B_0 m_q + {\cal O}(m_q^2)$ and $B_0$
the value of the quark condensate,
and include parameters (LECs) which can be determined by physical data, 
lattice QCD data or some theoretical assumptions~\cite{Pascalutsa07}.
The inclusion of the non analytic contributions on 
the baryon masses is fundamental for the comparison
between lattice QCD and physical data~\cite{Pascalutsa07,Young02a}.
In general, the linear extrapolation on $m_\pi$ is not a good approximation
for most observables, since non analytic contributions are not negligible, 
as discussed in Ref.~\cite{Pascalutsa05b,Pascalutsa06a,Gail06}
for the case of the $\gamma^\ast N \to \Delta(1232)$ transition.
This is further discussed in Section~\ref{sec-Delta-lQ2}.

As a rule, the form factors calculated in lattice QCD simulations
decrease monotonically with $Q^2$, the same way as the physical form factors.
One exception is the Dirac form factor for $J= \frac{1}{2}$.
The elastic form factor of neutral particles and
the transition form factors for $J^P= \frac{1}{2}^\pm$
vanish naturally when $Q^2=0$.
When compared with the physical form factors
the magnitude of the lattice QCD form factors
are in general reduced, and the falloff of the form factors
becomes slower when $m_\pi$ increases.
The results have been confirmed by existing simulations 
of electromagnetic form factors for the nucleon and the
$\gamma^\ast N \to \Delta(1232)$ transition~\cite{Alexandrou08a,Alexandrou06a}.

The results from lattice QCD can be interpreted through comparisons with
quark models in a regime where the pion/meson cloud effects are small.
For this purpose quark models can be extended to the lattice QCD regime.
This regime corresponds to lattice QCD simulations 
associated with large values of $m_\pi$ (say $m_\pi > 400$ MeV).
The extension of a quark model to lattice QCD should also take 
into account the change of properties in the lattice QCD regime
including the modifications of the masses 
and the couplings of the hadrons.

Most of the QCD states are unstable resonances that originate reactions
to final states of two or more particles.
The unstable states manifest as complex pole singularities of the scattering amplitudes,
and the very same dynamics that binds quarks and gluons into hadron resonances determines
also their decay into lighter hadrons.
Therefore, an analysis of these unstable states requires the understanding
of complex final state interactions  from robust multi-body reaction amplitudes
that respect the inviolable principles of quantum mechanics of unitarity and analyticity.
As it happens, a technical issue of lattice calculations is the extraction
of scattering amplitudes to these asymptotic states from the truncated finite spatial volume
and the discrete spectrum of QCD eigenstates obtained in a box.
Building upon the pioneer work of  the L\"uscher method, the review~\cite{Briceno18a}
relates non perturbative QCD results for observables obtained with finite
and infinite volume numerical calculations
(see also Refs.~\cite{Mai20a,Hansen21a}).
It also gives methods to obtain information from lattice QCD calculations about
a variety of resonance properties, including their masses and widths,
as well as their decay couplings and form factors.
A more recent work~\cite{Dawid23a}, explores three-particle amplitudes
for real energies below elastic thresholds and complex energies
in the physical and unphysical Riemann sheets, and it extracts positions
of three-particle bound-states that are in agreement with previous
finite-volume calculations.

\subsection{\it  Dyson-Schwinger methods \label{secDS}}

Together with lattice QCD calculations, Dyson-Schwinger methods belong
to the family of functional methods, which are propped up by the partition function path integral.
However, instead of obtaining the Green functions through a direct numerical way,
Dyson-Schwinger methods start by establishing the  equations of
motion of a quantum field theory - QCD for the case of hadron systems - 
by working out the relations between Green functions from the partition function.
Those relations form an infinite tower of integral equations that couple the different Green's
functions and $n$-point functions or vertices. 
By construction, the equations are non perturbative and sum an infinite number of diagrams.
The quarks get dressed by gluon exchanges, acquiring mass that runs with 4-momentum,
through dynamical chiral symmetry breaking~\cite{Nambu11a}.
In the limit of vanishing pion mass the chiral symmetry is enforced
by the consistency between the two and one body equations.
The quark mass evolution with decreasing $Q^2$ covers the
transition from the current quark mass region to the constituent quark mass region.

Hadron bound states and resonances are color singlets that correspond
to the poles of $n$-point functions.
Inserting a complete set of eigenstates of the QCD Hamiltonian produces
bound state poles, and gives the spectral decomposition of such $n$-point functions.
The baryon wave functions are the residues of the poles of the corresponding $n$-point functions.
The connected part of the $n$-point functions,
once amputated of their external $n$-legs, define the scattering $T$-matrix.
The Bethe-Salpeter equation for this matrix is then derived from the
classification and regrouping of graphs
entering the scattering matrix $T$
into irreducible and reducible classes.

The three-body Bethe-Salpeter equation is deceitfully simple,
but its solution is still challenging due to the number and complexity of
Lorentz invariant structures of the different amplitudes.
Quark-diquark reduction helps here, and leads to  Poincar\'e covariant Faddeev equations.
To justify such simplification the interquark interaction is subsumed
into effective diquark correlations.
Therefore gluons appear then only implicitly into quark and diquark propagators
and quark-diquark vertex functions. In the quark-diquark
simplification, the interaction between a quark and a diquark is a quark exchange,
that is iterated in all orders, where in every iteration step
the spectator quark and one quark inside the diquark exchange roles.

\begin{figure}[t]
\centerline{
\includegraphics[width=16cm]{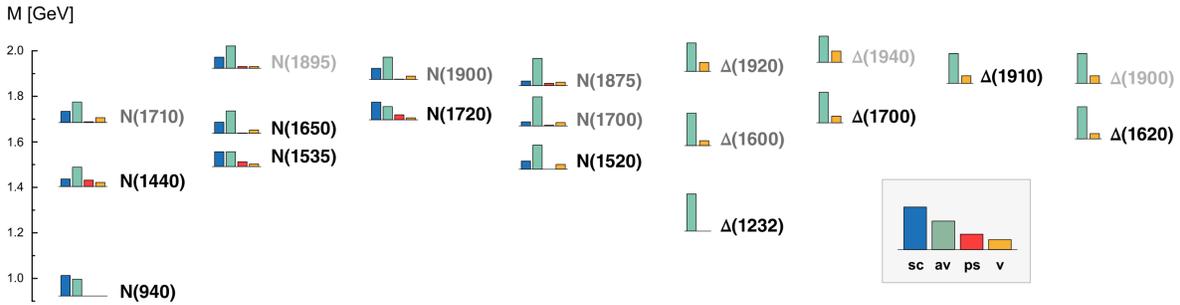}}
\caption{\footnotesize Baryon spectrum and weight of the different
diquark structures in the different baryons: scalar (sc);
axial-vector (av); pseudo-scalar (ps); vector (v).
Figure adapted from Ref.~\cite{Barabanov21}.
Courtesy of Gernot Eichmann.}
\label{fig-spectrum}
\end{figure}

The quark plus dynamical-diquark picture seems to work well
for the baryon structure~\cite{Eichmann16,Eichmann12,SAlepuz18},
and also throughout the whole rich baryon spectrum,
with the level ordering in the baryon spectrum affected
by diquark dynamics~\cite{Barabanov21}.
For instance, in beyond rainbow ladder Dyson-Schwinger calculations,
the mass difference between the Roper resonance and the  $J^P=\frac{1}{2}^-$ states:
$N(1535)\frac{1}{2}^-$ and  $N(1650)\frac{1}{2}^-$ is well reproduced
when in addition to the scalar and pseudovector diquark
structures a conveniently tuned (with 35$\%$ weight) 
contribution of the pseudoscalar diquark is considered.
Figure~\ref{fig-spectrum} illustrates the importance
of different diquark structures in the light baryon spectrum.

The Dyson-Schwinger results for every state $J^P$ comprise
far greater structural complexity
than quark model descriptions. $P$, $D$, $S$, $F$  waves and
their interferences are important for the observables,
and there is a crucial role of genuine relativistic structural components
as $P$ waves.
In these calculations the Roper and the first excited
negative parity states of the nucleon have substantial contributions
of relativistic orbital $P$-waves which are absent
in non relativistic quark models. They are originated by
lower components of the relativistic Dirac spinors 
of the three quarks.

Results and discussion of baryon electromagnetic transitions are found in
Refs.~\cite{Eichmann12,SAlepuz18,Nicmorus10,SAlepuz14,SAlepuz16,Segovia13a,Segovia14b,Segovia14,Segovia19b,Segovia15,Chen19a,Lu19}.
For detailed reviews see Refs.~\cite{Burkert19a,Eichmann16,Roberts94a}.
More references in~\cite{Roberts15a}.
In three-body calculations, since no quark-antiquark exchanges and correlations are considered 
one expects that the Dyson-Schwinger results  for the form factors
are similar to the results of constituent quark models 
at low $Q^2$. At large $Q^2$, however, Dyson-Schwinger equations 
are expected to describe the data 
associated to the transition form factors, since both quark masses and vertices run with $Q^2$.

\subsection{\it AdS/QCD--Holographic QCD   \label{secAdS-QCD}}

In the last decades it was demonstrated that string theory 
or gravity in anti-de-Sitter (AdS) space can provide 
a description of lower-dimensional (conformal) gauge theories (CFT) 
with strong coupling~\cite{Brodsky15a,Maldacena98,Witten98,Gubser98}.
This AdS/CFT correspondence can in certain conditions 
be applied to QCD-like theories~\cite{Witten98b,Sakai05,Klebanov00}.
Though the method is based on the semiclassical approximation of QCD,
it can be used to describe many properties of the hadronic systems due to its simplicity.

The AdS/CFT theories are divided in two main categories:
the top-down and bottom-up approaches.
The top-down approach is based on first principles 
and it is related to super symmetric strings 
based on D-brane physics~\cite{Sakai05,Nawa06a,Hong07a,Hashimoto08a,Hong08a}.
The bottom-up approach is more phenomenological
and derive the QCD properties in the confining
regime using 5D-fields in AdS space~\cite{Brodsky15a,Brodsky08a,Teramond09a,Karch06a,Gutsche12a}.

Most of explicit calculations of 
transition form factors using holography are based on the bottom-up approach.
Within the bottom-down approach, one can relate the results from AdS/CFT 
with the results from light-front dynamics based on a Hamiltonian 
that include the confining mechanism of
QCD~\cite{Brodsky15a,Brodsky08a,Teramond09a,Karch06a}.
The correspondence between AdS/QCD and light-front dynamics is a consequence 
of mapping the hadronic modes in AdS space and the Hamiltonian formulation 
of QCD in physical spacetime quantized on the light-front~\cite{Brodsky15a}.
The correspondence (duality) in known as 
light-front holography or holographic QCD.
In the limit of the massless quarks,
one can relate the AdS holographic variable $z$ 
with the impact separation $\zeta$, 
which measures the distance between constituent partons inside
the hadrons~\cite{Brodsky15a,Brodsky08a,Teramond09a,Karch06a}.

The light-front holography  has
been used to study the properties of the hadrons, 
such as the mass spectrum, parton distribution functions,
and structure form factors of mesons and baryons
using different types of confinements, 
including soft-wall and hard-wall potentials.
A review can be found in Ref.~\cite{Brodsky15a,Teramond12a}.

In the light-front holography formalism, one can represent the wave
functions of the hadrons using an expansion of Fock states
with a well defined number of partons (twist number $t$)~\cite{Gutsche12a,Teramond12a}. 
In the case of baryons, the first term corresponds 
to the three-quark state $(qqq)$ or the leading twist approximation ($t=3$).

Calculations of the nucleon elastic form factors 
where performed using different confining mechanisms 
in leading twist approximation~\cite{Hong08a,Teramond12a,Chakrabarti13a,Liu15a,Maji16a,Abidin09a}
and also with higher twist corrections~\cite{Gutsche12a,Sufian17,Gutsche18a}.
Holographic calculations of different 
$\gamma^\ast N \to N^\ast$ transition form factors 
can be found in 
Refs.~\cite{Teramond12a,Gutsche18a,Grigoryan09b,Gutsche13a,Roper-AdS1,Roper-AdS2,Gutsche20a,Lyubovitskij20a,Teramond12b,Fujii22a,Taghieva22a}.

A note about higher order Fock states 
using holographic QCD, 
in particular on the contributions from $q \bar q$ pairs:
in holographic QCD the substructure of the $q \bar q$ pair 
is neglected in first approximation, 
meaning  that the particles 
in the pair are not correlated as in a meson state, i.e.~one considers
two pointlike particles instead of an extended meson.
The consequence is that contributions of the 
$q \bar q$ pairs to the transition form factors give a 
much slower falloff with $Q^2$ 
than explicit calculations of 
meson cloud corrections to the 
three-quark baryon systems~\cite{NucleonGA-hol}.

The conclusion is that, on one hand, we must be careful 
in using holographic QCD models to estimate 
$q \bar q$ contributions to transitions form factors, on the other hand,
holographic QCD is, for its simplicity, 
a very promising method to estimate the contributions 
of the valence quark degrees of freedom at low $Q^2$,
for comparison with other estimates of the bare core effects.

\subsection{\it More on Quark Models  \label{sec-QuarkModels}}

In addition to the frameworks discussed in the previous sections
two other families of quark models that stand out due to the scope of their results (as shown in the next section).

One is the hypercentral quark model developed by Bijker,
Giannini and Santopinto and
collaborators~\cite{Giannini15,Giannini01a,Santopinto12}
use the combination of the Jacobi variables (\ref{eqJacobi})
in terms of hyperspherical coordinates (6D vector)
to derive a wave function for the variable
$x= \sqrt{{\bf \rho}^2 + {\bf \lambda}^2}$
with a confining potential $V(x)$ similar to
the Isgur-Karl model (see Section~\ref{secQM}).
Calculations of  both
the baryon spectrum and the transition amplitudes
between baryon states~\cite{Giannini15,Bijker94,Bijker96a,Bijker97a,Bijker00,Giannini01a,Bijker09a,Sanctis05,Santopinto12,Bijker16a} were accomplished.

Another relevant class of models is the light-front quark models
discussed in Refs.~\cite{Aznauryan07,Obukhovsky19a,Aznauryan12b,Aznauryan17a,Aznauryan15a}.
Light-front quark models are derived in the infinite momentum frame~\cite{NSTAR,Aznauryan07}
and include in first approximation only $qqq$ states.
The estimates are expected to be valid for $Q^2> 2$ GeV$^2$~\cite{Aznauryan07}.
The model from Aznauryan and Burkert in particular,
motivated by the results from the Dyson-Schwinger formalism,
takes into account the momentum dependence
of the quark masses phenomenologically.
This input contributes to the good description
of the data at large $Q^2$~\cite{Aznauryan12b,Aznauryan17a,Aznauryan15a,Aznauryan16a}.


\renewcommand{\theequation}{6.\arabic{equation}}
\setcounter{equation}{0}

\renewcommand{\thefigure}{6.\arabic{figure}}
\setcounter{figure}{0}

\section{Electromagnetic transitions: data and models \label{secData}}

In the present section we review the experimental results 
associated with the $\gamma^\ast N \to N^\ast$ transitions.
The results are compared with theoretical calculations 
based on different frameworks, and different
approaches are discussed.

More ample data exists for  $\Delta (1232) \frac{3}{2}^+$, 
$N(1440)\frac{1}{2}^+$, $N(1535)\frac{1}{2}^-$ and $N(1520) \frac{3}{2}^-$
resonances than for others.
The identification of the available data,
their sources and energy domains are presented on Tables~\ref{table-Data1} 
and \ref{table-Data2}.
Since presently available data 
is restricted to proton targets except for the neutron amplitudes 
at the photon point, in the following we present  results 
obtained with proton targets only, without an explicit reference. 

The $\Delta(1232)$ was the first nucleon resonance to be discovered. It is
the best known nucleon resonance with its isolated peak 
on the $\gamma N \to \pi N$ cross section clearly identified.
The following three resonances define the  second resonance 
region, where the individual effects are more difficult to disentangle.
The second known resonance, the $N(1440)\frac{1}{2}^+$ 
is more deceptive and its traces are better 
recognized in the two pion production channels
($N^\ast \to \pi \pi N$).
In turn, the $N(1535)\frac{1}{2}^-$ signatures appear more distinctively
in the decay on the $\eta N$ channel.  
Finally the $N(1520)\frac{3}{2}^-$ state is identified 
by combining  measurements on both $\pi N$ and $\pi \pi N$ decays.

In the following, we compare model results
for helicity amplitudes and form factor data.
The calculation of
transition form factors from the helicity amplitudes is possible only
when one has data for the transverse and scalar amplitudes for the same $Q^2$.
For most resonances, the available data is from
JLab/CLAS~\cite{CLAS09,CLAS12,CLAS16a,CLAS15,Aznauryan05b,Aznauryan05a,CLAS08}.
We recall from Section~\ref{sec-Expermental} that the analysis
of CLAS data is based on a combination 
of unitary isobar model and $t$-channel dispersion relations,
testing the consistency 
of the results~\cite{Aznauryan12a,CLAS09}.
Data analysis of MAID of different experiments 
are also available~\cite{Drechsel07,MAID2009,MAID2011,MAID-database,MAID-website}.
However, the MAID analyses differ from the CLAS analyses
for different $N^\ast$ states~\cite{CLAS09,CLAS16a,MAID2011}.
At the photon point, we consider the PDG result from 2022~\cite{PDG2022},
unless mentioned otherwise.
PDG presents the band for the helicity amplitudes
and lists the most relevant measurements.
In the comparison with theoretical models, 
these results should be taken with caution, 
since the central values and the errors can change year after year.
This variation is illustrated in Section~\ref{sec-N1535-lQ2} for 
the helicity amplitudes of the $N(1535)\frac{1}{2}^+$ resonance.
For the data on
helicity amplitudes or transition form factors
we consider also some relevant parametrizations 
and discuss their features.

There are four parametrizations that we will characterize here
in more detail: the MAID parametrizations~\cite{Drechsel07,MAID2009,MAID2011},
the Rational function parametrizations~\cite{Eichmann18},
the JLab parametrizations~\cite{JLab-website,Blin19a}
and the JLab-ST parametrizations~\cite{JLab-website,LowQ2param}.
The MAID parametrizations are expressed by polynomials 
and exponential functions for most $N^\ast$ states.
Their problems are the fast falloff at large $Q^2$
compared with the expected power law falloffs
and some incompatibilities to Siegert's theorem at low $Q^2$ 
(the reader may remember the discussion from Sections~\ref{sec-Siegert} and \ref{secMAID}).
The Rational function parametrizations 
use ratios between polynomials and are compatible 
with power law falloffs at large $Q^2$, as well as with the correlations 
between form factors at low $Q^2$.
The JLab parametrizations are empirical parametrizations
based mainly on the JLab data for the helicity amplitudes~\cite{JLab-website,JLab-database,Blin19a}.
The so-called JLab-ST parametrizations 
use the original JLab parametrizations, but are modified 
at low $Q^2$ in order to be made compatible with the pseudothreshold constraints.

The discussion of the following sections considers
models based on different degrees of freedom.
At low $Q^2$ one can compare the data 
with models that effectively incorporate meson-baryon degrees of freedom,
chiral EFT, and the large $N_c$ limit.
However, for intermediate and large $Q^2$,
relativistic models and
the valence quark degrees of freedom are favorable.
In that regime we expect that as the quark interaction becomes
weaker the baryons behave 
as systems of almost independent pointlike quarks and their
form factors are satisfactorily defined by power law falloffs,
as discussed in Section~\ref{sec-largeQ2}.
By contrast, at low $Q^2$, in the long wavelength limit,
one probes more peripheral regions and the 
picture of the baryons as three valence quark systems only is not appropriate.
In this case other effects, such as higher order Fock state contributions,
including gluon excitations and $q\bar q$ pairs should also taken into account.

Models based on meson-baryon degrees of freedom 
can include chiral theories, discussed in Section~\ref{secEFT},
or other effective meson-baryon interactions.
They consider meson interactions and meson-baryon 
loops, and correspond to coupled-channel reaction 
models where the interference of the channels is fundamental.
They do not include the structure of the baryons
explicitly, and the resonances are described as dynamically
generated from meson-baryon re-scattering.
As explained in Section~\ref{secDM}, in addition, dynamical coupled-channel 
reaction models may also decompose the baryon-photon
electromagnetic interaction  into a contribution from 
the bare core and a contribution from the meson-baryon states.
This decomposition is justified near the resonance pole  
when the non resonant background contributions
(associated to the Born terms) are subtracted~\cite{Pascalutsa07}.
The separation in two terms was discussed in Section~\ref{secDM}
and  helps to decide the domain of validity of the model assumptions.
For the  form factors the decomposition reads
\ba
F_i(Q^2) = F_i^{\rm B} (Q^2) +  F_i^{\rm MC} (Q^2), 
\hspace{1.5cm}
G_\alpha (Q^2) = G_{\alpha}^{\rm B} (Q^2) +  G_{\alpha}^{\rm MC} (Q^2), 
\label{eqGB-MC}
\ea 
where $i=1,2$ ($J=\frac{1}{2}$) or $\alpha =M,E,C$ ($J \ge \frac{3}{2}$), 
B is the contribution associated to the bare core
(valence quarks) and MC the contribution associated 
with the meson cloud (meson-baryon states).

The processes associated with the valence quark contributions 
(bare contributions) and the meson cloud contributions are represented
in Fig.~\ref{figPionCloud1} for the first orders.
The effect of the baryon core at low $Q^2$ 
can be estimated using frameworks based on quarks degrees of freedom 
or inferred from lattice QCD simulations~\cite{NSTAR,Aznauryan12a,N1440-1,JDiaz07b,Tiator04,Segovia15}.
Alternatively the magnitude of the meson cloud contribution 
can be extracted from the data based on theoretical estimates. The comparison of the magnitudes of 
these two main components gives clues for our understanding 
the nature of the $N^\ast$ resonance under study.
In addition to calculations based on valence quarks, 
we present here also the estimates of the 
meson cloud effects from the ANL-Osaka 
DCC model~\cite{Kamano16a,Kamano13a,Nakamura15a,Sato16a}.

The theoretical description of the $\gamma^\ast N \to N^\ast$ 
transition requires the description of the initial state $N$
and the final state $N^\ast$. 
Since we are focused on the properties of the nucleon resonances, 
description of the nucleon is assumed and 
we omit the discussion of the results of the nucleon elastic form factors.
Good discussions can be found  in many reviews, such as~\cite{Hyde04a,Perdrisat07,Arrington07a,Puckett18a}.

We start, then the discussion of the dominant
$\frac{1}{2}^\pm$ and  $\frac{3}{2}^\pm$ states in order.
At the end, we discuss less experimentally known states.


\subsection{\it $N(1440)\frac{1}{2}^+$ resonance \label{sec-N1440}}

The  $N(1440)\frac{1}{2}^+$ resonance, also known as Roper resonance, 
has the same quantum numbers of the nucleon, $N(939)\frac{1}{2}^+$, 
and has been interpreted in the context of 
the quark model framework as the first radial excitation 
of the nucleon~\cite{Burkert19a,Capstick00,Isgur79b}.
Contrarily to the $\Delta(1232)$, the Roper 
was not identified as a clear bump in the reaction cross section, 
but instead it was found in 
phase shifts analysis~\cite{Roper64a,Roper65a}. 
The Roper lies in the second resonance region where it
competes with the $N(1520)\frac{3}{2}^-$  
and $N(1535)\frac{1}{2}^-$ resonances, but the
contribution for the $\pi N$ channel
are overshadow in the presence of those other resonant states.

The properties of the Roper have been a long standing mystery.
The mass and the decay width are difficult to understand in the context of 
the quark model framework~\cite{Capstick00,Capstick86b}.
Quark models predict a large mass and
the decay ratios have significant
contributions from $\pi \pi N$ states, where the last 
is a mixture of $\pi  \Delta$ and $\sigma N$ channels~\cite{PDG2022}.
The effects of the $N(1440)$ and the decay on $\pi \pi N$ channel 
can be observed on $N N \to N N \pi \pi$ 
reactions~\cite{Hernandez02a,CELSIUS08a,ARuso98a,Cao10a}.
A difficulty for a long time was the reversed order between 
the $N (\frac{1}{2}^+)$ and $N (\frac{1}{2}^-)$ states 
in quark model calculations of the spectrum.
Most quark models predict that the
first negative parity resonance 
$N (\frac{1}{2}^-)$ is lighter than the first radial excitation
of the nucleon, $N(1440)\frac{1}{2}^+$, the Roper.  
Only recently, lattice QCD simulations 
had shown that the correct order is obtained when 
the pion masses became smaller than 300 MeV~\cite{Mathur05a,Mahbub10a}.
Dyson-Schwinger methods, predict the correct ordering of the Roper, $N(1520)\frac{3}{2}^-$  
and $N(1535)\frac{1}{2}^-$ states, and actually give the observed ordering
of states and good overall spectrum description of light
and strange baryons~\cite{Eichmann19a,Qin18a,Qin19a,Chen18a}.

Experimentally also, the picture of the $N(1440)$ structure became clearer 
in the last two decades with accurate measurements 
of the $\gamma^\ast N \to N(1440)$ transition 
helicity amplitudes 
at low, intermediate and large $Q^2$ 
at JLab/CLAS (up to $Q^2=4.2$ GeV$^2$)~\cite{Aznauryan12a,CLAS09,CLAS08}.
The new measurements were important to determine the
shape of the $\gamma^\ast N \to N(1440)$ helicity amplitudes, 
their definitive signs,
and to confirm that at large $Q^2$ the falloffs 
of the amplitudes are consistent with what is expected 
from a radial excitation of the 
nucleon~\cite{Aznauryan12a,Burkert19a,Aznauryan08a,Aznauryan07,CLAS09,CLAS12,CLAS16a,CLAS08}.
The signs of the amplitudes depend on the signs of 
the  pion couplings with the nucleon and the Roper ($g_{\pi N N}$ and $
g_{\pi N N^\ast}$)~\cite{Aznauryan12a,Aznauryan08a,Aznauryan07,Gelenava18a},
as discussed in Section~\ref{sec-amplitudes}.

The  $\gamma^\ast N \to N(1440)$ 
single pion electroproduction data from CLAS
in the range $Q^2= 0.3$--4.2 GeV$^2$~\cite{CLAS09} 
are in agreement with the MAID analysis~\cite{Drechsel07,MAID2009,MAID2011}
within the uncertainties~\cite{Aznauryan12a}.
The more recent CLAS double pion electroproduction data 
in the range $Q^2= 0.28$--1.3 GeV$^2$~\cite{CLAS12,CLAS16a,Mokeev09a}
confirms the magnitude of the CLAS single pion electroproduction 
in that range.

\subsubsection{\it Physical interpretation of the $N(1440)\frac{1}{2}^+$ state}

The first interpretations of the Roper are based on  
non relativistic $SU(6)$ quark models~\cite{Capstick00,Koniuk80}.
Different versions of quark models 
with different confinement mechanisms have been
used to estimate the properties of the Roper resonance, 
including the mass, transition magnetic moment, 
decay widths, and transition form factors.
These calculations are performed in non relativistic quark 
models~\cite{Capstick00,Giannini15,Santopinto12,Tiator04,Obukhovsky11a}, relativistic quarks 
models~\cite{N1440-1,N1710,Warns90,Weber90a,JDiaz04a,Dong99a,Kaewsnod22a}, 
light-front quark models~\cite{Aznauryan08a,Aznauryan07,Capstick95,Obukhovsky19a,Aznauryan12b,Aznauryan16a,Obukhovsky14a,Cardarelli97a}, 
and holographic QCD models~\cite{Teramond12a,Gutsche18a,Gutsche13a,Roper-AdS1,Roper-AdS2}.
Although some of these calculations reduce the mass of the Roper
below 1.7 GeV, the values are still larger than the measured value.
The discussion about the quark model calculations 
of the Roper mass can be found in 
Refs.~\cite{Crede13a,Burkert19a,Capstick00,Giannini15,Capstick86b,Glozman96,Giannini01a}.
In general the estimates of the mass are larger than the physical mass.
In the last years, the structure of the state 
was also studied using the Dyson-Schwinger 
formalism~\cite{Burkert19a,Eichmann19a,Wilson12,Segovia15,Chen19a,Segovia16a,Eichmann16b}.
In general the estimates of the mass are larger than the
physical mass ($M_R \simeq 1.7$--1.8 GeV)~\cite{Burkert19a,Segovia15}.
Reference~\cite{Eichmann19a}
gives values closer to the physical value, and the right relative ordering
of the Roper and the state $N(1535)\frac{1}{2}^-$.

The difficulties of the quark models in the description of 
the proprieties of the first $N(\frac{1}{2}^+)$ excited state
lead to a proliferation of proposals to describe 
the Roper system based on different physical mechanisms.
One of the first proposals was the interpretation 
of the Roper as a hybrid state composed of three quarks 
and a gluon [$(qqq)g$ state]~\cite{Li92a,Barnes83a,Capstick02a,Kisslinger95a}.
This proposal was discarded due to the magnitude and 
signs of the helicity amplitudes
when accurate CLAS data became available~\cite{Aznauryan12a,CLAS09}.
Another class of proposals interprets the $N(1440)$ resonance 
as a breathing mode of the nucleon within 
the frameworks of the quark-soliton and Skyrme models~\cite{Kaulfuss85a,Mattis85a}.
These descriptions of the $N(1440)$ state 
were also abandoned when the data showed
that at large $Q^2$ the Roper has the properties 
of a three valence quark system, 
and the falloff of the form factors is well
approximated by quark counting rules behavior.

Below $Q^2=2$ GeV$^2$ the $\gamma^\ast N \to N(1440)$ data cannot 
be described accurately by models 
based exclusively on the valence quark degrees of freedom.
In that regime, we need to consider corrections to the 
three valence quark picture, in particular models that take into account 
$q \bar q$, $(q \bar q)(q \bar q)$ states etc.~\cite{Li06c,Zou10a}, or meson cloud excitations.
There are different proposals that account for this extension.
The first quark models which include the $q \bar q$ excitations 
use the $^3P_0$ model where the initial photon is converted into a 
$q\bar q$ state to produce the final state 
$\pi N$~\cite{Koniuk80,JDiaz06a,Bijker16a,Li06c}.
In the same category are models which account for 
quark-meson interactions~\cite{Obukhovsky19a,Obukhovsky11a,Dong99a,Obukhovsky14a,Cano98a,Alberto01a,Meyer01,Dillig07a,Sekihara21a}, 
as cloudy bag models~\cite{Thomas84,Bermuth88,Golli18a,Golli08a,Golli09a,Chen08a}.
In general those models improve the description of the data below $Q^2=2$ GeV$^2$.
The $N(1440)$ and the $\gamma^\ast N \to N(1440)$ transition has also been 
described by chiral effective meson-baryon 
models~\cite{Bauer14a,Hernandez02a,Gelenava18a,Gegelia16a,Long11a} 
and meson-baryon coupled-channel models~\cite{Ronchen13a,Doring09a,Golli18a,Krehl00a,Schneider06a},
without explicit reference to quark content.
However, a pure molecular-type description 
of the Roper would lead to much softer 
transition form factors than observed 
experimentally~\cite{Burkert19a,Aznauryan08a}.

The dynamical coupled models in general, and the ANL-Osaka DCC model in particular, 
provide a hybrid description  of the Roper 
and of the $\gamma^\ast N \to N(1440)$ 
transition~\cite{Kamano16a,Burkert04,Kamano13a,Nakamura15a,Suzuki10a,JDiaz09a}.
While the bare core structure follows the structure of a $qqq$ system, 
the meson production mechanism modifies the baryons with an inclusion 
of a meson cloud associated with different decay channels, 
including $\pi N$, $\pi \pi N$, $\pi \Delta$, $\rho N$ and $\sigma N$~\cite{Suzuki10a}. 
The large mass of the Roper can be explained when we 
combine the calculations based on quark degrees of freedom 
with the ANL-Osaka DCC model~\cite{Suzuki10a}.
The analysis of the ANL-Osaka group concludes that 
a single bare pole for a mass $M_R \simeq 1.7$ GeV 
generates two different poles with masses $M_R \simeq 1.36$ GeV,
when the meson cloud dressing is taken into account, becoming
consistent with the observed mass of the Roper~\cite{Suzuki10a}.
The two poles appear in two different Riemann sheets 
and are associated with different decay channels, 
the physical and the unphysical $\pi  \Delta$ channel.
There is still same debate about either the poles represent 
different states or are the consequence of a single state 
in different physical or unphysical channels.
For more  detailed discussions the readers can consult 
Refs.~\cite{Ronchen13a,Arndt06a,Sarantsev08a,Capstick92,Arndt85a,Cutkosky90a}.
Nevertheless, the analysis of the  ANL-Osaka group
 shows how we can conciliate 
a large mass, when we consider the effects of valence 
quark degrees of freedom, 
with a smaller mass when we take into account 
meson cloud effects at the hadronic level.
Notice that by the effect of valence quarks 
we meant light undressed quarks, constituent quarks or 
quarks dressed by  gluon loops as in the Dyson-Schwinger framework.

The interplay of the effects associated with pure quarks 
and meson cloud dressing of the three-quark core 
is also visible when we analyze the lattice QCD data.
The properties of a radial excitation associated 
to a three-quark state are reproduced in 
lattice QCD simulations  -- where the zeros on the wave function 
are observed~\cite{DRoberts13a,DRoberts14a}.
The interpretation of the energy levels in lattice 
and their relation to the physical processes 
is a more complex task.
The connection between the lattice energy levels 
based on the L\"uscher method~\cite{Luscher91a,Luscher91b} 
indicates that the first $N\left(\frac{1}{2}^+\right)$ 
excitation cannot be interpreted just as a $qqq$ state, 
but resembles instead a dynamical generated resonance 
where $\pi N$ and other channels 
are strongly mixed~\cite{Wu18b,Liu17a,Kiratidis17a,Lang17a,Sun20a}.

The hybrid picture is advocated by dynamical coupled models
as described above, and this is also the case, of the non relativistic effective
field theory calculations that connect to the finite volume energy levels results
of lattice QCD, by evaluating the overlap of the lattice QCD states
with non local $(qqq)(q \bar q)$ operators and local three-quark interpolating fields.
The best reproduction of these overlaps from lattice QCD data 
by an Hamiltonian effective theory formulation 
is achieved with a combination of a three-quark radial excited state 
with a strong rescattering of the 
$\pi N$, $\pi \Delta$ and $\sigma N$ channels~\cite{Wu18b,Liu16a}.
This explanation is consistent with the
dynamical model/quark model analysis~\cite{Burkert19a,Suzuki10a,Kamano10}
mentioned above, and also with the interpretations of 
the lattice QCD results for the $N(1535)$ (strong three-quark state)
and $\Lambda (1405)$ (dominated by meson-baryon molecular states)~\cite{Wu18a}.
At large $Q^2$, in the regime where the meson-baryon effects are small,
it is the three-quark state component that dominates~\cite{Wu18a}.

Nevertheless, the scale dependence and continuum limit of the overlaps
with the quark interpolating fields may question decisive conclusions, and
on the other hand,  the falloff of the form factors at large $Q^2$ indicates
a $qqq$ quark structure. In addition,
different groups accomplished Dyson-Schwinger
calculations~\cite{Eichmann19a,Qin18a,Qin19a,Chen18a}
-- which are continuum approaches to the three valence quark bound-state problem
in quantum field theory -- indicating that the Roper channel
has the appearance of a radial excitation of the nucleon ground state.
In their turn those groups stress the importance  of contributions
from a rich structure of diquarks (scalar and pseudovector for the positive parity
and admixtures of even- and odd-parity diquarks for the negative parity cases)
in the structure of the $J=\frac{1}{2}^\pm$ states.
At present, and in summary, it thus seems fair to say that
the controversy on the structure of the Roper resonance is not yet fully decided.

To complete the picture, one notices that 
coupled-channel descriptions based on chiral EFT 
can also be used to understand the rule of the $\pi N$ 
and $\pi \pi N$ decays widths on the Roper decays~\cite{Bauer14a,Gegelia16a}.
Larger contributions of $\pi \pi N$ and 
$\sigma N$ channels are also found in the analysis from
Refs.~\cite{Sarantsev08a,JDiaz08a}.
The explicit inclusion of $q \bar q$ contributions 
in constituent quark models also improves the description 
of the Roper decay widths~\cite{JDiaz06a,Li06c}.
Calculations based on holographic QCD 
(AdS/QCD) can be found in
Refs.~\cite{Teramond12a,Gutsche18a,Gutsche13a,Roper-AdS1,Roper-AdS2,Fujii22a}.

\subsubsection{\it Helicity amplitudes \label{sec-Roper-Amps}}

\begin{figure}[t]
\centering

\includegraphics[width=3.2in]{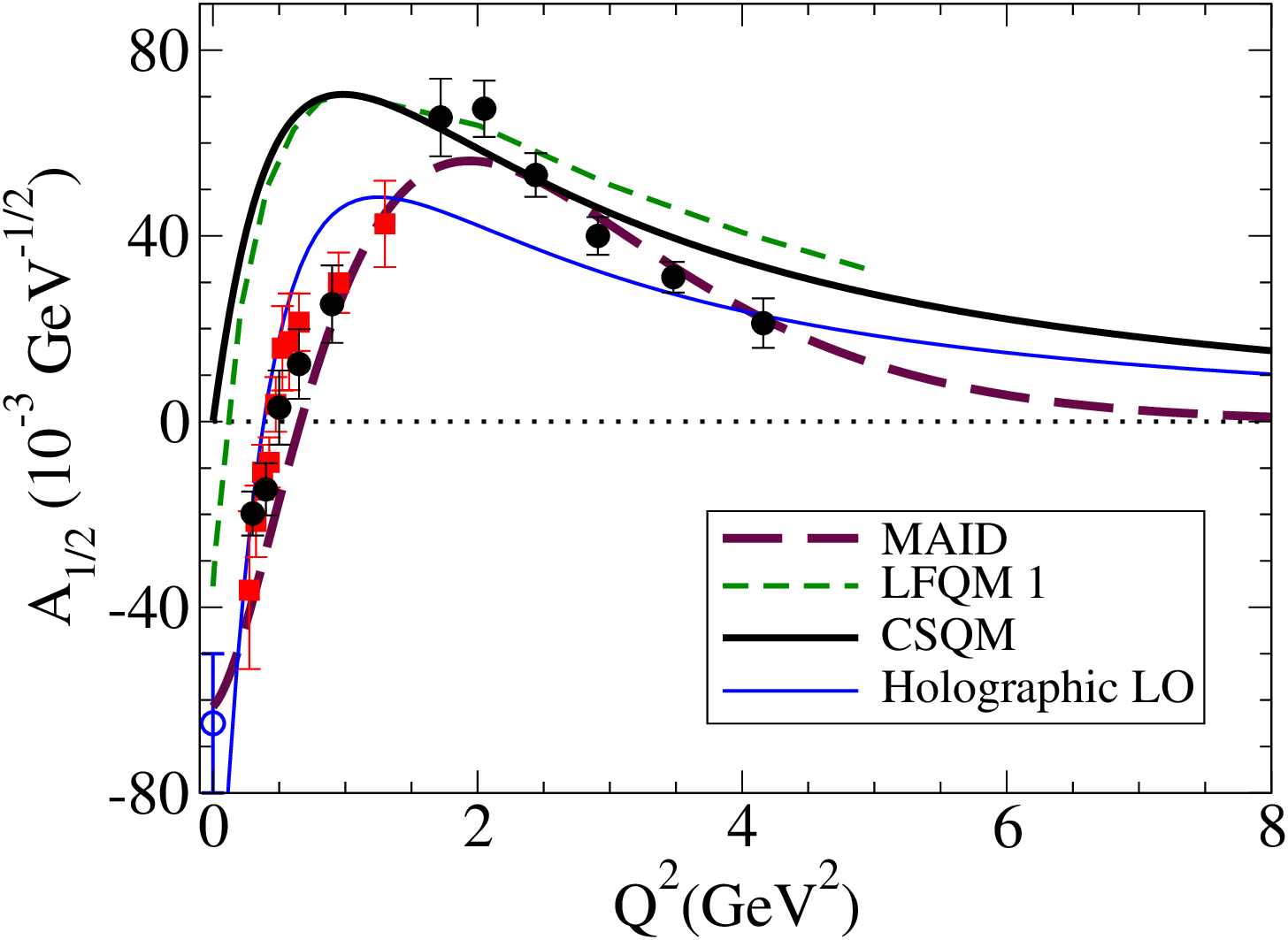} \hspace{1.cm}
\includegraphics[width=3.2in]{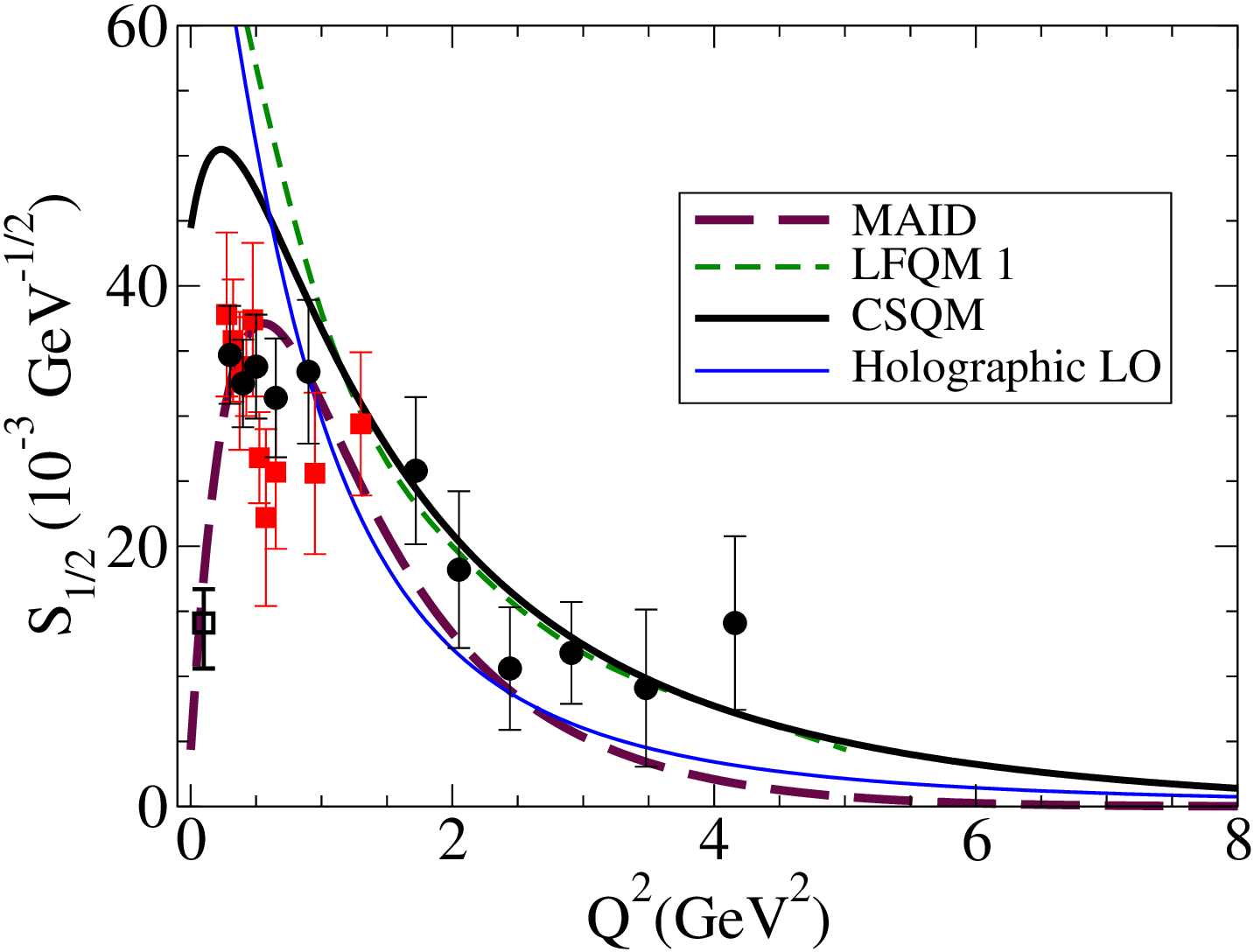} 
\caption{\footnotesize 
Calculations of the $\gamma^\ast N \to N(1440)$ helicity amplitudes that
focus on the bare quark contributions:
 light-front quark model (LFQM 1)~\cite{Aznauryan07}, 
covariant spectator quark model (CSQM)~\cite{N1440-1}, 
and Holographic QCD model in leading order (LO, leading twist)~\cite{Roper-AdS1}.
Comparison with the MAID parametrization~\cite{Drechsel07,MAID2009,MAID2011}.
The data are from JLab/CLAS, one pion production 
({\color{blue}{\Large $\bullet$}})~\cite{CLAS09}
and two pion production 
({\tiny {\color{red} $\blacksquare$}})~\cite{CLAS12,CLAS16a}
and PDG 2022 ({\large {\B $\bm \circ$}})~\cite{PDG2022}.
For $S_{1/2}$ we include also the MAMI data point at 
$Q^2=0.1$ GeV$^2$  ({\Large $\bm \square$})~\cite{Stajner17}.
\label{fig-Roper-Amps1}}

\bigskip
\vspace{.5cm}

\includegraphics[width=3.2in]{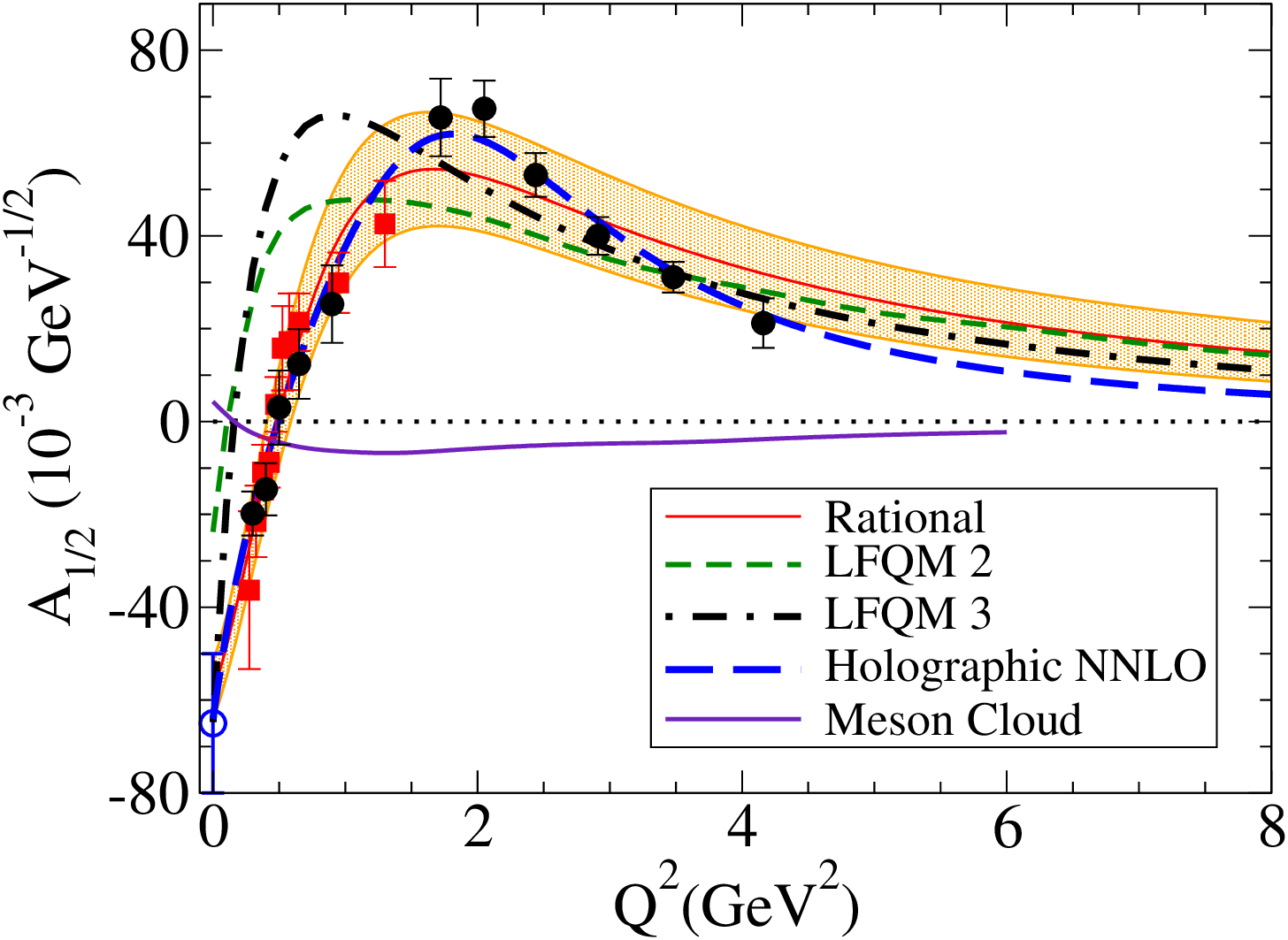} \hspace{1.cm}
\includegraphics[width=3.2in]{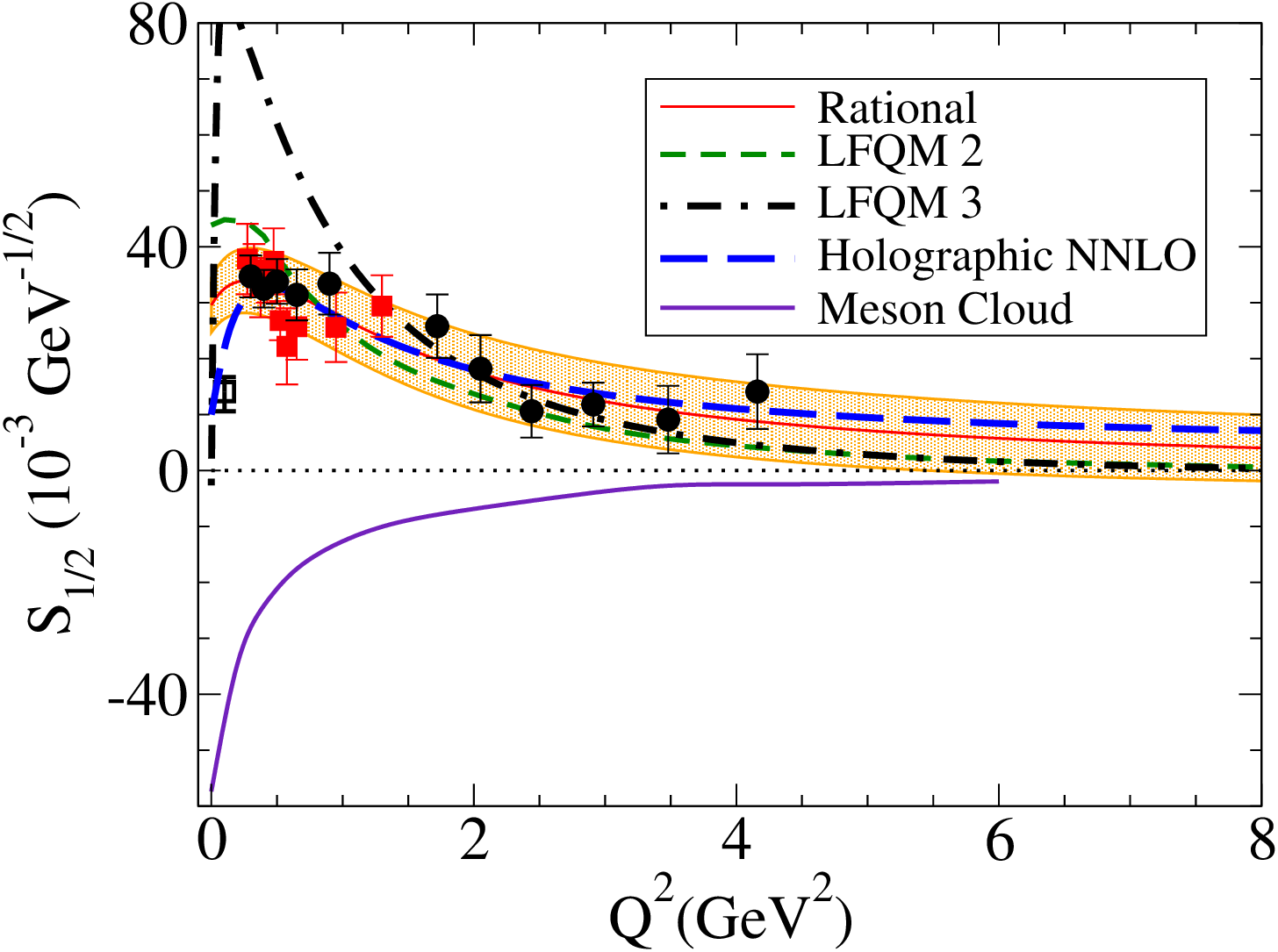} 
\caption{\footnotesize 
Calculations of the $\gamma^\ast N \to N(1440)$ helicity amplitudes with meson cloud contributions.
Calculations from LFQM 2~\cite{Aznauryan12b}, LFQM 3~\cite{Obukhovsky19a}
and Holographic QCD calculation in next-to-next leading order~\cite{Gutsche18a}.
Calculation of meson cloud contributions from 
ANL-Osaka DCC model~\cite{Kamano16a,Kamano13a,Nakamura15a,Sato16a} also shown, as well as
comparison with Rational function parametrization of Ref.~\cite{Eichmann18}. 
Data as in Fig.~\ref{fig-Roper-Amps1}.
Uncertainty bands calculated from the uncertainties of the form factor data.
\label{fig-Roper-Amps2}}
\end{figure}

We discuss now the $\gamma^\ast N \to N(1440)$  transition 
from the experimental and theoretical and point of view.
The data for the helicity amplitudes are presented in 
Figs.~\ref{fig-Roper-Amps1} and \ref{fig-Roper-Amps2},
in comparison with several model calculations. 
The displayed data includes the analysis of the single pion 
electroproduction~\cite{CLAS09} and double
pion electroproduction~\cite{CLAS12,CLAS16a} from JLab/CLAS.
Included in the graphs are also the PDG data at the photon point~\cite{PDG2022}
and the measurement of $S_{1/2}$ at $Q^2=0.1$ GeV$^2$ from MAMI~\cite{Stajner17}.
The low-$Q^2$ data is debated later here, after the discussion of the results 
in terms of the transition form factors. 
In the figures, we display also the MAID parametrization, 
which take into account  the world data~\cite{Drechsel07} 
and the Rational functions parametrization.
Not included in the figures are models which assume a $(qqq)g$ configuration~\cite{Li92a}.
These models predict the wrong sign for $A_{1/2}$ 
and a negligible $S_{1/2}$~\cite{Aznauryan12a}.

From Figs.~\ref{fig-Roper-Amps1} and \ref{fig-Roper-Amps2}, 
one concludes that the amplitude $S_{1/2}$ does not change sign with $Q^2$, while the amplitude $A_{1/2}$ 
is negative at low $Q^2$, and changes sign near $Q^2 \approx 0.5$ GeV$^2$.
The observation is valid also for the MAID (Fig.~\ref{fig-Roper-Amps1}) 
and Rational function (Fig.~\ref{fig-Roper-Amps2}) parametrizations.
The zero for the amplitude $A_{1/2}$ is a consequence of the zero of
the form factor $F_2$
[recall the relations (\ref{eqA12-b}) and (\ref{eqS12-b}), upper index]. 
As for the positive values of $S_{1/2}$ 
they are originated by the evolution of the form factors
$F_1$ and $F_2$, particularly the small magnitude of $F_2$ for large $Q^2$.
The form factors $F_1$ and $F_2$ are discussed in the next section.
The tendency of the amplitude $S_{1/2}$ at large $Q^2$, 
a slow or a fast falloff, will be tested by experiments for larger values of $Q^2$.

In Fig.~\ref{fig-Roper-Amps1}, we present results from calculations based 
on valence quark degrees of freedom.
We include calculations from the light-front quark model
from Aznauryan (LFQM 1)~\cite{Aznauryan07},  
the covariant spectator quark model (CSQM)~\cite{N1440-1},
and the AdS/QCD calculation in leading twist approximation from Ref.~\cite{Roper-AdS1}.
All model calculations assume that the $N(1440)$ is a three valence quark system.
The different models provide a fair description of the large-$Q^2$ data, 
an evidence that the Roper can indeed be interpreted as a system of three valence quarks, 
since the photon-quark coupling mechanisms are expected 
to dominate at sufficiently large $Q^2$.

The LFQM from Ref.~\cite{Aznauryan07} from 2007, follows an earlier 
work~\cite{Aznauryan82a}, long time before the CLAS data from 2009 became available, 
and provides a good prediction of the data for $Q^2 > 2$ GeV$^2$.
The calculations from the covariant spectator quark model 
are based on the model for the nucleon in Ref.~\cite{Nucleon} 
and uses a $N(1440)$ radial wave function which 
is fixed by the orthogonality\footnote{The approximation 
is valid when the pseudothreshold $Q^2= -(M_R-M)^2$ 
is not too far away from the photon point $Q^2=0$.
Under this approximation, the calculations 
are expected to be accurate for large $Q^2$.} 
with the nucleon without any further adjustable parameters~\cite{N1440-1,N1710}.
The estimates from the covariant spectator quark model are then true predictions.
The holographic QCD calculation from  Ref.~\cite{Roper-AdS1} is  
based on the formalism from Ref.~\cite{Gutsche13a}, 
but considers only the first Fock state ($qqq$ term).
The three couplings are adjusted by the nucleon data,
and the calculations are predictions of the large-$Q^2$ region.

In Fig.~\ref{fig-Roper-Amps2}, and for comparison to Fig.~\ref{fig-Roper-Amps1}, we show
calculations that include contributions from the meson cloud effects. 
The first observation on the two figures is that the
meson cloud screening effect of the quark core couplings lowers the curves of
the light-front quark models (Refs.~\cite{Obukhovsky19a,Aznauryan12b})
and that the estimate of the meson cloud contribution
from the ANL-Osaka DCC model (Meson Cloud)~\cite{Kamano16a,Kamano13a,Nakamura15a,Sato16a}
is also negative.
We can then conclude that, in general, the inclusion of 
meson cloud effect improves the description of the data,  
particularly below $Q^2=2$ GeV$^2$.

The LFQM from Ref.~\cite{Aznauryan12b} (LFQM 2) includes the meson cloud effects 
in the normalization of the amplitudes, and reduces
the valence quark contribution at large $Q^2$ with
the meson cloud contribution to  the Roper wave function being about 25\%.
In addition, LFQM 2 takes into account the momentum dependence 
of the quark masses as in the Dyson-Schwinger formalism~\cite{Aznauryan12b}.
The LFQM from Ref.~\cite{Obukhovsky19a} (LFQM 3) follows previous works~\cite{Obukhovsky14a},
and takes into account contributions from  $\sigma N$ states.
The holographic QCD model (Holographic NNLO) from Refs.~\cite{Gutsche18a,Gutsche13a} 
includes higher Fock states [$(qqq)g$ and $(qqq)(\bar q q)$]
where the independent couplings are adjusted by
the nucleon data and some $N(1440)$ or nucleon to $N(1440)$ transition data.
The calculation from Ref.~\cite{Gutsche13a}
is calibrated by the experimental result for $A_{1/2} (0)$.
The calculation presented in the figure is from Ref.~\cite{Gutsche18a},
where the couplings are adjusted by the helicity amplitude data.
The models LFQM 2 and Holographic NNLO, include meson cloud effects,
and improve the description of the data based only on valence quark degrees of freedom
(Fig.~\ref{fig-Roper-Amps1}).
One notices, however, that light-front quark models
due to its nature are more appropriate to the region
$Q^2 > 2$ GeV$^2$~\cite{Aznauryan07,Aznauryan17a}.

The second observation on Figs.~\ref{fig-Roper-Amps1} and \ref{fig-Roper-Amps2},
is that the different approaches predict
different asymptotic results for large $Q^2$ and possible data
at larger $Q^2$ may discriminate between them. 
For a more detailed comparison with earlier light-front quark 
models from Refs.~\cite{Capstick95,Weber90a,JDiaz04a} 
and with the gluonic model~\cite{Li92a}, we suggest the reader the
Refs.~\cite{Aznauryan12a,Aznauryan07,CLAS08,Cardarelli97a}.

On the magnitude of the meson cloud contributions,
estimated by the ANL-Osaka DCC model, displayed on Fig.~\ref{fig-Roper-Amps2},
we note that the ANL-Osaka DCC amplitudes are complex due to the opening
of the meson production channels, but for simplicity we present only the real parts. 
The  meson cloud contribution to $A_{1/2}$ is very small,
and the contribution for $S_{1/2}$ is large, differing in sign in the vicinity of $Q^2$.
However, it is important to mention that the imaginary parts 
of the meson contributions to  $A_{1/2}$ and $S_{1/2}$  
have a magnitude similar to the real parts.

We underline that as shown in Fig.~\ref{fig-Roper-Amps1} the MAID parametrization 
describes well the CLAS data in the range $Q^2=0$--4.2 GeV$^2$, its range of validity. 
Above $Q^2=5$ GeV$^2$ the parametrization 
falls off very fast with an exponential factor.
The Rational parametrization~\cite{Eichmann18} has the advantage 
of being compatible with the pseudothreshold constraints which are
$A_{1/2} \; \propto \; |{\bf q}|$ and $S_{1/2} \; \propto\; |{\bf q}|^2$,
and its slope at small $Q^2$ follows well the data.
The band of variation shown is estimated from the uncertainties
obtained from the form factor data.
The magnitude of the uncertainties is smaller for the transition form factor data
(see next section).
The large uncertainties in the amplitudes
are mainly a consequence of the conversion 
from transition form factors with the errors calculated in quadrature.
In the present case,  the inclusion of the uncertainties is important
to understand the limits of the parametrizations,
near the pseudothreshold, and also for their large-$Q^2$ predictions.

An important matter
is the expected shape of the helicity amplitudes 
or the transition form factors at low $Q^2$.
Comparing Figs.~\ref{fig-Roper-Amps1}  
and \ref{fig-Roper-Amps2}, one notices the differences 
between the MAID and the Rational parametrizations, 
particularly for the amplitude $S_{1/2}$.
Since  $S_{1/2}$ cannot be measured at $Q^2=0$,
one has to rely on data near $Q^2=0$. 
One can nevertheless analyze the limit  of (\ref{eqS12-b})
\ba
S_{1/2} (0)= (M_R^2 -M^2) \sqrt{\frac{\pi \alpha K}{M M_R}} 
\left[\left.\frac{d F_1}{d Q^2} \right|_{0} - \frac{F_2(0)}{(M_R + M)^2} \right],
\label{eqS120}
\ea
where $\left. \frac{d F_1}{d Q^2}\right|_{0}$ represents
the first derivative of $F_1$ at $Q^2=0$, 
and can be inferred from the expansion 
$F_1(Q^2) \simeq \left. \frac{d F_1}{d Q^2}\right|_{0} Q^2$.
The sign of $S_{1/2} (0)$ is then determined by $\left. \frac{d F_1}{d Q^2}\right|_{0}$
and by the term on $F_2(0)$, which is positive.

The Rational parametrization suggests that $S_{1/2}(0)$ is large and positive, 
$S_{1/2}(0) \simeq 30 \times 10^{-3}$ GeV$^{-1/2}$, while the data from MAMI 
and the MAID parametrization point to a smaller value.
But, as we will see in the next section on the form factor data,
a smaller values for $S_{1/2} (0)$ may imply 
negative values for  $\left. \frac{d F_1}{d Q^2} \right|_{0}$, and consequently 
an atypical shape for $F_1$ near $Q^2=0$,  
The determination of the magnitude of $S_{1/2}$ below $Q^2=0.3$ GeV$^2$ 
is then crucial.
Notice also that a consequence of the pseudothreshold constraints 
is that both amplitudes, $A_{1/2}$ and  $S_{1/2}$,
must vanish in the limit $Q^2= -(M_R-M)^2$. 
In the Rational parametrization the indications of the inflection
of the amplitude $S_{1/2}$ is clear already near $Q^2=0$, but
as for $A_{1/2}$ the inflection can be seen only below $Q^2=0$, 
as documented in Ref.~\cite{Eichmann18}.

In this presentation we avoid the discussion of 
calculations based on chiral EFTs, which are valid below $Q^2=0.6$ GeV$^2$~\cite{Bauer14a,Gelenava18a}
due to the lack of data below $Q^2=0.3$ GeV$^2$.
We report, nevertheless that the chiral model from Mainz~\cite{Bauer14a} 
provides a good description of the data 
once the low energy constants are fixed.

\begin{figure}[t]
\centering

\includegraphics[width=3.2in]{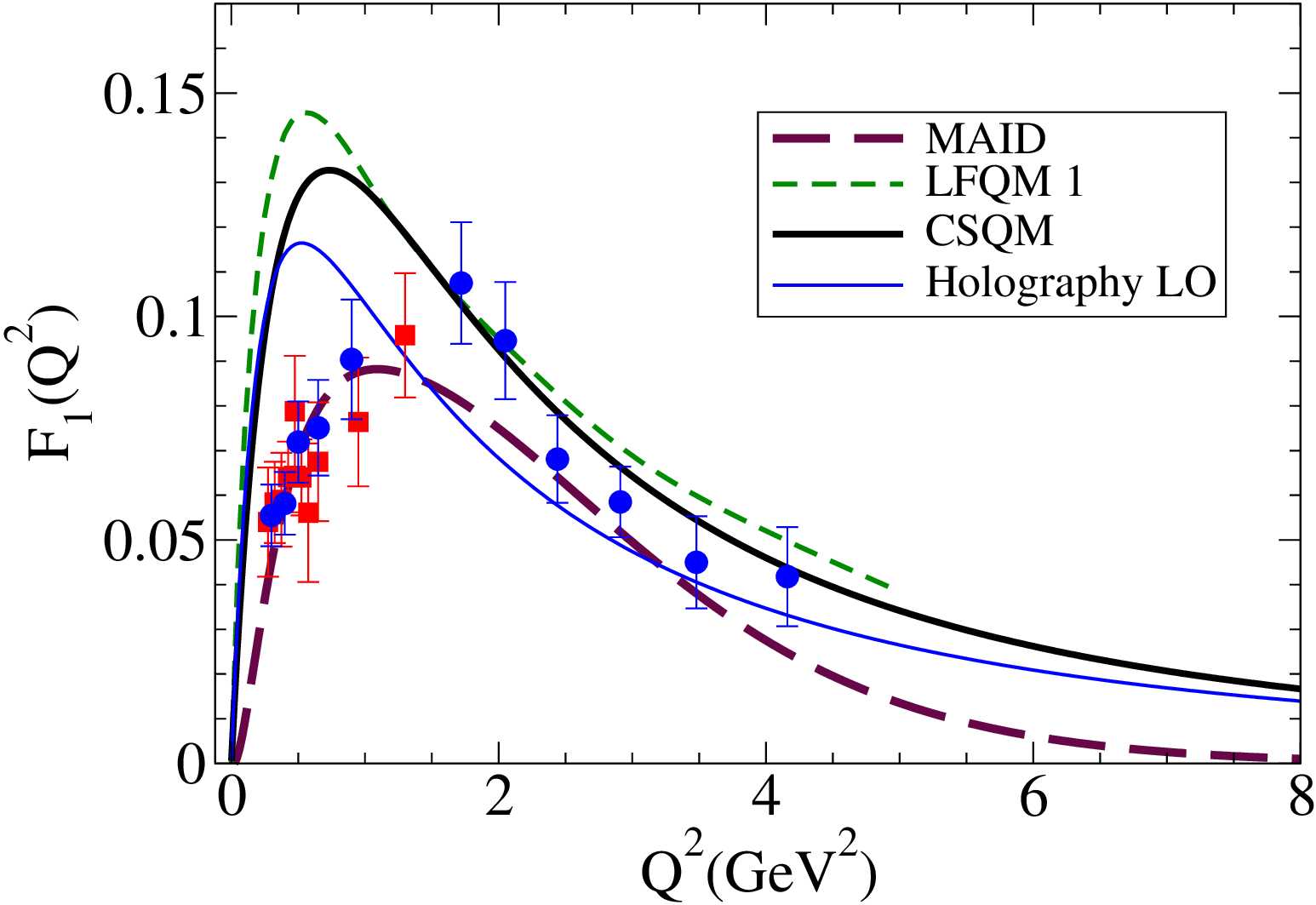} \hspace{1.cm}
\includegraphics[width=3.2in]{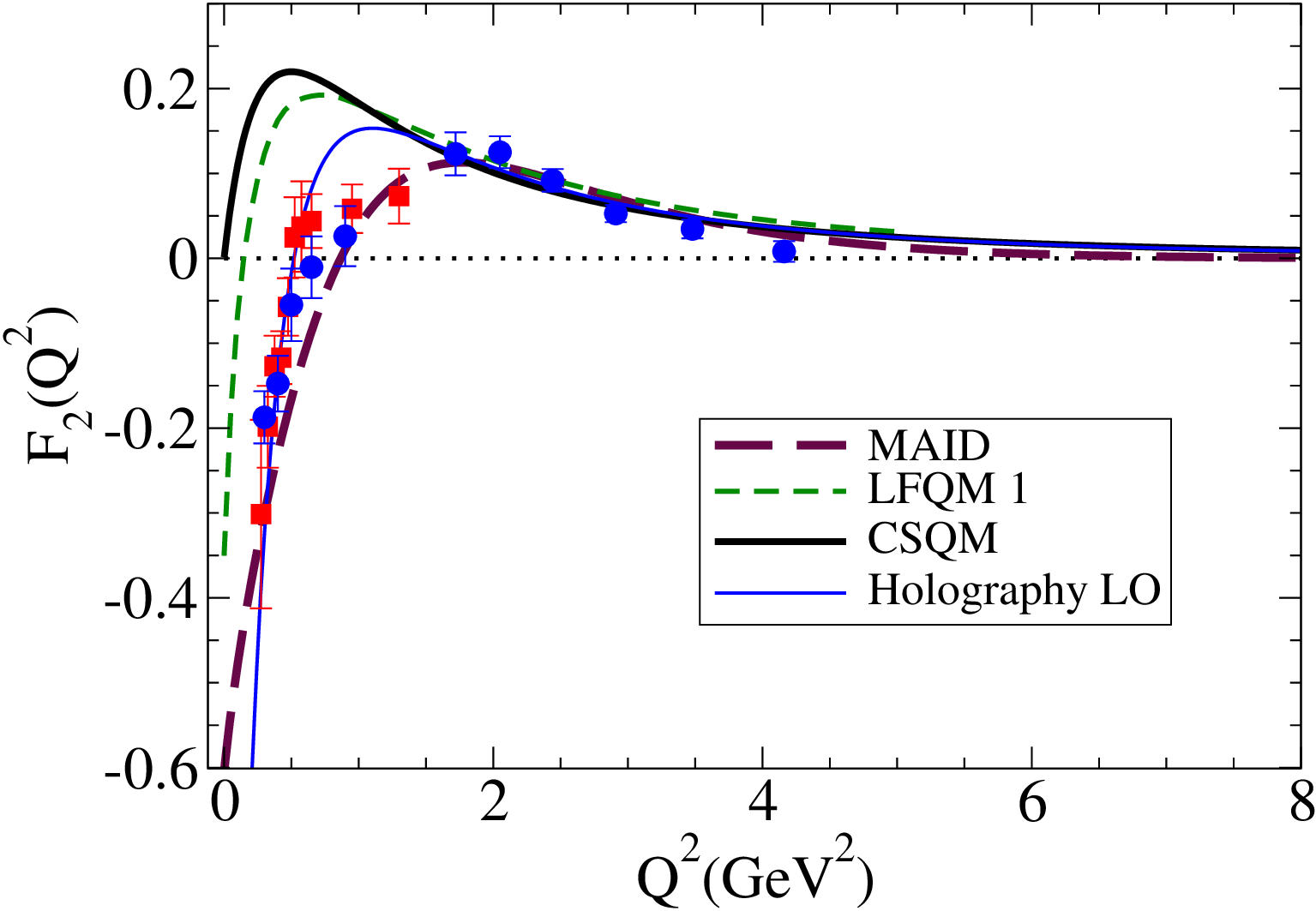} 
\caption{\footnotesize 
Calculations of $\gamma^\ast N \to N(1440)$
transition form factors.
Focus on the bare contributions.
The figure includes calculations of light-front quark model  (LFQM 1)~\cite{Aznauryan07}, 
covariant spectator quark model (CSQM)~\cite{N1440-1}, 
and Holographic QCD model in leading order (LO, leading twist)~\cite{Roper-AdS1}.
Comparison with the MAID parametrization~\cite{Drechsel07,MAID2009,MAID2011}.
The data are from JLab/CLAS, one pion production 
({\color{blue}{\Large $\bullet$}})~\cite{CLAS09}
and two pion production 
({\tiny {\color{red} $\blacksquare$}})~\cite{CLAS12,CLAS16a},
and PDG 2022 ({\large {\B $\bm \circ$}})~\cite{PDG2022}.
\label{fig-Roper-FF1}}

\bigskip
\vspace{.5cm}

\includegraphics[width=3.2in]{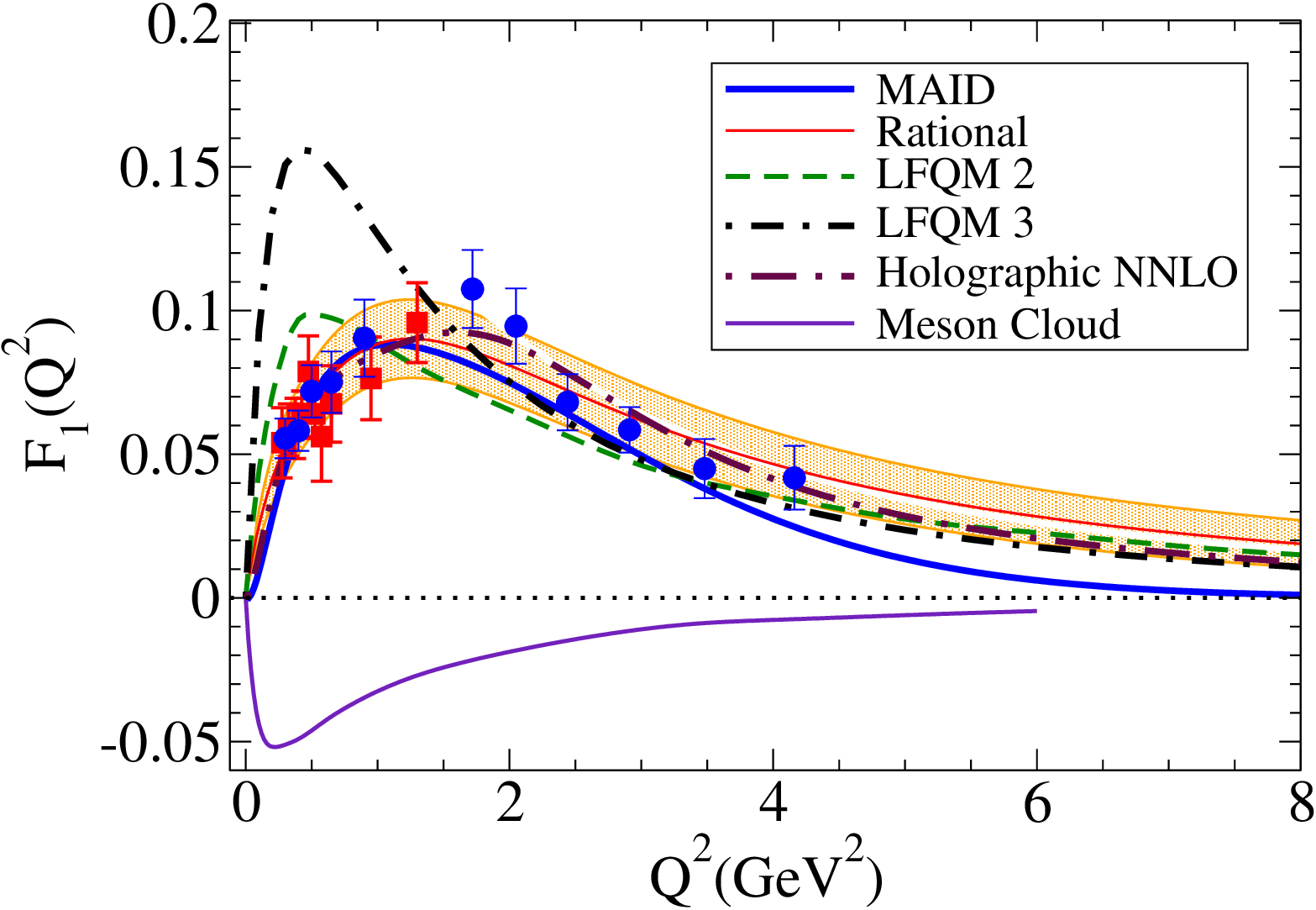} \hspace{1.cm}
\includegraphics[width=3.2in]{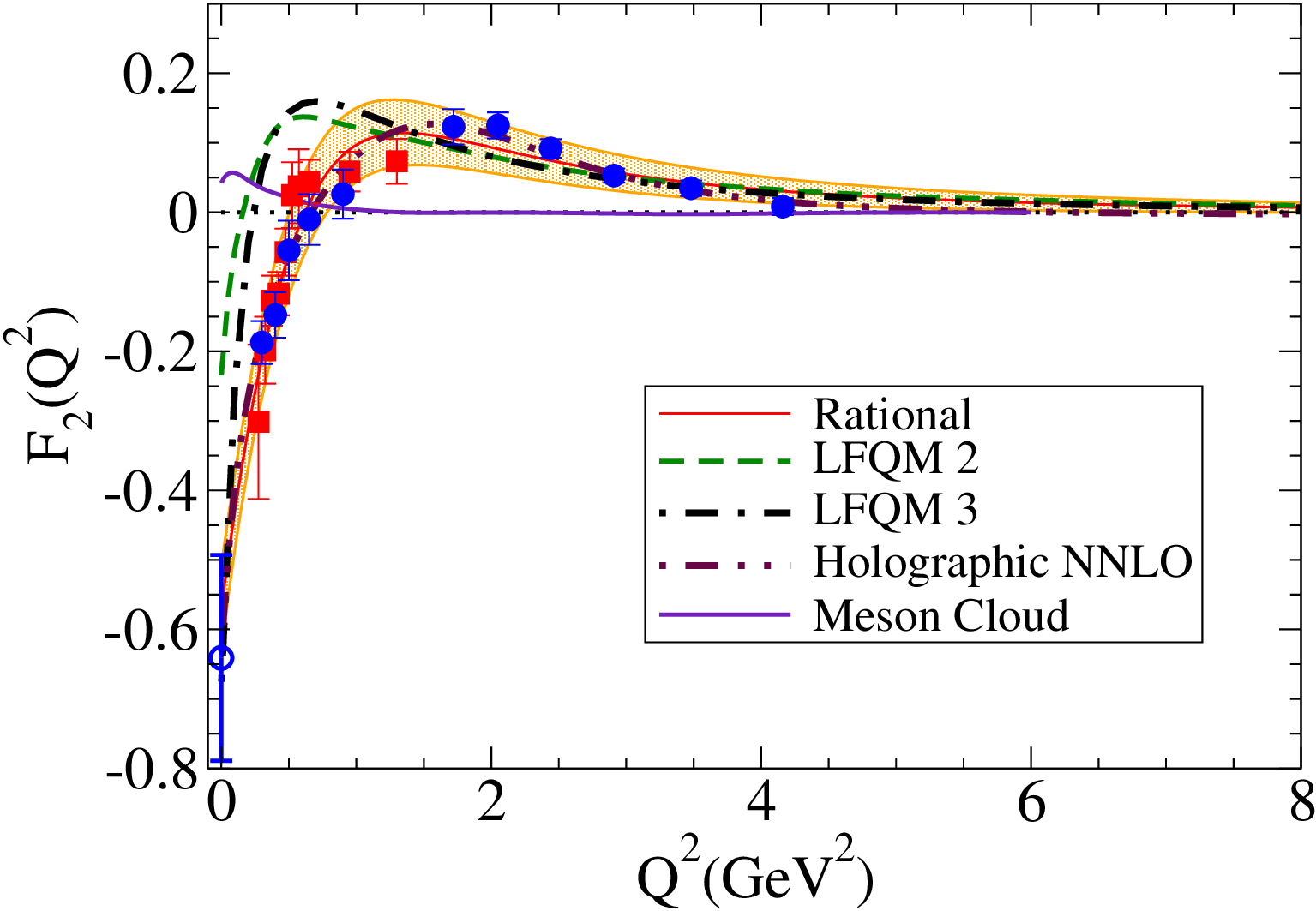} 
\caption{\footnotesize 
Calculations of $\gamma^\ast N \to N(1440)$
transition form factors.
Calculations from LFQM 2~\cite{Aznauryan12b}, LFQM 3~\cite{Obukhovsky19a}
and holographic QCD calculation in next-to-next leading order~\cite{Gutsche18a}
Comparison with the Rational parametrization~\cite{Eichmann18}.
Include calculation of meson cloud contributions from 
ANL-Osaka DCC model~\cite{Kamano16a,Kamano13a,Nakamura15a,Sato16a}.
Data as in Fig.~\ref{fig-Roper-FF1}. Uncertainty bands calculated from the uncertainties
of the form factor data.
\label{fig-Roper-FF2}}
\end{figure}

\subsubsection{\it Transition form factors  \label{sec-Roper-FFs}}

The conversion of the helicity amplitudes to the transition form factors
is presented in Figs.~\ref{fig-Roper-FF1} and \ref{fig-Roper-FF2},
based on the relations from Section~\ref{sec-spin12}.
Notice that, contrary to the helicity amplitudes, 
the form factors $F_1$ and $F_2$ are independent 
and uncorrelated functions.

The form factor data indicate that $F_1$ is positive 
and $F_2$ has a zero near $Q^2 \simeq 0.5$ GeV$^2$.
Zeros at the form factors are expected for states that 
are radial excitations of a ground state~\cite{Burkert19a,Eichmann16}.
The zero of $F_2$ was found in 
lattice QCD simulations for $m_\pi \simeq 400$ MeV~\cite{Lin08,Lin12a}.

The differences between the MAID and the Rational parametrizations 
become more clear in the form factor representation.
The faster falloff of MAID for large $Q^2$ is very evident in the graph for $F_1$
from  Fig.~\ref{fig-Roper-FF1}, and the difference in 
the $Q^2$-dependence near $Q^2=0$  of  both representations is also visible.
In the MAID parametrization $F_1$ behaves 
as $F_1 \; \propto \; Q^2 (Q^2 - Q_0^2)$ where $Q^2_0 =0.023$ GeV$^2$.
The MAID parametrization for $F_1$ resembles then $F_1 \; \propto \; Q^4$ near $Q^2=0$. 
The consequence of the form $F_1 \; \propto \; Q^2 (Q^2 - Q_0^2)$
is that $\left. \frac{d F_1}{d Q^2}\right|_{0} \propto \; -Q^2_0$
becoming  negative for MAID in the close vicinity of the origin
(not evident in the figure due to the scale),
while a simpler description of the data suggests instead
a positive derivative for $F_1$. 
There are at the moment, no physical justification
for the existence of a zero of $F_2$ near $Q^2=0$, and 
it comes as a consequence of the uncorrelated parametrization  
of the amplitudes $A_{1/2}$ and $S_{1/2}$.

The form factors results justify the signs of $A_{1/2}\;  \propto \; (F_1 + F_2)$ 
and $S_{1/2} \; \propto \;  (F_1 - \tau F_2)$ shown in the data of the previous section.
$A_{1/2}$ is negative at low $Q^2$ and became positive at large $Q^2$, 
when both form factors are positive.
The function $S_{1/2}$ is positive at low-$Q^2$ because $F_1$ and $- \tau F_2$ 
are both positive.
Notice, however, that we cannot exclude the possibility 
that $S_{1/2}$ becomes negative at large $Q^2$, 
if the positive term $\tau F_2$ became larger than $F_1$.
This possibility appears in the graph for $S_{1/2}$
in Fig.~\ref{fig-Roper-Amps2}, 
where one notices that negative values for $Q^2 > 6$ GeV$^2$ are within
the uncertainty band
obtained with the Rational parametrization.
In contrast, the last data point ($Q^2 = 4.2$ GeV$^2$) seems to
suggest that the amplitude is not falling fast enough to become negative.

The zero for the amplitude $A_{1/2} \;  \propto \; (F_1 + F_2)$  in 
Figs.~\ref{fig-Roper-Amps1} and  \ref{fig-Roper-Amps2}
is naturally explained because $F_1$ and $F_2$ have different 
signs at low $Q^2$. The zero of $A_{1/2}$ is below the zero of $F_2$.
The exact position of the zero, however, 
depends on the meson cloud contributions, since 
those contributions change the shape of $F_1$ and $F_2$ 
at small $Q^2$~\cite{Burkert19a,Segovia19b,Segovia15,Chen19a}.

The Roper transition form factors have also been 
calculated within the Dyson-Schwinger 
formalism~\cite{Wilson12,Segovia15,Chen19a,Segovia16a}.
Recent calculations have a trend similar to the model 
LFQM 2~\cite{Aznauryan12b} and overestimate the data below $Q^2=2$ GeV$^2$~\cite{Segovia15}.
In that region the comparison with the data improves when 
one considers that meson cloud effects may account for 
20\% of the transition reducing the magnitude of $F_1$ 
to close to the measured data~\cite{Burkert19a,Segovia16a}.
Also, the calculated $F_2$  approaches the data but with a different shape.
The meson cloud contributions inferred by the Dyson-Schwinger 
calculations are similar to the ANL-Osaka estimate for $F_1$.
But in its contribution to $F_2$ it differs in sign from the ANL-Osaka estimate,
suggesting that the meson cloud 
contributions may affect $F_1$ and $F_2$  differently.
The quark core of the Roper is augmented by a meson cloud, which also
reduces its quark core mass by about 20$\%$, working out also the
Roper mass puzzle in constituent quark
models treatments~\cite{Burkert19a,Segovia15}.
In Ref.~\cite{Segovia19b}
Segovia makes the breaking down of Roper form factors into the several contributions corresponding to 
the relevant diquark correlations within the baryon. The isoscalar-scalar and isovector-pseudovector diquarks are dominant in the nucleon and in the Roper, as also seen
by Eichmann and Fischer~\cite{Eichmann19a}. 
Interestingly, the results of of Ref.~\cite{Segovia19b}
show for large $Q^2$ the scattering of the photon off
a diquark is almost as important as its scattering off
an uncorrelated bystander dressed-quark.

Overall, one can conclude that models which take into account 
meson cloud effects are in general closer to the 
helicity amplitude and form factors data.
Another observation is that light-front quark models 
may generate large results at low $Q^2$ 
and are more appropriated to region $Q^2 >2$ GeV$^2$~\cite{Aznauryan07}.
The calculations for $F_1$ may be the exception due to
the small magnitude associated with the form $F_1 \; \propto \; Q^2$.

\subsubsection{\it AdS/QCD calculations}

We expand here a bit more on the calculations based on light-front holography
(see Section~\ref{secAdS-QCD}) and their results.

The $\gamma^\ast N \to N(1440)$ transition 
was one of the first transitions to be estimated 
using light-front holography (Holographic QCD). 
The holographic wave function became simpler 
for the $J^P= \frac{1}{2}^+$ systems where the baryon 
can be decomposed into a quark-diquark system~\cite{Gutsche13a,Teramond12b}.

The first holographic study of the 
$\gamma^\ast N \to N(1440)$ transition 
by Teramond and Brodsky~\cite{Teramond12a,Teramond12b} 
derived the following parameter free 
expression for the Dirac form factor
\ba
F_1(Q^2) = 
\frac{\sqrt{2}}{3} \frac{\frac{Q^2}{m_\rho^2}}{
\left(1+ \frac{Q^2}{m_\rho^2} \right) \left(1+ \frac{Q^2}{m_{\rho'}^2} \right)
\left(1+ \frac{Q^2}{m_{\rho''}^2} \right) },
\label{eqRoper-AdS-F1}
\ea
where $m_\rho$, $m_{\rho '}$ and  $m_{\rho ''}$ are 
the masses of the $\rho$ meson and its first excitations.
Notice the simplicity of the expression (\ref{eqRoper-AdS-F1}), where
all the parameters are expressed in terms of physical masses.

\begin{figure}[t]
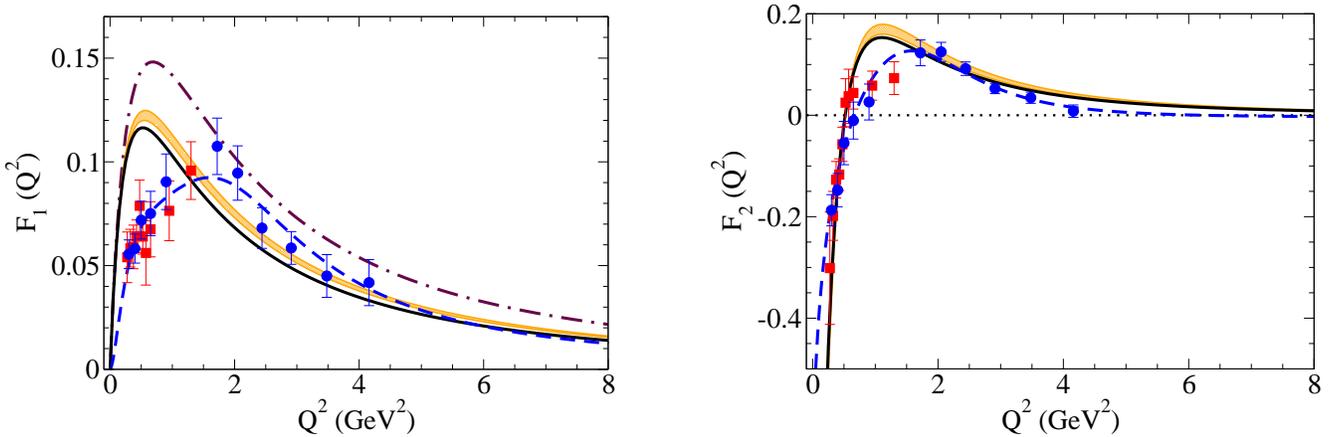

\begin{center}
\includegraphics[width=3.2in]{F1R-N1440} \hspace{1.cm}
\includegraphics[width=3.2in]{F2R-N1440} 
\caption{\footnotesize 
Estimates of $\gamma^\ast N \to N(1440)$ 
transition form factors based on light-front holography.
The dashed-dotted line represent the result 
from Eq.~(\ref{eqRoper-AdS-F1})~\cite{Teramond12a,Teramond12b}.
The dashed line is the twist-5 estimate (NNLO) from 
Ref.~\cite{Gutsche18a}.
The band represents the calculation from Ref.~\cite{Roper-AdS1},
and the solid line the analytic expressions valid for $Q^2 > 2$ GeV$^2$
from Ref.~\cite{Roper-AdS2}. 
Data as in Fig.~\ref{fig-Roper-FF1}.
\label{fig-Roper-AdS}}
\end{center}
\end{figure}

Calculations of the $\gamma^\ast N \to N(1440)$ transition 
form factors $F_1$ and $F_2$ considering $(qqq)g$ 
and $(qqq) q \bar q$ states ($t=4,5$) 
are performed in Refs.~\cite{Gutsche18a,Gutsche13a}.
The authors conclude that the $qqq$ component 
is the dominant contribution to the transition.
One notices, however, that similar parametrizations are obtained 
in leading order and higher orders when the parameters  
are adjusted by static properties
(masses and photocouplings)~\cite{Gutsche13a}.
One has then to be careful in the interpretation 
of the leading order estimates, depending on 
the way the parameters are fixed.
In the twist-5 calculation from Ref.~\cite{Gutsche18a}
(NNLO approach) the couplings are adjusted to
the form factor data.

The transition form factors are also estimated 
at the lowest level in Ref.~\cite{Roper-AdS1}, 
where the free parameters 
are fixed by the nucleon elastic form factor data
above $Q^2= 1.5$ GeV$^2$, to avoid contamination 
$q \bar q$ and $(q \bar q)(q \bar q)$ effects.
A simplified parametrization of the 
transition form factors in terms of the 
nucleon, Roper and $\rho$ meson masses valid 
for $Q^2 > 2$ GeV$^2$ was proposed in Ref.~\cite{Roper-AdS2}.

In Fig.~\ref{fig-Roper-AdS}, we compare 
the calculations from Refs.~\cite{Teramond12a,Gutsche18a,Roper-AdS1,Roper-AdS2}
with the form factor data.
In general for large $Q^2$ the models provide an approximate description 
of the data for $F_1$, and slightly overestimate  the data for $F_2$,
as most quark models.
The estimate from Teramond and Brodsky gives the largest 
contribution to $F_1$.
The band represents the results from Ref.~\cite{Roper-AdS1}
within the uncertainties provided by the nucleon elastic form factor data.
The calculations which include gluon and 
$q \bar q$ contributions improve the description 
of $F_1$ at low $Q^2$ (see dashed line, fit to the data)~\cite{Gutsche18a}.
This may be an indication that meson cloud contributions 
are in fact important for the Dirac form factor.
It is also interesting to notice that the holographic 
estimates for $F_2$ below $Q^2=1$ GeV$^2$ are
in good agreement with the data, 
suggesting that meson cloud contributions may be small 
for this function.
[We omit here the photon point, which is more sensitive to the parametrizations].
For the function $F_2$, one notices also  
the gap between models and data in the range $Q^2=1$--2 GeV$^2$.
Since the data in this window are exclusively 
from two pion production, additional data 
are necessary to confirm if the differences 
are due to models or limitations of the data.

To finish the section, we recall the discussion from Section~\ref{secAdS-QCD},
according with holographic QCD can be very useful as an estimate 
of the contributions from the bare core, but that corrections 
to the leading order should be taken with care.

\subsubsection*{\it Short notes}

In summary, the Roper is today clearly identified as the first radial 
excitation of the nucleon, on basis of state-of-the-art lattice QCD
and Dyson-Schwinger calculations.
Most calculations based on valence quarks 
predict similar falloffs of the Roper form factors in the large-$Q^2$ region.
In addition parametrizations of the form factor data consistent with quark counting rules 
provide a better description of the helicity amplitudes and 
transition amplitudes at large $Q^2$.
This gives support to the
view that constituent quark models continue to be a valuable part
of the strong interaction toolkit to guide 
first principle QCD calculations.
Finally, the low-$Q^2$ data of the form factors is better explained when we consider 
meson cloud dressing associated to $\pi N$ and $\pi \pi N$ channels.


\subsection{\it $N(1535)\frac{1}{2}^-$ resonance \label{sec-N1535}}

The $N(1535)\frac{1}{2}^-$ is a resonance of the 
second resonance region and the lightest negative parity 
nucleon resonance. 
It is the chiral partner of the nucleon $N(939)\frac{1}{2}^+$.
If the chiral symmetry is exact the nucleon and 
$N(1535)$ would have the same mass~\cite{Capstick00}. 

Most of the data about the $\gamma^\ast N \to N(1535)$ transition 
come from $\eta N$ decay channel (JLab/Hall C and the
CLAS)~\cite{Denizli07,Dalton09,Armstrong99,Thompson01,Aznauryan05b} 
and $\pi N$ decay channel (CLAS)~\cite{CLAS09} 
in the period of 1999-2009 (see Table~\ref{table-Data1}).
The data extracted from $\eta$ production
determines the transverse amplitude $A_{1/2}$ 
in the range $Q^2=$0.2--4.0 GeV$^2$ 
and $Q^2= 5.7$ and $7.0$ GeV$^2$~\cite{Dalton09},
under the assumption that 
the longitudinal amplitude $S_{1/2}$ is negligible.
The data from CLAS provide 
the first results for the longitudinal amplitude $S_{1/2}$
for $Q^2=$0.3--4.2 GeV$^2$~\cite{CLAS09}.
The results from CLAS demonstrate the dominance
of the amplitude $A_{1/2}$ and the slow falloff
of the amplitude~\cite{Aznauryan12a}.
The analysis of the data presumes that the $\pi \pi N$
branching ratio is negligible (near 1\%)~\cite{PDG2022}
and that the $\pi N$ and $\eta N$ branching 
ratios are almost equal~\cite{Aznauryan12a,CLAS09}
\ba
\beta_{\pi N} \simeq 0.485, 
\hspace{1.5cm}
\beta_{\eta N} \simeq 0.46.
\ea

An interesting aspect about the $N(1535)\frac{1}{2}^-$ 
is its vicinity in mass with another $N(\frac{1}{2}^-)$ state 
the $N(1650)\frac{1}{2}^-$ resonance.  
The two resonances differ in their decay modes.
Within the constituent quark model picture 
the $N(1535)$ and $N(1650)$ resonances can be represented 
as a mixture of two different configurations 
with relative angular momentum $L=1$  
with core spin  $S=\frac{1}{2}$ and $S=\frac{3}{2}$ 
and negative parity~\cite{Capstick00,Burkert-SQTM}
\ba
& &
\left|N(1535)  \right>  = \cos \theta_S \left| N^2 \; \frac{1}{2}^- \right> 
-   \sin \theta_S \left| N^4 \; \frac{1}{2}^- \right>, \nonumber \\
& &
\left|N(1650)  \right>  = \sin \theta_S \left| N^2\;  \frac{1}{2}^- \right> 
+   \cos \theta_S \left| N^4 \; \frac{1}{2}^- \right>.
\label{eqMixt-N1535}     
\ea
The notation $\left| N^{2S +1} \; \frac{1}{2}^- \right>$
is used to represent combinations of states with $S=\frac{1}{2}$ and $S=\frac{3}{2}$~\cite{Capstick00}.
The mixture of the two states is defined by the mixing angle 
$\theta_S$ which is a consequence of the $SU(6)$ breaking 
due to the color hyperfine interaction 
between quarks, and can be generated by different sources:
one-gluon exchange, one-pion exchange or 
Goldstone-boson exchange~\cite{Capstick00,Isgur78a,Chiang03a,He03a}.
The label $S$ is derived from the spectroscopic notation
$S_{11}$ associated with the $S$-wave pion on the $\pi N$ decay.
The angle can be estimated by the decay properties of 
the two nucleon resonances~\cite{Isgur77a,Hey75a}.
The present estimate is $\theta_S \simeq - 31^\circ$~\cite{Aznauryan12a}.

Traditionally, quark models with a confining potential
predict that the mass of the $N(1535)$ resonance
is smaller than the mass of the 
$N(1440)$~\cite{Capstick00,Isgur78a,Glozman96b,Giannini91}.
The same order appears in lattice QCD simulations with $m_\pi > 500$ MeV.
Only when the pion masses became smaller 
than 300 MeV, we recover the 
order of the physical masses~\cite{Mathur05a}.
These results suggest that the quark-antiquark contributions 
or meson cloud effects are important for the 
description of the physical properties of 
the $N(1535)$ and the  
$\gamma^\ast N \to N(1535)$ transition~\cite{An09a,JDiaz06a,Golli11a,Liu16a,Meyer01,Zhao02,Liu06a,Guo22a}.

The $N(1535)$ resonance can also be described by 
effective chiral meson-baryon models~\cite{Chiang03a,Liu06a,Doring10a,Kaiser95a,Inoue02a,Bruns11a,Garzon14a,Khemchandani13a}
and meson-baryon coupled-channel models~\cite{Jido08a,Ronchen13a,Doring09a,LambdaStar,Krehl00a,Schneider06a,Nacher99a,Nieves01a,Hyodo08a}.
In these formalisms the $N(1535)$ and $N(1650)$ resonances 
can be dynamically generated by one or more 
baryon-meson states, and the pole positions 
and decay rates can be estimated. 
The contributions to the transition currents 
can be regarded as meson cloud effects 
since the photon interacts with baryon-meson systems.
Calculations based on the chiral unitary model conclude 
that the resonance is mainly based on baryon-meson components,
although genuine quark states can help to improve the agreement 
with the data~\cite{Jido08a,Hyodo08a}.
Also coupled-channel dynamical models have been applied to the description 
of the $N(1535)$ resonance  and the 
$\gamma^\ast N \to N(1535)$ transition~\cite{Drechsel07,JDiaz09a,Kamano09a,Tiator04,Golli11a,Chen03a}.
Some dynamical models include processes that 
can be interpreted as contributions from 
valence quarks to the transitions 
or bare core contributions~\cite{Kamano16a,Kamano13a,Nakamura15a,JDiaz09a}.

The  $\gamma^\ast N \to N(1535)$ transition is a
very interesting reaction from the theoretical point of view, 
because in spite of the variety of frameworks tested,
there is at the moment, no model 
that describes the measured helicity transition amplitudes 
$A_{1/2}$ and $S_{1/2}$, or the transition form factors $F_1$ and $F_2$, 
in the full range of $Q^2$.

\subsubsection{\it Helicity amplitudes \label{sec-N1535-Amps}}

The analysis of the $\gamma^\ast N \to N(1535)$ transition
has traditionally been done  in terms of the 
helicity amplitudes $A_{1/2}$ and $S_{1/2}$.
Along the years different calculations of the helicity amplitudes
were developed.
There are calculations based on constituent quark models, 
starting with the Karl-Isgur model~\cite{Capstick00,Isgur78a,Isgur77a},
non relativistic quark models~\cite{Santopinto12,Meyer01,Aiello98a}, 
relativistic quark models~\cite{Warns90,Konen90a},
light-front quark models~\cite{Pace99a,Capstick95,Obukhovsky19a,Aznauryan12b,Aznauryan17a,Pace00a},
and quark models which also take into 
account meson cloud effects~\cite{Helminen02a,An09a,Golli11a,Zhao02}.
In the class of the models based on the valence quark degrees of freedom 
there are also calculations within the frameworks of 
QCD sum rules~\cite{Braun09,Anikin15a,Aliev13a}
and AdS/QCD~\cite{Gutsche20a,Bayona12a}.

\begin{figure}[t]
\centering 

\includegraphics[width=3.25in]{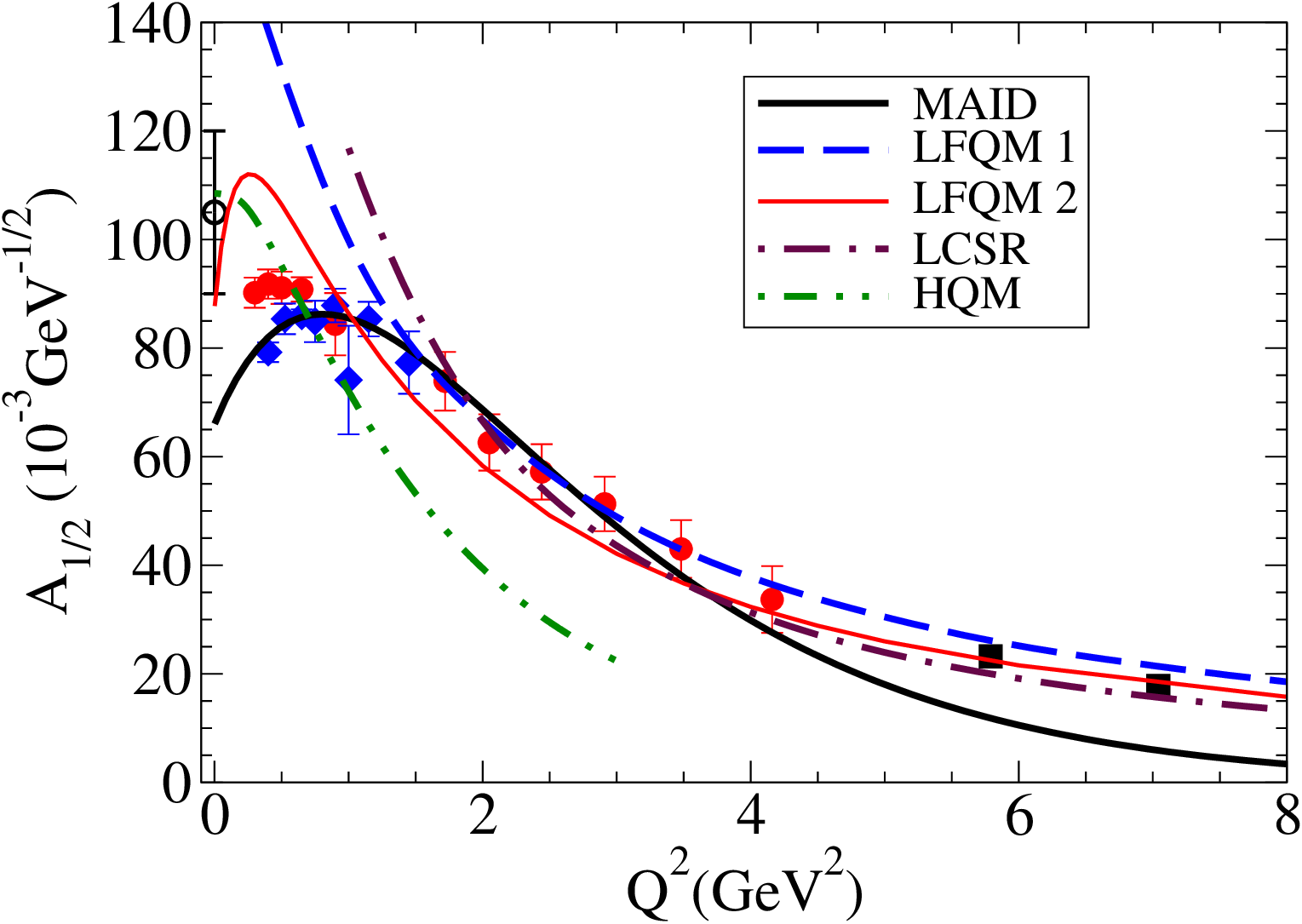} \hspace{.5cm}
\includegraphics[width=3.25in]{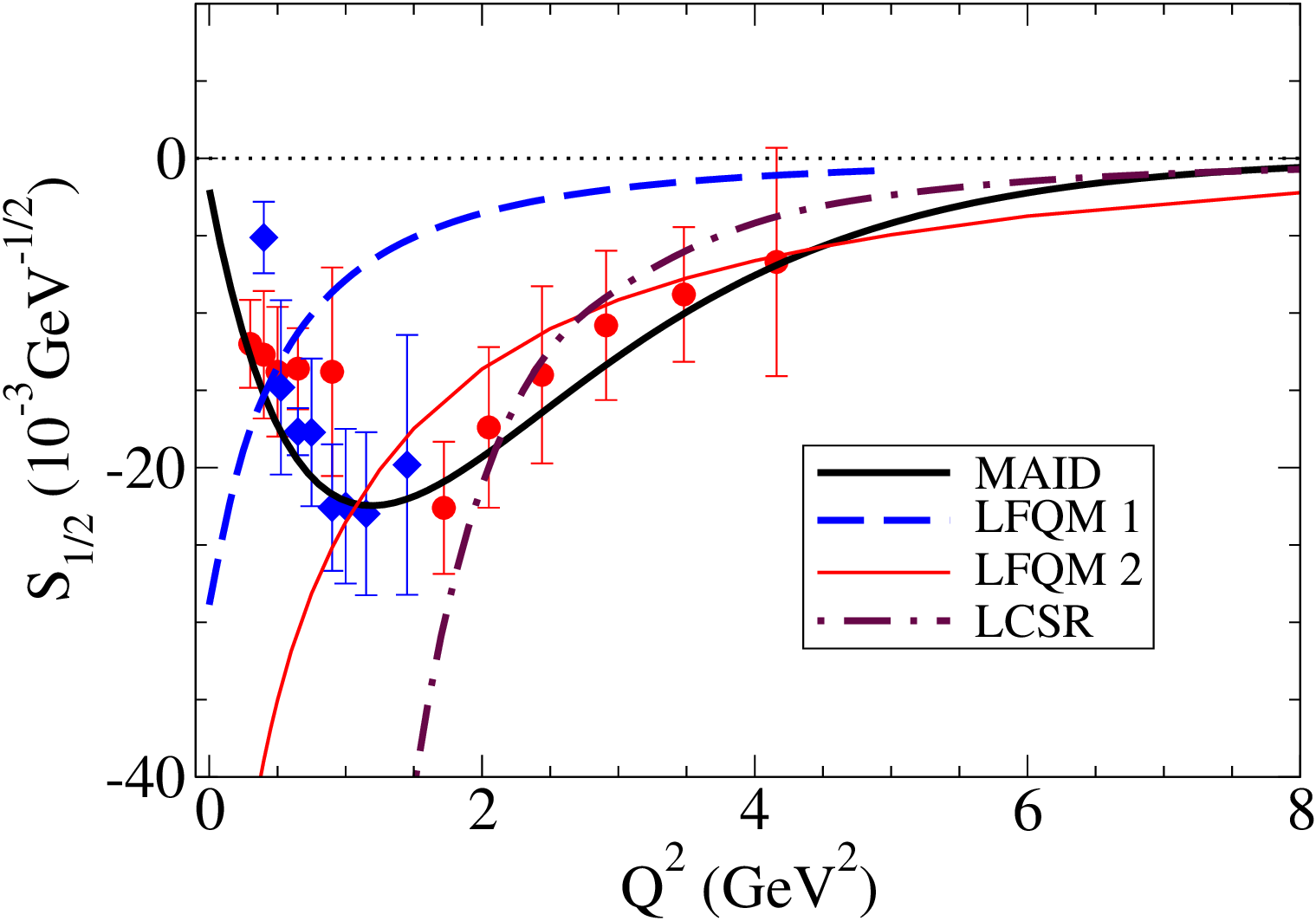} 
\caption{\footnotesize 
$\gamma^\ast N \to N(1535)$ helicity amplitudes. 
Comparison with the light-front quark models 
LFQM 1~\cite{Aznauryan17a}, LFQM 2~\cite{Obukhovsky19a}, 
the hypercentral quark model~\cite{Santopinto12,Aiello98a}  
and light-cone sum rules calculations LCSR~\cite{Anikin15a}. 
The MAID parametrization is also included.
The data are from JLab/CLAS ({\Large {\color{red} $\bullet$}})~\cite{CLAS09}, 
JLab/Hall C ({\scriptsize $\blacksquare$})~\cite{Dalton09} and 
PDG 2022 ({\Large ${\bm  \circ}$})~\cite{PDG2022}.
We present also the data analysis of MAID 
 ({\Large {\B $\blackdiamond$}})~\cite{Drechsel07,MAID-database}.
\label{figAmps-N1535-1}}

\bigskip
\vspace{.5cm}

\includegraphics[width=3.25in]{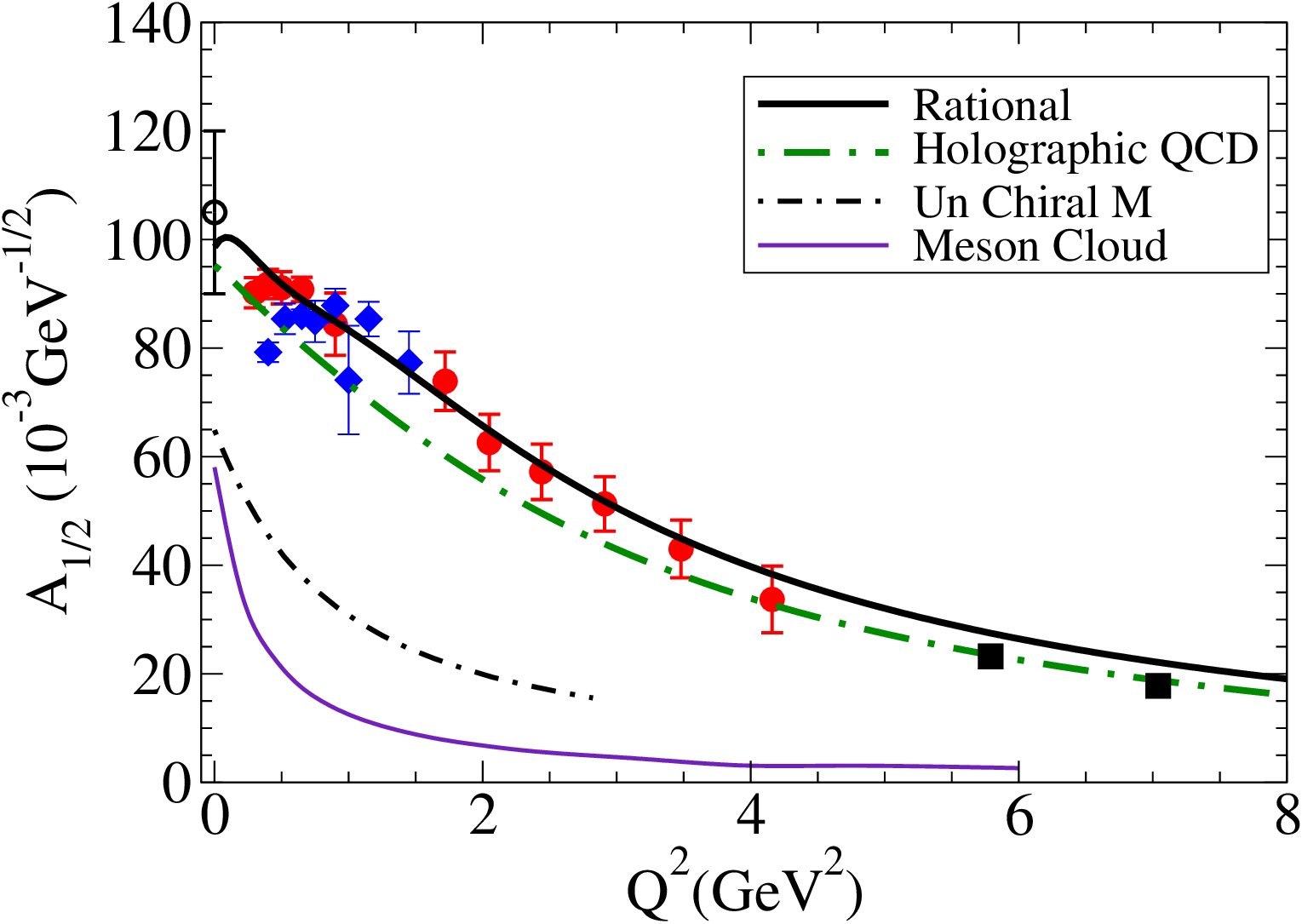} \hspace{.5cm}
\includegraphics[width=3.25in]{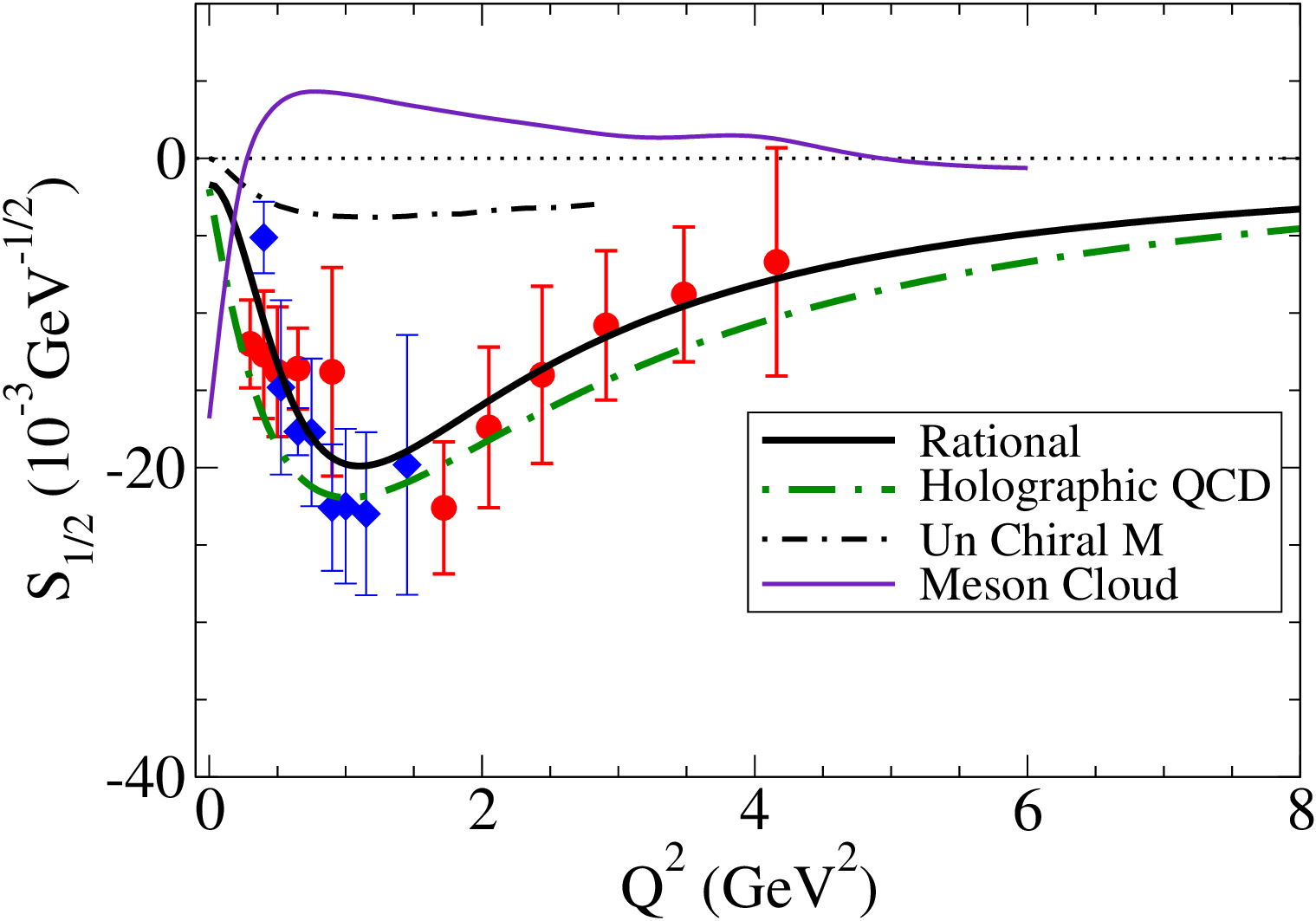} 
\caption{\footnotesize 
$\gamma^\ast N \to N(1535)$ helicity amplitudes.
Comparison with the Unitary Chiral Model~\cite{Jido08a},  
the holographic QCD model from Ref.~\cite{Gutsche20a}
and the Rational parametrization from Ref.~\cite{Eichmann18}.
The estimates are also compared with the 
meson cloud contributions calculated by 
ANL-Osaka DCC model~\cite{Kamano16a,Kamano13a,Nakamura15a}. 
Data as in Fig.~\ref{figAmps-N1535-1}.
\label{figAmps-N1535-2}}

\end{figure}

In Fig.~\ref{figAmps-N1535-1}, we compare results from different models
based on valence quark degrees of freedom. 
In particular, we present the calculations of the light-front 
quark models from Aznauryan and Burkert~\cite{Aznauryan17a} (LFQM 1) 
and Obukhovsky et al.~\cite{Obukhovsky19a} (LFQM 2),
the hypercentral quark model (HQM) from Santopinto,
Giannini et al.~\cite{Santopinto12,Aiello98a} (for $A_{1/2}$), 
and calculations from light-cone sum rules 
from Anikin, Braun et al.~\cite{Anikin15a}.
The LFQM 1 calculations take into account the normalization associated 
to meson cloud effects relevant at low $Q^2$.
The suppression of the electromagnetic amplitudes from the extended meson cloud 
for  the nucleon and the $N(1535)$ is about 20\%.
The LFQM 2 calculations include a mixture of about 10\% 
of meson cloud associated with the $K \Lambda$ channel~\cite{Obukhovsky19a}.
A slightly better description of the amplitudes can be obtained
when we omit the last contribution.
The inclusion of the $\Lambda K$  states provides, however,   
a better description of the form factors data, as will be seen in the next section.
Earlier calculations based on light-front quark models
with similar properties can be found in Refs.~\cite{Pace99a,Capstick95,Pace00a}.
The light-cone sum rule calculations correspond to 
the next-to-leading order in 
the distribution amplitudes 
for the strong coupling constant ($\alpha_s$)~\cite{NSTAR,Anikin15a,Braun06b}.
These calculations take advantage of the similarity 
of the distribution amplitudes of the $N(\frac{1}{2}^+)$ 
and $N(\frac{1}{2}^-)$ states in the 
light-cone sum rules framework~\cite{NSTAR,Braun06b}.
The coefficients of the distribution amplitudes are 
calibrated using lattice QCD simulations~\cite{Braun06b} 
and physical data~\cite{Anikin15a}.
The calculations include quark states with no angular momentum 
and $p$-state quarks. Since they are based on calibrations of the light-cone distribution amplitudes 
in the intermediate $Q^2$~\cite{Anikin15a,Braun06b}, they are valid for the region $Q^2 >2$ GeV$^2$.
Results from  next-to-leading order~\cite{Anikin15a} calculations
include significant contributions from $p$-state quarks and 
improve the description of the data for $S_{1/2}$, compared
to the leading order~\cite{Braun06b}.

The models are compared with JLab/CLAS data~\cite{CLAS09} for both amplitudes, with JLab/Hall C data
for $Q^2 > 4$ GeV$^2$ for the amplitude $A_{1/2}$~\cite{Dalton09},
and with the MAID parametrization~\cite{Drechsel07,MAID2009,MAID2011}.
Additional data for the amplitude $A_{1/2}$ in Refs.~\cite{Denizli07,Armstrong99,Thompson01} from 
JLab is based on the $\eta$ production.
It assumes $S_{1/2}=0$ and for this reason that data are omitted here.
The large-$Q^2$ data from JLab/Hall C~\cite{Dalton09} 
are included here because the suppression of $S_{1/2}$ is more justified for large $Q^2$.
In the present case, we include also data from the MAID analysis 
in the region  $Q^2=1$--2 GeV$^2$, where there are no CLAS data.

From Fig.~\ref{figAmps-N1535-1}, we conclude that 
quark models provide a fair description of the $A_{1/2}$ data only above $Q^2=2$ GeV$^2$.
For the amplitude $S_{1/2}$, LFQM 1 produces a too small magnitude in that region
while LFQM 2 provides a good description of the data for $Q^2 > 1$ GeV$^2$.
At low $Q^2$, light-front quark models overestimate the magnitude of the data.
The fast falloff from the MAID parametrizations is the consequence of the range of the data used 
in the calibration ($Q^2 <5$ GeV$^2$) and the type 
of the parametrizations (exponential falloffs).
For comparison, the Rational parametrization
is shown in Fig.~\ref{figAmps-N1535-2}.

In order to infer the effect of corrections to the results from a three-quark structure only,
in Fig.~\ref{figAmps-N1535-2}, we show model calculations which take into account 
meson cloud dressing effects or explicit 
$q \bar q$ states on the baryon wave functions.
In the figure, we present calculations from holographic QCD (AdS/QCD) from Ref.~\cite{Gutsche20a},
the unitary chiral model~\cite{Jido08a} 
and the meson cloud contributions determined by the 
ANL-Osaka DCC model~\cite{Kamano16a,Kamano13a,Nakamura15a,Sato16a}.
The AdS/QCD calculation takes into account the $(qqq)g$ and $(qqq)\bar q q$ contributions
to the masses and transition amplitudes~\cite{Gutsche20a}.
The mixture coefficients between the different Fock states are calibrated by the data.
From the comparison with the data, one can conclude that 
the additional contributions, including 
 $(qqq) (\bar q q)$ degrees of freedom  improve a description based exclusively on valence quarks.
The calculations of the unitary chiral model from Ref.~\cite{Jido08a} 
describe the $N(1535)$ state as a dynamically generated resonance,  
and the contributions can be interpreted as pure meson cloud effects.
The amplitudes calculated by the
unitary chiral and ANL-Osaka models 
have imaginary parts, due to the opening of the meson production channels.
For simplicity, we present only the results associated to the real part.

The results from Fig.~\ref{figAmps-N1535-2} suggest   
that the meson cloud effects may generate important contributions 
to the helicity amplitudes at low $Q^2$, 
and also that those contributions are suppressed when $Q^2$ increases.
The analytic properties and constraints of the helicity amplitudes at low $Q^2$
are discussed in more detail in Section~\ref{sec-N1535-lQ2}.
In the figure, we include also the Rational parametrizations from Ref.~\cite{Eichmann18}.
These parametrization are obtained considering rational 
functions for the form factors $F_1$ and $F_2$ 
which are adjusted to the CLAS data below 4.2 GeV$^2$.
Compared to the MAID parametrizations, 
one observes much softer falloffs with $Q^2$, 
similar to LFQM 2 and the Holographic QCD model.
In general, models which include meson cloud effects 
provide estimates closer to the data at low $Q^2$.
This is true for the cloudy bag model from Ref.~\cite{Golli11a} 
and for models that take into account
$(qqq)q \bar q$ states~\cite{Helminen02a,An09a,Liu16a,Liu06a}.
The contribution of a state $(qqq) s \bar s$, in particular,  
can help to understand the large $\eta N$ and 
$K \Lambda$--$K \Sigma$ decay branching ratios~\cite{Liu06a}.

For the discussion of the nature of the meson cloud contributions
it is also important to take into account the helicity amplitudes 
associated with the experiments with neutron targets, 
which are known only at the photon point.
The combination of the data for proton and neutron targets at $Q^2=0$: 
$A_{1/2}^p (0) = 0.105\pm0.015$ GeV$^{-1/2}$ 
and $A_{1/2}^n (0) = -0.075\pm0.020$ GeV$^{-1/2}$~\cite{PDG2022},
points out for a dominance of the isovector component in the 
$\gamma^\ast N \to N(1535)$ transition.
Also the meson cloud contributions appear 
to be dominated by the isovector contributions,
according with ANL-Osaka DCC calculations~\cite{Kamano16a,Kamano13a,Nakamura15a}
and meson cloud estimates based on the LFQM 1~\cite{Aznauryan17a}.

Overall, we can conclude the light-front quark models 
and the light-cone sum rules provide a good description of 
the data above 2 GeV$^2$.
The AdS/QCD model from Ref.~\cite{Gutsche20a} 
describe well all ranges of $Q^2$ when the $(qqq)g$ 
and $(qqq) q \bar q$ contributions are included and 
the coefficients adjusted to the helicity amplitude data
(see Fig.~\ref{figAmps-N1535-2}).

\subsubsection{\it Transition form factors \label{sec-N1535-FF}}

\begin{figure}[t]
\centering

\includegraphics[width=3.in]{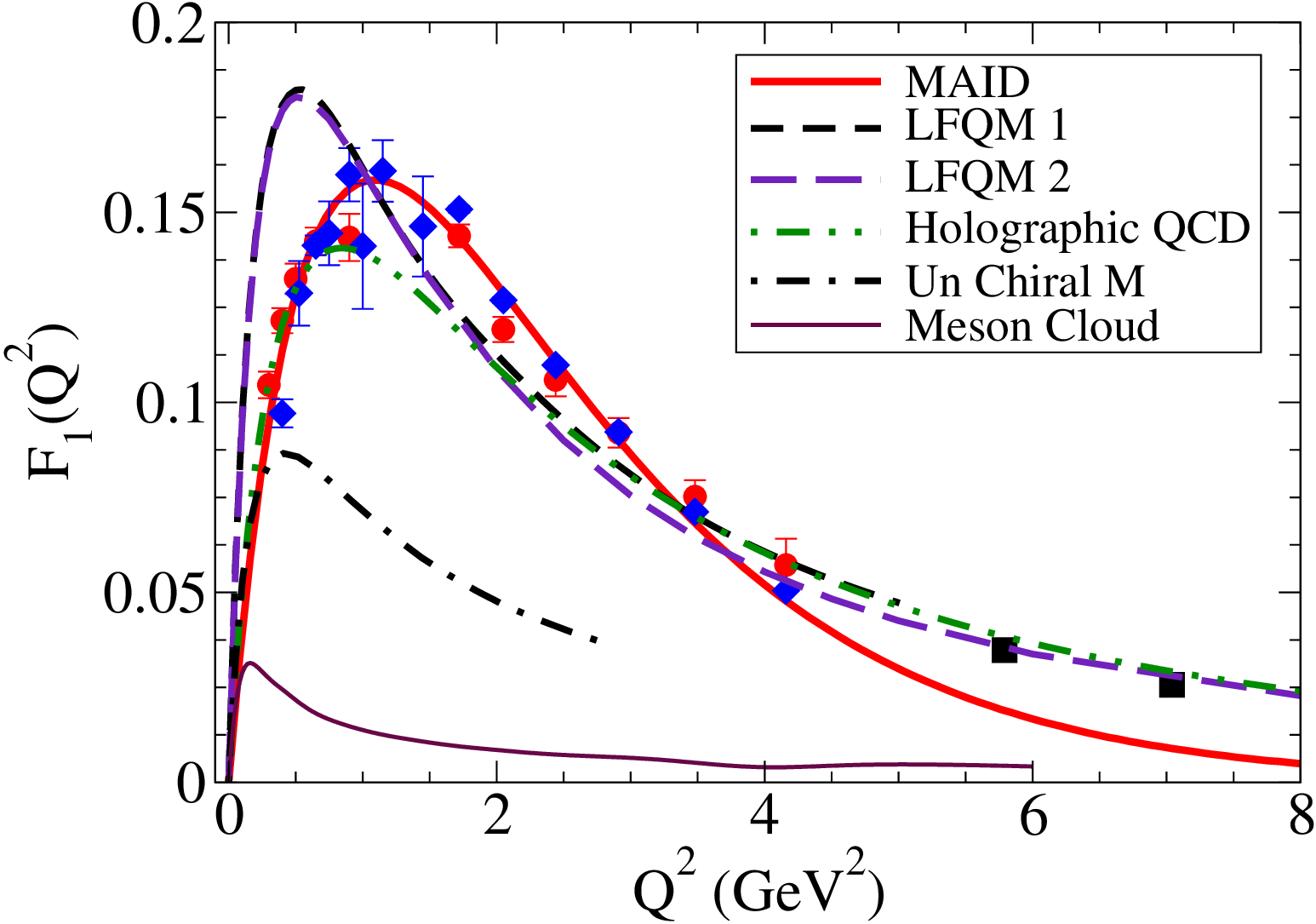} \hspace{.5cm}
\includegraphics[width=3.in]{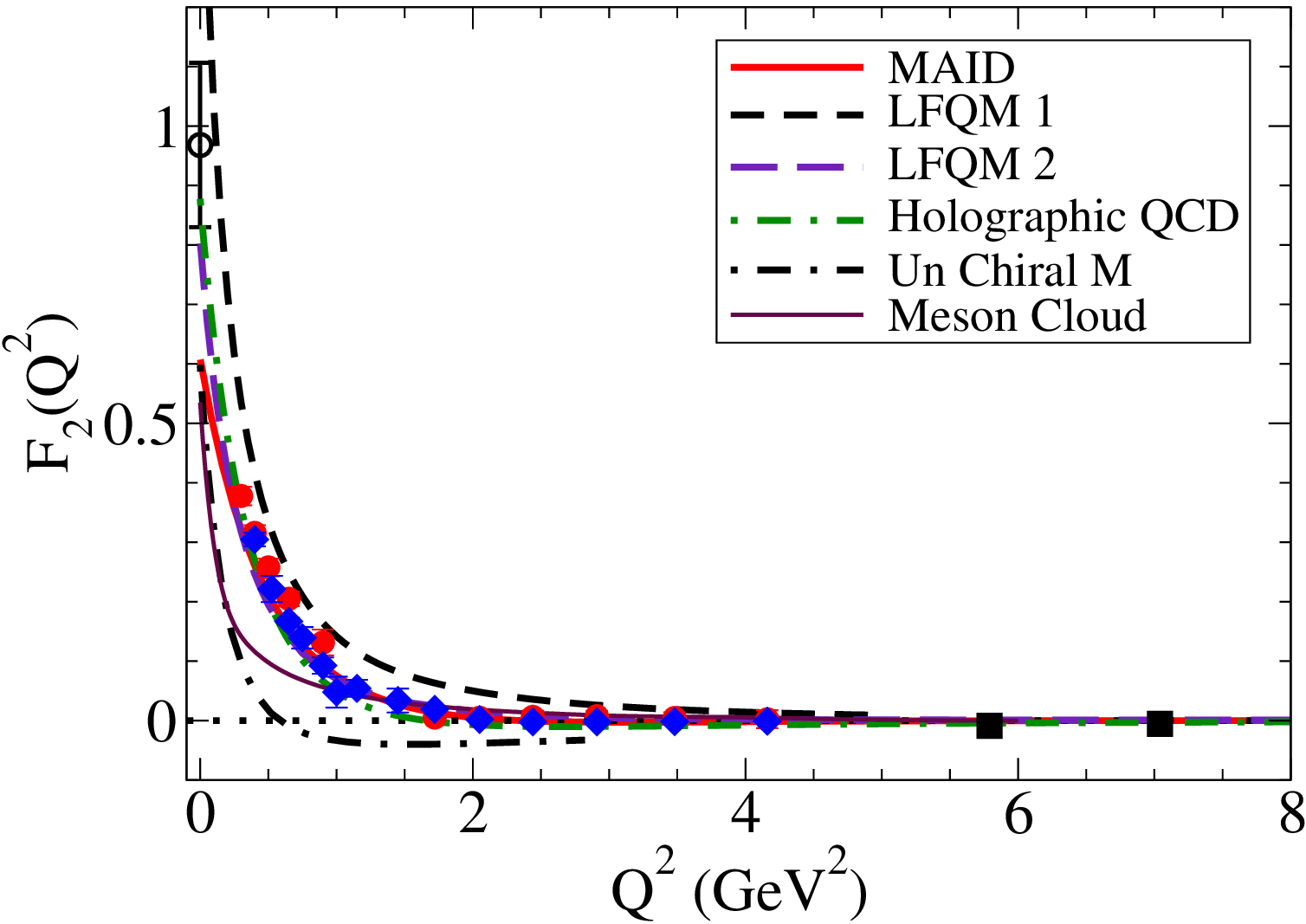} 
\caption{\footnotesize 
$\gamma^\ast N \to N(1535)$ transition form factors.  
Calculations of from LFQM 1~\cite{Aznauryan17a}, 
LFQM 2~\cite{Obukhovsky19a}, 
holographic QCD~\cite{Gutsche20a}, 
unitary chiral model~\cite{Jido08a} and 
light-cone sum rules~\cite{Anikin15a}. 
The estimates are also compared with the 
meson cloud contributions calculations of the 
ANL-Osaka DCC model~\cite{Kamano16a,Kamano13a,Nakamura15a}. 
The data are from JLab/CLAS ({\Large {\color{red} $\bullet$}})~\cite{CLAS09}, 
JLab/Hall C ({\scriptsize $\blacksquare$})~\cite{Dalton09} and 
PDG 2022 ({\Large ${\bm  \circ}$})~\cite{PDG2022}.
We present also the data analysis of MAID 
 ({\Large {\B $\blackdiamond$}})~\cite{Drechsel07,MAID-database}.
\label{figF1F2-N1535}}

\bigskip
\vspace{.5cm}

\includegraphics[width=3.in]{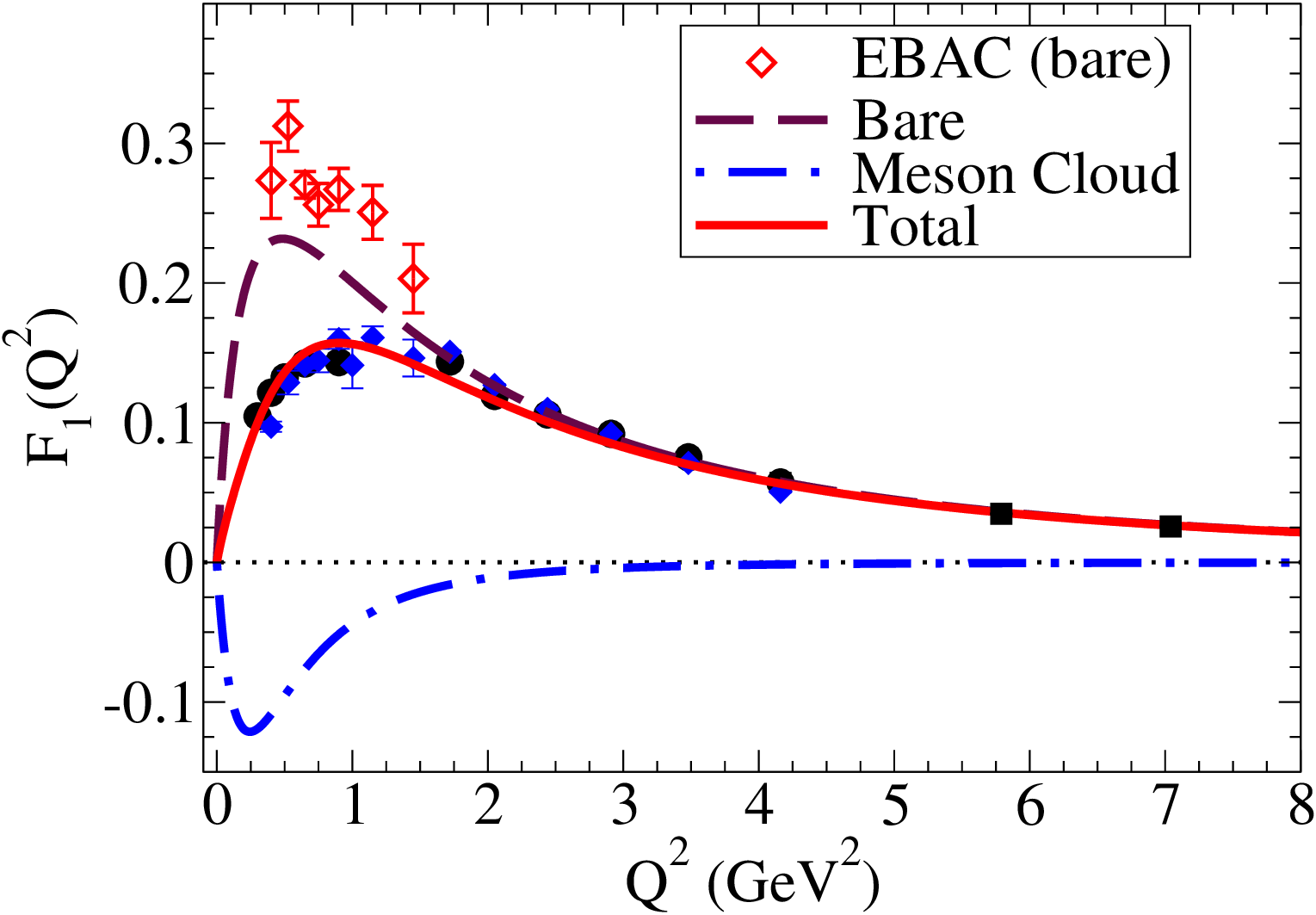} \hspace{.5cm}
\includegraphics[width=3.in]{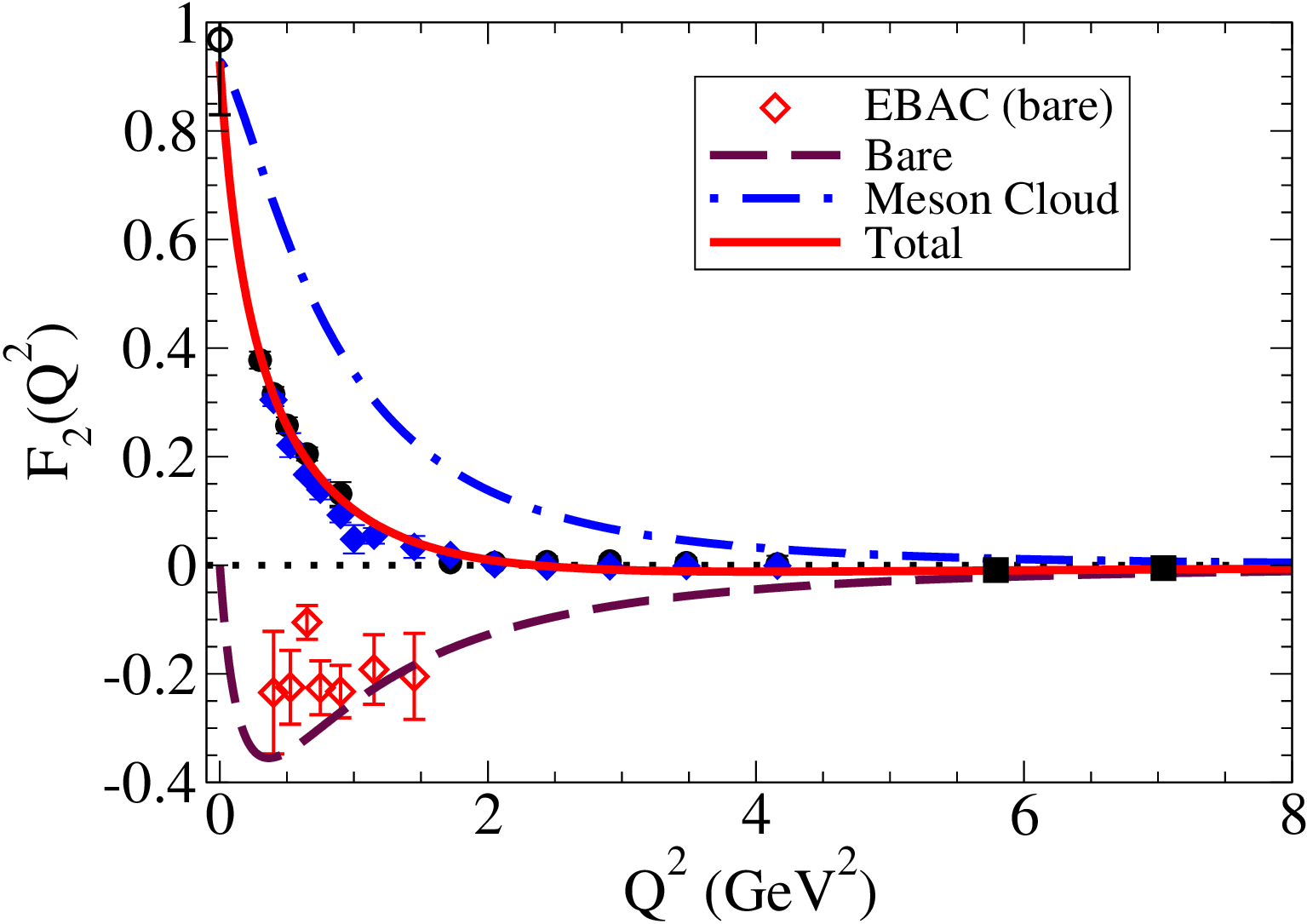} 
\caption{\footnotesize 
$\gamma^\ast N \to N(1535)$ transition form factor. 
Results form the covariant spectator quark model~\cite{N1535-TL}.
Data as in Fig.~\ref{figF1F2-N1535}.
The EBAC data are from Refs.~\cite{JDiaz09a,KamanoPC}
\label{figF1F2-N1535-2}}
\end{figure}

In Figs.~\ref{figF1F2-N1535} and \ref{figF1F2-N1535-2}, we present 
calculations of the transition form factors $F_1$ and $F_2$.
The form factors are calculated using the relations 
(\ref{eqF1-b})--(\ref{eqF2-b}) that invert the relations 
for the helicity amplitudes in terms of form factors.
The data for the region $Q^2 > 4.2$ GeV$^2$ is the result 
of JLab/Hall C experiments~\cite{Dalton09},   
where the amplitude $A_{1/2}$ is determined under the assumption that $S_{1/2}=0$.

In Fig.~\ref{figF1F2-N1535}, we include models 
discussed in Fig.~\ref{figAmps-N1535-1}: LFQM 1~\cite{Aznauryan17a} 
and LFQM 2~\cite{Obukhovsky19a}, the light-cone sum rules~\cite{Anikin15a},
the Holographic QCD model~\cite{Gutsche20a}, as well as 
the unitary chiral model~\cite{Jido08a} and 
the meson cloud contributions determined by the 
ANL-Osaka DCC model~\cite{Kamano16a,Kamano13a,Nakamura15a,Sato16a}.
As in the case of the helicity amplitudes the AdS/QCD model 
provides a good description of the transition form factors.
This result is also a consequence of the fit of the model to the data.
The light-front quark models overestimate the data for $F_1$ at low $Q^2$.
LFQM 2 describes well the $F_2$ data.
The meson cloud contributions have large contributions to $F_2$ 
near $Q^2=0$, and are suppressed when $Q^2$ increases, as expected.
The meson cloud contributions are much more significant to larger $Q^2$ values
for $F_2$ than for $F_1$.

The $\gamma^\ast N \to N(1535)$ transition form factors have also been 
calculated within the Poincar\'e-covariant Faddeev equation framework 
using a contact interaction and the baryons as quark-diquark systems~\cite{Raya21a}.
The quarks are dressed by interaction with the gluon as in the
Dyson-Schwinger framework 
and the diquarks are fully dynamical.
Although the contact interaction limits the predictions for intermediate $Q^2$,
the calculations have the advantage of being algebraic and can be 
used as input for more sophisticated studies.
The calculations lead to a bare mass in the range 1.7--1.8 GeV 
and a fair description of the $F_1$ data when 
the anomalous magnetic moment of the dressed quark 
is taken into account~\cite{Wilson12}.
The form factor $F_2$ is underestimated at low $Q^2$, 
and overestimated at large $Q^2$.

We focus now on the discussion of the results for $F_2$.
Notice that the data drops off to zero within the error bars,
when $Q^2 > 2$ GeV$^2$.
The consequences of this result, assumed to be valid for larger values of $Q^2$, 
are discussed in the Section~\ref{sec-N1535-LQ2}, dedicated to the large-$Q^2$ region.
There are two simple interpretations of the result $F_2 \simeq 0$, but
which of the two scenarios does happen is not known yet:
Do the valence quark and the meson cloud contributions per se
fall off so fast that there are both negligible 
above a certain value of $Q^2$?
Or is there a cancellation between the 
valence quark and meson cloud contributions 
for intermediate values of $Q^2$?

In Fig.~\ref{figF1F2-N1535-2}, we present 
the results of the covariant spectator quark 
model, in the semirelativistic approximation~\cite{N1535-1,N1535-TL,SemiR},
discussed in Section~\ref{secCSQM}.
Reference~\cite{N1535-TL} shows that the result
$F_2 \simeq 0$ is a consequence of the cancellation 
between the valence quark and meson cloud contributions 
for intermediate $Q^2$. In that study of the $N(1535)$ Dalitz decay, 
the reactions with the neutron target are also shown to be as relevant 
as the reactions with proton target, demonstrating the importance of models for both isospin states.
The calculations are in the figure compared with the data
and with previous EBAC estimates of the baryon core 
(no meson cloud contributions) from Refs.~\cite{JDiaz09a,KamanoPC}.
The bare contribution to the form factors 
$F_1$ and $F_2$ are parameter free calculations,
since the parameters associated with the $N(1535)$ wave function are correlated 
to the parameters of the nucleon radial wave function in the 
semirelativistic approximation~\cite{N1535-TL,SemiR}.
Notice the closeness between this parameter free model estimate 
and the EBAC estimate below 2 GeV$^2$.
Notice also the agreement of the covariant spectator quark model
with the data for $F_1$ for $Q^2 > 5$ GeV$^2$, in a region  
where the meson cloud effects are expected to be negligible. 
These results support the use of the covariant spectator quark model
to parametrize the valence quark contributions 
to the transition form factors.
To obtain an accurate description of the data at 
low $Q^2$, say below 2 GeV$^2$, one needs to 
include an effective parametrization of the meson cloud effects.
This parametrization is developed using the 
knowledge that the meson cloud contributions 
are dominated by isovector contributions, as discussed above.
The parametrization of the meson cloud presented 
in Fig.~\ref{figF1F2-N1535-2} assumes also falloffs 
compatible with pQCD calculations (see Section~\ref{secPQCD}).

\begin{figure}[t]
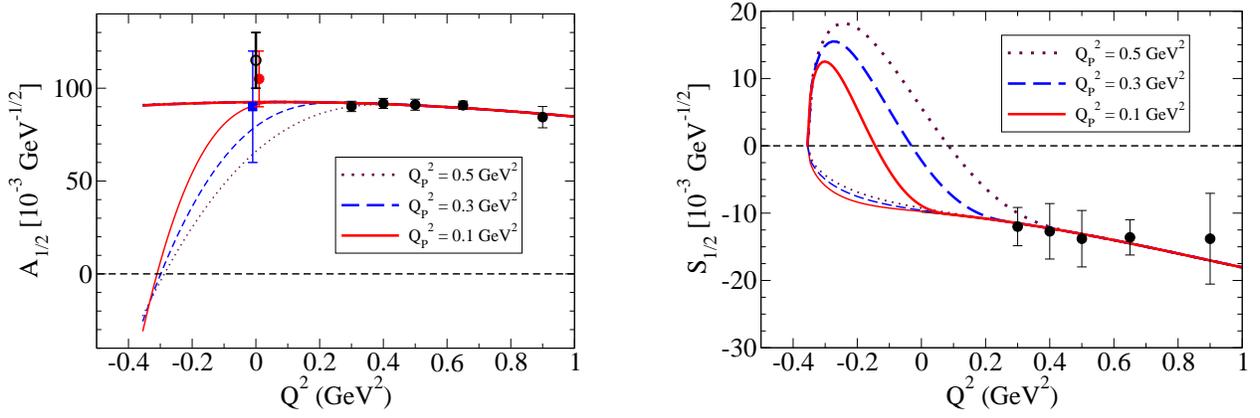

\begin{center}
\includegraphics[width=3.in]{A12-N1535-v4} \hspace{1.cm}
\includegraphics[width=3.in]{S12-N1535-v4} 
\caption{\footnotesize 
$\gamma^\ast \to N(1535)$ transition.
Discussion about the low-$Q^2$ region and the constraints of 
Siegert's theorem.
The figure illustrates the uncertainty in the related shapes
and signs of the two amplitudes in the vicinity of the photon point ($Q^2=0$)
and at the pseudothreshold point ($Q^2=-(M_R-M_N)^2$).
At $Q^2=0$, we include data from PDG 2012 ({\Large {\B $\blackdiamond$}})~\cite{PDG2012},
PDG 2016 ({\Large ${\bm  \circ}$})~\cite{PDG2016} and PDG 2022 
({\Large {\color{red}$\bullet$}})~\cite{PDG2022}.
The values are shifted from $Q^2=0$ for a better visualization.
The finite-$Q^2$ data are from JLab/CLAS~\cite{CLAS09}.
Thick and thin lines on both the left and right panels are
two sets of consistent extrapolations of the two related amplitudes into the timelike region. 
\label{figN1535-ST}}
\end{center}
\end{figure}

\subsubsection{\it Low-$Q^2$ region \label{sec-N1535-lQ2}}

There are two important points for discussion 
on the data and behavior of the functions describing the $\gamma^\ast N \to N(1535)$ transition 
in the low-$Q^2$ region:
the quality of the data (in range, precision and accuracy) 
and the impact of the low-$Q^2$ constraints
near the pseudothreshold $Q^2= -(M_R-M)^2$, introduced in Section~\ref{sec-Siegert}.

First, the amplitude $A_{1/2}$ is not very well known at the photon point.
The PDG results for $A_{1/2}(0)$ cover a large range of values reported by
different groups~\cite{PDG2022,PDG2012,PDG2016},
revealing that accuracy has still to be improved.
The central value of the PDG, based on 
the Breit-Wigner photon decay amplitudes, has been changing
from 2012~\cite{PDG2012} to 2016~\cite{PDG2016} and 2022~\cite{PDG2022}.
The $A_{1/2}(0)$ data from PDG 2012, PDG 2016 and PDG 2022 
are presented in Fig.~\ref{figN1535-ST}, showing changes with time.
Second, another limitation of the helicity amplitude data,
also common to other transitions, is the lack of data within the interval 
between $Q^2=0$ and 0.3 GeV$^2$ which prevents precise
extrapolations of the data down to the photon point or below.

The atypical shape of the $\gamma^\ast N \to N(1535)$ helicity
amplitudes and the possible conflict with Siegert's theorem
were already reported by Tiator 
in the context of the MAID parametrizations~\cite{Tiator16},
as discussed in Section~\ref{sec-Siegert}.
We discuss here the problem based on available low-$Q^2$ data.

In Fig.~\ref{figN1535-ST}, we present the CLAS data~\cite{CLAS09} 
below $Q^2=1$ GeV$^2$ in comparison with different parametrizations 
of the data compatible with Siegert's theorem~\cite{LowQ2param}.
In the spacelike region ($Q^2 \ge 0$) we represent
the JLab parametrization~\cite{JLab-website,Blin19a} above a point $Q_P^2$.
There are considered three cases: $Q_P^2=0.1$, 0.3 and 0.5 GeV$^2$.
Below the point $Q_P^2$, we present extensions of the parametrization
to the region $-(M_R-M)^2  \le Q^2 \le Q_P^2$ consistent with $A_{1/2} \; \propto \; 1$, 
$S_{1/2} \; \propto \; |{\bf q}|$ and 
$A_{1/2} = \sqrt{2} (M_R -M) S_{1/2}/ |{\bf q}|$, near
the pseudothreshold $|\bold{q}| = 0$ or  $Q^2= -(M_R-M)^2$.
These conditions are called the pseudothreshold constraints 
(Section~\ref{sec-Siegert}, Table~\ref{table-Siegert1}).
Each value of $Q_P^2$ defines then an extension to the timelike region.

The presented parametrizations (JLab-ST) were derived in Ref.~\cite{LowQ2param},
where it is proposed a method to determine extensions
of parametrizations for low $Q^2$, consistent with both the data and Siegert's theorem.
The figure illustrates also the uncertainty in the knowledge of the shape
of the two helicity amplitudes near $Q^2=0$.

First one calculates three extrapolations of the $A_{1/2}$ data down
to the pseudothreshold point, beyond three different limit points $Q_P^2=0.5$, 0.3 and 0.1 GeV$^2$
by assuming that the amplitude $A_{1/2}$ is smooth and varies slowly.
The thick red line on the left panel depicts all the three extensions
-- all the three lines overlap below $Q^2=0$.
The corresponding three lines for the amplitude $S_{1/2}$ are then obtained from the relation 
$A_{1/2} \; \propto \; S_{1/2}/|{\bf q}|$.
They are shown on the right panel also by the three thicker lines.
This fast variation of $S_{1/2}$ below $Q_P^2$  
is necessary  for the relation between $A_{1/2}$ and $S_{1/2}$ to be valid.
In this case both $A_{1/2}$ and $S_{1/2}$
are positive to the right of the pseudothreshold point.

Inversely, if one assumes  that instead all the three extensions
of $S_{1/2}$ down to the timelike region are smooth and vary slowly~\cite{LowQ2param}
(thin lines on the right panel),
then it is the corresponding extrapolations of $A_{1/2}$  that change more drastically when $Q^2$ 
decreases: this is shown by the three thin lines in the left panel
that are the partners of the thin lines for $S_{1/2}$ on the right panel.
In this case near the pseudothreshold
$A_{1/2}$ and $S_{1/2}$ are both negative to right of the pseudothreshold point.

This serves to illustrate that more precise data for $A_{1/2}(0)$, 
or accurate data for both amplitudes, in the range $Q^2=0$--0.3 GeV$^2$, are needed to
enable us to decide about shape and sign of the two amplitudes below the photon point,
and the values of $S_{1/2}$ near $Q^2=0$.

The more recent PDG estimate (red data point from PDG 2022) favors
the solutions represented by the thick lines:
a smooth or slow variation of $A_{1/2}$ and a rapid variation
for $S_{1/2}$, and positive signs of both to the right of 
the pseudothreshold point.
This behavior needs to be confirmed 
by measurements of the amplitude $S_{1/2}$ below $Q^2=0.3$ GeV$^2$.
Notice, however, that the red thin lines corresponding  
to $Q^2_P =0.1$ GeV$^2$ is still compatible with the $A_{1/2} (0)$ most recent data
(red data point).
The sensitivity of the parametrizations of the data 
to the low-$Q^2$ data was also noted in the calculations from Ref.~\cite{N1535-TL}, 
presented in Fig.~\ref{figF1F2-N1535-2}.

The parametrizations of the amplitudes $A_{1/2}$ and $S_{1/2}$ can be 
translated into parameterizations of the  form factors $F_1$ and $F_2$.
In this case, however, the pseudothreshold constraints
are not so explicitly manifest in the behavior of the curves.
The consequences are that $F_1$ and $F_2$ are finite at the pseudothreshold
($F_1$, $F_2 = {\cal O}(1)$).
More details can be found in Ref.~\cite{Siegert-N1535}.

To conclude, Fig.~\ref{figN1535-ST} shows the need for
precise and accurate measurements in the range $Q^2=0$--0.3 GeV$^2$ to determine
the shape of the helicity amplitudes near the photon point.
It shows also that parameterizations
of the two helicity amplitudes in the low-$Q^2$ region have to take
into account physical correlations between these amplitudes.

\subsubsection{\it Large-$Q^2$ region \label{sec-N1535-LQ2} }

We discuss now the large-$Q^2$ region.
We divide this section in two parts.
We start with the discussion of an empirical relation
between the amplitudes $A_{1/2}$ and $S_{1/2}$ 
for $Q^2 > 2$ GeV$^2$. Afterwards we discuss the dependence of $A_{1/2}$ 
and $F_1$ at very large $Q^2$ according to estimates from pQCD.

In Section~\ref{sec-N1535-FF}, it was mentioned that 
the measured data for $F_2$ suggest that, 
within the error bars, the Pauli form factor vanishes for $Q^2 > 2$ GeV$^2$.
The consequence of the relation $F_2 \simeq 0$
is according to Eq.~(\ref{eqF2-b}) that the 
amplitudes $S_{1/2}$ and $A_{1/2}$ are proportional:
using the expression 
$|{\bf q}| \simeq  (M_R + M) \sqrt{1 + \tau} \frac{Q}{2M_R}$,
valid for $Q^2 \gg (M_R-M)^2 \simeq 0.36$ GeV$^2$, one obtains~\cite{N1535-2}
\ba
S_{1/2} \simeq - \frac{\sqrt{1 + \tau}}{\sqrt{2}} 
\frac{M_R^2 -M^2}{ 2 M_R Q} A_{1/2}.
\label{eqS12-N1535}
\ea

\begin{figure}[t]
\begin{center}
\includegraphics[width=3.8in]{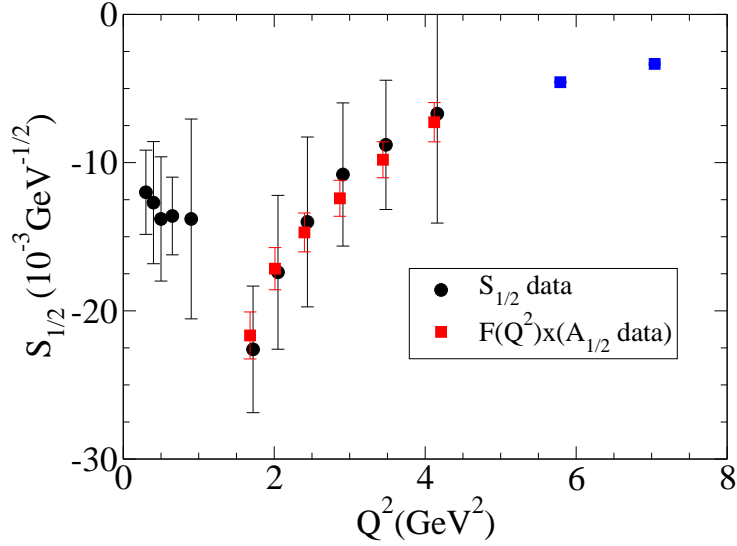} 
\caption{\footnotesize
$\gamma^\ast N \to N(1535)$ transition.
Test of the large-$Q^2$ region proportionality term in (\ref{eqS12-N1535}) for $Q^2 > 2$ GeV$^2$;
$F(Q^2) = - \frac{\sqrt{1 + \tau}}{\sqrt{2}} \frac{M_R^2 -M^2}{ 2 M_R Q}$.
The black bullets ({\Large $\bullet$}) represent the CLAS data 
for $A_{1/2}$~\cite{CLAS09}.
The red squares ({\scriptsize {\color{red} $\blacksquare$}}) 
represent the amplitude $S_{1/2}$ calculated by Eq.~(\ref{eqS12-N1535}).
The blue squares  ({\scriptsize {\color{blue} $\blacksquare$}}) 
are predictions for $S_{1/2}$ for large $Q^2$
based on the $A_{1/2}$ data from JLab/Hall C~\cite{Dalton09}.
 \label{fig-S12-N1525-LQ2}}
\end{center}
\end{figure}

The relation (\ref{eqS12-N1535}) is tested 
in Fig.~\ref{fig-S12-N1525-LQ2} against the 
one pion production CLAS data~\cite{CLAS09}.
The estimates of $S_{1/2}$ are in good agreement with 
the $S_{1/2}$ data up to 4 GeV$^2$. 
For more definitive conclusions it is necessary to improve 
the precision of the $S_{1/2}$ data and extend the range 
of the measurements for higher $Q^2$.
We can also apply Eq.~(\ref{eqS12-N1535}) to predict the amplitude $S_{1/2}$
for larger $Q^2$ using the results of the amplitude $A_{1/2}$
from Dalton et al.~\cite{Dalton09} for $Q^2=5.7$ and 7.0 GeV$^2$,
as exemplified in Fig.~\ref{fig-S12-N1525-LQ2}.

Assuming that Eq.~(\ref{eqS12-N1535}) holds 
for arbitrary large values of $Q^2$, we conclude that 
the ratio $S_{1/2}/A_{1/2}$ converges to 
$- \frac{1}{\sqrt{2}} \frac{M_R -M}{2 M_R} \simeq -0.13$ 
($\tau \gg 1$)~\cite{N1535-2}.
Note, however, that this approximation  is good 
only when $Q^2 \gg (M_R +M)^2 \simeq 6.1$ GeV$^2$,
meaning that the convergence is expected to be very slow.
The ratio $S_{1/2}/A_{1/2} \simeq -0.13$ can also be used 
to justify the approximation which neglects $S_{1/2}$ 
in the measurements of the transition cross sections.

An important point for this discussion is whether $F_2$ really vanishes or not, having 
just a small magnitude compared with $F_1$. 
If it really vanishes, there is a deviation from the pQCD result $F_2 \; \propto \; 1/Q^6$.
More definitive conclusions can be obtained from
the ratio $Q^2F_2/F_1$ at large $Q^2$ (convergence to small constant or to zero).

\begin{figure}[t]
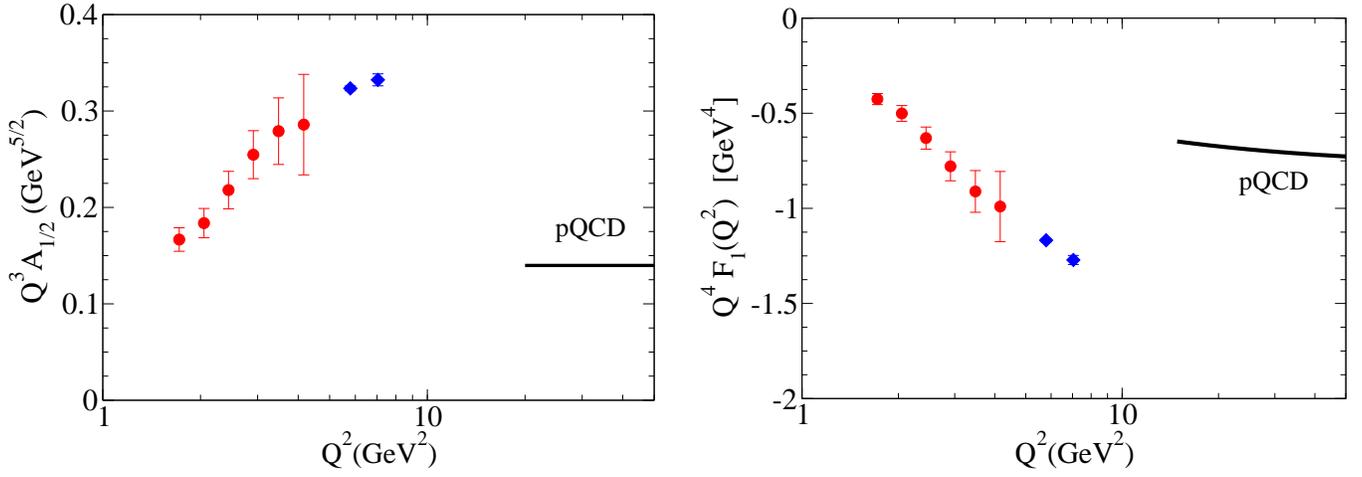
  
\begin{center}
\mbox{
\includegraphics[width=3.4in]{Q3A12-N1535-v1} \hspace{.3cm}
\includegraphics[width=3.4in]{Q4F1-N1535-v1}}
\caption{\footnotesize
pQCD estimates for the $\gamma^\ast N \to N(1535)$ transition.
{\bf Left panel:} 
Data for $Q^3 A_{1/2}$ compared with pQCD result. 
{\bf Right panel:} 
Data for $Q^4 F_1$ compared with pQCD result for $F_1$.
The data are from CLAS~\cite{CLAS09} 
({\scriptsize {\color{red} $\blacksquare$}}) 
and JLab/Hall C~\cite{Dalton09} ({\Large {\B $\blackdiamond$}}).
\label{fig-N1535-pQCD}}
\end{center}
\end{figure}

The asymptotic behavior for the amplitude $A_{1/2}$
has been calculated by C.~Carlson et al.~\cite{Carlson98a,Carlson88a}
within pQCD, and using distribution amplitudes derived 
from QCD sum rules.
The authors calculate the limit $Q^3 A_{1/2}$ 
using different parametrizations for the nucleon 
and $N(1535)$ distribution amplitudes.
The upper limit of those estimates gives~\cite{N1535-1,Carlson98a}
\ba
Q^3 A_{1/2} = e \sqrt{\frac{M}{M_R^2-M^2}} \beta,
\ea
where $\beta = 0.58$ GeV$^3$.
The comparison of the more recent data for large $Q^2$
with this result is presented 
on the left panel from Fig.~\ref{fig-N1535-pQCD}.
It is clear in the graph that the pQCD calculation underestimates
the present data, showing that the pQCD limit is not reached
in the $Q^2$ region of the data available.

As for the form factors $F_1$ and $F_2$,
we expect, according with pQCD arguments, 
$F_1 \; \propto \; 1/Q^4$ and  $F_2 \; \propto \; 1/Q^6$,
apart logarithmic corrections.
One can then use Eq.~(\ref{eqA12-b}) and by
neglecting the $F_2$ term one concludes that 
\ba
Q^4 F_1 = 
- \sqrt{\frac{2 M^2 Q^2}{(M_R + M)^2 + Q^2}} \beta. 
\ea 
The comparison between the data and pQCD estimate 
is presented on the right panel from Fig.~\ref{fig-N1535-pQCD}.
In the present case the deviation of the pQCD estimate 
from a flat line is the consequence of 
the conversion factor $\sqrt{Q^2/Q_+^2}$, 
and illustrate that finite corrections to a constant  
are expected, unless in the region $Q^2 \gg (M_R + M)^2$.

From Fig.~\ref{fig-N1535-pQCD}, one can conclude that 
either we are still very far away from the pQCD scaling region,
or the pQCD calculations are underestimations.
Notice that a similar situation happens 
in the case of the  $\gamma^\ast N \to \Delta(1232)$ 
transition form factors, in particular to the 
electric quadrupole form factor, as will be seen in Section~\ref{sec-Delta-LLQ2}. 
New data from the JLab-12 GeV upgrade, or future extensions 
in the energy range of the measurements~\cite{Carman23a} may help
to determine the large-$Q^2$ dependence of the  $\gamma^\ast N \to N(1535)$
amplitudes and form factors.

\subsubsection*{\it Short notes}

The $\gamma^\ast N \to N(1535)$ data obtained 
in the last decades suggest that the transition 
can be described for large $Q^2$ based mainly on the valence quark degrees 
of freedom dominated by $p$-wave quark states or quark-diquark $P$-states.
Light-front quark models with meson cloud contributions 
and light-cone sum rules describe well the data.

Data at low-$Q^2$ favor models that consider the interplay 
between systems of three quarks and significant 
meson cloud contributions. 
These models predict relevant meson cloud contributions at moderate $Q^2$.
The form factor $F_2$ in particular is very sensitive 
to the valence quark physics and the meson cloud physics,
even in the relatively large-$Q^2$ region.

\subsection{\it $\Delta(1232)\frac{3}{2}^+$  resonance \label{sec-D1232}}

The $\Delta(1232)$, the first nucleon resonance to be known,
is very clearly identified by the its $\pi N$ decay
(with a branching ratio of 99.4\%). 
The radiative decay $\Delta(1232) \to \gamma N$ accounts only for 0.6\%
of the decays.
The third possible decay $\Delta(1232) \to e^+ e^- N$ 
has a branching ratio of $4.2 \times 10^{-5}$~\cite{PDG2022}.

Non relativistic $SU(6)$ symmetry-based constituent quark models interpret
the $\gamma^\ast N \to \Delta (1232)$ transition 
as a spin-flip of a single quark when hit by the photon,
originating a new baryon with spin $3/2$, 
the $\Delta(1232)$~\cite{NSTAR,Aznauryan12a,Pascalutsa07,Beg64}.
In those simple quark model calculations transitions of $s$-wave quark
wave functions to $d$-wave 
quark states are forbidden~\cite{Pascalutsa07,Isgur82} by the interaction with photons,
and  therefore the electric quadrupole $G_E$ and the Coulomb 
quadrupole  $G_C$ form factors defined in Section~\ref{sec32} vanish identically.
Only excited $l>0$ quark states in the 
nucleon and/or $\Delta(1232)$ wave functions lead to non zero
but small results for the quadrupole form factors 
$G_E$ and $G_C$~\cite{Isgur82,Glashow79,Becchi65,Krivoruchenko91,Buchmann01}.   
The $\gamma^\ast N \to \Delta (1232)$ transition can thus
by regarded as a magnetic transition where the 
magnetic dipole form factor $G_M$ dominates. This dominance of     
the magnetic dipole form factor is indeed seen in the data
at low and intermediate squared momentum transfer,
within the range  $0 \le Q^2 < 8$ GeV$^2$. %
The consequence of this dominance 
is that in first approximation of $SU(6)$, 
we can neglect at low and intermediate $Q^2$
the contributions to the electric form factor ($G_E \simeq 0$).
From Eq.~(\ref{eqAmp32p}) it follows that the amplitudes $A_{1/2}$ and $A_{3/2}$
are correlated functions according to
$A_{3/2} \simeq \sqrt{3} A_{1/2} \simeq - {\frac {\sqrt{3}}{4  F_{1+}}} \, G_M$.

Non zero results for the form factors $G_E$ and $G_C$
indicate a deviation of the $\Delta(1232)$
from a spherical shape~\cite{Stave08,Buchmann01,Alexandrou09b,Delta-Shape}.
This deviation can be the consequence of high orbital angular momentum of the quarks,
meson cloud effects or relativistic effects~\cite{Eichmann16,Glashow79,Buchmann01},
as discussed later.

Traditionally, the quadrupole form factors 
of the nucleon electroexcitation to the $\Delta(1232)$
are represented through their electromagnetic ratios
with the leading magnetic dipole form factor,
\ba  
R_{EM} (Q^2) = - \frac{G_E(Q^2)}{G_M(Q^2)}, 
\hspace{1.5cm}
R_{SM} (Q^2) = - \frac{|{\bf q}|}{2 M_\Delta}\frac{G_C(Q^2)}{G_M(Q^2)}, 
\label{eqEMC-RMC}
\ea
where $|{\bf q}|$ is the magnitude of the photon 3-momentum 
at the $\Delta(1232)$ rest frame.  
Some authors use also EMR and CMR to represent 
$R_{EM}$ and $R_{SM}$, respectively~\cite{Pascalutsa07}.

The  $\gamma^\ast N \to \Delta$ transition 
magnetic form factor $G_M$ at $Q^2=0$ determines the
magnetic moment defined
as~\cite{Pascalutsa07,Tiator03} 
\ba
\mu_{N \Delta} = \sqrt{\frac{M_\Delta}{M}} G_M(0) \mu_N,
\label{eq-mu}
\ea
where $M_\Delta$ is the $\Delta (1232)$ mass. Similarly,
the transition electric quadrupole moment can be defined as
\ba
Q_{N\Delta} = - \frac{6}{M K} \sqrt{\frac{M_\Delta}{M}} G_E(0),
\ea
where $K= \frac{M_\Delta^2-M^2}{2 M_\Delta}$, as before.
$Q_{N\Delta}$ is expressed in fm$^2$ or the inverse of square mass.

In alternative to the Jones and Scadron form factors 
$G_\alpha$ ($\alpha=M,E,C$) some authors use also 
the Ash form factors $\overline G_\alpha$~\cite{Drechsel07,Pascalutsa07,Ash67}, defined as
\ba
\overline G_\alpha (Q^2) = \frac{G_\alpha(Q^2)}{\sqrt{1 + \frac{Q^2}{(M_\Delta + M)^2}}}.
\label{eqGM-Ash}
\ea 
The Ash form factors have then a faster falloff with $Q^2$.
For very large $Q^2$, when
$Q^2 \gg (M_\Delta + M)^2$, $\overline G_\alpha  (Q^2) \simeq \frac{M_\Delta + M}{Q}  G_\alpha (Q^2)$.

The current experimental information about the electromagnetic structure 
of the $\Delta(1232)$ at the photon point is 
the following~\cite{PDG2022,Tiator03}
\ba
G_M(0) = 3.02 \pm 0.03,
\hspace{1cm}
R_{EM}(0) = -2.5\pm0.5 \%. 
\label{eqGM0}
\ea
The result for $G_M(0)$ is an extrapolation 
from the finite $Q^2$ data~\cite{Tiator03}.
The result for $R_{EM}(0)$ is the average of PDG entries~\cite{PDG2022}.
Combining both results, we obtain
\ba
G_E(0)= 0.076\pm 0.015.
\ea 
\noindent
The PDG result~\cite{PDG2022} for the quadrupole electric moment is 
\ba
Q_{N\Delta} = - 0.087 \pm0.017 \; \mbox{fm}^2.
\ea
More precise results can be obtained 
from a better analysis of $R_{EM}(0)$~\cite{Tiator03}.  

We give also here the helicity amplitudes 
at the photon point~\cite{PDG2022}
\ba
A_{1/2} (0) = -0.135\pm0.007 \; \mbox{GeV}^{1/2},
\hspace{1cm}
A_{3/2} (0) = -0.255\pm0.007 \; \mbox{GeV}^{1/2}.
\ea
These PDG results for the transverse amplitudes at $Q^2=0$ 
are averages over results from different experiments.
For that reason the PDG values 
should not be used to calculate $G_M(0)$ and $G_E(0)$~\cite{Pascalutsa07}.

To end this summary on the $\gamma^\ast N \to \Delta (1232)$ transition properties,
we illustrate how quark models describe the  $\gamma^\ast N \to \Delta (1232)$ 
transition magnetic moment. The calculation with $s$-wave quark wave functions 
in the baryon wave functions is given by~\cite{Beg64} 
\ba
\mu_{N \Delta} = \frac{2 \sqrt{2}}{3} \mu_p, 
\ea
where $\mu_p$ is the proton magnetic moment in 
nuclear magneton units $\mu_N \equiv \frac{e}{2M}$.
Importantly, from (\ref{eq-mu}) one concludes that such a
simple calculation gives the theoretical value 
\ba
G_M(0) = 2.3  \hspace{.3cm} [SU(6) \; \mbox{model}]
\label{eqGM-QM}
\ea
that, by comparing with (\ref{eqGM0}), underestimates $G_M(0)$ by about 25\%.      
More sophisticated calculations to fill in this gap between experiment
and quark model calculations will be reported in the next sections.

\subsubsection{\it Transition form factor data \label{sec-DeltaFF}}

\begin{figure}[t]
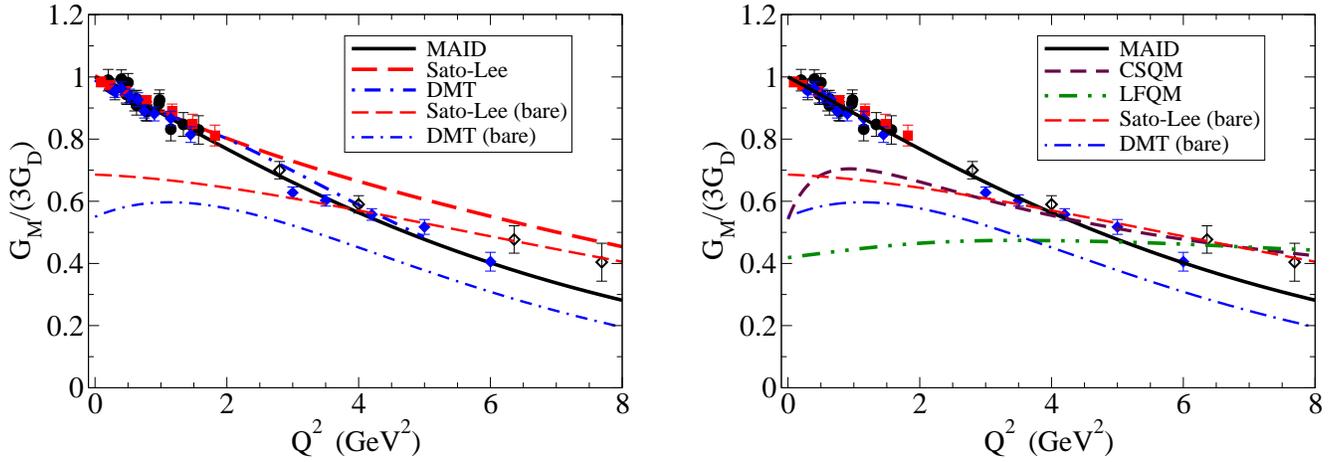

\begin{center}
\mbox{
\includegraphics[width=3.25in]{GM-total6} \hspace{.7cm}
\includegraphics[width=3.25in]{GM-total5} }
\caption{\footnotesize
$\gamma^\ast N \to \Delta(1232)$ magnetic form factor $G_M$ normalized by $3 G_D$.
$G_D= \left(\frac{\Lambda_D^2}{\Lambda_D^2 + Q^2} \right)^2$ with $\Lambda_D^2 \simeq 0.71$ GeV$^2$.
The parametrization MAID 2007~\cite{Drechsel07} is included for reference.
{\bf Left panel:} 
Comparison with results of the Sato-Lee~\cite{SatoLee01} and DMT~\cite{Kamalov01a} models.
{\bf Right panel:} 
Comparison with calculation of the covariant spectator quark model (CSQM)~\cite{LatticeD} 
and the light-front quark model from Aznauryan and Burkert~\cite{Aznauryan15a}.
The data are from DESY ({\Large $\bullet$})~\cite{DESY68}, 
SLAC  ({\scriptsize {\color{red} $\blacksquare$}})~\cite{SLAC75},
JLab/CLAS ({\Large {\B $\blackdiamond$}})~\cite{CLAS09,JLab-database}
and JLab/Hall C  ({\Large ${\bm \diamond}$})~\cite{Villano09,Frolov98}.
The results of the bare parametrizations for the models 
Sato-Lee and DMT are also presented. 
\label{figGMtotal}}
\end{center}
\end{figure}

The first measurements of the $\gamma^\ast N \to \Delta(1232)$
magnetic dipole transition form factor were 
performed at DESY (1968)~\cite{DESY68} and SLAC (1975)~\cite{SLAC75}.
Those measurements suggested that $G_M(0) \simeq 3$ consistently
with the recent results of (\ref{eqGM0}).
The observed falloff of $G_M$ with $Q^2$ 
suggest that it would be more convenient to compare the data with the
dipole form factor form $G_D$ compatible with the first 
proton and neutron electric and magnetic 
form factor measurements at SLAC~\cite{HalzenMartin,Hyde04a,Perdrisat07,Litt70},
with $G_D= \left(\frac{\Lambda_D^2}{\Lambda_D^2 + Q^2} \right)^2$,
and $\Lambda_D^2 \simeq 0.71$ GeV$^2$.
The normalization of $G_M$ by $3 G_D$ generates a softer dependence on $Q^2$ 
and facilitates the study of $G_M$ in the low- and large-$Q^2$ regions.

The recent available data on the $G_M$ form factor, 
properly normalized by $3G_D$, is presented in Fig.~\ref{figGMtotal}
in comparison with the MAID parametrization~\cite{Drechsel07}.
In the left panel the results are compared with the 
Sato-Lee~\cite{SatoLee01,JDiaz07b} and DMT~\cite{Kamalov02,Kamalov01a} dynamical models.
In the graph, we also include the bare parametrizations of the Sato-Lee and DMT models.
These bare parameterizations are attempts to describe
in an effective way the contributions from 
the valence quarks for the baryon cores
(the reader may revisit Section~\ref{secDM} for details) which at $Q^2=0$
underestimate $G_M$ in about 25\%, as we saw already with simple $SU(3)$ quark model estimates.
Actually, it is a general feature of constituent quark models results to fall short of  
the experimental magnitude of $G_M$ at low $Q^2$,
independently of the adopted confinement mechanism~\cite{Aznauryan12a,Pascalutsa07}.

\begin{figure}[t]
\centering

\includegraphics[width=2.8in]{REM-v10a} \hspace{.3cm}
\includegraphics[width=2.75in]{REM-v10b} \hspace{.2cm}
\includegraphics[width=1.1in]{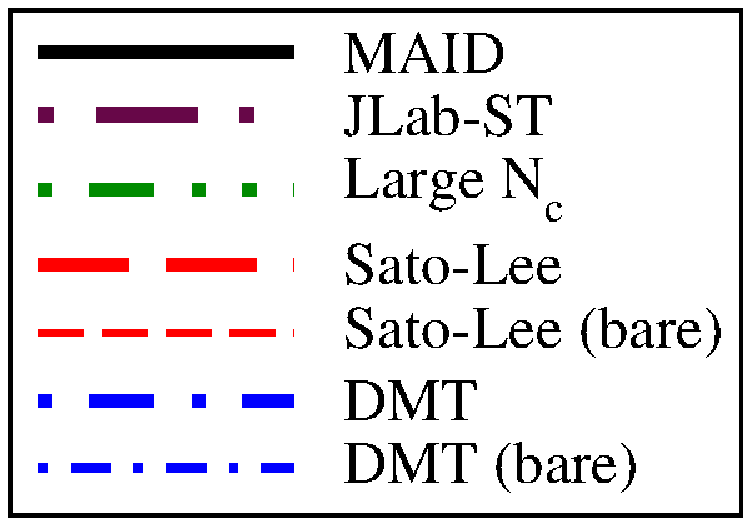} 
\caption{\footnotesize
$\gamma^\ast N \to \Delta(1232)$ transition. 
Quadrupole ratio $R_{EM}$ as function of $Q^2$.
The data are from ELSA ({\scriptsize $\blacksquare$})~\cite{ELSA97}, 
MAMI ({\B {\Large ${\bm  \circ}$}})~\cite{Beck00a,Stave08,Sparveris13}, 
MIT-Bates ({\Large ${\bm \diamond}$})~\cite{MIT-Bates03,Sparveris05a},
JLab/Hall A ({\Large $\blackdiamond$})~\cite{Blomberg16},
JLab/CLAS ({\Large {\color{red} $\bullet$}})~\cite{CLAS09} and 
JLab/Hall C ({{\B \Large $\blackdiamond$}})~\cite{Villano09,Frolov98}.
The dots represent the horizontal and the vertical axis.
Models and parametrizations explained in the main text.
\label{fig-Delta-REM}}

\bigskip

\vspace{.5cm}

\includegraphics[width=2.8in]{RSM-v10a} \hspace{.3cm}
\includegraphics[width=2.75in]{RSM-v10b} \hspace{.2cm}
\includegraphics[width=1.1in]{label1B} 
\caption{\footnotesize
$\gamma^\ast N \to \Delta(1232)$ transition. 
Quadrupole ratio $R_{SM}$ as function of $Q^2$.
The data are from ELSA ({\B {\Large ${\bm  \circ}$}})~\cite{Pospischil01,Stave08}, 
MIT-Bates ({\Large ${\bm \diamond}$})~\cite{MIT-Bates03,Sparveris05a},
JLab/Hall A ({\Large ${\bm \blackdiamond}$})~\cite{Blomberg16},
JLab/CLAS ({\Large {\color{red} $\bullet$}})~\cite{CLAS09} and   
JLab/Hall C ({{\B \Large $\blackdiamond$}})~\cite{Villano09,Frolov98}.
The dots represent the horizontal and the vertical axis.
Models and parametrizations explained in the main text.
\label{fig-Delta-RSM}}

\end{figure}

The gap between quark model estimates and the empirical data
shows the limitations of the interpretation of baryons 
as systems composed of three valence quarks,
and motivates the decomposition (\ref{eqGB-MC}).
In the case of the $\Delta(1232)$, with a mass very close to
the $\pi N$ channel threshold, this decay channel (99.4\%) dominates,
the baryon is extended by the pion cloud around its core and we can consider that the
{\it meson cloud} is reduced mainly  to {\it pion cloud}, replacing $G_M^{\rm MC}$ by $G_M^{\pi}$.

However, at large $Q^2$ after a certain value of $Q^2$, one expects the valence quark 
effects to dominate~\cite{Burkert04,Kamalov01a,Kamalov99a,SatoLee96,SatoLee01}.
The bare core contributions 
are then parameterized using effective functions that reproduce 
 the asymptotic falloffs at large $Q^2$. 
The pion cloud contributions are obtained by switching on the coupling of effective pions
with the nucleon and the $\Delta(1232)$ systems, which
provides an extra strength to the form factors near $Q^2=0$ [See Fig.~\ref{figPionCloud1}].
The calibration of the pion cloud contributions
presented in Fig.~\ref{figGMtotal} is based 
mainly on the data below 4 GeV$^2$ from the
$\Delta^+(1232) \to \pi^0 p$ and  $\Delta^+(1232) \to \pi^+ n $ data~\cite{Kamalov01a,SatoLee01}.

In addition to the magnetic dipole form factor $G_M$ 
there are also today precise data available for 
the electric  and Coulomb quadrupole form factors, represented 
in terms of the ratios $R_{EM}$ and $R_{SM}$, defined by Eqs.~(\ref{eqEMC-RMC}).
The  JLab/Hall C~\cite{Villano09} data for large $Q^2$, 
particularly the data for $Q^2=6.4$ and 7.7 GeV$^2$,
are extracted from the unpolarized cross section 
with poor statistics for $G_E$, and different results may be obtained 
using different analyses, such as the one proposed by Ref.~\cite{CLAS09} (JLab/CLAS). 
Figs.~\ref{fig-Delta-REM} and \ref{fig-Delta-RSM} show 
the electromagnetic ratios, in percentage, at low $Q^2$ (left panel) 
and in the range $Q^2=0$--8 GeV$^2$ (right panel), which
confirms that they are in fact small.  In the figures we include, besides
the MAID parametrization~\cite{Drechsel07},
the JLab-ST parametrization~\cite{JLab-website,JLab-database,LowQ2param}  
extended till the pseudothreshold $Q^2 \simeq -0.09$ GeV$^2$.
From the comparison of the bare contributions of the dynamical models with the data,  
one concludes that the valence quark degrees of freedom effects are much more reduced 
for $G_E$ and $G_C$ than for $G_M$ at low $Q^2$.
We also included in Figs.~\ref{fig-Delta-REM}
and \ref{fig-Delta-RSM}, 
parametrizations based on large $N_c$ calculations by Buchmann, 
Pascalutsa and Vanderhaeghen~\cite{Pascalutsa07a,Buchmann04a}.
As discussed in the following sections, the large $N_c$ results can be 
interpreted as a simulation of the pion cloud effects
(see Section~\ref{secLargeNc}) and they
also suggests that, contrary to $G_M$, $G_E$ and $G_C$ 
may be dominated by pion cloud effects 
at low $Q^2$~\cite{Pascalutsa07a,Siegert3,Blomberg16,Buchmann04a,Siegert1}.

A general conclusion of this section is that the description of the data 
for the  $\gamma^\ast N \to \Delta(1232)$ transition is obtained within dynamical 
coupled-channel reaction framework. This is the case of the Sato-Lee and DMT models
shown in Fig.~\ref{figGMtotal}.
A closing remark is that the Sato-Lee and DMT bare or quark-core estimates 
at low $Q^2$ are compatible with most quark model 
calculations~\cite{Blomberg16,JDiaz07b,Siegert1}.
Next sections presents results from some of these models.

\begin{table}[t]
\begin{center}
\begin{tabular}{l   c  c }
\hline
\hline
           &    $G_M(0)$  & $G_E(0)$ \\
\hline
\hline
Data              & 3.02  &  0.076 \\
                  &       &         \\
$SU(6)$ symmetry  & 2.3   &  0.0 \\
SL (bare)~\cite{JDiaz07b,SatoLee01}         & 2.0   &  0.025    \\
DMT (bare)~\cite{Kamalov99a}      & 1.65  &  $-0.008$     \\
Non relativistic QM~\cite{Isgur82,Koniuk80}  & 2.2 & $-0.002$ \\
Relativistic QM~\cite{Capstick90}         & 2.3  & $-0.003$ \\
Hypercentral QM~\cite{Aiello96} & 2.0  &  0.003  \\   
QM+ MEC~\cite{Buchmann97}      &  2.2  &  0.075 \\
MIT~\cite{Donoghue75}          &  2.2  &        \\
                               &       &        \\
CBM~\cite{Bermuth88}           &  2.1  & 0.041   \\
CBM~\cite{Fiolhais96}          &  2.4  & 0.042 \\
CBM ($R=0.9$ fm)~\cite{Lu97}   &  2.5  &     \\
CBM ($R=0.8$ fm)~\cite{Lu97}   &  2.7  &      \\
Chiral QM~\cite{Faessler06}    &  2.9  & 0.090 \\
\hline
\hline
\end{tabular}
\end{center}
\caption{\footnotesize 
$\gamma^\ast N \to \Delta(1232)$ transition. 
Calculations of $G_M$ and $G_E$ at the photon point from different frameworks.
At the top we include estimates based exclusively on quarks.
At the bottom, after the MIT estimate, we present calculations 
that take into account effects associated with meson cloud dressing.
\label{table-GM0-GE0}}
\end{table}

\subsubsection{\it The role of valence quarks and the meson cloud \label{sec-QM-MC}}

We divide this discussion in two parts.
First we analyze the results for magnetic dipole form factor $G_M$.
After that we analyze the results for the electric quadrupole ($G_E$)
and Coulomb quadrupole ($G_C$) form factors.

\subsubsection*{\it Discussion of results for $G_M$}

The $SU(6)$ results of Eq.~(\ref{eqGM-QM}) for $G_M(0)$
assumes that in the limit $Q^2=0$ 
(when $|{\bf q}| \simeq 0.26$ GeV),
the overlap of the nucleon and $\Delta(1232)$ radial wave functions 
is maximum, providing an upper limit to the $G_M$ estimate.
But in general, when $|{\bf q}| = \frac{M_\Delta^2 -M^2}{2 M_\Delta} > 0$,
the overlap of the radial wave functions deviates from 
unity by a term proportional to $|{\bf q}|$,
reducing even more the estimate from  Eq.~(\ref{eqGM-QM}).


Relativistic quark model calculations correct the non relativistic $SU(6)$ quark model 
estimate to $G_M(0)$ $\simeq 2$~\cite{JDiaz07b}, 
a value used as a
reference to the bare parametrization of the Sato-Lee model 
and other dynamical models~\cite{SatoLee96,SatoLee01}.
Different classes of constituent quark models 
provide similar results for $G_M$ at low $Q^2$. 
A list of estimates of $G_M(0)$ from different representative valence quark models
is presented in the upper part of Table~\ref{table-GM0-GE0}.
Estimates of $G_E(0)$ are also included in the last column for future discussion.
From the table, one confirms that estimates from quark core contributions
(upper part) are about $G_M(0) \approx 2.2$.
In particular we note that the  result of the MIT bag model 
when the pion production at the bag surface is not taken 
into account is $2.2$~\cite{Thomas84,Donoghue75}.
However, the lower part of Table~\ref{table-GM0-GE0}
shows larger values for a family of cloudy bag models.
When meson cloud effects are considered, as in cloudy bag models (CBM), 
the results can rise above $G_M(0)= 2.3$.
The same happens when we consider chiral quark models~\cite{Pascalutsa07,Faessler06},
where the effect of the chiral symmetry on quark 
properties (mass, anomalous magnetic moments, etc.),
improves the description of the data at low $Q^2$.
In summary, one obtains values closer to 
the physical value for $G_M(0)$ by considering baryon and  meson degrees of freedom together.

In the right panel of Fig.~\ref{figGMtotal}, 
we have compared the $G_M$ data with the quark model calculations and 
also with the bare estimates of the Sato-Lee and DMT models.
We discuss here also the covariant spectator quark model (CSQM),
(see Section~\ref{secCSQM}) and the light-front quark 
model (LFQM) from Aznauryan and Burkert, 
which takes into account the dependence of the quark mass on $Q^2$~\cite{Aznauryan15a}.
In this last calculation the valence quark contribution 
of the LFQM is adjusted to the $Q^2 > 5$ GeV$^2$ data, taking into account 
effectively the pion cloud effects 
on the normalization of the $\Delta(1232)$ 
wave function~\cite{Aznauryan15a}.
Similar results are obtained with other quark models.
Examples are the hypercentral quark model~\cite{Sanctis05,Santopinto12}, 
the LFQM from Capstick and Keister~\cite{Capstick95}
and the Diaz and Riska~\cite{JDiaz07b,JDiaz05},
and calculations based on the Bethe-Salpeter equation~\cite{Ronniger13}. 
From the magnitude of the quark model estimates 
and from the bare estimates of the dynamical models,
one can conclude that, the bare contributions dominate above 3 GeV$^2$.
This result is a consequence of the fast falloff of the pion cloud contributions:
from the discussion of Section~\ref{sec-largeQ2}, one infers that the 
falloff of the meson cloud contributions goes like $1/Q^8$,
while the bare contributions have a $1/Q^4$ behavior.
Notice in any case that the magnitude of the pion cloud contributions 
is fundamental to fill the gap between  bare and dressed contributions. 
There are also estimates of transition form factors based on 
soliton models~\cite{Kim20,Ledwig10a,Ledwig09a}.
The estimates at low $Q^2$ are comparable with the data but
the falloffs are in general slower that the suggested by
the physical data~\cite{DecupletDecays2,Kim20}.

There are also calculations of transition form factors based
on Generalized Parton Distributions (GPDs).
In this approach elastic and inelastic interactions of baryons with photons 
are described using structure functions that separate the mechanism  
of the interacting parton (quark) from the role of 
the remaining partons (quarks, antiquarks and gluons)~\cite{Pascalutsa07,Goeke01a}.
These structure functions can be measured in deeply virtual Compton scattering 
$\gamma^\ast N \to \gamma N^\ast$, where the initial (virtual) photon 
has a very large virtuality $Q_h^2$.
The GPDs depend on both transitions $N \to N$ or $N \to \Delta$
and on three kinematic variables:
the fraction of momentum of the quark (Bjorken $x$ variable),
the transverse component of 
the Compton scattering transfer momentum 
and the virtuality of the transition $Q_h^2$~\cite{Pascalutsa07}.
The applications of the GPD formalism to the 
$\gamma^\ast N \to \Delta$ transition uses sum rules
that relate the GPD functions with the nucleon elastic and
$\gamma^\ast N \to \Delta$ form factors.
The estimates are expected to be valid for large $Q^2$, 
where the single quark-photon interaction dominates~\cite{Pascalutsa07,Stoler03}. 
Some applications use $\gamma^\ast N \to \Delta$ and $\gamma^\ast N \to N$  
large $N_c$ relations based on the isovector character of 
the transition (see Section~\ref{secLargeNc}).
The calculations provide a good description of $G_M$ at large $Q^2$, 
but the obtained results lack strength at low $Q^2$~\cite{Burkert04,Stoler03}.
The results are improved when Regge parametrizations 
of the GPDs are considered~\cite{Guidal05a}. 
The GPDs formalism has also been used to estimate 
the impact of the two-photon exchange contributions to 
the $\gamma^\ast N  \to \Delta(1232)$ transition form factors~\cite{Pascalutsa06b}.


The limitation of the calculations based on valence quark models  
in the description of $G_M$ can be illustrated in some frameworks.
Within the covariant spectator quark model (Section~\ref{secCSQM}),
if one takes into account only 
the contributions associated with the valence quark degrees of freedom,
one can conclude that when the nucleon and $\Delta(1232)$ systems 
are dominated by quark-diquark $S$-state configurations,
the quadrupole form factors vanish ($G_E=0$ and $G_C=0$).
In that case, one obtains for the magnetic 
dipole form factor~\cite{NDeltaD,NSTAR2017,NDelta,DecupletDecays}
\ba
G_M^{\rm B}(Q^2) =  \frac{4}{3\sqrt{3}} 
\frac{M}{M_\Delta + M} \left[
f_{1-} (Q^2)+ \frac{M_\Delta + M}{2 M} f_{2-}(Q^2) 
\right] \int_k \psi_\Delta(P_\Delta,k) \psi_N(P_N,k),
\label{eqGM-bare}
\ea
where $f_{i-} (Q^2)$ $(i=1,2)$ are the
Dirac and Pauli isovector constituent 
quark form factors and $\psi_\Delta$, $\psi_N$ 
are the $\Delta$  and nucleon radial wave functions, respectively.
We assume here that the radial wave functions are non negative functions.
In the limit $Q^2=0$ the Cauchy-Schwarz inequality
for integrals\footnote{In the $\Delta$ rest frame,
we can use the Cauchy-Schwarz inequality for integrals, 
valid for real radial functions, to write~\cite{NDeltaD,DecupletDecays}
\ba
\left(\int_k \psi_\Delta (P_\Delta,k)\psi_N(P_N,k)\right)^2 \le  
\left(\int_k [\psi_\Delta(P_\Delta,k)]^2\right) 
\left( \int_k [\psi_N(P_N,k)]^2 \right). 
\nonumber
\ea
Since the $\Delta$ radial wave function is normalized to unit,
when $P_\Delta=(M_\Delta,0,0,0)$, one has $\int_k [\psi_\Delta(P_\Delta,k)]^2=1$.
As for the nucleon radial wave function, one has by construction
$\psi_N (P_N, k) \le \psi_N(\bar P_N,k)$, 
where $\bar P_N=(M,0,0,0)$ corresponds to the case $|{\bf q}|=0$.
Thus $\int_k [\psi_N(P_N,k)]^2 \le \int_k [\psi_N(\bar P_N,k)]^2 =1$,
according with the normalization of $\psi_N$.
One can then write $\int_k \psi_\Delta \psi_N \le 1$, 
where the equality stands only for the case $|{\bf q}|=0$,
equivalent to the limit $M_\Delta = M$.} 
shows that the overlap integral of the two radial wave functions
satisfies the relation
$\int_k \psi_\Delta \psi_N \le 1$.
When we include the values 
of the quark isovector form factors at $Q^2=0$ in Eq.~(\ref{eqGM-bare})
one obtains $G_M(0) \le 2.07$, below the experimental value 
from Eq.~(\ref{eqGM0})~\cite{LatticeD,NDeltaD,NDelta}.
This estimate is just an upper limit (see footnote), and lower underestimations of $G_M$
can be obtained~\cite{DecupletDecays}.
The two independent parameters of the $\Delta(1232)$ 
radial wave function (see Section~\ref{secCSQM})
are adjusted in order to reproduce the EBAC results.

\begin{figure}[t]
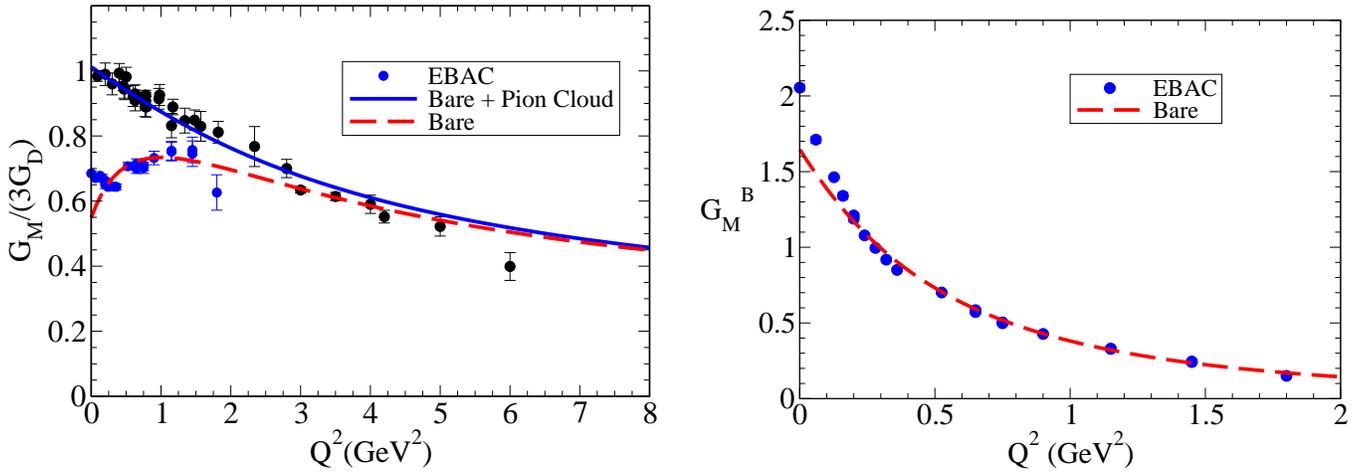

\begin{center}
\mbox{
\includegraphics[width=3.4in]{GM-CSQM-v1} \hspace{.3cm}
\includegraphics[width=3.4in]{GM-EBAC}}
\caption{\footnotesize
$\gamma^\ast N \to \Delta(1232)$ transition. 
{\bf Left panel:} 
Results from the covariant spectator quark model in the $S$-state approximation
for the magnetic form factor~\cite{NDelta}. 
Data from DESY, SLAC and JLab~\cite{Frolov98,CLAS06b,DESY68,SLAC75,DESY72}.
{\bf Right panel:} Comparison of the fit of $G_M^B$ with the EBAC results
(contribution from the core)~\cite{JDiaz07b}.
The fit varies only the two parameters of the 
$\Delta$ radial wave functions~\cite{LatticeD,NDelta}.
\label{figGM-CSQM1}}
\end{center}
\end{figure}

The results from the covariant spectator quark model 
for the magnetic form factor in the $S$-state approximation, 
are presented in Fig.~\ref{figGM-CSQM1} together with the EBAC estimate
of the bare contributions (blue bullets). 
The figure shows that the bare component of the model describes the data only at large $Q^2$.
This confirms the previously discussed general result
that valence quark based contributions are 
insufficient to describe the low-$Q^2$ data.

To fix the problem, within the covariant spectator quark model
it is considered a simplified parametrization 
of the pion cloud contributions $G_M^\pi$
using a product of two dipoles 
$G_M^\pi (Q^2) = 3 \lambda_\pi \left( \frac{\Lambda_\pi^2}{\Lambda_\pi^2 + Q^2}
\right)^2 G_D$,
where  $\lambda_\pi =0.441$, $\Lambda_\pi^2= 1.53$ GeV$^2$
and $G_D$ is the nucleon dipole form factor~\cite{LatticeD,NDeltaD,DecupletDecays}.
This parametrization is motivated by 
the representation of the $G_M$ data, normalized by $3G_D$, 
by the expected falloff at large $Q^2$
for a $(qqq)(q \bar q)$ system
(proportional to $1/Q^8$, see Section~\ref{secPQCD})
and by the shape of the function $G_M - G_M^B$, where $G_M$ 
is a parametrization of the data~\cite{NDelta,Timelike}.
The result of the combination of the bare valence quark contributions with 
the pion cloud parametrization just described is also presented 
in the left panel of Fig.~\ref{figGM-CSQM1}, normalized by  $3G_D$.
The inclusion of the pion cloud parametrization 
improves the description at low $Q^2$, 
without destroying the description in the large-$Q^2$ region.
In more recent works the pion cloud parametrization was improved
by taking a decomposition into two terms associated 
to the two diagrams shown in Fig.~\ref{figPionCloud1}:
the coupling with the pion, where we include an 
explicit dependence on the pion physical form factor
$F_\pi (Q^2)$, which is well known experimentally,
and the coupling with the intermediate baryon state. 
These improvements were inspired and led to the possibility of extending 
the model to the octet baryon to decuplet baryon transitions
and to the timelike region ($\Delta$ Dalitz decay: 
$\Delta \to e^+ e^- N$)~\cite{Timelike,Timelike2,DecupletDecays2,DecupletDecays3}.

The limitations of the description of the $G_M$ data 
by valence quark degrees of freedom were explicitly illustrated here
naturally
within the context of the covariant spectator quark model formalism.
But similar results are obtained by other constituent quark models, 
even if they comply with somewhat different
overlaps between the radial wave functions of the initial and final states.
The results of diverse constituent quark models are
compiled on the upper part of Table~\ref{table-GM0-GE0}
indicates that different quark models,
calibrated by the nucleon data, can lead to similar results.

\begin{figure}[t]  
\begin{center}
\mbox{
\includegraphics[width=2.7in]{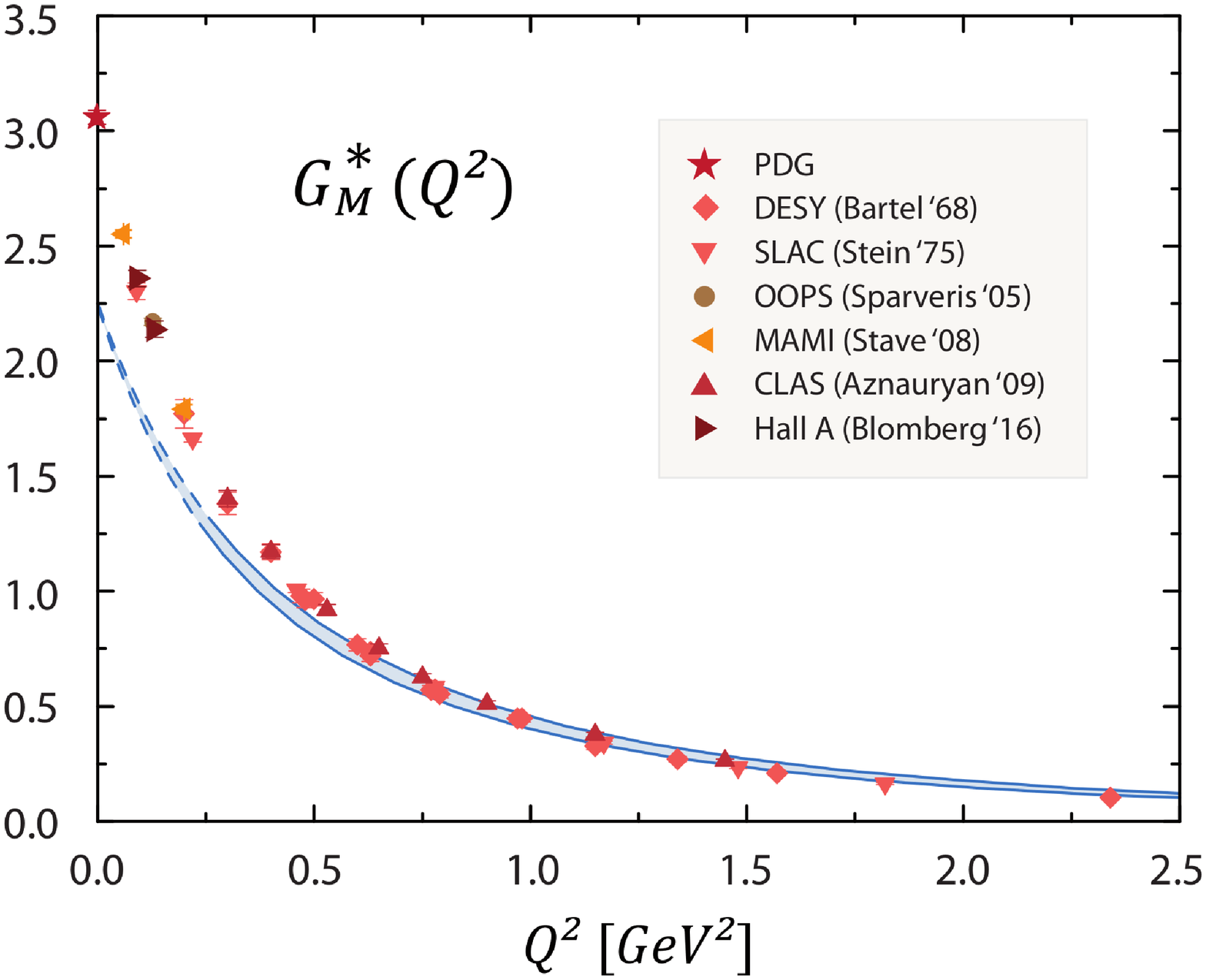}} \hspace{.5cm}
\includegraphics[width=3.5in]{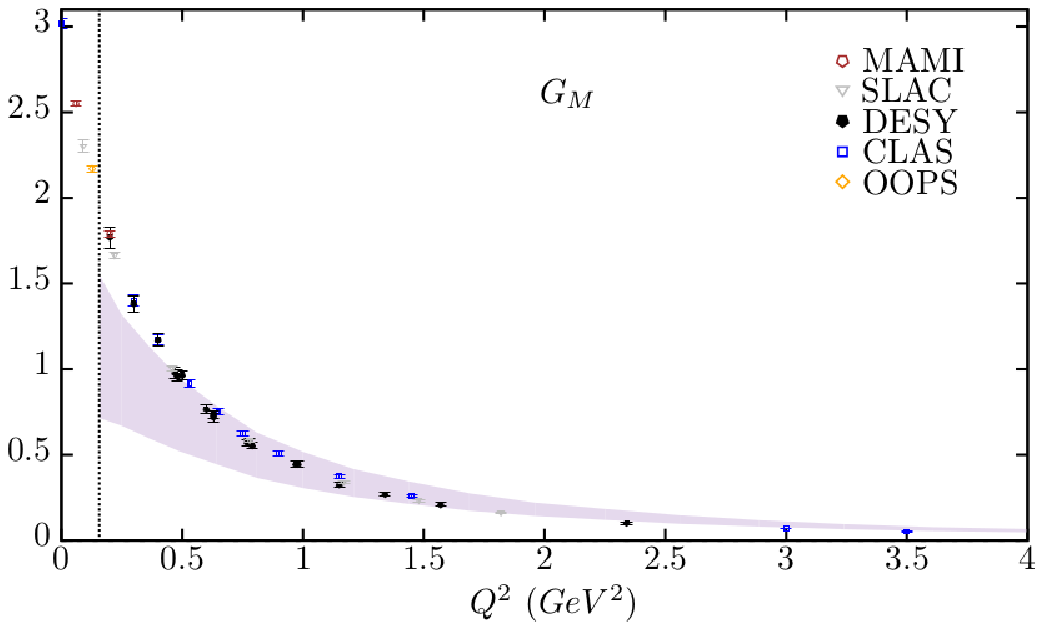}
\caption{\footnotesize
$\gamma^\ast N \to \Delta(1232)$ transition. 
Jones and Scadron form factor $G_M^\ast \equiv G_M$ with 
the Dyson-Schwinger equation formalism.
{\bf Left panel:} results from Refs.~\cite{Eichmann16,Eichmann12}.
{\bf Right panel:} results from Ref.~\cite{SAlepuz18}.
The data are from DESY~\cite{DESY68,DESY72}, SLAC~\cite{SLAC75},
OOPS/MIT-Bates~\cite{Sparveris05a}, MAMI~\cite{Stave08}, 
JLab/CLAS~\cite{CLAS09} and JLab/Hall A~\cite{Blomberg16}.
Courtesy of Gernot Eichmann and Christian Fischer. 
\label{fig-GM-DSE}}
\end{center}
\end{figure}

Dyson-Schwinger calculations for $G_M$ from the Graz~\cite{Eichmann16,Eichmann12} 
and Giessen~\cite{Eichmann16,Eichmann12,SAlepuz18} groups are presented in Fig.~\ref{fig-GM-DSE}.
Similar results for $G_M$ are obtained by the Argonne group~\cite{Segovia14b,Lu19}.
The underprediction of $G_M$ at low $Q^2$  by quark models discussed above
is also observed in these calculations based on the Dyson-Schwinger equations
for three dynamical quarks.
The calculations from Refs.~\cite{Eichmann16,Eichmann12,SAlepuz18} 
use the quark-diquark approximation to the nucleon and $\Delta(1232)$
systems, and also throughout the whole rich baryon spectrum.
This is proven to be a good approximation 
to the three-body calculations~\cite{Eichmann16,SAlepuz18},
indicating that diquarks are valid strong-interaction quasiparticle
degrees of freedom at the hadronic scale in the interior of baryons.
Calculations from the Argonne group
in a first stage~\cite{Segovia14b,Segovia16a} used 
a contact interaction, but more recently the contact interaction
was replaced by a QCD-kindred interaction, 
using quark and diquark dressed propagators~\cite{Lu19,Segovia20b}.
In Fig.~\ref{fig-GM-DSE},
the underestimation of the data below 0.5 GeV$^2$ is evident.
In light of what phenomenological quark models
and dynamical coupled channel models tell us on the importance of the pion cloud,
we can understand the lack of strength near $Q^2=0$
of the Dyson-Schwinger calculations also, since so far they do not include $(qqq)(q \bar q)$ states.


\subsubsection*{\it Discussion of results for $G_E$ and $G_C$}

Although smaller than the dominant magnetic form factor,
the two quadrupole form factors $G_E$ and $G_C$ of the electroexcitation of the $\Delta (1232)$
are an interesting source of structure and deformation information.
Experiments can access only the spectroscopic quadrupole moment of the baryons
(average of the spin projections). This is zero for spin $\frac{1}{2}$ 
particles, meaning that for the nucleon the intrinsic quadrupole moment cannot 
be measured~\cite{Gross06a,Nucleon,Fixed-Axis,Buchmann01,Buchmann07a,Kvinikhidze06,Miller03}.
As for the $\Delta(1232)$, direct measurements of the elastic form factors, 
which can give direct information about the shape of the system,
are difficult due to the short particle 
lifetime~\cite{PDG2022,DeltaFF1,DeltaFF2,Bernstein03,Delta-Shape,Alexandrou09a}.
Quark model calculations of the quadrupole form factors $G_E$ and $G_C$
have to go beyond valence quark $s$-states in the baryon wave functions, and
consider deviations of valence quark distributions from spherically symmetric distributions.
This makes the two sub-leading form factors interesting manifestations
of the deformation of the nucleon and $\Delta(1232)$ from spherical 
shape~\cite{Isgur82,Pascalutsa05b,Stave06,Stave08,Blomberg16,Bernstein03,Delta-Shape},
by opening the possibility of asymmetric components
of the wave functions being extracted from the two $\gamma^\ast N \to \Delta(1232)$ 
quadrupole form factors~\cite{Pascalutsa07,Stave08,Sparveris13,Buchmann01}.

Unfortunately, contributions from
$d$-wave quark states to the nucleon and/or $\Delta(1232)$ radial 
wave functions provide only a small fraction (10--20\%) of the measured 
magnitudes~\cite{Isgur82,Pascalutsa05b,JDiaz07b,Bernstein03,Buchmann01},
as seen for example in the results for $G_E(0)$ 
in the upper part of Table~\ref{table-GM0-GE0}, and displayed in Figs.~\ref{fig-Delta-REM} 
and \ref{fig-Delta-RSM}  for the bare quark parametrizations of $G_E$ and $G_C$ 
in the Sato-Lee model.
Non vanishing  $R_{EM}$ and $R_{SM}$ form factors  
are also manifestations of another type of dynamical deformation: 
non valence quark-antiquark pairs as degrees of freedom that contribute
to pion cloud extensions of the baryon quark cores~\cite{Buchmann01,Buchmann97}.
In other words, contributions to $G_E$ and $G_C$ are generated when
the pion cloud dressing of the nucleon and $\Delta(1232)$ is taken into account.

In this picture the nucleon and the $\Delta(1232)$ are interpreted
respectively as symmetric three-quark core states of spin
$S=\frac{1}{2}$ and $S=\frac{3}{2}$, 
surrounded by $p$-wave pion states, 
and both systems are deformed~\cite{Buchmann01}.
In the nucleon case (prolate shape) the pion cloud is oriented in the spin direction ($\hat z$)
and in the $\Delta(1232)$ case (oblate shape)  transversely to
the spin direction~\cite{Bernstein03,Buchmann01}.
From the possible contributions to the quadrupole form factors, 
pion cloud and $d$-wave quark states contributions, 
the pion cloud is the dominant effect and accounts for  
about 80-90\% of the total~\cite{Bernstein03,Buchmann01,Buchmann07a}.
The magnitude of the pion cloud contributions to the  $G_E$ and $G_C$
form factors was already inferred in Figs.~\ref{fig-Delta-REM}
and \ref{fig-Delta-RSM},
from the comparison of bare contributions with the data.
The pion cloud contributions to the quadrupole form factors 
derived within the large $N_c$ limit are discussed in Section~\ref{secLargeNc}.

The electric $G_E$ and Coulomb $G_C$ quadrupole form factors and the ratios $R_{EM}$ and $R_{SM}$
have been also calculated in the Dyson-Schwinger equation
framework~\cite{Eichmann16,Eichmann12,SAlepuz18,Segovia13a,Segovia14,Lu19,Segovia16a,Segovia20b}.
Notice, however, that the ratios are calculated 
using the Dyson-Schwinger result for $G_M$, which enhances 
the ratios at low $Q^2$, since $G_M$ in Dyson-Schwinger equations underestimate 
the physical function in that region.  
However, calculations based on the quark-diquark
approximation~\cite{Eichmann16,Eichmann12,SAlepuz18}
suggest that the results above $Q^2=1$ GeV$^2$ are similar 
to the physical results for the quadrupole form factors, 
and for the ratios\footnote{Above $Q^2=1$ GeV$^2$, the pion cloud 
effects are substantially reduced in $G_E$, $G_C$ and $G_M$, 
and the Dyson-Schwinger quark-diquark calculations 
provide a good approximation to the physical functions.
As a consequence the calculations of the ratios $R_{EM}$ and $R_{SM}$
are also similar to the measured ratios.~\cite{SAlepuz18}.}.
Below $Q^2=1$ GeV$^2$, one has different behaviors for $R_{EM}$ and $R_{SM}$.
The results for $R_{SM}$ underestimate the data at low $Q^2$, consistent
with the expected dominance of the pion cloud effects. 
The results for $R_{EM}$ are closer to the physical data, due to significant contributions from 
$p$-wave quark states to the form factor $G_E$~\cite{Eichmann12,SAlepuz18}.
These calculations suggest that the main contribution to $G_E$
are not pion cloud effects, but the lower components 
of the quark spinors, a relativistic effect absent in non relativistic quark model calculations.
Dyson-Schwinger calculations based on contact interaction tend
to give large contributions~\cite{Segovia13a,Segovia14,Segovia16a}
to $G_E$ and $G_C$ at low $Q^2$ and a fast falloff with $Q^2$~\cite{Segovia16a}.
More recent calculations using a QCD-kindred interactions, 
also based on the quark-diquark approximation~\cite{Lu19,Segovia20b} 
predict an underestimation of the data for $R_{EM}$ at low and intermediate $Q^2$.
The results for $R_{SM}$ are similar to the results 
in the quark-diquark approximation from Refs.~\cite{Eichmann12,SAlepuz18}.

The quadrupole form factors  $G_E$ and $G_C$, have been also evaluated in lattice QCD.
Lattice QCD simulations for large pion masses
(a regime where the pion cloud effects are suppressed)
in the region $Q^2 > 1$ GeV$^2$ 
show that the valence quark contributions to $G_E$ and $G_C$
although small, they are also still important.
This topic is discussed in the following section.
The explicit comparison of the data for these two observables with
results of quark models is presented in Section~\ref{sec-Delta-lQ2},
where we compare different frameworks 
at low $Q^2$ ($Q^2 < 0.6$ GeV$^2$).


\subsubsection*{\it Short notes}

A general feature of constituent quark models is the underestimation of $G_M(0)$.
This result is relevant. It is obtained also in
relativistic Dyson-Schwinger dynamical quark models.
Within the covariant spectator quark model 
an upper limit for $G_M(0)$ is obtained (Eq.~(\ref{eqGM-bare}))
that is consistent with the results of several quark models. 
In the following sections, we show that it
is also consistent with lattice QCD simulations 
when we consider large pion masses. This consistency indicates that the origin
for the missing magnitude of the theoretical quark model results
at low $Q^2$ relates to pion cloud effects in that region,
and it has an operative value: it implies that
constituent quark models are to be tested, compared
and calibrated best against lattice QCD simulations
in the large $m_\pi$ regime, since only in
that regime pion cloud effects are suppressed.
This will be discussed in the next section.

The non-leading electric and Coulomb form factors, $G_E$ and $G_C$ would potentially
inform us on the importance of small components in the baryons wave function.
However, the information on the relative importance of contributions from non-spherical 
$l>0$ components in the wave functions and the pion cloud effects is model dependent.
This leads us at this point also to the need of resorting to lattice QCD data and
in particular to a regime of large pion masses, to
incorporate model independence, which opens up for the next section.

\subsubsection{\it Lattice QCD results \label{sec-latticeQCD-Delta} }

Lattice QCD simulations provided results for $\gamma^\ast N \to \Delta(1232)$
transition form factors 
for different pion mass values and
in different approximations 
(quenched, partial quenched and unquenched/full QCD).
The first calculations, by Leinweber, Draper and Woloshyn~\cite{Leinweber93},
for $m_\pi \simeq 0.6$--1.0 GeV suggested already that 
the electric quadrupole form factor was small compared to 
the magnetic  dipole form factor [$R_{EM}= (-3 \pm 8)\%$].
Later, lattice QCD simulations were
performed by the Nicosia-MIT group in quenched 
and full QCD approximations, within the range 
$Q^2=0$--2.5 GeV$^2$~\cite{Alexandrou08a,Alexandrou04a,Alexandrou05a,Alexandrou11a}.

\begin{figure}[t]
\begin{center}
\mbox{
\includegraphics[width=2.0in]{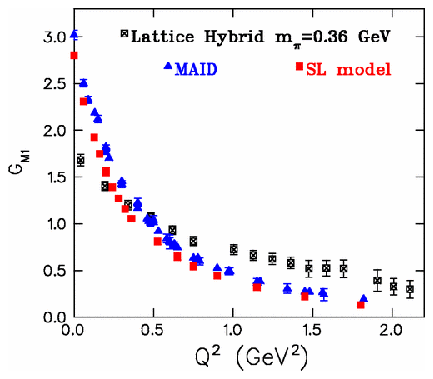} 
\includegraphics[width=2.62in]{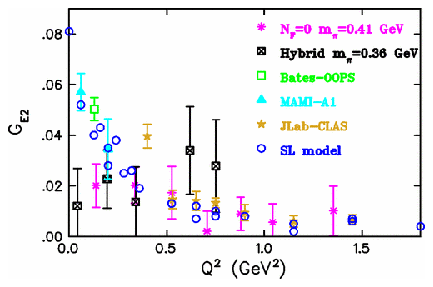}  
\includegraphics[width=2.44in]{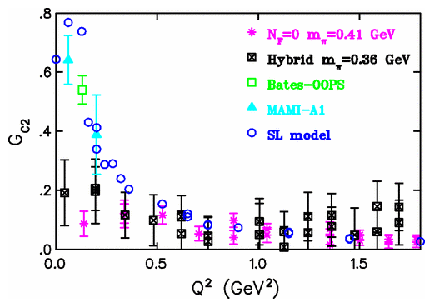} }
\caption{\footnotesize 
Lattice QCD results for the $\gamma^\ast N \to \Delta(1232)$ transition form factors
$G_{M1} \equiv G_{M}$, $G_{E2}  \equiv G_E $ and $G_{C2} =G_C$~\cite{Alexandrou08a}.
Left: $G_M$ compared with the data analysis from MAID 2007. 
The results of the hybrid action are similar to 
quenched results with $m_\pi = 411$ MeV.
Center and right: $G_E$ and $G_C$ compared with 
data pre-2009 from MIT-Bates, MAMI and JLab/CLAS. 
In all cases the results are compared to 
the analysis of the Sato-Lee model~\cite{JDiaz07b}.
Courtesy of Constancia Alexandrou.
Reprinted with permission from
\href{https://journals.aps.org/prd/abstract/10.1103/PhysRevD.77.085012}{C.~Alexandrou,
G.~Koutsou, H.~Neff, J.~W.~Negele,
W.~Schroers and A.~Tsapalis, Phys.~Rev.~D 77, 085012 (2008).} \\
Copyright (2008) by American Physical Society.
\label{fig-Lattice-NDelta}}
\end{center}
\end{figure}

The most complete study so far, includes quenched 
($m_\pi= 411$, 490 and 563 MeV)
and full QCD simulations ($m_\pi= 353$, 384, 498, 509, 594 and 691 MeV)
for $G_M$, $G_E$ and $G_C$~\cite{Alexandrou08a}.
The results show that 
the unquenched effects are within statistical errors
in the pion mass range 0.5–-0.8 GeV~\cite{Pascalutsa07,Alexandrou08a}.
The results for $G_M$ are relatively precise at low $Q^2$ (with uncertainties of about 2\%) 
but became more  inaccurate when $Q^2$ increases.
The quadrupole form factors results
are less precise due to their smaller magnitude.
The comparison of the lattice QCD data 
with the physical data for $G_M$, $G_E$ and $G_C$ is shown in Fig.~\ref{fig-Lattice-NDelta}.
This comparison shows that in general lattice QCD form factors 
underestimate the physical data near $Q^2=0$, and 
that the falloffs of the lattice results are 
slower than the physical form factors.
In general, the falloffs became slower when $m_\pi$ increases.
The lattice QCD results for $G_M$ are consistent 
with the results obtained from quark models at low $Q^2$, in a regime of $m_\pi > 400$ MeV, 
where the meson cloud effects are small.
For the quadrupole form factors, $G_E$ and $G_C$, 
the lattice results became comparable with experimental
data for $Q^2 \approx 0.5$ GeV$^2$.

Given that the pion cloud effects are suppressed for large unphysical
pion masses,
we can link now the bare contributions of the
covariant spectator quark model
to the lattice QCD regime data (see discussion in Section~\ref{secCSQM}).
The covariant spectator quark model wave function for the $\Delta(1232)$, 
 is composed by a dominant $S$-state and two small $D$ states, 
associated the core spin states $S=\frac{3}{2}$ ($D3$) 
and $S=\frac{1}{2}$ ($D1$)~\cite{LatticeD,NDeltaD}
\ba
\Psi_\Delta =
N \left[ \Psi_S  + a \Psi_{D3}  + b \Psi_{D1}  \right],
\ea
where $a$ and $b$ are two small mixture coefficients,
and $N$ is the normalization constant.

\begin{figure}[t]
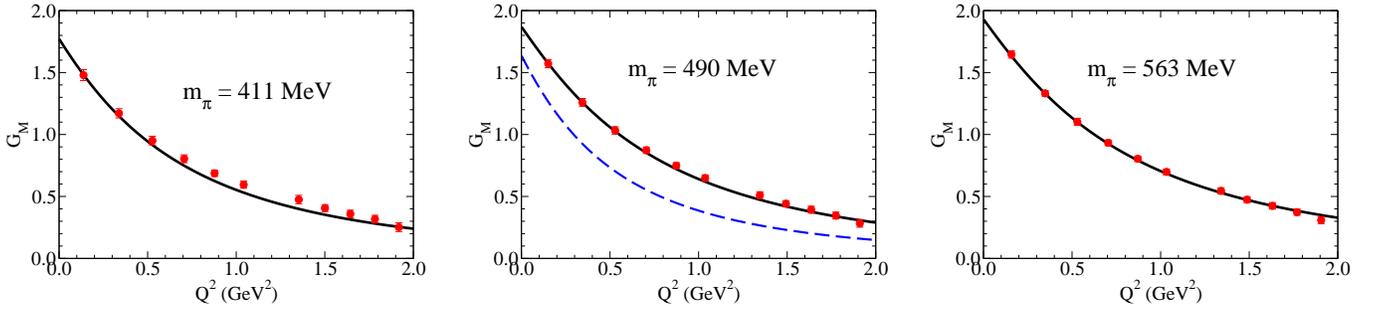

\begin{center}
\mbox{
\includegraphics[width=2.2in]{GM411a} \hspace{.3cm}
\includegraphics[width=2.2in]{GM490a}  \hspace{.3cm}
\includegraphics[width=2.2in]{GM563a} }
\caption{\footnotesize
$\gamma^\ast N \to \Delta(1232)$ transition. 
Covariant spectator quark model (solid line) compared to lattice QCD data~\cite{LatticeD}.
The dashed-line indicate the extrapolation 
to the physical limit (bare contributions).
The quenched lattice QCD data are from Ref.~\cite{Alexandrou08a}.
\label{figGM-lattice}}
\end{center}
\end{figure}

The results for the valence quark contribution for $G_M$ 
in the $S$-state approximation ($a=b=0$) were already presented in Fig.~\ref{figGM-CSQM1}.
Figure~\ref{figGM-lattice} shows the results of the 
covariant spectator quark model when extended to the lattice QCD regime
for different values of $m_{\pi}$.
Notice in Fig.~\ref{figGM-lattice} the excellent agreement
with the lattice QCD data for all values of $m_\pi$.

The calculations use the parametrization determined
by the previous fit to the EBAC estimates of the bare core, based 
on the dominance of the $S$-state, combined with the $D$-state components,
that contribute to the $G_E$ and $G_C$ form factors~\cite{Siegert3,LatticeD,Siegert1}.
The admixture coefficients and the $D3$ and $D1$ radial wave function 
range parameters are fitted to the lattice QCD data for $G_E$ and $G_C$~\cite{LatticeD}.
The calibration uses only unquenched lattice data above 400 MeV,
in order to avoid contamination from meson cloud effects,
and thus to be independent of extra assumptions and parameters. 
From the fit one obtains an admixture of 0.72\%
for both $D$ states~\cite{LatticeD,Siegert1}.
From this one can conclude that a model with $D$ states is
still dominated by the $S$-state $\Delta(1232)$ wave function.
The present parametrization for the bare contribution
at the physical point is similar to the result
displayed on the right panel of Fig.~\ref{figGM-CSQM1}.
Thus, the complete calculations and calibration with the lattice data
preserve the good description of the $G_M$ data
after the pion cloud contributions are included,
and provide a calibration of the $D$-states.
We further conclude that the covariant spectator quark model
$S$-state calibration is consistent with two estimates 
of form factors: 
the lattice QCD simulations with large values of $m_\pi$ and 
the bare quark contribution of quark models in the physical regime~\cite{Siegert3,LatticeD}
(see right panel of Fig.~\ref{figGM-CSQM1}).

\begin{figure}[t]
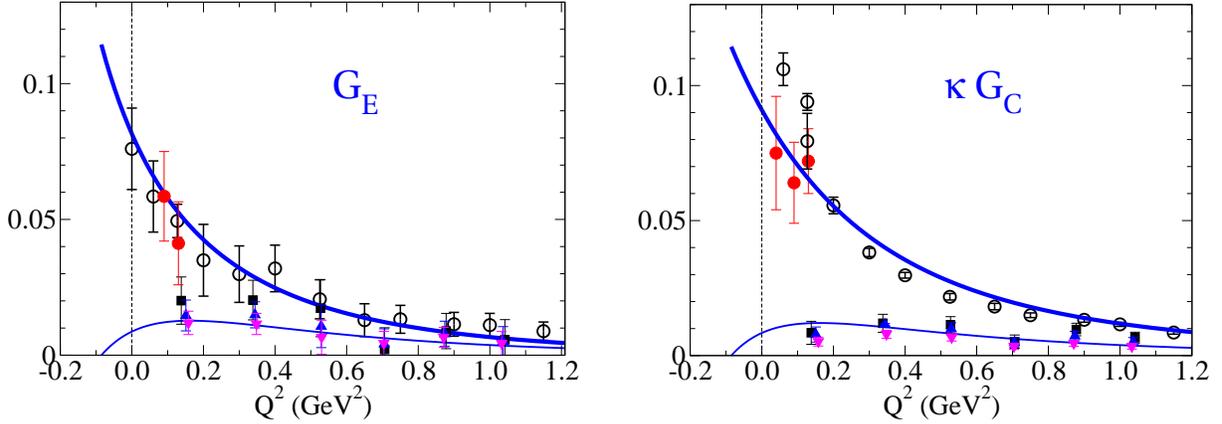

\begin{center}
\includegraphics[width=3.in]{GE-v3.eps} \hspace{.5cm}
\includegraphics[width=3.in]{GC-v3.eps} 
\caption{\footnotesize
$\gamma^\ast N \to \Delta(1232)$ transition.
Electric and Coulomb quadrupole form factor $G_E$  and $G_C$ 
(multiplied by $\kappa = \frac{M_\Delta -M}{2 M_\Delta}$) 
compared with the data and the covariant spectator quark model~\cite{Siegert3}
constrained by (quenched) lattice QCD data 
from Ref.~\cite{Alexandrou08a}
for $m_\pi = 411$, 490 and 563 MeV (represented 
by squares, upper triangles and lower triangles, respectively).
The thick line represents the combination of bare an pion cloud contributions
and the thin lines the bare contributions. 
Data from PDG, JLab/CLAS, MAMI and 
MIT-Bates~\cite{PDG2022,CLAS09,Pospischil01,MIT-Bates01,MIT-Bates03,Sparveris05a,Sparveris13} 
({\Large $\bm \circ$}) and JLab/Hall A~\cite{Blomberg16} ({\color{red}{\LARGE $\bullet$}}).
In the conversion of the $R_{EM}$ and $R_{SM}$ data to the $G_E$ and $G_C$
we use the MAID parametrization of $G_M$~\cite{Drechsel07}.
\label{fig-GEGC-CSQM}}
\end{center}
\end{figure}

We turn now to the quadrupole form factors $G_E$ and $G_C$.
In Fig.~\ref{fig-GEGC-CSQM}, we present the bare contributions (thin lines)
determined by the fit to the lattice QCD data
in comparison with the physical data.
The figure confirms that for the quadrupole form factors
the bare contribution accounts also only for a fraction of 
the measured form factors at low $Q^2$.
The two graphs feature the gap between the bare
contributions and the total
(bare quark plus pion cloud, given by the thick lines), which is close to the physical data.
What are the pion cloud contributions shown in the figure?
Following Eq.~(\ref{eqGB-MC}) the decomposition
of the quadrupole form factors $G_E$ and $G_C$ form factors
into valence quark and pion cloud components,
would allow us to extract the meson cloud effects
(defined by construction as the missing strength of both form factors)
from both the experimental and lattice data.
Nevertheless, in the case of the $\gamma^\ast N \to \Delta(1232)$ transition, 
the necessary pion cloud contributions can be estimated 
using large $N_c$ relations, discussed in detail in Section~\ref{secLargeNc}.
The large $N_c$ limit relations have no adjustable parameters, 
apart from the connection with the neutron electric form factor,
which is really a strong motivation of the interpretation
in terms of the pion cloud contribution.
In Fig.~\ref{fig-GEGC-CSQM}, we combine the valence quark contributions (thin lines)
with these large $N_c$ pion cloud parametrizations
from Eqs.~(\ref{eqGEGC-LNC}) and  (\ref{eqBeta}),
to obtain the final result for the form factors $G_E$ and $\kappa G_C$ (thick lines),
where $\kappa= \frac{M_\Delta -M}{2 M_\Delta}$.
We also show the lattice QCD data used
in the fits of the valence quark contributions of the model\footnote{Notice, however,
that the thin line does not represent the direct fit 
to the lattice data, but are the results of the extrapolation of the lattice parametrization 
to the physical limit~\cite{LatticeD}.}.


The use of $\kappa G_C$ instead of $G_C$ is done 
to simplify the discussion of the quadrupole form factors at low $Q^2$,
and to make the connection with Siegert's theorem (see Section~\ref{sec-Siegert}) that establishes 
that at the pseudothreshold, the lowest value of $Q^2$ in the graphs, 
the results of the two functions converge, $G_E = \kappa G_C$.
It can be shown that bare contributions for $G_E$ and  $G_C$ 
are proportional to the angular integral of $Y_{20} (z)$,
as a consequence of the orthogonality of 
the $\Delta(1232)$ $D$-states with the nucleon $S$-state~\cite{NDeltaD,Siegert1}.
Near the pseudothreshold the overlap integrals 
of the radial wave functions are proportional to $|{\bf q}|$
and vanish in the limit $|{\bf q}|=0$.
The corollary of this result is that the bare contributions 
to $G_E$ and $G_C$ vanish at the pseudothreshold~\cite{Siegert1}.
This property is responsible by the soft behavior 
of the bare contributions near $Q^2=0$ and the	 smooth turning point,
a necessary condition to obtain zero contributions at 
the point $|{\bf q}|=0$.
Notice that the smooth transition from the pseudothreshold ($Q^2 \simeq -0.09$  GeV$^2$)
and the region $Q^2 \simeq 0.3$ GeV$^2$ is also observed in  
the lattice QCD data.  For future reference, it is important to keep in mind that since the 
bare contributions vanish at the pseudothreshold 
(as shown in Fig.~\ref{fig-GEGC-CSQM})
the constraints at the pseudothreshold
are transferred to the pion cloud contributions. 


When we overlap the two graphs of Fig.~\ref{fig-GEGC-CSQM}, we conclude that 
$G_E \simeq \kappa G_C$ is a good approximation 
to the bare contributions for finite  $Q^2$~\cite{Siegert3,Siegert1,GlobalParam}.
A close relation, which can be derived from the large $N_c$ limit is 
$G_E (Q^2) = \frac{M_\Delta^2 -M^2}{4 M_\Delta^2} G_C (Q^2)$~\cite{Pascalutsa07a} 
(see Section~\ref{secLargeNc} for more details)
and deviates from $G_E (Q^2)\simeq \kappa G_C (Q^2)$ by a term of the order $1/N_c^2$ 
of about 12\%\footnote{Notice that 
  $\frac{M_\Delta^2-M^2}{4 M_\Delta^2} =
  \kappa \left( 1 - \frac{M_\Delta -M}{2 M_\Delta}\right) \simeq \kappa$,
a correction of the order $1/N_c^2$.}.
Some models, like the Sato-Lee model, use the relation 
$G_E = \frac{M_\Delta^2 -M^2}{4 M_\Delta^2} G_C$ to estimate the bare contribution of $G_C$ 
from the result for $G_E$~\cite{SatoLee01,JDiaz07b}.
These relations can also be used to justify the order 
of magnitude between $G_C$ and $G_E$: $G_E \approx 0.1 \; G_C$,
for any component (bare or pion cloud) near $Q^2=0$.

For future reference, we point out that according to the results 
from Fig.~\ref{fig-GEGC-CSQM}  the bare contributions became comparable 
to the physical data for $Q^2 > 0.9$ GeV$^2$ for $G_E$ 
and for $Q^2 > 1.1$ GeV$^2$ for $G_C$.

\subsubsection*{\it Short notes}

From Figs.~\ref{fig-Lattice-NDelta}, \ref{figGM-lattice} (central panel)
and  \ref{fig-GEGC-CSQM} one learns that lattice QCD calculations underestimate
the physical form factors of the $\gamma^\ast N \to \Delta(1232)$
transition at low $Q^2$.  The use of lattice QCD data and/or inspiring general
approaches as the large $N_c$ limit to constrain more detailed models
as the covariant spectator quark model, illustrates the need to combine different approaches
to interpret experimental data in light
of different and complementary physical mechanisms.

\subsubsection{\it Low-$Q^2$ region \label{sec-Delta-lQ2}}

We focus now on the low-$Q^2$ region.
Since the magnitude of $G_M$ was already discussed in detail 
in the previous sections
we look here for the quadrupole form factors $G_E$ and $G_C$, 
and the electromagnetic ratios ($R_{EM}$ and $R_{SM}$).
Also, we do not represent in this section the
results from lattice QCD simulations,
because the functions $G_E$ and $G_M$ were already discussed
in the previous section, and also because the ratios
are misleading when calculated in terms of a
lattice QCD function $G_M$ that underestimates the physical function $G_M$
in the low-$Q^2$ region, as pointed out already.
The direct comparison of the lattice QCD ratios with the physical 
ratios would give the impression that the physical and lattice QCD functions 
$R_{EM}$ and $R_{SM}$ are comparable at low $Q^2$.

On the other hand chiral EFTs have a special role
in this section on the low-$Q^2$ limit since chiral EFTs 
are based on expansions powers of $Q^2$ normalized by 
the chiral symmetry breaking scale $\Lambda_{\chi }\simeq 1$ GeV
(see Section~\ref{secEFT}).
Once calibrated the parameters of the chiral EFT (low energy constants) 
the models can be used to predict the $Q^2$-dependence of the 
transition form factors within the limits of the $Q^2$-expansions. 
Chiral EFTs are also important in the analysis 
of the lattice QCD data where extrapolations of the pion mass 
to the  physical point ($m_\pi =138$ MeV) or to the chiral limit ($m_\pi=0$), 
for fixed values of $Q^2$ are needed. 
These extrapolations must be done with care, since 
in some frameworks $G_E$ or $G_C$ diverge in 
the chiral limit~\cite{Pascalutsa07,Pascalutsa06a,Pascalutsa08a,Butler93,Gail06}.

\begin{figure}[t]
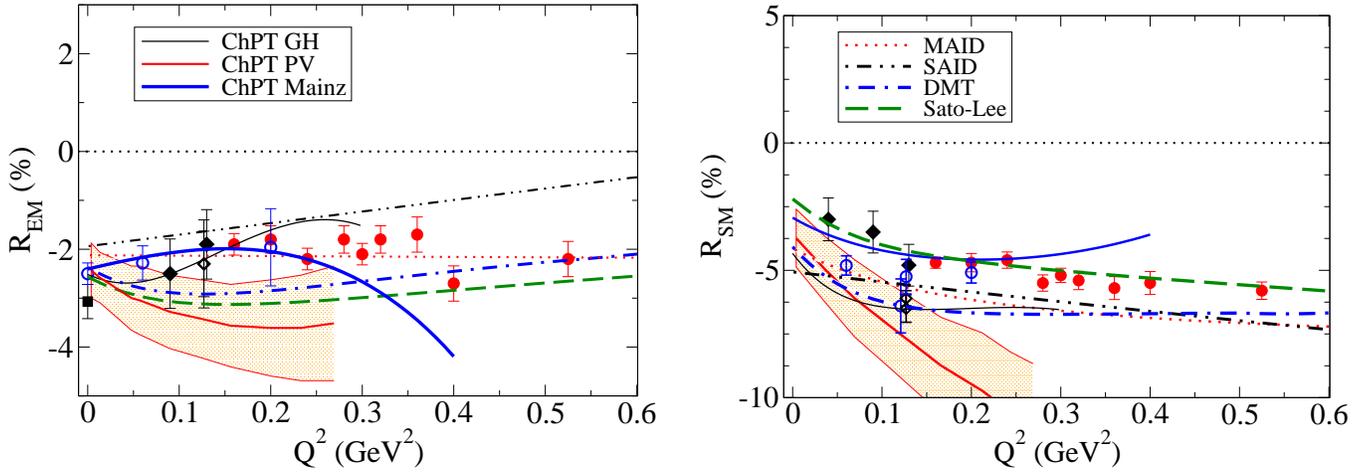

\begin{center}
\mbox{
\includegraphics[width=3.4in]{REM-ChPT-v4} \hspace{.3cm}
\includegraphics[width=3.4in]{RSM-ChPT-v4}}
\caption{\footnotesize 
$\gamma^\ast N \to \Delta(1232)$ electromagnetic ratios at low $Q^2$.
In both panels we include calculations of chiral EFTs,
dynamical models and parametrizations of the data.
The chiral EFTs curves are defined in the box of the left panel;
the dynamical models and phenomenological parametrization
curves are defined on the box of the right panel.
Chiral EFTs calculations are from Gail and Hemmert (GH)~\cite{Gail06} 
and Pascalutsa and Vanderhaeghen (PV) (within the error band)~\cite{Pascalutsa05b,Pascalutsa06a}
and the Mainz group~\cite{Hilt18}. 
The empirical parametrizations are from MAID~\cite{Drechsel07} and  
SAID~\cite{SAID-website}. 
Dynamical models are from Sato-Lee~\cite{SatoLee01}
and DMT~\cite{Kamalov01a}.  
The data are from ELSA ({\scriptsize $\blacksquare$})~\cite{ELSA97}, 
MAMI ({\B {\Large ${\bm  \circ}$}})~\cite{Beck00a,Stave08,Sparveris13}, 
MIT-Bates ({\Large ${\bm \diamond}$})~\cite{MIT-Bates03,Sparveris05a},
JLab/Hall A ({\Large $\blackdiamond$})~\cite{Blomberg16} and
JLab/CLAS ({\Large {\color{red} $\bullet$}})~\cite{CLAS09}.
\label{fig-REM-RSM-LQ2}}
\end{center}
\end{figure}

In Fig.~\ref{fig-REM-RSM-LQ2}, we compare the data below 0.6 GeV$^2$ with several models, 
including the chiral EFTs from Refs.~\cite{Pascalutsa06a,Pascalutsa08a,Gail06}.
The figure shows results from chiral theories together with results from, 
dynamical models (Sato-Lee and DMT),
and the popular MAID~\cite{Drechsel07,MAID-database} and SAID~\cite{SAID-website,Arndt02}
parametrizations of the data.
(The reader may recall from Figs.~\ref{fig-Delta-REM} and \ref{fig-Delta-RSM}
that the curves from the Sato-Lee~\cite{SatoLee96,SatoLee01} 
and DMT~\cite{Kamalov01a,Kamalov99a} take into account the effects of the pion cloud dressing 
in the low-$Q^2$ region).
Quark model calculations provide contributions 
of the order of $\pm 1\%$  to the electromagnetic
ratios~\cite{Pascalutsa07,Isgur82,Sanctis05,Glashow79,Becchi65,Krivoruchenko91,Buchmann01,Capstick90}.
The reader can consult Refs.~\cite{Pascalutsa07,Stave08,Bernstein07a,JDiaz07b}
for details of the comparison.

Several chiral EFTs have been applied to the study of the $Q^2$-dependence 
of the $\gamma^\ast N \to \Delta(1232)$ transition form factors. 
In Fig.~\ref{fig-REM-RSM-LQ2}, we present the results 
from Pascalutsa and Vanderhaeghen (PV)~\cite{Pascalutsa05b,Pascalutsa06a,Pascalutsa08a} 
based on an expansion in the scales $(M_\Delta -M)/\Lambda_{\chi}$
and $m_\pi/\Lambda_{\chi}$ ($\delta$- expansion),
and the  calculations of Gail and Hemmert (GH)~\cite{Gellas99,Gail06}
based on the so-called {\it small scale expansion} or $\epsilon$-expansion, 
where $\epsilon$ can represent  $(M_\Delta -M)/\Lambda_{\chi}$ or 
$m_\pi/\Lambda_{\chi}$ (only one scale is considered).
The free parameters of the $\delta$-expansion are fixed 
by constraints at $Q^2=0$~\cite{Pascalutsa06a}. 
The parametrization of the $\epsilon$-expansion used 
also the low-$Q^2$ data for $G_M$~\cite{Gail06}.
In some cases the accuracy of the estimate can be improved, 
including an extra $Q^2$-dependence on some functions associated 
with higher order corrections 
in the chiral expansion~\cite{Pascalutsa05b}.
For a discussion about the formal differences between 
the $\delta$- and  $\epsilon$- expansions check Ref.~\cite{Pascalutsa07}.
In the graphs shown here we include also the chiral calculation 
from the Mainz group (by Hilt et al.), calibrated by the more recent 
data for the quadrupole form factors~\cite{Hilt18}.
The curves for the chiral EFTs calculations in Fig.~\ref{fig-REM-RSM-LQ2}
denote the different $Q^2$ upper limits of validity of the different expansions shown.
These limits are 0.3 GeV$^2$ for GH, 0.27 GeV$^2$ for PV,
and 0.4 GeV$^2$ for the Mainz group.
For the calculations of Pascalutsa and Vanderhaeghen 
a theoretical error band estimated from the magnitude 
of the next leading order contributions~\cite{Pascalutsa05b,Pascalutsa06a,Pascalutsa08a}
is also shown.
Notice in the graphs the good agreement of the chiral EFTs with the $R_{EM}$ data.
As for $R_{SM}$, we can conclude that, 
except for the Mainz group, the models are a bit below 
the more recent data from JLab/Hall A~\cite{Blomberg16} 
for $Q^2 < 0.15$ GeV$^2$, and the PV 
estimate deviates more from the data  with the error band broadening
as $Q^2$ increases, up to the validity limit of the expansion.

Chiral EFTs can also be compared with lattice QCD calculations
for a given value of $m_\pi$ for fixed $Q^2$~\cite{Pascalutsa05b,Gail06}, 
as far as $m_\pi$ is not too far from the physical value (say $m_\pi <  $ 500 MeV), 
in order to use $m_\pi/\Lambda_\chi$ as expansion parameter.
In the comparison with the lattice QCD simulations it is important 
to take into account the modification of the baryon masses 
in the lattice regime~\cite{Pascalutsa07}.
The modifications on the baryon masses can be calculated in chiral EFTs
taking into account the self-energy contributions 
originated from meson cloud
dressing~\cite{Pascalutsa07,Pascalutsa05b,Pascalutsa06a,Bernard01a,Young02a,Young03a,Leinweber04a}.

An excellent example of the importance of chiral EFTs to compare physical data 
to lattice QCD data is provided by the calculations 
from Pascalutsa and Vanderhaeghen
for $R_{EM}$ and $R_{SM}$, around $Q^2=0.1$ GeV$^2$~\cite{Pascalutsa05b,Pascalutsa06a,Pascalutsa08a}.
In that region, there are measurements from  
MAMI and MIT-Bates~\cite{Pospischil01,MIT-Bates01,MIT-Bates03,Sparveris05a} for $Q^2 \simeq 0.12$ GeV$^2$
and lattice QCD simulations from Alexandrou et al.~for 
$Q^2=0.1$ GeV$^2$~\cite{Alexandrou04a}.
A linear extrapolation of the quenched lattice QCD data suggested that
the physical function $R_{SM}$ near $Q^2=0.1$ GeV$^2$ 
should be $R_{SM} (0.1 \; \mbox{GeV}^2) \simeq - 1\%$,
while the measured data at the time 
gave $R_{SM}  (0.13 \; \mbox{GeV}^2) \simeq - (6.0 \mbox{--} 6.5) \%$,  
with errors of about 0.5--1.0\%.
It was the extrapolation provided by the Pascalutsa and Vanderhaeghen
calculation~\cite{Pascalutsa05b,Pascalutsa06a,Pascalutsa08a},
with a relevant non analytic contribution 
that showed that the lattice data was indeed consistent with the physical data.
Also for $R_{EM}$ a non analytic dependence was shown important to 
make the connection between lattice QCD
simulations and physical data~\cite{Pascalutsa07}.

For completeness we mention that meanwhile it was realized
that the measurements from MAMI and MIT-Bates 
were overestimated. The most recent determinations 
decrease the magnitude of the data to 
$R_{SM} (0.13 \; \mbox{GeV}^2) = - (4.8 \pm 0.8) \%$ (JLab/Hall A)~\cite{Blomberg16}.
The main differences come for lower values of $Q^2$: 0.04 and 0.09 GeV$^2$.
The result for the lowest value of $Q^2$ is now 
$R_{SM} (0.04 \; \mbox{GeV}^2) = - (3.50 \pm 0.88) \%$ (JLab/Hall A)~\cite{Blomberg16}.
The reduction in magnitude of the $R_{SM}$ data, 
suggests that the large $N_c$ relation for $Q^2 \simeq 0$,
$R_{EM} \simeq R_{SM}$, is valid, a result  
that has been challenged by the data since 
the first accurate measurements, two decades ago.
This topic is discussed in the next section in the context 
of the large $N_c$ limit.

New lattice QCD simulations with fixed values of $Q^2$ below $0.1$ GeV$^2$, 
may be necessary to understand the new data 
from JLab/Hall A.
The values of $m_\pi$ must become closer to the physical pion mass 
in order to satisfy the limits of the chiral expansions; 
and the unquenched processes must be taken into account, 
since the assumption that the pion/meson cloud effects 
can be neglected, is no longer valid. 
New lattice QCD simulations and/or new low-$Q^2$ data 
can also be used to recalibrate the chiral EFTs mentioned above, 
and provide increasingly more accurate predictions
of $G_E$, $G_C$, $R_{EM}$ and $R_{SM}$ near $Q^2=0$.

\begin{figure}[t]
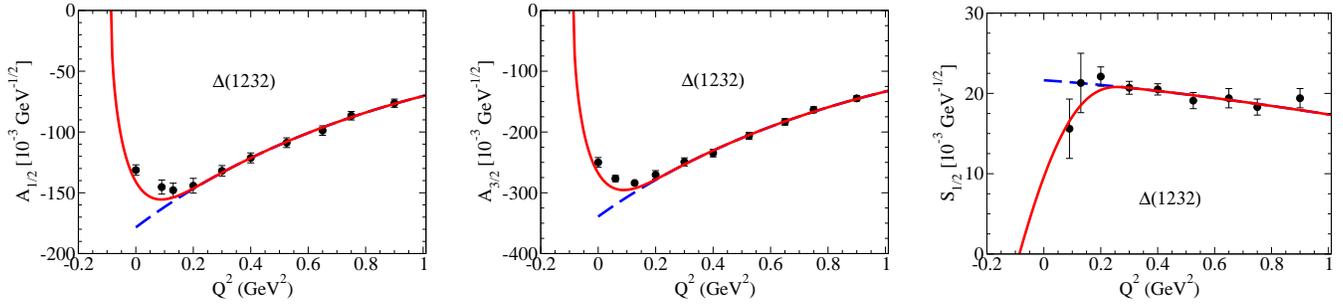

\begin{center}
\mbox{
\includegraphics[width=2.2in]{A12-D1232} \hspace{.3cm}
\includegraphics[width=2.2in]{A32-D1232}  \hspace{.3cm}
\includegraphics[width=2.1in]{S12-D1232} }
\end{center}
\caption{\footnotesize 
$\gamma^\ast N \to \Delta(1232)$ helicity amplitudes. 
JLab parametrization for the $\Delta(1232)$ amplitudes~\cite{JLab-website} 
(dashed blue line), and parametrization (JLab-ST) corrected below $Q^2=0.3$ GeV$^2$ 
by the pseudothreshold constraints (solid red line)~\cite{LowQ2param}. 
The data are from PDG 2022~\cite{PDG2022}, MAMI~\cite{Stave08,Sparveris13},  
MIT-Bates~\cite{MIT-Bates03,Sparveris05a}, JLab/Hall A~\cite{Blomberg16}
and JLab/CLAS~\cite{CLAS09}. 
\label{fig-Siegert-Delta1}}
\end{figure}

To finish the discussion about the transition form factors near $Q^2=0$, 
we recall the discussion in Section~\ref{sec-Siegert}
about the constraints on the form factors $G_E$ and $G_C$ 
and the helicity amplitudes $A_{1/2}$, $A_{3/2}$ and $S_{1/2}$, 
near the pseudothreshold. 
In Fig.~\ref{fig-Siegert-Delta1}, 
we present the JLab-ST parametrizations~\cite{JLab-website,LowQ2param} for these amplitudes,
discussed in Figs.~\ref{fig-Delta-REM} and \ref{fig-Delta-RSM} for
the transition form factors, at intermediate and large $Q^2$.
The solid line provides a good description of the low-$Q^2$ data, 
and exhibits a smooth approximation to the pseudothreshold,  
consistent with the expected leading dependence on $|{\bf q}|$: 
$A_{1/2}$, $A_{3/2} \; \propto \; |{\bf q}|$  and $S_{1/2} \; \propto \; |{\bf q}|^2$
(Section~\ref{sec-Siegert}, Table~\ref{table-Siegert1}). 
The main conclusion is then that, we cannot ignore the constraints 
on the helicity amplitudes near the pseudothreshold (limit $|{\bf q}|=0$),
and also that naive parametrizations of the data may fail in the verification 
of those constraints and in the description of the low-$Q^2$ data~\cite{LowQ2param}.
The signature of the pseudothreshold constraints is manifest 
in the turning points observed in the three 
$\gamma^\ast N \to \Delta(1232)$ helicity amplitudes.

\subsubsection*{\it Short notes}

In the small-$Q^2$ region: i) chiral EFT models
define extrapolations of lattice data
to the physical limit providing a prime example of the importance
of combining different approaches to make progress in hadron physics.
ii) data parametrizations of helicity amplitudes that are not guided
by Siegert's theorem may lead to unphysical behavior.

\subsubsection{\it Large $N_c$ limit   \label{secLargeNc}}

In a pure $SU(6)$ symmetric model 
the electric form factor of the neutron is identically zero ($G_{En} \equiv 0$),
and the $\gamma^\ast N \to \Delta (1232)$ 
electric and Coulomb quadrupole
form factors also identically zero ($G_E \equiv 0$ and $G_C \equiv 0$).
Equivalent results are obtained in the large $N_c$ limit, a limit where the baryons
are infinitely massive  and exact $SU(2 N_f)$
spin-flavor symmetry is obtained, with $N_f$ the number of light quark flavors.
This property implies that in this approach no model dependent interaction
and form of wave function are necessary to generate many of the results
that are otherwise obtained by spin-flavor $SU(6)$ symmetric quark models.
In the large $N_c$ limit the baryon masses, $M$ and $M_\Delta$ scale 
with $N_c$, while the mass difference is $M_\Delta -M= {\cal O}(1/N_c)$~\cite{Pascalutsa07a,Jenkins02}.

Although the large $N_c$ limit gives results only for $Q^2=0$, 
it can be extended to finite $Q^2$
using the low-$Q^2$ expansion
\ba
G_{En} (Q^2) \simeq -  \frac{1}{6} r_n^2 Q^2,
\label{eqGEn-lQ2}
\ea
where $r_n^2 = -0.1161 \pm 0.0022$ fm$^2$~\cite{PDG2022}
is the neutron electric square radius.

In the leading order of large $N_c$, 
the magnetic form factor is $G_M(0) = {\cal O}(N_c^0)$,
while $G_E (0)$ and $G_C(0)$ correspond to contributions 
${\cal O}(1/N_c^2)$~\cite{Jenkins94a,Jenkins02}.
For the transition magnetic moment, 
we obtain up to $1/N_c^2$ corrections~\cite{Jenkins94a}
\ba
\mu_{N \Delta} = \frac{1}{\sqrt{2}}(\kappa_p - \kappa_n), 
\hspace{1.5cm}
G_M(0) \simeq 2.62,
\label{eqGM-LNC}
\ea
where $\kappa_p = 1.793$ and $\kappa_n= -1.913$ are 
the nucleon anomalous magnetic moments
($\kappa_V = \kappa_p -\kappa_n$ defines the isovector anomalous magnetic moment).
The underestimation of the data by these results is about 13\%, 
an error consistent with terms of the order $1/N_c^2$.

Calculations based on the large $N_c$ limit have shown that~\cite{Pascalutsa07a}
\ba
G_E^\pi (Q^2) = \left( \frac{M}{M_\Delta}\right)^{3/2} \frac{M_\Delta^2 - M^2}{2 \sqrt{2}}
\frac{\tilde G_{En} (Q^2)}{ 1 + \beta Q^2},
\hspace{1.cm}
G_C^\pi (Q^2) = \left( \frac{M}{M_\Delta}\right)^{1/2} \sqrt{2} M_\Delta  M  \tilde G_{En} (Q^2),
\label{eqGEGC-LNC}
\ea
where $\tilde G_{En} (Q^2) \equiv G_{En}(Q^2)/Q^2$.

The relation for $G_C$ was first derived by Buchmann and collaborators
within a constituent quark model with two-body 
exchange currents~\cite{Buchmann97}.
These two-body exchange currents include 
diagrams with creation of quark-antiquark
pairs induced by pion exchanges
between quarks~\cite{Buchmann04a,Buchmann97,Buchmann02b,Buchmann02c,Grabmayr01a}
the reason why they are classified as pion cloud effects.
An equivalent relation was later derived by
Buchmann et al.~within the large $N_c$ limit formalism~\cite{Buchmann02b,Buchmann02c}.
The relation for $G_E$ with $\beta=0$ was obtained
by Pascalutsa and Vanderhaeghen
in the large $N_c$ limit~\cite{Pascalutsa07a}.
In the large $N_c$ limit $G_E$ and $G_C$ are described 
by quark-antiquark effects terms 
of the order $1/N_c^2$, while $G_M$  
is described by leading order terms ${\cal O}(N_c^0)$
associated to valence quark effects~\cite{Jenkins02}.
This explains the use of the  upper index $\pi$ in Eq.(\ref{eqGEGC-LNC})
which specifies that they are originated by a specific type of meson cloud (MC),
as introduced in the notation of Eq.~(\ref{eqGB-MC}).
As the large $N_c$ limit calculations, calculations of quark models
with $SU(6)$ symmetry also lead to results for $G_{En}$, $G_E$ and $G_C$ 
proportional to $r_n^2$ and they also come as effects of
pion cloud/quark-antiquark dressing 
of the quark cores~\cite{Pascalutsa07a,Buchmann04a,Buchmann97,Buchmann02b,Buchmann02c,Grabmayr01a}.

On the other hand quark models which break $SU(6)$ can provide also 
non zero contributions to $G_E$ and $G_C$  due 
to small $d$-wave quark states on the baryon wave function,
but in this case they correspond to the bare quark contributions
to quadrupole form factors ($G_E^{\rm B}$ and $G_C^{\rm B}$)
in the notation of the decomposition of Eq.~(\ref{eqGB-MC}).
However, these contributions are small when compared with
the experimental data for $G_E$ and $G_C$ 
and the estimates $G_E^\pi$ and $G_C^\pi$ from
Eqs.~(\ref{eqGEGC-LNC})~\cite{Pascalutsa07,Buchmann01}:
recall from Section~\ref{sec-Delta-lQ2} that the valence quark contributions to $G_E$ and $G_C$ 
are of the order of 10\%.

Combining the two relations (\ref{eqGEGC-LNC}) one concludes that, near $Q^2=0$
\ba
\frac{1}{ 1 + \beta Q^2} G_E^\pi (Q^2) = \frac{M_\Delta^2 -M^2}{4 M_\Delta^2} G_C^\pi (Q^2).
\label{eqGEGC-LNC4}
\ea
Note also that, when taking $\beta=0$, one gets
$G_E = \frac{M_\Delta^2 -M^2}{4 M_\Delta^2} G_C$, 
discussed already at the end of Section~\ref{sec-latticeQCD-Delta},

The corollary of Eq.~(\ref{eqGEGC-LNC4}) is that 
in the limit $Q^2=0$, the functions $R_{EM}$, $R_{SM}$, defined by Eq.~(\ref{eqEMC-RMC}),
satisfy 
\ba
R_{EM}(0) = R_{SM} (0),
\label{eqREMRSM0}
\ea
apart terms of the order $1/N_c^2$~\cite{Pascalutsa07a}.
The previous relation has been challenged by the data, 
because till recently the lowest $Q^2$ measurement of $G_C$ appeared to be compatible with 
$R_{SM}(0) \simeq - 5\%$~\cite{Aznauryan12a,Pascalutsa07,Pascalutsa06a,Stave08,Blomberg16,Bernstein03,Siegert1}, while $R_{EM}(0)  \simeq - 2.5\%$~\cite{PDG2022}.
However, the most recent measurement of $R_{SM}$ near $Q^2=0$ 
is consistent with (\ref{eqREMRSM0}), within the error bars~\cite{Siegert3,Blomberg16},
with the main differences in the new data came for lower values of $Q^2$: 0.04 and 0.09 GeV$^2$.
The result for the lowest value of $Q^2$ available is now 
$R_{SM} (0.04 \; \mbox{GeV}^2) = - (3.50 \pm 0.88) \%$ (JLab/Hall A)~\cite{Blomberg16}.
Notice that (\ref{eqREMRSM0}) is a limit 
defined at the photon point, and that the two functions differ significantly 
for finite $Q^2$, including at the pseudothreshold, where $R_{EM} \to \mbox{const}$
and $R_{SM} \; \propto \; |{\bf q}| \to \;  0$.

From Eq.~(\ref{eqGEn-lQ2}) the quadrupole form factors 
depend on the parametrization used for the function $G_{En}(Q^2)$.
One of the most popular ones is 
the Galster parametrization~\cite{Galster71}
\ba
G_{En}(Q^2) = - \mu_n \frac{ a \tau_{\scriptscriptstyle N}}{ 1 + d\, \tau_{\scriptscriptstyle N}} G_D,
\label{eqGalster}
\ea
where $\mu_n$ is the nucleon magnetic moment, 
$\tau_{\scriptscriptstyle N}= \frac{Q^2}{4 M^2}$, $G_D$ is the dipole form factor and
$a$, $d$ are two dimensionless parameters.
The parameter $a$ can be determined by the neutron square radius $r_n^2$,
through $a= \frac{2M^2}{3 \mu_n} r_n^2$  ($a=0.9$),
while $d$ is determined by $r_n^2$ and the fourth momentum of the $Q^2$
expansion of $G_{En}$: $r_n^4$~\cite{Grabmayr01a}.
Since $r_n^4$ cannot be directly measured,  $d$ is usually interpreted 
as a free parameter, adjusted by the $G_{En}$ data.
Typical values for $d$ are $d=1.75$ or 
$d=2.8$~\cite{Siegert3,Buchmann04a,Grabmayr01a,Buchmann09}.
The value of $d$ controls the asymptotic value of $G_C$ for large $Q^2$,
in case the term (\ref{eqGEGC-LNC}) is the dominant contribution
at large $Q^2$~\cite{Buchmann04a}.
At low and intermediate $Q^2$ the value
$d=2.8$ provides the best approximation to the data~\cite{Siegert3,Siegert1}.

\begin{figure}[t]
\begin{center}
  \includegraphics[width=5.8in]{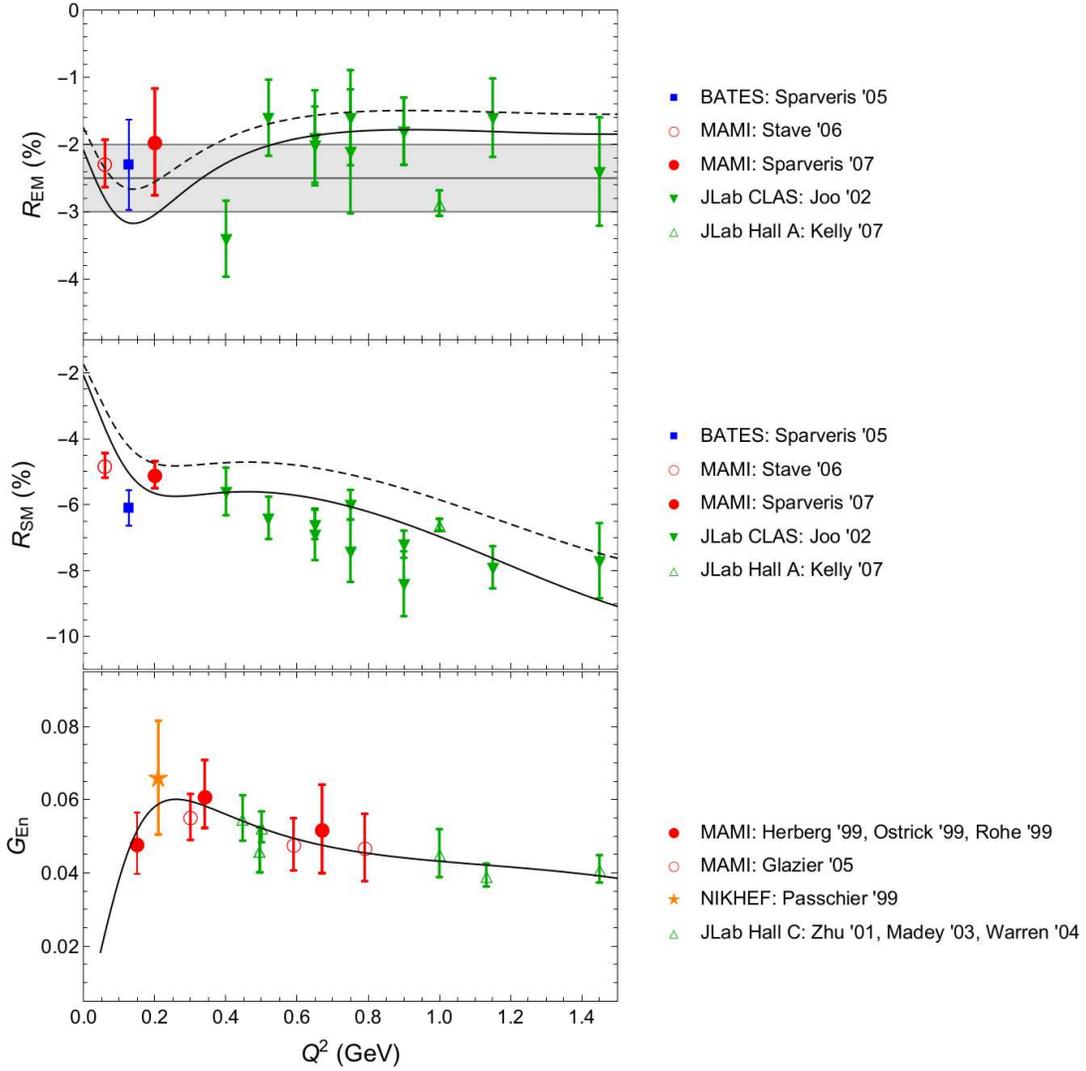}
\caption{\footnotesize 
$\gamma^\ast N \to \Delta(1232)$ transition. 
The solid lines are calculations of $G_{En}$, $R_{EM}$ and $R_{SM}$ from 
Pascalutsa, Vanderhaeghen and Hagelstein~\cite{Hagelstein17a}.
The calculations use the $G_{En}$ parametrization from Ref.~\cite{Bradford06a},
(\ref{eqGEGC-LNC}) 
with $\beta=0$ for the quadrupole form factors 
and $G_M \simeq \sqrt{2} F_{2p}$. 
The dashed lines are calculations with $G_M (Q^2) = \frac{3.02}{\kappa_p} F_{2p} (Q^2)$, 
where $G_M$ is normalized by the measured value of $G_M(0)$.
The data for $R_{EM}$ and $R_{SM}$ are from MIT-Bates~\cite{Sparveris05a}, MAMI~\cite{Stave06,Sparveris07},
JLab/CLAS~\cite{CLAS02} and JLab/Hall A~\cite{Kelly05b}.
The data for $G_{En}$ are from MAMI, NIKHEF and 
JLab/Hall C. Compilation taken from Ref.~\cite{Bradford06a}.
Courtesy of Vladimir Pascalutsa.
\label{figLargeNc1}}
\vspace{-0.8cm}
\end{center}
\end{figure}

Notwithstanding this phenomenological result, the range of validity
of the large $N_c$ limit relations (\ref{eqGEGC-LNC})
between the functions $G_{En}$, $G_E$ and $G_C$ their applicability
is restricted in principle to the $Q^2 \ll 1$ GeV$^2$ region
since they are based on the low-$Q^2$ expansion (\ref{eqGEn-lQ2})~\cite{Pascalutsa07a}.
In addition, near the pseudothreshold when $Q^2 = -(M_\Delta -M)^2 < 0$,
Eqs.~(\ref{eqGEGC-LNC}) 
are not compatible with Siegert's theorem: $G_E = \frac{M_\Delta -M}{2 M_\Delta} G_C$,
when $\beta=0$
(Section~\ref{sec-Siegert}, Table~\ref{table-Siegert1}),
violating the theorem in a term $1/N_c^2$~\cite{Siegert3,Siegert1}.
However, one obtains a consistent description of  Siegert's theorem 
when one uses~\cite{Siegert3}
\ba
\beta = \frac{1}{2 M_\Delta( M_\Delta - M)}.
\label{eqBeta}
\ea
The results of the parametrizations (\ref{eqGEGC-LNC}) for 
$G_E^\pi$ and $G_C^\pi$ with $\beta$ fixed by this relation were the ones
represented in Fig.~\ref{fig-GEGC-CSQM}, combined with the 
covariant spectator quark model estimates
for the bare contributions (thin lines).

When the relations (\ref{eqGEGC-LNC}) are combined with a 
parametrization of the data for $G_M$ for the determination
of the ratios $R_{EM}$ and $R_{SM}$, the large $N_c$ parametrizations 
of $G_E$ and $G_C$ lead to underestimations of the data for both $R_{EM}$ and $R_{SM}$.
The gap between the estimates and the data was shown
in Figs.~\ref{fig-Delta-REM} and  \ref{fig-Delta-RSM}
(dotted-dotted-dash lines) for the case $\beta=0$.
The underestimation of the large $N_c$ relations
has been recently discussed in the literature~\cite{Siegert3,Blomberg16,Atac21a}.
The results from Figs.~\ref{fig-Delta-REM} and  \ref{fig-Delta-RSM}
show  that extra contributions of the order of 10-20\% 
may improve the description of the data at low $Q^2$.

To determine the ratios  $R_{EM}$ and $R_{SM}$ from 
(\ref{eqGEGC-LNC})  different estimates for $G_M$ are considered in the literature.
Pascalutsa and Vanderhaeghen~\cite{Pascalutsa07a} also use the case $\beta=0$ but
generalize $G_M$ to finite $Q^2$ 
using $G_M (Q^2) \to  [F_{2p}(Q^2) -F_{2n}(Q^2)]/\sqrt{2}$, 
where $F_{2N}$ ($N=p,n$) are the nucleon Pauli form factors.
More recently, the estimate of $G_M$ was modified to 
$G_M (Q^2) \to  \sqrt{2} F_{2p}(Q^2)$, taking advantage 
of another large $N_c$ relation: $F_{2n}(Q^2)= - F_{2p}(Q^2)$~\cite{Hagelstein17a}.
Buchmann and collaborators consider instead and also for $\beta=0$
a relation between the transition form factor $G_M$ and the neutron 
magnetic form factor $G_M (Q^2)= - \sqrt{2} G_{Mn} (Q^2)$~\cite{Buchmann04a,Buchmann09}.
Combining  (\ref{eqGEGC-LNC}) for $G_E$, $G_C \; \propto \; \tilde G_{En}$, 
with this last relation, one concludes that $R_{EM}$  can be determined by
$G_{En}/G_{Mn}$, which is known 
as the $SU(6)$ result for $R_{EM}$~\cite{Buchmann09,Atac21a}. 
The estimates based on $G_{Mn}$  increase $R_{EM}$ and $R_{SM}$ by 10--15\%,
improving the agreement with the data 
because the function $G_M$ is below the data~\cite{Siegert2}.
Concerning the parametrizations
used for $G_{En}$~\cite{Kelly04a,Friedrich03,Platchkov90,Kaskulov04,Gentile11}
several choices lead to very similar results~\cite{GlobalParam}.
Inverting the procedure, the correlations between $G_E$, $G_C$ and $G_{En}$ 
have also been used to obtain more accurate parametrizations 
of the neutron electric form factor $G_{En}$~\cite{GlobalParam,Atac21a}.

\begin{figure}[t]
\begin{center}
\includegraphics[width=3.6in]{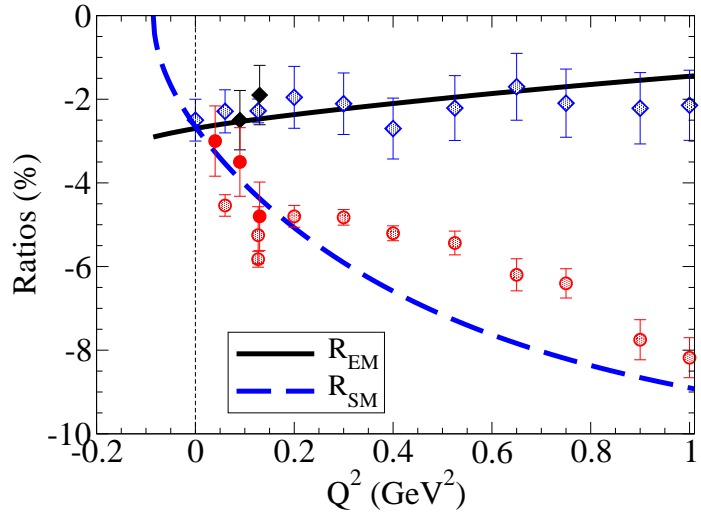}
\end{center}
\caption{\footnotesize
$\gamma^\ast N \to \Delta(1232)$ transition.
Comparison of results for $R_{EM}$ and $R_{SM}$ 
between the pseudothreshold and $Q^2=1$ GeV$^2$, specially showing
the constraint $R_{EM} =R_{SM}$ at the photon $Q^2=0$ point. 
Model from Refs.~\cite{Siegert3,LatticeD}. 
The data are from PDG 2022~\cite{PDG2022}, MAMI~\cite{Stave08,Sparveris13},  
MIT-Bates~\cite{MIT-Bates03,Sparveris05a}, JLab/Hall A~\cite{Blomberg16}
and JLab/CLAS~\cite{CLAS09}.
For a discussion of the data check Ref.~\cite{Siegert3}. 
\label{figREM-RSM}}
\end{figure}

Figure~\ref{figLargeNc1} shows
the predictions from Pascalutsa and Vanderhaeghen (solid line)~\cite{Pascalutsa07a} 
based on the relation $G_M (Q^2) \simeq  \sqrt{2} F_{2p}(Q^2)$ 
and on the $G_{En}$ parametrization from Ref.~\cite{Bradford06a}.
The enhancement of the ratios due to parametrization 
used for $G_M$ near $Q^2=0$ (correction of about 20\%)
contributes to the good agreement to the pre-2016 data. 
When the function $G_M$ is modified 
to reproduce the value $G_M(0) \simeq 3.02$ (dashed lines)
one obtains an underestimation of the $R_{EM}$ and $R_{SM}$ 
of the order of 20\%.

For a more detailed discussion of the ratios $R_{EM}$ and $R_{SM}$ near $Q^2=0$,
we present in Fig.~\ref{figREM-RSM} the data below $Q^2= 1$ GeV$^2$,
including the recent JLab/Hall A data from Ref.~\cite{Blomberg16}.
The data are compared with the calculation of the 
covariant spectator quark model from Ref.~\cite{Siegert3},
discussed already in Section~\ref{sec-latticeQCD-Delta}
(see also Fig.~\ref{fig-GEGC-CSQM}).
The data and the model results are consistent with the Eq.~(\ref{eqREMRSM0})
within the experimental error bars.
Numerically the model gives $R_{SM}(0) - R_{EM} (0) = 0.05\%$~\cite{Siegert3}.

Now, and to set a relative scale for their domain of validity,
how much does the behavior of the relations (\ref{eqGEGC-LNC})
deviate from the expected pQCD behavior for large $Q^2$?
According to  pQCD analysis of the leading order dependence of 
the electric and Coulomb quadrupole form factors (see Section~\ref{sec-largeQ2})
one should expect $G_E^{\rm B} \; \propto \; 1/Q^4$ and   $G_C^{\rm B} \; \propto \; 1/Q^6$.
When we take into account that the contributions from
the meson cloud excitations ($q \bar q$ contributions) 
should be suppressed by extra global factor $1/Q^4$,
the asymptotic contributions of the pion cloud terms 
behaves as $G_E^{\pi} \; \propto \; 1/Q^8$ and $G_C^{\pi} \; \propto \; 1/Q^{10}$.
But from (\ref{eqGEGC-LNC}), 
$G_C^{\pi} \; \propto \; 1/Q^6$. How to conciliate this with the pQCD expected result
$G_C^{\pi} \; \propto \; 1/Q^{10}$?
Naturally,
the relations (\ref{eqGEGC-LNC}), both for $G_E^\pi$ and $G_C^\pi$, 
are in fact valid only in the low-$Q^2$ region, and should be modified 
above a certain value of $Q^2$.
The replacement $G_C^\pi \to G_C^\pi /( 1 + Q^2/\Lambda_C^2)^2$,
where $\Lambda_C^2$ is a cutoff to be determined phenomenologically,
conciliating $G_C^\pi$ with the expected asymptotic behavior at large $Q^2$~\cite{Siegert3,Siegert1}.
Also for $G_E^{\pi}$, we can expect modifications at large $Q^2$, 
although the falloff at large $Q^2$ is already consistent with the expected $1/Q^8$.
The consequence of the modified form is that $G_C^\pi$ is suppressed, 
and that the bare component $G_C^{\rm B} \; \propto \; 1/Q^6$
dominates at large $Q^2$.
The dominance of the bare contributions over the pion cloud contributions 
is corroborated by Fig.~\ref{fig-GEGC-CSQM} from
the comparison of the bare contributions with the data above 1 GeV$^2$.
However, phenomenologically, if the original relation for
$G_C^\pi$ is justified for arbitrary large $Q^2$, 
we recover the Buchmann estimate for $R_{SM}$
at large $Q^2$~\cite{Buchmann04a,Buchmann09}:
$R_{SM} (Q^2) \simeq \frac{1}{4} \frac{M}{M_\Delta} \left(- \frac{a}{d} \right)$,
a constant determined by the parameters of the Galster formula
(\ref{eqGalster}) given above.

\subsubsection*{\it Short notes}

Both the large $N_c$ limit calculations and quark models with $SU(6)$
symmetry including two-body exchange currents lead to relations between the
$\gamma^\ast N \to \Delta(1232)$ $G_E$ and $G_C$ form factors
and the neutron electric form factor $G_{En}$. 
The physical process behind these connections are effects of
pion cloud/quark-antiquark dressing of the quark cores.
The large $N_c$ limit relations establish that at the photon point
$R_{EM}(0) = R_{SM} (0)$ apart $1/{N_c}^2$ corrections.
Although this constraint was challenged for a while by data,
it is consistent with the
JLab/Hall A new data obtained at very low $Q^2$.

\subsubsection{\it Large $Q^2$ and the onset of pQCD \label{sec-Delta-LLQ2}}

The asymptotic dependence of the 
$\gamma^\ast N \to \Delta(1232)$ transition 
form factors and helicity amplitudes 
has been calculated by 
C.~Carlson et al.~\cite{Carlson86a,Carlson98a,Carlson88a} 
using different forms for the nucleon and 
$\Delta(1232)$ distributions amplitudes.
Some of those distribution amplitudes 
were proposed by Chernyak-Zhitnisky (CZ)~\cite{Chernyak84a}
and King-Sacharadja (KZ)~\cite{King86a} based on the QCD sum rules formalism.
Using the distribution amplitudes it is possible 
to calculate the asymptotic result for the 
leading order helicity amplitude, $A_{1/2}$~\cite{Carlson98a}
\ba
\frac{Q^3}{M^3} |A_{1/2}| = 0.022 \; \mbox{GeV}^{-1/2} \;  \mbox{(CZ)}, 
\hspace{1.4cm}
\frac{Q^3}{M^3} |A_{1/2}| = 0.036 \; \mbox{GeV}^{-1/2} \;  \mbox{(KS)}.
\label{eqA12ass}
\ea 
These results are affected by uncertainties 
associated to the QCD sum rules framework 
in the determination of the distribution amplitudes.
The authors state that the uncertainties can change 
the estimate in 2 or 3 times more or less~\cite{Carlson98a}.

The  asymptotic result for $|A_{1/2}|$ can be used to 
calculate the asymptotic result for $G_M$.
Some care is necessary in the conversion because, 
we need to take into account that at large $Q^2$: $G_M = -G_E$,
the pQCD result from Eq.~(\ref{eqGEGM-J3/2}).
Using the relation at large $Q^2$ and Eq.~(\ref{eqAmp32p}) for $A_{1/2}$
one obtains the asymptotic relation 
$G_M \simeq - F_{1+} A_{1/2}$, 
where $F_{1+} \simeq 1.66 \frac{M}{Q}$ in GeV$^{1/2}$.
Combining these results with Eqs.~(\ref{eqA12ass}), we get
\ba
\frac{Q^4}{M^4} G_M = 0.037 \; \mbox{(CZ)}, \hspace{1.6cm}
\frac{Q^4}{M^4} G_M = 0.059 \; \mbox{(KS)}.
\ea
As a term of comparison, the estimated 
for the proton magnetic moment is~\cite{Carlson98a}
\ba
\frac{Q^4}{M^4} G_{Mp} \simeq 1.28. 
\ea
In these analyses, we ignore logarithmic corrections.

\begin{figure}[t]
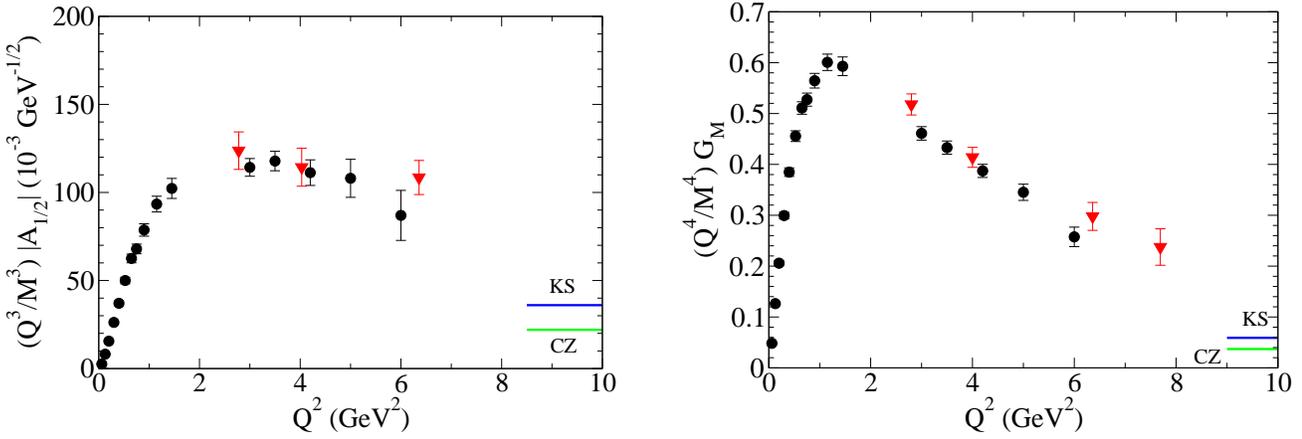

\mbox{
\includegraphics[width=3.2in]{Q3A12-D1232} \hspace{.6cm}
\includegraphics[width=3.2in]{Q4GM-D1232}  }
\caption{\footnotesize
Asymptotic form of 
$\gamma^\ast N \to \Delta(1232)$ 
transition amplitudes and form factors.
{\bf Left panel:} 
Absolute value of the amplitude $A_{1/2}$ scaled by the factor $Q^3/M^3$.
{\bf Right panel:}
Magnetic dipole form factor $G_M$ normalized by $Q^4/M^4$.
The asymptotic result from 
C.~Carlson et al.~\cite{Carlson98a} 
using the CZ and KS distribution amplitudes 
are displayed at large $Q^2$.
The data are from JLab/CLAS ({\Large $\bullet$})~\cite{CLAS09} 
and JLab/Hall C 
({\Large {\color{red}$\blacktriangledown$}})~\cite{Villano09,Frolov98}. 
\label{fig-Scalling-D1232}}
\end{figure}

The experimental tests of the possible scaling relations are presented 
in Fig.~\ref{fig-Scalling-D1232} for the magnitude of 
$A_{1/2}$ and for the magnetic form factor $G_M$.
On the left panel, one can notice an approximate scale 
of $|A_{1/2}|$ with $1/Q^3$, within the large uncertainties of the data.
The asymptotic level for the CZ and KS are presented at the end of the scale of $Q^2$.
The test for $G_M$ is presented on the right panel.
The graph for $Q^4 G_M$ is not so conclusive 
about the scaling $G_M \; \propto \; 1/Q^4$.
However, the magnitude is closer to the asymptotic value.

The complete investigation of the $\gamma^\ast N \to \Delta(1232)$
asymptotic amplitudes requires also the test of the relations
$S_{1/2} \; \propto \; 1/Q^3$ and $A_{3/2} \; \propto \; 1/Q^5$.
The results for $S_{1/2}$ are compatible with the expected falloff
(see Fig.~\ref{fig-Amps-LQ2}, Section~\ref{sec-TransitionFF}).
As for the amplitude $A_{3/2}$, the test of scaling
at the present range of $Q^2$ is premature
due to the failure of the 
expected large-$Q^2$ relation $G_M =- G_E$.
The magnitude observed presently to $G_E$ is about 
two orders of magnitude smaller than $G_M$.
This result suggests that the range of validity 
of the asymptotic region is achieved only for much larger values of $Q^2$.
The asymptotic region for the  $\gamma^\ast N \to \Delta(1232)$ transition 
may be beyond the range of a possible JLab-22 GeV upgrade~\cite{Carman23a}.

There are also estimates of the Coulomb quadrupole form factor $G_C$ 
based on pQCD arguments.
Idibi et al.~calculated the Breit frame amplitude $G_0$ 
in leading order pQCD~\cite{Idilbi04}.
The result is then used to calculate  $R_{SM}$.
The final expression, calibrated by 
the JLab/Hall C data~\cite{Frolov98}, takes the form
\ba
R_{SM} (Q^2) \simeq - c \; \frac{|{\bf q}|}{Q^2} \log^2 \left( \frac{Q^2}{\Lambda^2}\right), 
\label{eqRSM-pQCD}
\ea
where $c=0.013$ and $\Lambda = 0.25$ GeV.
The relation (\ref{eqRSM-pQCD}) provides a good description 
of the data up to $Q^2=4$ GeV$^2$.
Under discussion is whether the range of scaling for $G_C$
is valid for such a small scale of the measured $Q^2$ region~\cite{Idilbi04}.

\subsubsection*{\it Short notes}

The available data for the $\gamma^\ast N \to \Delta(1232)$
transition form factors have falloffs with increasing $Q^2$ that go along with
power laws consistent  with the outset of the asymptotic pQCD behavior
only far beyond the $Q^2$ region experimentally scrutinized so far.
Even without the experimental validation of the falloffs pQCD power laws,
the observed magnitude of $G_E$ is about 
two orders of magnitude smaller than $G_M$, 
providing evidence that the measured region is well away
of the asymptotic region, where with the relation $G_M= - G_E$ is expected to hold.


\subsection{\it $N(1520)\frac{3}{2}^-$ resonance \label{sec-N1520}}

The $N(1520)\frac{3}{2}^-$ is a resonance 
of the second resonance region characterized 
by negative parity and by three helicity amplitudes.
The resonance is clearly identified in the 
$\pi N$ and $\pi \pi N$ channels.
The branching ratios of the decays are about 
60\% for $\pi N$ and 40\% for $\pi \pi N$~\cite{PDG2022}.

The first data related to the $\gamma^\ast N \to N(1520)$
come from the experiments at DESY and NINA in 
the 70' and 80'~\cite{DESY75a,NINA74a}, but
accurate data in a wide region of $Q^2$ became available 
only with CLAS measurements around 2009~\cite{CLAS09}.
Before that the scalar amplitude was unknown.
Measurements of the full set of the helicity 
amplitudes at CLAS in the range $Q^2=0.3$--4.2 GeV$^2$ 
using $\pi N$ and $\pi \pi N$ 
decays~\cite{CLAS09,CLAS12,CLAS16a,Aznauryan05a}
provide important information about the 
electromagnetic properties of the transition 
at low and intermediate $Q^2$.
There are 
relevant discrepancies between the CLAS 
and MAID analysis of the helicity amplitudes~\cite{N1520SL}.

Similarly to the $N(1535)\frac{1}{2}^-$ 
and $N(1650)\frac{1}{2}^-$ resonances, 
the  $N(1520)\frac{3}{2}^-$ and $N(1700)\frac{3}{2}^-$ resonances 
can be considered partner resonances, 
since they can be interpreted as 
mixtures of two configurations that couple
relative angular momentum $L=1$ with 
core spins $S=\frac{1}{2}$ or $S=\frac{3}{2}$~\cite{Aznauryan12a,Capstick00,Burkert-SQTM}
\ba
& &
\left|N(1520)  \right>  = \cos \theta_D \left| N^2\;  \sfrac{3}{2}^-\right> 
-   \sin \theta_D \left| N^4\; \sfrac{3}{2}^- \right>, \nonumber \\
& &
\left|N(1700)  \right>  = \sin \theta_D \left| N^2 \; \sfrac{3}{2}^-\right> 
+   \cos \theta_D \left| N^4 \; \sfrac{3}{2}^- \right>, 
\label{eqMixt-N1520}
\ea
where we use the notation $\left| N^{2S +1} \; \frac{3}{2}^- \right>$     
and $\theta_D$ is a mixture angle.
Since the mixture angle $\theta_D$ is small 
($\theta_D \simeq 6^\circ$, $\cos \theta_D \simeq 0.99$)~\cite{Burkert-SQTM},
$N(1520)\frac{3}{2}^-$ is almost an $S=\frac{1}{2}$ pure state 
and $N(1700)\frac{3}{2}^-$ is almost an $S=\frac{3}{2}$ pure state.
The label $D$ comes from the spectroscopic notation 
$D_{13}$ for the $D$-wave pion on the $N^\ast$ decay.
From the theoretical point of view $N(1520)\frac{3}{2}^-$
can be interpreted as a resonance dominated by the core spin $S=\frac{1}{2}$
configuration. 
In an exact $SU(6)$ limit 
$N(1520)\frac{3}{2}^-$ and $N(1700)\frac{3}{2}^-$ 
would correspond to the same state
(degenerated states)~\cite{Capstick00}.

\subsubsection{\it Helicity amplitudes and transition form factors \label{sec-N1520-Amps}}

Calculations of the $\gamma^\ast N \to N(1520)$ 
transition amplitudes have been performed 
using non relativistic quark 
models~\cite{Koniuk80,Bijker96a,Santopinto12,Aiello98a,Close72a} and 
relativistic quark models~\cite{Ronniger13a,Capstick95,N1520SL,SemiR,Aznauryan12b,Aznauryan17a,Warns90,Merten02a}.
There are also calculations based on the cloudy bag model~\cite{Golli13a},
light-cone sum rules~\cite{Aliev14a}
and AdS/QCD~\cite{Lyubovitskij20a}.
The model calculations can be compared with the 
ANL-Osaka DCC calculations of the meson cloud contributions
that are extracted by setting the bare contributions 
to zero~\cite{Kamano16a,Kamano13a,Nakamura15a,Sato16a}.

\begin{figure}[t]
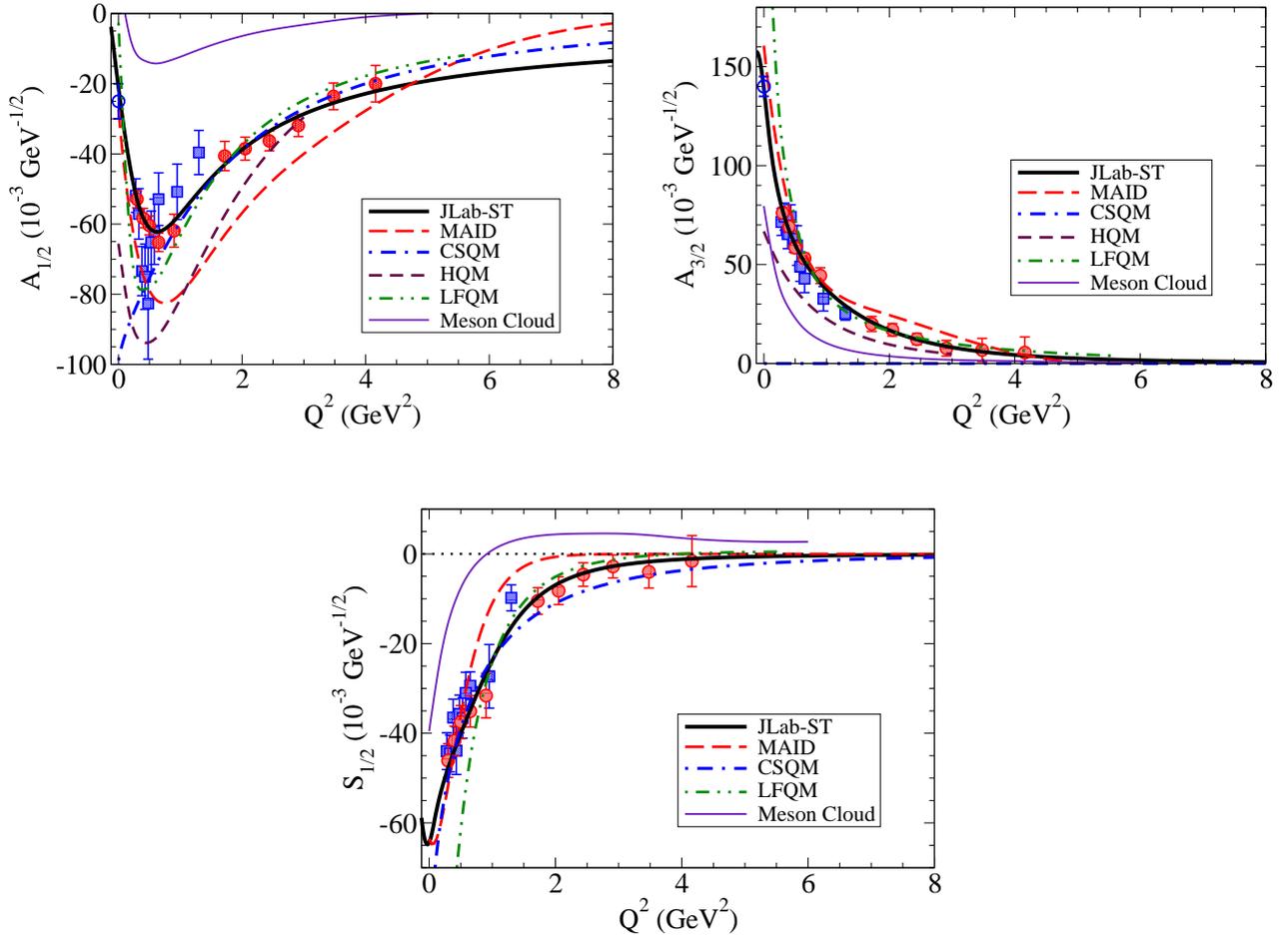

\begin{center}
\mbox{
\includegraphics[width=3.2in]{A12-N1520-v3} \hspace{.3cm}
\includegraphics[width=3.2in]{A32-N1520-v3}}  \\
\vspace{1.0cm}
\includegraphics[width=3.2in]{S12-N1520-v3}  
\end{center}
\caption{\footnotesize 
$\gamma^\ast N \to N(1520)$ helicity amplitudes.
The model calculations are from the covariant spectator quark model (CSQM)~\cite{SemiR}, 
hypercentral quark Model (HQM)~\cite{Santopinto12,Aiello98a} and LFQM~\cite{Aznauryan17a}.
The results are compared with the MAID~\cite{MAID2009} 
and JLab-ST parametrizations~\cite{JLab-website,LowQ2param}.
The data are from CLAS: one pion production 
({\color{red} {\LARGE $\bullet$}})~\cite{CLAS09} 
and two pion production 
({\scriptsize {\color{blue} $\blacksquare$}})~\cite{CLAS12,CLAS16a}, 
and PDG 2022 ({\Large {\B $\bm \circ$}})~\cite{PDG2022}. Meson cloud contribution from the
ANL-Osaka DCC model.
\label{figAmps-N1520-1}}
\end{figure}

In Fig.~\ref{figAmps-N1520-1}, we show the CLAS data for the helicity amplitudes
from one pion electroproduction~\cite{CLAS09}, 
two pion electroproduction~\cite{CLAS12,CLAS16a}
and the PDG result at $Q^2=0$~\cite{PDG2022}. 
The data extracted from CLAS assumes the  branching ratio of 
60\% for $\pi N$ decay and 40\% of $\pi \pi N$~\cite{CLAS09}.
The figure show as well model calculations
and parametrizations of the data.
Notice that at the photon point the magnitude of 
$A_{1/2}$ is small and the magnitude of $A_{3/2}$ is very large.
In contrast $A_{1/2}$ dominates  $A_{3/2}$ for large $Q^2$.
The figure also shows the MAID~\cite{MAID2009,MAID2011} 
and JLab-ST parametrizations~\cite{JLab-website,LowQ2param} of the helicity data.
The deviation of the MAID parametrization from the 
CLAS data is manifest for the amplitudes $A_{1/2}$ and $S_{1/2}$.

Figure~\ref{figAmps-N1520-1} includes
the calculations from three models:
the covariant spectator quark model (CSQM) from Ref.~\cite{SemiR},
the hypercentral quark model (HQM)~\cite{Santopinto12,Aiello98a} and the 
LFQM from Aznauryan and Burkert~\cite{Aznauryan17a}.
The last one takes into account the momentum dependence of the quark mass while
the first two are based exclusively on fixed constituent valence quark masses.
The LFQM assumes a contribution of about 15\% from the meson cloud 
which is included in the normalization of the $N(1520)$ wave function~\cite{Aznauryan17a}.
Both covariant spectator quark model and LFQM provide 
a good description of the data for the amplitudes $A_{1/2}$ and $S_{1/2}$.
For the amplitude $A_{3/2}$, 
the hypercentral quark model underestimates the data at small $Q^2$ 
while the covariant spectator quark model result vanishes, and LFQM
is in better agreement with the data.
Only LFQM
predicts a substantial contribution for $A_{3/2}$ near $Q^2=0$.
However, LFQM is expected
to be accurate only for $Q^2 > 2$ GeV$^2$~\cite{Aznauryan07,Aznauryan17a}.

The preliminary conclusion is that in general 
models based on quark degrees of freedom only underestimate the magnitude of $A_{3/2}$,
predicting only one third or one half 
of the measured amplitude at $Q^2=0$~\cite{Ronniger13a,Golli13a,Warns90,Merten02a}.
An unquenched quark model~\cite{Bijker09a}
which takes into account the $q\bar q$ contributions 
suggest that those effects are significant, 
and can explain the gap to the measured amplitude.
This result indicates that the amplitude $A_{3/2}$ may have significant 
contributions from meson cloud effects not explicitly
included in the quark model calculations.
The calculations from ANL-Osaka DCC model of the meson cloud contribution 
to the amplitude are also displayed in the figure and
confirm this interpretation~\cite{Kamano16a,Kamano13a,Nakamura15a,JDiaz08a},
suggesting that the meson cloud effects  contribute to 
about 50\% of the amplitude~\cite{CLAS16a}.
In summary and in general quark models describe 
well the amplitudes $A_{1/2}$ and $S_{1/2}$, 
while meson cloud effects are more relevant for the amplitude $A_{3/2}$.
Another important conclusion from the study 
of the $\gamma^\ast N \to N(1520)$ transition is that 
the meson cloud contributions are dominated by 
isovector contributions~\cite{N1520TL,Aznauryan12b,Aznauryan17a}.

\begin{figure}[t]
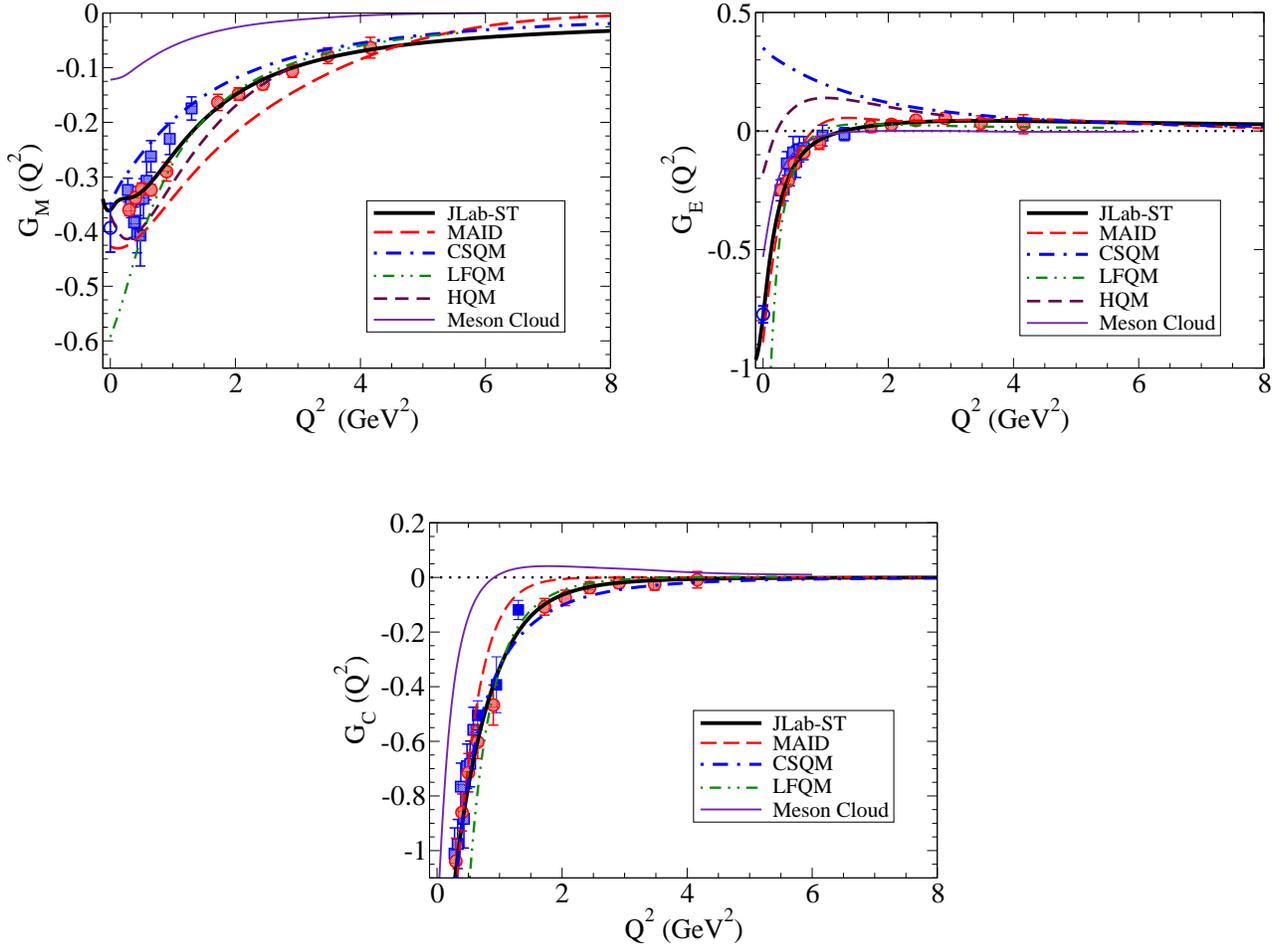

\begin{center}
\mbox{
\includegraphics[width=3.2in]{GM-N1520-v2} \hspace{.3cm}
\includegraphics[width=3.2in]{GE-N1520-v2} } \\ 
\vspace{1.cm}
\includegraphics[width=3.2in]{GC-N1520-v2}  
\end{center}
\caption{\footnotesize 
$\gamma^\ast N \to N(1520)$ transition form factors.
The model calculations are from the covariant spectator quark model (CSQM)~\cite{SemiR}, 
hypercentral quark model (HQM)~\cite{Santopinto12,Aiello98a} and LFQM~\cite{Aznauryan17a}.
The results are compared with the MAID~\cite{MAID2009} 
and JLab-ST parametrizations~\cite{JLab-website,LowQ2param}.
The data are from CLAS: one pion production 
({\color{red} {\LARGE $\bullet$}})~\cite{CLAS09} and
two pion production ({\scriptsize {\color{blue} $\blacksquare$}})~\cite{CLAS12,CLAS16a}, 
and PDG 2022 ({\Large {\B $\bm \circ$}})~\cite{PDG2022}. Meson cloud contribution from the
ANL-Osaka DCC model.
\label{figFF-N1520-1}}
\end{figure}

In Fig.~\ref{figFF-N1520-1}, we present the results from the $\gamma^\ast N \to N(1520)$
form factors, as defined in Section~\ref{sec-TransitionFF}.
The form factors have a more regular or monotonic
behavior with $Q^2$ than the helicity amplitudes.
The form factor $G_M$ may be the exception, since the function 
is expected to have a turning point near or below $Q^2=0$, 
in order to satisfy the condition $G_M=0$ 
at the pseudothreshold~\cite{LowQ2param,Siegert2}
(see Section~\ref{sec-Siegert}, Table~\ref{table-Siegert1}).
The monotonous shape of the form factors contrasts 
with the amplitude $A_{1/2}$ whose curvature is imposed by its suppression near $Q^2=0$.
This can be understood noticing that 
for states $J^P= \frac{3}{2}^-$ from Eq.~(\ref{eqAmp32m}): 
$A_{1/2} \; \propto \; \sqrt{Q_+^2} (G_E - 3 G_M)$ 
and that near $Q^2=0$, $|G_E - 3 G_M|$ is small,
as seen in Fig.~\ref{figFF-N1520-1}.
The enhancement of the amplitude for $Q^2 > 0$ 
is a consequence of small differences between 
the $G_E$ and $3 G_M$, amplified by the factor 
$\sqrt{1 + Q^2/(M_R+M)^2}$.

From Fig.~\ref{figFF-N1520-1} we conclude also that the quark models:
the covariant spectator quark model and LFQM 
provide a good description of the multipole form factor data for $Q^2 >2$ GeV$^2$.
The covariant spectator quark model
for large values of $Q^2$ approaches the data, but fails at small $Q^2$ for $G_E$.
This is the consequence of the model giving $A_{3/2}=0$: because of the factors
have the form $G_M \;\propto \; (A_{3/2}/\sqrt{3} - A_{1/2})$ and $G_E  \; \propto \; (\sqrt{3} A_{3/2} + A_{1/2})$,
according to Eqs.~(\ref{eqGM1m}) and (\ref{eqGE1m}).
The suppression of $A_{3/2}$ affects
more $G_E$  than $G_M$ near $Q^2=0$, but implies
for the whole range of $Q^2$ the asymptotically valid relation
[recall that $A_{3/2} \; \propto (G_E + G_M)$]~\cite{NSTAR2017,N1520TL,SemiR}.
We note that covariant spectator quark model predictions are based on the semirelativistic approximation
a parametrization with minimum number of parameters mostly fixed by 
the nucleon form factor data~\cite{Nucleon,SemiR}.

The form factor representation obtained in this section allows us
to judge quickly on the relation (\ref{eqGEGM-J3/2}) between 
the functions $G_E$ and $G_M$ at large $Q^2$: $G_E \simeq -G_M$.
But a more detailed discussion of the behavior of the form factors
at large $Q^2$ is presented in the next section.

\subsubsection*{\it Short notes}

Near $Q^2=0$ the  amplitude $A_{3/2}$ is much larger than the amplitude $A_{1/2}$
for the $\gamma^\ast N \to N(1520)$ transition.
The low-$Q^2$ region has large contributions from $q \bar{q}$ states or 
from isovector meson cloud effects.
This prevents a good description of
the transition by valence quark models in the low-$Q^2$ region.
Nevertheless such models, as the covariant spectator quark model,
describe the data well in the large-$Q^2$ region,
as a consequence of the relation $A_{3/2} \; \propto \; (G_E + G_M)$,
and the observation that $A_{3/2}$ is suppressed comparatively 
to $A_{1/2}$ for very large $Q^2$.

\subsubsection{\it Large-$Q^2$ region}

\begin{figure}[t]
\begin{center}
\includegraphics[width=2.1in]{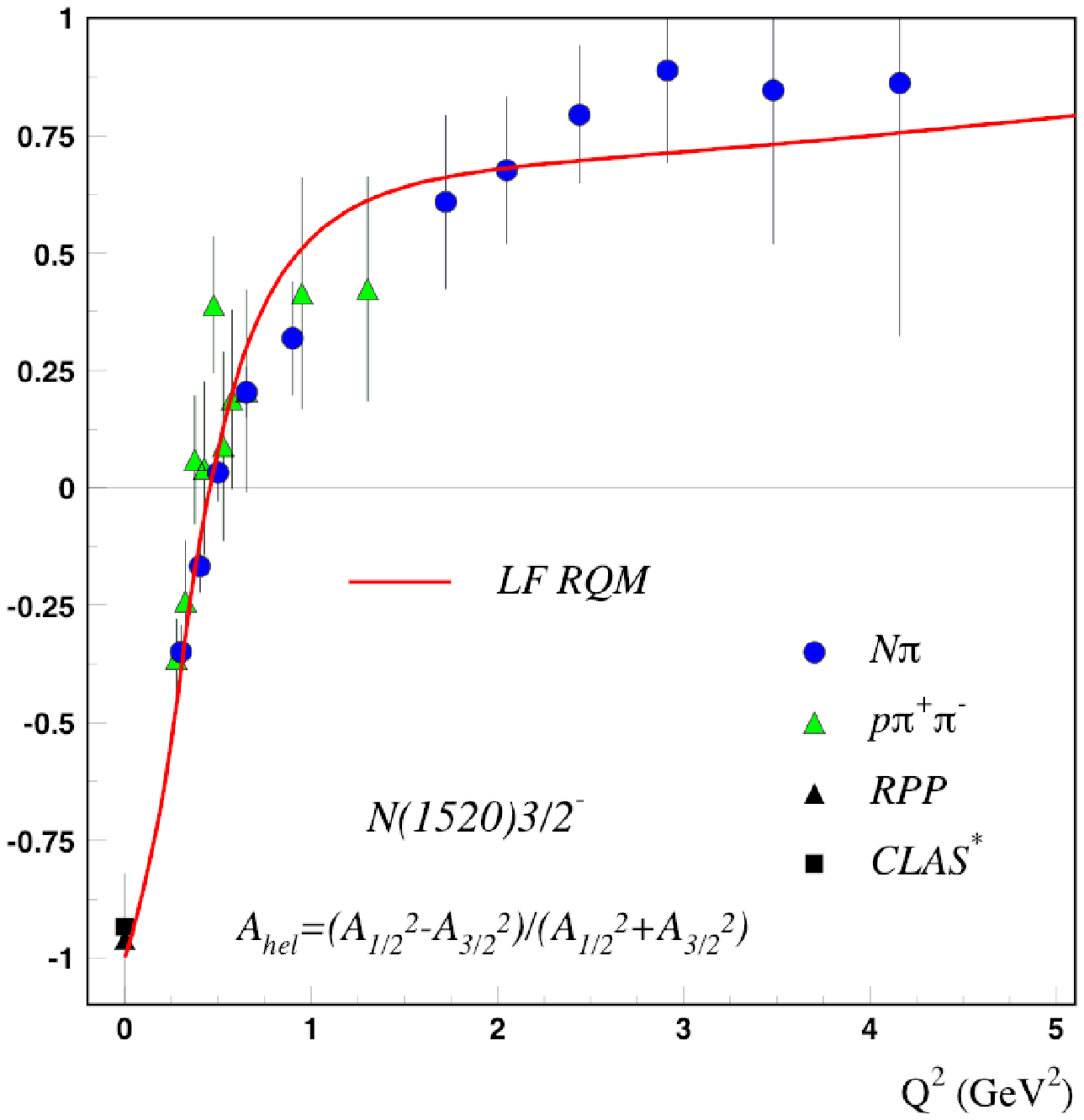} \hspace{0.3cm}
\includegraphics[width=3.in]{GMGE-N1520-GD} 
\caption{\footnotesize
Two illustrations of the asymptotic large-$Q^2$ regime for the
$\gamma^\ast N \to N(1520)$ transition.
{\bf Left panel:} Function $A_{\rm hel}$ from Eq.~(\ref{eqAhel}) 
compared with Light Front Relativistic Quark Model
from Aznauryan and Burkert~\cite{Gross22a}.
The data are from JLab/CLAS~\cite{CLAS09,CLAS16a} and PDG~\cite{PDG2022}.
{\bf Right panel:} 
$\gamma^\ast N \to N(1520)$ transition form factors normalized 
by the dipole function $G_D=\left(1 + \frac{Q^2}{\Lambda_D^2} \right)^{-2}$, 
where $\Lambda_D^2= 0.71$ GeV$^2$.
Results compared with the JLab-ST parametrization~\cite{JLab-website,LowQ2param}.
The data are from PDG~\cite{PDG2022} and CLAS~\cite{CLAS09}.
Figure from the left panel: Courtesy of Volker Burkert and Inna Aznauryan. 
Reprinted with permission from
\href{https://link.springer.com/article/10.1140/epjc/s10052-023-11949-2}{F.~Gross, E.~Klempt, S.~J.~Brodsky, A.~J.~Buras, V.~D.~Burkert, G.~Heinrich and K.~Jakobs, \textit{et al.}
Eur. Phys. J. C \textbf{83}, 1125 (2023).}
Copyright (2023) by Springer.
\label{fig-N1520-GEGM}}
\end{center}
\end{figure}

The study of the asymptotic dependence 
of the  $\gamma^\ast N \to N(1520)$ structure functions
can be done using the comparison between the falloffs 
of the amplitudes $A_{1/2}$ and $A_{3/2}$.
Those falloffs are estimated for large $Q^2$ 
as $A_{1/2} \; \propto \; 1/Q^3$ and $A_{3/2} \; \propto \; 1/Q^5$,
as discussed in Section~\ref{sec-largeQ2}.

The comparison between the transverse amplitudes
can be done using the helicity  asymmetry~\cite{Aznauryan12a,CLAS09}
\ba
A_{\rm hel} = \frac{A_{1/2}^2 - A_{3/2}^2}{A_{1/2}^2+ A_{3/2}^2}
\label{eqAhel}.
\ea
Near the photon point one has $A_{\rm hel} (0) \simeq -1$
since $A_{3/2}$ dominates for small $Q^2$.
For larger values of $Q^2$, $A_{3/2}$ has a stronger falloff ($\propto \; 1/Q^5$)
than $A_{1/2}$ ($ \propto \;1/Q^3$) and the ratio  approaches $A_{\rm hel} (Q^2) \simeq 1$.
The large-$Q^2$ limit can be followed using $x= \frac{A_{3/2}^2}{A_{1/2}^2}$,
with $x\; \propto \; 1/Q^4$ for large $Q^2$.
One obtains then the asymptotic form 
$A_{\rm hel} (Q^2) \simeq \frac{1-x}{1+x} \simeq 1 - 2 x$ for large $Q^2$,
justifying the fast convergence of $A_{\rm hel}$ to one.
The function  $A_{\rm hel}$ is displayed on the left panel 
of Fig.~\ref{fig-N1520-GEGM}.
The single pion and double pion production data
is compared with the light front quark model from
Aznauryan and Burkert~\cite{Gross22a}.

The previous analysis is based on the property that $A_{3/2}$ decreases
the magnitude very fast, 
which is equivalent to $A_{3/2}$ being negligible for large $Q^2$.
We can now look for this result using the transition form factors.
The previous asymptotic relations for $A_{1/2}$ and $A_{3/2}$
can be translated into $G_E \; \propto \; 1/Q^4$, $G_M \; \propto \; 1/Q^4$
and $G_E + G_M \; \propto \; 1/Q^6$, meaning that $G_M= - G_E$ for large $Q^2$
(see Section~\ref{sec-largeQ2}).

This behavior can also be tested using available parametrizations of the data.
The results are presented in the right panel of Fig.~\ref{fig-N1520-GEGM}.
where we compare 
the form factors $G_M$ and $-G_E$ normalized by the dipole form factor, up to $Q^2= 8$ GeV$^2$.
The figure shows that the data associated with the highest values of $Q^2$
are consistent with $G_M = - G_E$ within the error bars, consistently with $|A_{1/2}| \gg |A_{3/2}|$
or $|A_{3/2}| \simeq 0$ at large $Q^2$.
The same trend is suggested by the JLab-ST parametrization, derived from the CLAS data,
extended here for larger values of $Q^2$.
One can notice, however, that the trend of $G_E$ and $G_M$ scaling with the dipole form factor,
is not manifest yet in the represented region, since the results deviate from a flat line.
Some models and parametrizations present a faster scaling with $G_D$~\cite{N1520SL}.
In the graph, the slow falloff of $G_M + G_E$ with $Q^2$ is also visible. 
The form factor analyses show that their convergence 
to the asymptotic behavior may be slower that 
the inferred from the analysis of the helicity amplitudes 
based on Eq.~(\ref{eqAhel}).
Future data for higher $Q^2$, including results from JLab-12 GeV upgrade 
may confirm if the convergence of  $G_M$ to  $- G_E$ is fast or slow.
For theory, the asymptotic behavior provides constraints on theoretical
models based on valence quark degrees of freedom
which should at very large $Q^2$ be consistent with 
the asymptotic relations between form factors.

\subsubsection*{\it Short notes}

This previous analysis shows that properties of the observables in
the large-$Q^2$  region may be tested in a region available by present experiments.
In this aspect, $\gamma^\ast N \to N(1520)$ transition 
is very different from the $\gamma^\ast N \to \Delta(1232)$,
since in that case the experiments are further away of the scale 
of convergence for the relation $G_M = -G_E$.

%

\subsection{\it Other $\frac{1}{2}^+$ resonances \label{sec-N12p}}

In addition to the state $N(1440)\frac{1}{2}^+$ 
other states 
$N\left(\frac{1}{2}^+\right)$ are worth to discuss: the $N(1710) \frac{1}{2}^+$ state,
a four star state,
whose form factors were measured for the first time in 2015 
at JLab/CLAS~\cite{CLAS15}, and also the  $N(1880) \frac{1}{2}^+$,
at the moment still
a three star resonance~\cite{PDG2022}.

\begin{figure}[t]
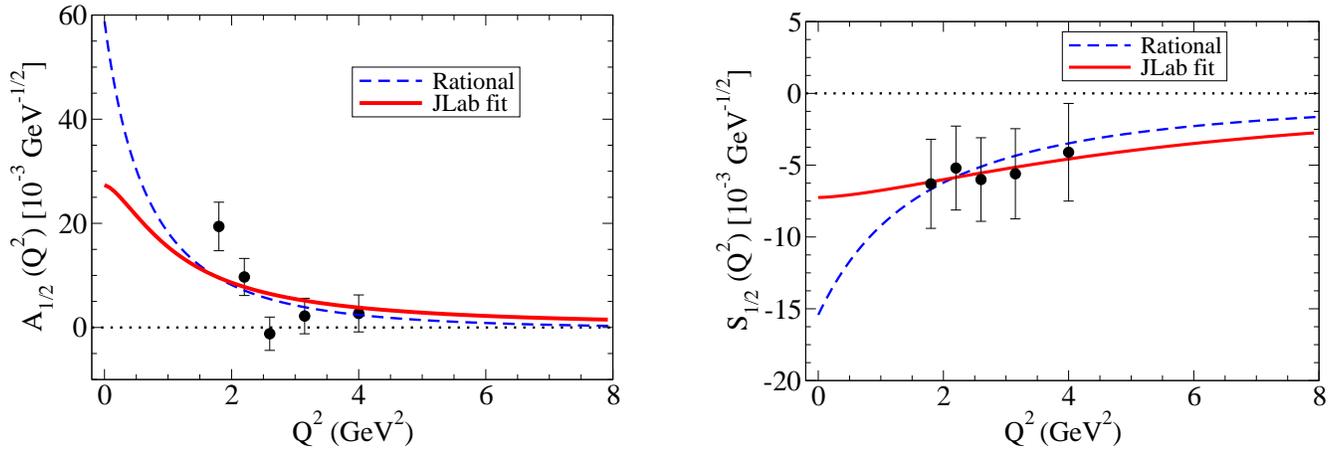

\begin{center}
\includegraphics[width=3.2in]{AmpA12-N1710} \hspace{1.cm}
\includegraphics[width=3.2in]{AmpS12-N1710} 
\caption{\footnotesize 
Parametrizations of the $\gamma^\ast N \to N(1710) \frac{1}{2}^+$ 
helicity amplitudes.
The JLab parametrization is from Ref.~\cite{JLab-website}.
The Rational fraction parametrization is from Ref.~\cite{Eichmann16}. 
The data are from JLab/CLAS~\cite{CLAS15}.
PDG data at $Q^2=0$ are not included.
\label{fig-N1710}}
\end{center}
\end{figure}

\subsubsection{\it $N(1710)\frac{1}{2}^+$ resonance}

While the Roper, $N(1440)\frac{1}{2}^+$  resonance can be interpreted
as the first radial excitation of the nucleon,
the $N(1710)\frac{1}{2}^+$
cannot be interpreted in the 
context of the $SU(6)\otimes O(3)$ quark model
just as the second radial excitation of the nucleon.
According to the $SU(6)\otimes O(3)$ classification the
$N(1710)\frac{1}{2}^+$ corresponds to a mixed symmetry state with $L^P=0^+$
(see chapter ``Quark Model'' in PDG 2022~\cite{PDG2022}).

The electromagnetic structure of the   $N(1710)\frac{1}{2}^+$
was probed at JLab/CLAS in the range $Q^2=1.8$--4.0 GeV$^2$~\cite{CLAS15}.
The results for the two helicity amplitudes are presented in Fig.~\ref{fig-N1710}.
The data was extracted from the $\pi N$ decays 
under the assumption that the decay fraction
was $\beta_{\pi N} \simeq 0.15$.
In the figure is visible the lack of data in the range $Q^2=0$--1.8 GeV$^2$.
More accurate determinations of the $N(1710)\frac{1}{2}^+$
helicity amplitudes may be obtained in a near future 
from eta production, since the $\eta N$ channel 
has a larger branching ratio $\beta_{\eta N} \approx 0.3$~\cite{PDG2022}.
Also the $\pi \pi N$ channel can be used 
to extract data at low $Q^2$~\cite{PDG2022,Mokeev22a,Mokeev20a}.
There are measurements at the photon point for $A_{1/2}$ 
but there is no consensus yet about the sign and magnitude~\cite{PDG2022}.
The PDG estimate has been changing along the years.
The parametrizations represented in the figure were also
constrained by the estimate of $A_{1/2}(0)$ at the time.
In these conditions, one can only conjecture about 
the low-$Q^2$ behavior of the helicity amplitudes.
The large-$Q^2$ data displayed in Fig.~\ref{fig-N1710} 
is consistent with the predictions from the 
hypercentral quark model~\cite{Santopinto12,CLAS15}.
Other estimates can be found in Refs.~\cite{Ronniger13a,Capstick95,Lyubovitskij20a,Taghieva22a}.

We notice that the $N(1710)\frac{1}{2}^+$ has also been predicted 
by models which relate it to the Roper~\cite{Suzuki10a,Golli08a}.
The best example is the EBAC/ANL-Osaka analysis, 
which interprets the two states originated from 
a bare state with a mass near 1.8 GeV~\cite{Suzuki10a}, with
the difference between the states coming from different channels taking part in their meson dressing.

\subsubsection{\it $N(1880)\frac{1}{2}^+$ resonance}

The properties of the $N(1880)\frac{1}{2}^+$ are 
less well known, as expected from a three star resonance.
Therefore, the calculations of the $\gamma^\ast N \to N(1880)$ 
transition amplitudes are scarce.
Calculations of the $\gamma^\ast N \to N(1880)$ amplitudes
can be found in Ref.~\cite{Ronniger13a}.
The calculations of the transition amplitudes 
are simplified if the $N(1880)\frac{1}{2}^+$
can be interpreted as the second radial excitation 
of the nucleon and the first radial excitation of the Roper. 
In that case the spatial wave function is symmetric and the 
transition amplitudes and form factors can be 
estimated using the correlations with the ground state 
and first radial excitation.
The calculations based on the covariant spectator 
quark model~\cite{N1440-1,Nucleon}, suggest that the helicity 
amplitudes for the $\gamma^\ast N \to N(1880)$ 
and the $\gamma^\ast N \to N(1440)$ transitions
have the same magnitude for $Q^2 > 6$ GeV$^2$~\cite{NSTAR2017,N1710}.

%

\subsection{\it  Other $\frac{3}{2}^+$ resonances \label{sec-N32p}}

\begin{figure}[t]
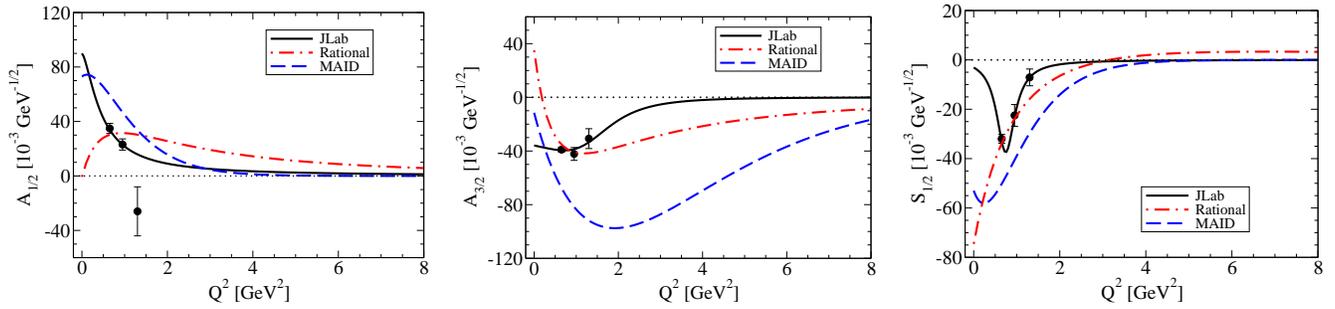

\begin{center}
\includegraphics[width=2.2in]{A12-N1720} \hspace{.1cm}
\includegraphics[width=2.2in]{A32-N1720} \hspace{.1cm}
\includegraphics[width=2.2in]{S12-N1720} 
\caption{\footnotesize 
$\gamma^\ast N \to N(1720) \frac{3}{2}^+$ helicity amplitudes.
The JLab parametrization is from Ref.~\cite{JLab-website}.
The Rational parametrization is from Ref.~\cite{Eichmann16}.
The data are from JLab/CLAS~\cite{Mokeev20a}.
\label{figN1720}}
\end{center}
\end{figure}

There are two $J^P= \frac{3}{2}^+$ states worthy of discussion,
the $N(1720) \frac{3}{2}^+$ and $\Delta(1600) \frac{3}{2}^+$ resonances.

\subsubsection{\it $N(1720)\frac{3}{2}^+$ resonance \label{secNS32p}}

The $N(1720)\frac{3}{2}^+$ resonance has been interpreted 
in the context of the $SU(6)\otimes O(3)$ symmetry 
as a state of the multiplet $[56,2_2^+]$, associated to a symmetric $J=\frac{3}{2}$ spin state 
with angular momentum $L=2$~\cite{PDG2022}.
The state decays predominantly into the $\pi \Delta$ channel. 
The decay to $\pi N$ is only about 10\%.
The $\gamma^\ast N \to N(1720)$ was measured recently at JLab/CLAS 
using the $\gamma^\ast p \to \pi^+ \pi^- p$ reaction in the range 
$Q^2=0.65$--1.30 GeV$^2$~\cite{Mokeev20a}.
The results are presented in Fig.~\ref{figN1720}.
The transverse amplitudes at the photon point are still imprecise,
since PDG selects experiments with very different results~\cite{PDG2022}.

In Fig.~\ref{figN1720}, we present also the MAID,  
JLab and Rational parametrizations of the data.
The MAID parametrization is based on the MAID analysis of the database 
including the JLab/CLAS data~\cite{MAID2009,MAID2011}.
From the comparison we can conclude that there is not  
a clear trend of the data, and that the low-$Q^2$ behavior 
is still largely undetermined.
The analyses of CLAS~\cite{Mokeev20a}, MAID~\cite{MAID2009,MAID2011}
and PDG group~\cite{PDG2022} made different assumptions
about the $N(1720)$ decay branching ratios 
on the different decay channels.
Different results may be obtained in a near future if
the branching ratios are modified by more accurate measurements
near $W \simeq 1.7$~\cite{Mokeev22a,Mokeev20a}.
Calculations of the transition amplitudes have been performed in quark model 
frameworks~\cite{Santopinto12,Ronniger13a}
and within AdS/QCD~\cite{Lyubovitskij20a}.
Except for the holographic model that is adjusted to the data,
no model describes the signs of the three amplitudes.

It is worth noticing that the classification based on the 
$SU(6)\otimes O(3)$ symmetry predicts more $N\left(\frac{3}{2}^+ \right)$ states
with different structures, that can be members of the multiplets
$[70,2^+]$, $[70,0^+]$ or $[20,1^+]$, and have a mass 
not much larger than 1.7 GeV~\cite{Capstick00,Isgur78a,Isgur79a,Capstick02a,Zhao06a,Close90a}.
There is then the possibility that more states $N\left(\frac{3}{2}^+ \right)$
may be detected in a narrow region of $W$.
In the region $W >$ 1.6 GeV the correspondence  
of the $N(J^P)$ and $\Delta(J^P)$ states in terms of the mass 
is sometimes tentative (see chapter ``Quark Model'' in PDG 2022~\cite{PDG2022}).

Recently there were some evidences of a state $J^+ = \frac{3}{2}^+$ 
near the $N(1720)\frac{3}{2}^+$ discussed above, from the analysis 
of the $\gamma^\ast p \to \pi^+ \pi^- p$ data~\cite{Mokeev20a,Kamano13a,CLAS03a}.
The properties of this state are different from the first $
N(1720)\frac{3}{2}^+$ state since it has different couplings
to the $\pi \pi N$ channels~\cite{Aznauryan12a}.
Helicity amplitudes have been extracted from the 
$\gamma^\ast N \to \pi \pi N$ data for $Q^2=0$ 
and finite $Q^2$~\cite{Mokeev20a,CLAS19a}.
At the moment this new state has no PDG classification.
More definitive conclusions may be obtained 
from a coupled-channel analysis, including 
$\pi N \to \pi N$, $\gamma^\ast N \to \pi N$, 
$\gamma^\ast N \to \pi  \pi N$ and $\pi N \to \pi \pi N$ data~\cite{Aznauryan12a}.

\subsubsection{\it $\Delta(1600)\frac{3}{2}^+$ resonance}

The $\Delta(1600)$ has been discussed in the literature 
because it can be interpreted in a valence quark model framework, 
as the radial excitation of the $\Delta(1232)$.
The $\Delta(1600)$ can then be regarded as 
the equivalent to the Roper in the isospin $\frac{3}{2}$ sector.
Although it is four star resonance, the data associated 
with the $\gamma^\ast N \to \Delta(1600)$ transition 
is at the moment restricted to the photon point.
The state decays predominantly in the $\pi \Delta(1232)$ 
($\sim 70\%$), $\pi N(1440)$  ($\sim 20\%$) and 
$\pi N$ ($\sim 15\%$)~\cite{PDG2022}.

Calculation of the $\gamma^\ast N \to \Delta(1600)$ transition 
amplitudes and form factors can be found in
Refs.~\cite{Ronniger13a,Capstick95,Golli19,Delta1600,Lu19,Aznauryan15a,Aznauryan16a}.
The first measurements of the $\gamma^\ast N \to \Delta(1600)$ 
helicity amplitudes for finite $Q^2$ from JLab experiments
were reported recently~\cite{CLAS23a},
following the progress in the analysis of the states from the second and third 
resonance region~\cite{CLAS16a,CLAS15}.
The results from Ref.~\cite{CLAS23a} suggest that, as in the case of the $\Delta(1232)$, 
the transition to the $\Delta(1600)$ is dominated
by the magnetic form factor $G_M$.
The electric form factor is compatible with zero.

\subsection{\it $\frac{1}{2}^-$ and $\frac{3}{2}^-$ resonance
from the multiplet $[70,1_1^-]$  \label{secSQTM2}}

In this section, we discuss some model calculations for the 
states $N(1650) \frac{1}{2}^-$,  $N(1700) \frac{3}{2}^-$,
 $\Delta(1620) \frac{1}{2}^-$ and  $\Delta(1700) \frac{3}{2}^-$.
Some of the calculations are based on the single quark transition model (SQTM)
discussed in Section~\ref{secSQTM}.
The reason to look at these states is that $SU(6)$ breaking
due to the color hyperfine interaction between quarks 
 mixes the states $N(1535)$-$N(1650)$ 
and the states $N(1520)$-$N(1700)$, respectively
(see Sections~\ref{sec-N1535} and \ref{sec-N1520}).
Subsequently the $\gamma^\ast N \to N^\ast$ current,  
where $N^\ast$ is a member of the multiplet $[70,1_1^-]$,
can be expressed in terms of three 
amplitudes $A$, $B$ and $C$, and two mixture 
angles of the $N\left( \frac{1}{2}^-\right)$ ($\theta_S$)  
and $N\left( \frac{3}{2}^-\right)$ ($\theta_D$) states, 
defined by Eqs.~(\ref{eqMixt-N1535}) and (\ref{eqMixt-N1520}).
Tables with explicit expressions 
can be found in Refs.~\cite{Aznauryan12a,Burkert04,SQTM}.
A consequence of the SQTM relations is that
the amplitudes for the states $N(1650)\frac{1}{2}^-$ and $N(1700)\frac{3}{2}^-$
are obtained from the amplitudes for 
$N(1535)\frac{1}{2}^-$ and $N(1520)\frac{3}{2}^-$ replacing $\cos \theta$ by $\sin \theta$, 
respectively, where $\theta$ can be $\theta_S$ ($J^P = \frac{1}{2}^-$) 
or $\theta_D$ ($J^P = \frac{3}{2}^-$)\footnote{In simple words, 
we obtain the amplitudes of the heavy states 
multiplying the amplitudes of the light states 
by $\frac{\sin \theta}{\cos \theta}$.}.
One expects then that the amplitudes for $N(1650)$ and $N(1700)$ 
are considerably smaller than for their counterparts.

The basic idea of the calculations shown here is then that from the parametrizations 
of the amplitude $A_{1/2}$ for the $N(1535)$ and of the amplitudes 
$A_{1/2}$, $A_{3/2}$ for the $N(1520)$,
we can determine the functions $A$, $B$ and $C$,
and predict the transverse 
amplitudes for the states 
$N(1650) \frac{1}{2}^-$,  $N(1700) \frac{3}{2}^-$,
$\Delta(1620) \frac{1}{2}^-$ and  $\Delta(1700) \frac{3}{2}^-$.
In principle, we could also estimate the 
amplitudes $A_{1/2}$ and $A_{3/2}$ of the state $N(1675)\frac{5}{2}^-$,
also a member of the multiplet.
Unfortunately this state does not mix with the other states 
in single quark interactions for proton targets~\cite{Burkert-SQTM}.
For neutron targets, however,  the amplitudes 
can be calculated within the SQTM.
The state  $N(1675)\frac{5}{2}^-$ is discussed in Section~\ref{sec-N52m}.

The input for the SQTM should be exclusively based 
on valence quark contributions,
which dominate the intermediate and large-$Q^2$ region.
However, in practice the results are improved when 
we use more accurate descriptions of the 
$N(1535)$ and $N(1520)$ data at low $Q^2$, where $q\bar{q}$
or meson cloud effects enlarge the baryons,
as in the calculations discussed in previous sections with meson cloud
dressing effects and applied here to these higher mass states.
The reader may also remember how relevant is for low $Q^2$ that empirical parametrizations 
of the data for the amplitudes $A_{1/2}$ and $A_{3/2}$ are be avoided, 
unless they are compatible with the pseudothreshold constraints,
as discussed in Section~\ref{sec-Siegert}.

\begin{figure}[t]
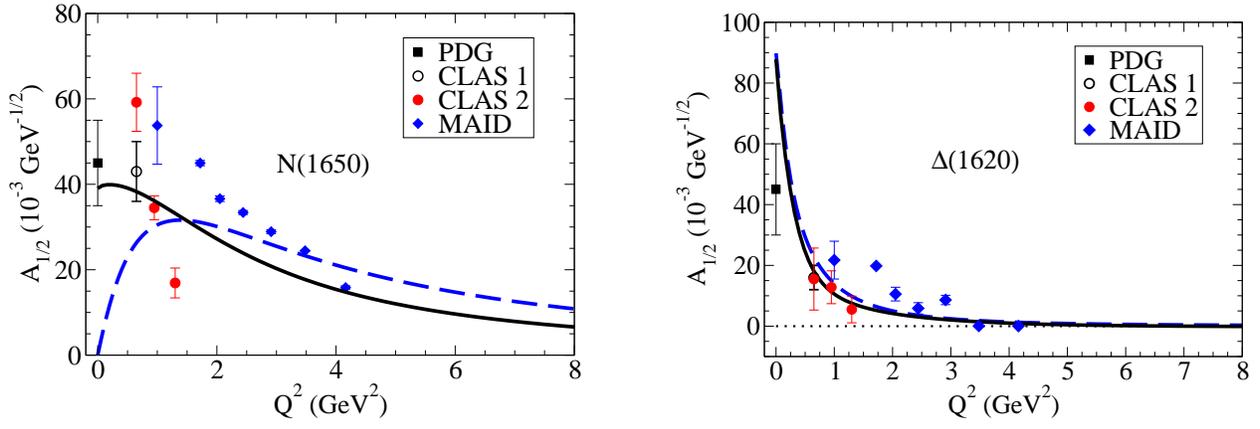

\begin{center}
\includegraphics[width=3.in]{A12-N1650} \hspace{1.cm}
\includegraphics[width=3.in]{A12-D1620} 
\caption{\footnotesize 
$\gamma^\ast N \to N(1650) \frac{1}{2}^-$ and $\gamma^\ast N \to \Delta(1620) \frac{1}{2}^-$
transition amplitude $A_{1/2}$.
Calculations based on the SQTM.
The solid line represents the calculation from Ref.~\cite{Aznauryan12a}
and the dashed line represents the calculation from Ref.~\cite{SQTM}. 
The CLAS 1 data are from Ref.~\cite{Aznauryan05a},
the CLAS 2 data are from Refs.~\cite{CLAS16a,Mokeev14a},
the MAID data are from Refs.~\cite{MAID2011,MAID-database,MAID-website},
and the $Q^2=0$ data are from PDG 2020~\cite{PDG2020}.
\label{figN1650}}
\end{center}
\end{figure}

\begin{figure}[t]
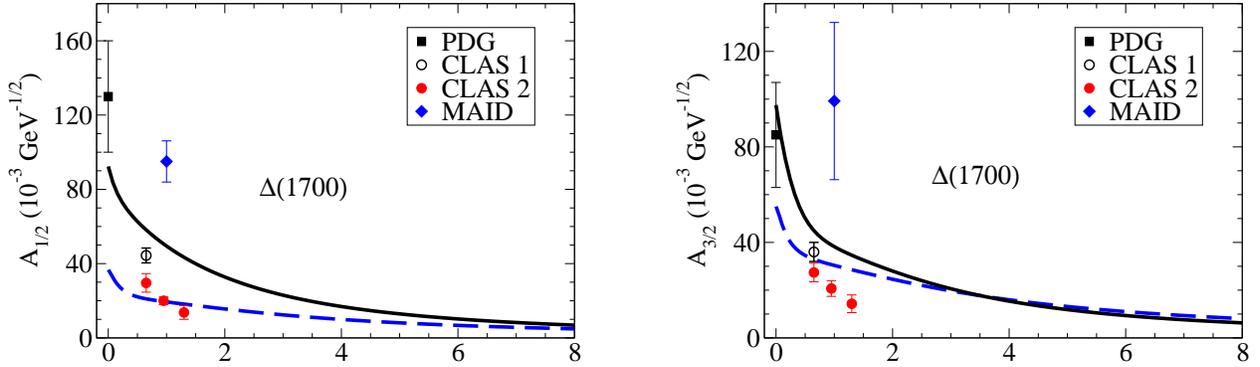

\begin{center}
\includegraphics[width=3.in]{A12-D1700} \hspace{1.cm}
\includegraphics[width=3.in]{A32-D1700} 
\caption{\footnotesize 
$\gamma^\ast N \to \Delta(1700) \frac{3}{2}^-$ transverse transition amplitudes.
Calculations based on the SQTM. The solid line represents the calculation
from Ref.~\cite{Aznauryan12a}
and the dashed line represents the calculation from Ref.~\cite{SQTM}. 
Data as described in Fig.~\ref{figN1650}.
\label{figD1700}}
\end{center}
\end{figure}

The amplitudes $A_{1/2}$ for the states $N(1650)$ and $\Delta(1620)$
are represented in Fig.~\ref{figN1650}.
The figure shows that magnitude of the amplitude $A_{1/2}$ for  $N(1650)$
is smaller comparatively to $N(1535)$:
\mbox{$A_{1/2} (0) \approx 100 \times 10^{-3}$} GeV$^{-1/2}$,
as expected from the previous discussion.
The amplitudes for $\Delta(1700)$ are in Fig.~\ref{figD1700}.
The calculations are from Refs.~\cite{Aznauryan12a,SQTM}.
In Figs.~\ref{figN1650} and \ref{figD1700}, 
we present two calculations of transverse amplitudes 
based on the SQTM formalism.
In the first calculation, Aznauryan and Burkert 
used parametrizations of the JLab data~\cite{Aznauryan12a}
combined with the angles $\theta_S= - 32^\circ$ 
and $\theta_D= 14^\circ$.
The second calculation uses the results 
of the covariant spectator quark model for $N(1535)$,  
$N(1520)$ and the angles $\theta_S= - 32^\circ$,   $\theta_D = 6^\circ$.
In the last case, only the bare contributions are taken into account,
except for the explicit parametrization 
of the amplitude $A_{3/2}$ for the $N(1520)$.
The data on Figs.~\ref{figN1650} and \ref{figD1700}, 
come mainly from Refs.~\cite{CLAS16a,Mokeev14a} 
from two pion production and from MAID~\cite{MAID2011,MAID-database,MAID-website}.
Data from JLab can also be found in the database~\cite{JLab-database}.

The discussion about the $N(1700)\frac{3}{2}^-$, a three-star state, 
is omitted due to the limited available data.
Apart the result from the combined analysis of one pion and two pion 
electroproduction data from CLAS for 
$Q^2=0.65$ GeV$^2$~\cite{Aznauryan05a},
one has only the PDG result for $Q^2=0$.
The PDG results for $A_{1/2}$ and $A_{3/2}$ have been changing along the years,
and their sign has been stable only since 2015~\cite{PDG2022,PDG2014}.

Based on the conditions from the SQTM, one expects 
the estimates presented in Figs.~\ref{figN1650} and \ref{figD1700}
to be valid for $Q^2 > 1.5$ GeV$^2$ (intermediate and large $Q^2$).
From the figures, we can conclude that the two model predictions
are comparable with the data, 
although there are still differences 
between the data from different analysis.
The calculations for the $\Delta(1620)$, in particular, 
suggest a very fast falloff for the amplitude $A_{1/2}$ close to $1/Q^5$~\cite{SQTM}, 
an apparent deviation from the main rule $A_{1/2} \; \propto \; 1/Q^3$, 
discussed in Section~\ref{sec-largeQ2}.
Unfortunately, at the moment the available data are very scarce, 
and almost nonexistent for $Q^2> 2$ GeV$^2$.

Explicit calculations of the $N(1650) \frac{1}{2}^-$,  $N(1700) \frac{3}{2}^-$
amplitudes based on the states $\left| N^4 \; \frac{1}{2}^- \right>$
and  $\left| N^4 \; \frac{3}{2}^- \right>$, omitted on SQTM, 
are presented in Ref.~\cite{Aznauryan17a}.
More model calculations of the states 
$N(1650) \frac{1}{2}^-$,  $N(1700) \frac{3}{2}^-$,
$\Delta(1620) \frac{1}{2}^-$ and  $\Delta(1700) \frac{3}{2}^-$
can be found in 
Refs.~\cite{Santopinto12,Ronniger13a,Pace99a,Capstick95,Golli11a,Golli13a,Lyubovitskij20a,Aznauryan17a,Merten02a}.
The previous references include also estimates of 
the scalar amplitude $S_{1/2}$, not discussed here because 
the SQTM estimates are restricted to transverse amplitudes.

Future data from the JLab 12 GeV-upgrade will 
be fundamental to test model calculations for the states $N(1650)$,
$N(1700)$, $\Delta(1620)$ and $\Delta(1700)$.

\subsubsection*{\it Short notes}

$SU(6)$ breaking
due to the color hyperfine interaction between quarks in single quark transition model
(SQTM) allows models and parametrizations 
  calibrated in the second resonance region
to be applied and make reasonable predictions of the transition amplitudes
for higher mass states of the third resonance region.
More data in this region is necessary for precise information
on the power law of the falloff of the amplitudes,
as well as their behavior near the photon point.

\subsection{\it $\frac{5}{2}^+$ resonances \label{sec-N52p}}

\begin{figure}[t]
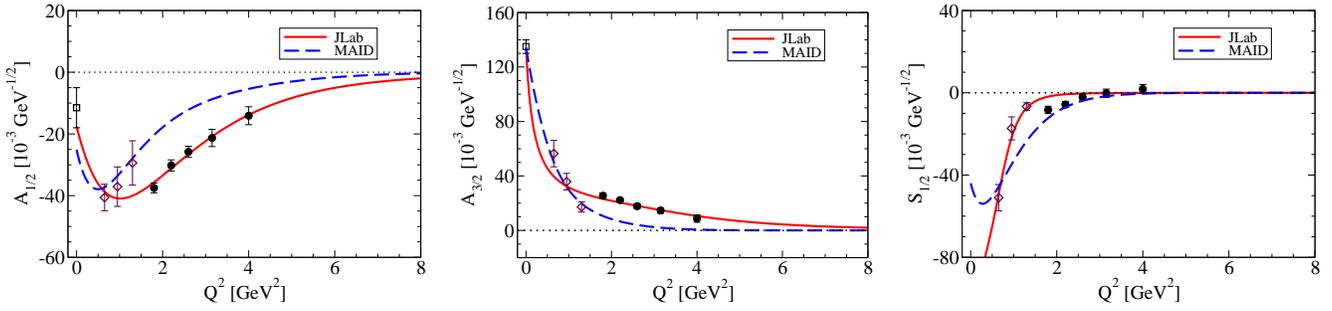

\begin{center}
\includegraphics[width=2.2in]{A12-N1680} \hspace{.1cm}
\includegraphics[width=2.2in]{A32-N1680} \hspace{.1cm}
\includegraphics[width=2.2in]{S12-N1680} 
\caption{\footnotesize 
$\gamma^\ast N \to N(1680)\frac{5}{2}^+$ helicity amplitudes.
The JLab parametrization is from Ref.~\cite{JLab-website},
and the MAID parametrization (2011) is from Ref.~\cite{MAID2009,MAID2011}. 
The data are from MAID~\cite{MAID2011,MAID-database}, preliminary double-pion electroproduction 
from CLAS~\cite{Mokeev14a} and single-pion electroproduction from CLAS (2015)~\cite{CLAS15}
and from PDG 2022~\cite{PDG2022} (for $Q^2=0$).
\label{figN1680}}
\end{center}
\end{figure}

The $N(1680)\frac{5}{2}^+$ resonance is the first known $J^P = \frac{5}{2}^+$ resonance 
and can be identified by the decays into the 
$\pi N$ ($\sim 65\%$), but has also important $\pi \pi N$ ($\sim 40\%$)
contributions originated by intermediate $\pi \Delta$,
$\rho N$ and $\sigma N$ decay channels~\cite{PDG2022}.
Besides an isolated measurement from JLab/CLAS 
in 2005 for $Q^2=0.6$ GeV$^2$ from 
single- and double-pion electroproduction data~\cite{Aznauryan05a},
there are preliminary JLab/CLAS (2014) results 
from double-pion electroproduction data~\cite{Mokeev14a},
and the data analysis from MAID from several pion electroproduction
experiments up to $W=1.7$ GeV~\cite{MAID2011,MAID-database}.
This analysis has used in MAID 2011 parametrization 
of nucleon resonances.
More recently $\gamma^\ast N \to N(1680)$ 
transition amplitudes were measured in a wide region of $Q^2$
(from 1.8 to 4.0 GeV$^2$) at JLab/CLAS from 
the one pion electroproduction~\cite{CLAS15}.
The data results are presented in Fig.~\ref{figN1680},
together with the MAID~\cite{MAID2009,MAID2011} and 
JLab parametrizations~\cite{JLab-website}.
The JLab parametrization is based 
on the low-$Q^2$ data from MAID, double-pion production data from CLAS~\cite{MAID2011,Mokeev14a},
below 1.5 GeV$^2$ and the recent one pion electroproduction CLAS data~\cite{CLAS15}.

There are predictions of the  $\gamma^\ast N \to N(1680)$ 
transition amplitudes based on non relativistic quark models~\cite{Santopinto12,Li90a} and
relativistic quark models~\cite{Ronniger13a,Merten03a}.
There are also VMD parametrizations~\cite{Vereshkov10a}
based on the MAID 2011 analysis and 
parametrizations based on holographic QCD~\cite{Lyubovitskij20a}.

The models from Ronniger et al.~\cite{Ronniger13a}, 
Li and Close~\cite{Li90a}, and  
Merten et al.~\cite{Merten03a} 
predict similar magnitudes for $A_{1/2}$ and $A_{3/2}$ above $Q^2=2$ GeV$^2$,
but underestimate the amplitude $A_{3/2}$~\cite{CLAS15}.
The model from Ronniger et al.~\cite{Ronniger13a} 
and the hypercentral quark model~\cite{Santopinto12} provide a good 
description of the $A_{1/2}$ data including the low-$Q^2$ region.
The hypercentral quark model~\cite{Santopinto12} 
provides as well a good approximation to the amplitude $S_{1/2}$ 
at large $Q^2$~\cite{CLAS15}.
In general quark models predict a dominance of the $A_{1/2}$ 
over $A_{3/2}$ for the transition which is not confirmed by the
data which exists only up to 4 GeV$^2$.



\subsection{\it $\frac{5}{2}^-$ resonances \label{sec-N52m}}

The last nucleon resonance to be discussed here is the $N(1675)\frac{5}{2}^-$.
The resonance decays predominantly to $\pi N$ ($\sim 40\%$),  
but was also significant contributions from 
the $\pi \Delta$ ($\sim 30\%$) and $\sigma N$ ($\sim 5\%$)
channels~\cite{PDG2022}.
In the $SU(6) \otimes O(3)$ symmetry group the state 
is part of the $[70,1_1^-]$ multiplet, discussed in Section~\ref{secSQTM2}.

As mentioned earlier, in the SQTM framework the $\gamma^\ast N \to N(1675)$ helicity amplitudes
are suppressed in the case the proton targets
(due to the Moorhouse rule~\cite{Burkert-SQTM,Moorhouse66a},
on contributions of the $S_{3q}=\frac{3}{2}$
quark spin states to the transition amplitudes).
The small contributions to the $N(1675)$ amplitudes are then 
the consequence of $SU(6)$ breaking due to 
the hyperfine interaction between quarks~\cite{Santopinto12,Aznauryan15b}.
Notice, however, that the SQTM predicts significant contributions 
for the amplitudes associated with neutron targets,
which is supported by the 
magnitudes of the transverse amplitudes 
measured at the photon point~\cite{PDG2022,Burkert-SQTM,Aznauryan17a}.

\begin{figure}[t]
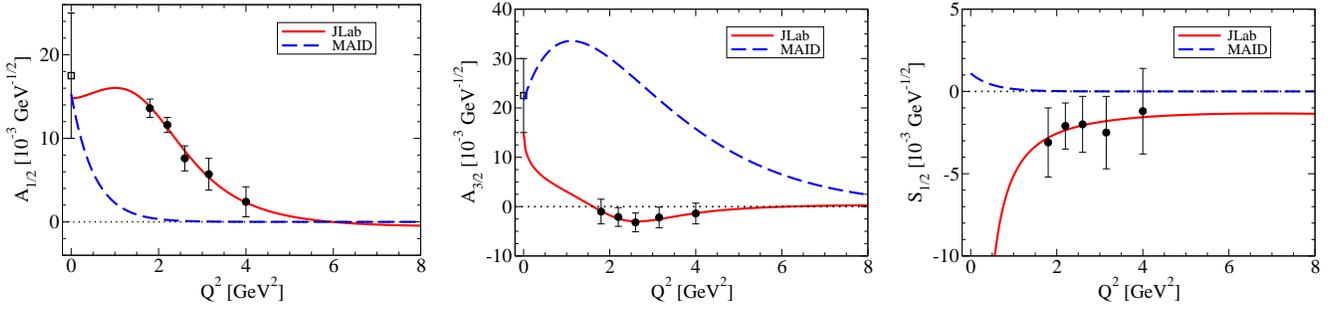

\begin{center}
\includegraphics[width=2.2in]{A12-N1675} \hspace{.1cm}
\includegraphics[width=2.2in]{A32-N1675} \hspace{.1cm}
\includegraphics[width=2.2in]{S12-N1675} 
\caption{\footnotesize 
$\gamma^\ast N \to N(1675) \frac{5}{2}^-$ helicity amplitudes.
The JLab parametrization is from Ref.~\cite{JLab-website},
and the MAID parametrization (2011) is from Ref.~\cite{MAID2009,MAID2011}. 
The data are from JLab/CLAS (2015)~\cite{CLAS15} and PDG 2022~\cite{PDG2022} (for $Q^2=0$).
\label{figN1675}}
\end{center}
\end{figure}

The first data on the $\gamma^\ast N \to N(1675)$ 
transition amplitudes was obtained by the MAID analysis from MAMI, ELSA 
and JLab data~\cite{MAID2009,MAID2011,MAID-database}.
More recently the transition amplitudes were measured at JLab/CLAS 
in the $\pi N$ channel in the range $Q^2=1.8$--4.0 GeV$^2$~\cite{CLAS15}.
The new data and the PDG selection for $Q^2=0$ are presented in Fig.~\ref{figN1675}.
The experimental results are compared with the JLab parametrization~\cite{JLab-website},
and the MAID parametrization, derived from the MAID 2011 analysis~\cite{MAID2011}.
Figure~\ref{figN1675} confirms the small magnitude 
of the helicity amplitudes compared with states 
discussed in the previous sections.

Consistently with the SQTM, quark models predict is general
only small contributions to the transition amplitudes,  
and underestimate the data.
This is valid for the (non relativistic) hypercentral quark model~\cite{Santopinto12}
and for the relativistic model of Merten et al.~\cite{Merten03a}, 
and for the light-front quark model from Aznauryan and Burkert~\cite{Aznauryan17a}.
The comparison of the models with the data can be found in Refs.~\cite{CLAS15,Aznauryan17a}.
Compared to the SQTM model, relativistic models include 
contributions which violate $SU(6)$, and may provide additional strength 
to the helicity amplitudes.
The $SU(6)$ violation is, however, small~\cite{Aznauryan17a}.
The significant underestimation associated with 
some quark models has been interpreted as a manifestation 
that the amplitudes, particularly the transverse amplitudes,
may be dominated by meson cloud effects~\cite{CLAS15,Aznauryan17a}.
This interpretation is supported by the magnitude 
of the meson cloud contributions obtained within the EBAC/ANL-Osaka dynamical 
coupled-channel model~\cite{Kamano16a,JDiaz08a,JDiaz09a,Aznauryan17a}.
The comparison of the magnitude of the EBAC/ANL-Osaka 
calculation with the data can be found in Ref.~\cite{Aznauryan15b}
for the amplitudes $A_{1/2}$ and $A_{3/2}$.
The relativistic quark model from Ronniger and Metsch~\cite{Ronniger13a} 
based on the Bethe-Salpeter equation
reproduce the experimental magnitude and sign of 
the three amplitudes at large $Q^2$.
The holographic QCD model from Ref.~\cite{Lyubovitskij20a}
provide a parametrization of the data 
compatible with the expected large-$Q^2$ behavior.

More data is necessary at low $Q^2$ in order 
to infer the shape of the $\gamma^\ast N \to N(1675)$ 
helicity amplitudes at intermediate $Q^2$,
to draw mode definitive conclusions about the $N(1675)$ state~\cite{Aznauryan15b}.

\subsection{\it Transverse densities \label{sec-Trans}}

The interpretation of form factors as Fourier transforms of
charge distributions is not general, and it was
criticized~\cite{Alexandrou09b,Delta-Shape,Kelly02a}
because it does hold only non relativistically. Relativistic
boost corrections increase with $Q^2/M^2$ and therefore make
that interpretation unclear. Such boost effects can be negligible
for the small binding energies and large masses of nuclei, but
for the smaller hadron masses the corrections become relevant specially
for large $Q^2$ and can even affect the charge radii extraction.
An alternative formalism to study the electromagnetic structure 
of the nucleon resonances and the $\gamma^\ast N \to N^\ast$ 
transitions is the transverse charge
density formalism~\cite{Alexandrou09b,Burkardt02a,Miller10a},
or simply the transverse density formalism, where the interpretation
of form factors is made more general and unambiguous.

Instead of looking for the transition current $J^\mu$ in a given frame
($N^\ast$ rest frame, Breit frame or Lab frame), 
the infinite momentum frame is considered and the transverse charge distributions is defined
The infinite moment frame is characterized by a very large 
longitudinal momentum along $\hat z$, $P_z$, of the initial and final baryons.
In these conditions we can write, 
using the light-front notation~\cite{Drell70a}
\ba
p'= \left(E', \sfrac{1}{2}{\bf q}_\perp, P_z\right),
\hspace{1cm}
p= \left(E ,- \sfrac{1}{2}{\bf q}_\perp, P_z\right),
\ea
and 
\ba
P^+ = E' + E = 2 P_z + {\cal O}\left(\frac{1}{P_z}\right), \hspace{1cm}
q= \left( \frac{M_R^2 -M^2}{2 P_+},  {\bf q}_\perp,0 \right),
\label{eqIMF2}
\ea
where $E'= P_z + {\cal O}\left(\sfrac{1}{P_z}\right)$ and
$E= P_z + {\cal O}\left(\sfrac{1}{P_z}\right)$. 
In the infinite momentum frame, one has $Q^2 = {\bf q}_\perp^2$, 
reducing the analysis from 3-dimensions to 
a 2-dimension plane, transverse to direction of the baryons,
where ${\bf q}_\perp$ lies.
To calculate the transverse amplitudes, 
defined in the plane transverse to $\hat z$,  
one has to project the electromagnetic current into states of  
the transverse spins ${\bf S}_\perp^\prime$ (final state) 
and ${\bf S}_\perp$ (initial state). 
These means that, instead of the light-front helicity 
states defined in the direction of ${\bf p}^\prime$ and ${\bf p}$,
we are interested in the spin projections along ${\bf S}_\perp^\prime$ 
($s_\perp^\prime$) and ${\bf S}_\perp$  ($s_\perp$).
Once defined the transverse spin states, 
we can project the light-front current, $J^+ = J^0 + J^3$,
into the light-front states $\left| P^+, \pm {\bf q}_\perp, s_\perp \right>_B$,
of a baryon $B$ with transverse spin projection $s_\perp$.

The transverse densities are 
defined by two-dimensional Fourier transform~\cite{Tiator09a,Carlson08a}
\ba
\rho_T^{N N^\ast} ({\bf b}) 
= \int \frac{d^2 {\bf q}_\perp}{(2 \pi)^2} 
e^{- i {\bf q}_\perp \dot {\bf b}} 
\sfrac{1}{2 P_+} \,
\frac{}{}_{N^\ast \! \!}\left< P^+, \sfrac{1}{2}{\bf q}_\perp, s_\perp \right| J^+(0)
\left| P^+, - \sfrac{1}{2}{\bf q}_\perp, s_\perp \right>_N.
\label{eqTrans1}
\ea
The upper indices $N$ and $N^\ast$ label the initial and final states.
Equation (\ref{eqTrans1}) transforms the function of ${\bf q}_\perp$ 
into a function of the impact parameter ${\bf b}$, 
and it is defined on the plane $(b_x,b_y)$ for a given 
value of $s_\perp$.
Since ${\bf S}_\perp$ is defined in the transverse plane, we need to specify 
the direction of ${\bf S}_\perp$.
It is customary to choose ${\bf S}_\perp$ with the direction 
of $\hat x$ or $\hat z$~\cite{Alexandrou09a,Carlson08a}.
The transverse densities can be extended naturally for 
elastic transitions of spin $1/2$ and 
the spin $1/2$ particles~\cite{Alexandrou09b,Miller07a}.

These $\rho_T^{N N^\ast} ({\bf b})$ distribution allows us 
to observe asymmetries on the $\gamma^\ast N \to N^\ast$ 
transitions and also in elastic transitions ($N$, $\Delta(1232)$, etc.).
Studies of transverse densities have been performed for 
the $\Delta(1232)$, $N(1440)$, $N(1535)$ and $N(1520)$ 
transitions~\cite{MAID2011,Tiator09a,Carlson08a}, 
as well as for the  $\Delta(1232)$ 
elastic transition~\cite{Alexandrou09b,Alexandrou09a}.  
A very significant part of the developed work used the 
MAID parametrizations~\cite{Drechsel07,MAID2009,MAID2011}.
The transverse densities representation can also be used to 
analyze the flavor decomposition and study the $u$ and $d$ quark 
distributions in the impact parameter plane~\cite{Burkardt02a,Miller10a,Tiator09a,Carlson08a}.
For a discussion of the transverse density results 
for nucleon resonances we recommend~\cite{MAID2011,Tiator09a,Carlson08a}.

Two extra notes are:
\begin{itemize}
\item 
The transverse densities come from integrals of
the transition form factors ($F_1$, $F_2$ or 
$G_M$, $G_E$, $G_C$) as functions of $Q^2$~\cite{Tiator09a,Carlson08a}.
Thus, in a sense, transverse densities use the information already included in the form 
factors expressed in terms of $Q^2$.
\item 
The region of the small transverse distances depend   
on the parametrizations of the large-$Q^2$ regions,
which are at the moment limited by the existing data 
and type of the extrapolation for large $Q^2$.
\end{itemize}
Thus data in a larger $Q^2$ region is necessary to reduce 
the uncertainties for small $|{\bf b}|$~\cite{Aznauryan12a}.
Parametrizations compatible with pQCD falloffs 
are preferable to parametrizations with faster falloffs.


\renewcommand{\theequation}{7.\arabic{equation}}
\setcounter{equation}{0}

\renewcommand{\thefigure}{7.\arabic{figure}}
\setcounter{figure}{0}

\section{Electromagnetic structure of baryons with heavy quarks
\label{secBaryons}}

In the previous sections, we restrict our discussion 
to baryon systems composed exclusively on quarks $u$ and $d$ (light quarks),
with emphasis to the nucleon excitations 
(transition $\gamma^\ast N \to N^\ast$).
To finish the review about the electromagnetic structure 
of baryons, we discuss here briefly studies on 
baryons which include one or more heavy quarks.
In this context, we count the strange quark abusively  
as heavy quark, when the correct description would be non light quark.

In the following we discuss then $\gamma^\ast B \to B$ 
and $\gamma^\ast B \to B^\ast$ transitions, 
where $B$ and $B^\ast$ represent baryons with at least a non light quark.
We start with elastic transitions, followed by inelastic transitions.
At the end we describe recent developments 
on the study of the  $\gamma^\ast B \to B$ 
and $\gamma^\ast B \to B^\ast$ transitions in the timelike region 
which can bring some light about the structure of heavy baryons 
(baryons with heavy quarks), which are unaccessed in spacelike 
(electron scattering on baryons).

The first estimates of the electromagnetic properties 
of the heavy baryons are based on 
non relativistic quark 
models~\cite{Capstick00,Isgur78a,Koniuk80,Isgur79b,Giannini15,Giannini89,Isgur80,Isgur77a,Santopinto15a}.
Some calculations of static properties for 
octet baryons, decuplet baryons, 
octet to decuplet transitions 
and other transitions have been performed 
using some symmetries including $SU(6)$ spin-flavor,  
collective $U(7)$ model, $U$-spin, large $N_c$ limit 
and chiral perturbation theory~\cite{Bijker96a,Bijker09a,Jenkins94a,Santopinto15a,Santopinto10a,Jenkins91a,Buchmann02c,Kubis01,Ledwig11a,Keller12,Buchmann03a,Buchmann08a,Buchmann19}.

\subsection{\it Elastic transitions}

There are almost no empirical information about 
the electromagnetic structure of hyperons and heavy baryons,
due to their short lifetime.
The available information is restricted 
to the magnetic moments of same baryon octet members 
($\Lambda$, $\Sigma^+$, $\Sigma^-$, $\Xi^0$ and $\Xi^-$)
and decuplet baryon members ($\Omega^-$)~\cite{PDG2022}.
One has also a single measurement 
of the $\Sigma^-$ electric charge radius from 
the SELEX collaboration~\cite{PDG2022,SELEX01a}.

A note about the $\Delta(1232)$ is in order, 
since there are measurements for 
the $\Delta^{++}$ and $\Delta^+$ magnetic moments,
although with significant error bars, 
based on several methods~\cite{PDG2022,DeltaFF0,DeltaFF2,PDG2016,Kotulla02}.
Calculations of $\Delta(1232)$ elastic form factors 
can be compared with 
lattice QCD simulations~\cite{Alexandrou09b,Alexandrou09a,Boinepalli09a}
or chiral extrapolations of those calculations.

Since the $Q^2$-dependence of the elastic form factors 
of baryons with heavy quarks cannot be tested directly,
one has to rely on theoretical estimates.
However, there are not many predictions 
associated with the elastic form factors.
Calculations of octet baryon form factors based 
on quark models and soliton models 
can be found in Refs.~\cite{Omega,Octet2,Medium,Ledwig09a,Jakob93,Kubis99a,Silva05,Kim19}.
Estimates based on the Dyson-Schwinger framework can be found in 
Refs.~\cite{Nicmorus10,SAlepuz16,Segovia14b,SAlepuz13}.
The available calculations can be compared with lattice QCD 
simulations for the octet baryon 
and decuplet baryon~\cite{Alexandrou09a,Leinweber93,Boinepalli09a,Lin09a,Alexandrou10a,Boinepalli06a}.

The $\Omega^-$ baryon is a special case, 
since simulations at the physical strange quark mass are possible at present,
and meson cloud effects, dominated in principle by the kaon 
are expected to be small due the chiral 
suppression of heavy mesons~\cite{Omega,HyperonFF,Omega2,Omega3}.

\subsection{\it Inelastic transitions}

Experimentally one has today information 
about two types of transitions associated to radiative decays:
the single transition between octet baryon members 
($\Sigma^0 \to \gamma \Lambda$), and 
three transitions between the  baryon decuplet and the baryon octet:
$\Delta \to \gamma N$, $\Sigma^{0\ast} \to \gamma \Lambda$ and 
$\Sigma^{+ \ast} \to \gamma \Sigma^+$.
The radiative decay widths give us the transition
magnetic moment and $G_M(0)$, apart from the sign.
One has then information about the  $\Sigma^0 \to \gamma \Lambda$,
$\Delta \to \gamma N$, $\Sigma^{0\ast} \to \gamma \Lambda$ and 
$\Sigma^{+ \ast} \to \gamma \Sigma^+$
transition magnetic moments~\cite{PDG2022,DecupletDecays2}.
Recent developments suggest that the 
$\Sigma^{0\ast} \to \gamma \Sigma^0$ magnetic moment 
may also be measured in a near future~\cite{DecupletDecays2,DecupletDecays3,HADES21a}.

To the previous discussion, we can add the weak radiative decays,
when there is a change of flavor between in initial and final states.
Tabled at PDG~\cite{PDG2022} are the decays of the hyperons:
$\Lambda \to \gamma n$, $\Sigma^+ \to \gamma p$, 
$\Xi^0 \to \gamma \Lambda$
and $\Xi^- \to \gamma \Sigma^-$.
From the previous cases, two Dalitz decays
have been measured: $\Sigma^+ \to \mu^+ \mu^- p$
and  $\Xi^0 \to e^+ e^- \Lambda$~\cite{PDG2022}.

Except for the $\gamma^\ast N \to \Delta(1232)$ 
there are then almost no experimental information about 
the octet baryon to decuplet baryon transitions,
except for the value of $G_M(0)$ for the cases mentioned above.
The connection between the $\gamma^\ast N \to \Delta$ transition 
and the reaming transitions based on the $SU(3)$ flavor symmetry 
and related symmetries have been used to estimate 
the magnetic moments of the transitions:
$\gamma^\ast \Sigma^0  \to \Sigma^{0 \ast}$, $\gamma^\ast \Sigma^-  \to \Sigma^{- \ast} $,
$\gamma^\ast \Xi^0   \to \Xi^{0 \ast}$ and $\gamma^\ast \Xi^- \to \Xi^{- \ast} $.
In an exact $SU(3)$ flavor symmetry or in $U$-spin symmetry 
the decays  $\Sigma^{- \ast} \to \gamma \Sigma^-$ and $\Xi^{- \ast} \to \gamma \Xi^-$
are forbidden, meaning that the magnetic moment vanishes 
(zero decay width)~\cite{Keller12}.
Since the $SU(3)$ flavor symmetry is violated in nature 
(the $s$ quark is heavier than quarks $u$ and $d$),
theoretical calculations predict very small but non zero results 
for the magnetic transition moments~\cite{DecupletDecays2}.
A discussion about the results of transitions between the
baryon octet and the baryon decuplet and relevant references 
can be found in Refs.~\cite{DecupletDecays2,DecupletDecays,Keller12}.

Calculations of $\gamma^\ast \Lambda \to \Sigma^0$ transition 
form factors are presented in Refs.~\cite{Octet4,SAlepuz18,Granados17}.
The shape of the form factors resembles the results 
for $\gamma^\ast N \to N\left(\frac{1}{2}^+\right)$ 
since they are also constrained by the condition $F_1(0)=0$.
As in the elastic transitions, we can define
the electric $G_E=F_1 + F_2$
and magnetic $G_M= F_1 - \tau F_2$ form factors.
There are similarities between the
$\gamma^\ast \Lambda \to \Sigma^0$ transition form factors 
and baryon octet form factors of neutral baryons,
like the neutron for instance~\cite{Octet2,Medium,Octet4}.

In Refs.~\cite{Octet2Decuplet,DecupletDecays2,Kim20,Junker20}, 
one can find calculations of  baryon octet to 
baryon decuplet transition form factors based on 
quark models, soliton models and chiral perturbation theory.
Dyson-Schwinger results for the octet baryon to decuplet baryon transitions
are given in Ref.~\cite{SAlepuz18}.
The results from Refs.~\cite{DecupletDecays2,SAlepuz18} 
have a similar shape for $Q^2 > 1 $ GeV$^2$.
Differences below that range are due to inclusion 
of meson cloud contributions in Ref.~\cite{DecupletDecays2} 
similarly to the case of the $\gamma^\ast N \to \Delta(1232)$ transition.

Inelastic transitions probe the timelike kinematic region ($q^2 =- Q^2 >  0$)
of the nucleon excitations $N^\ast$.
This is the region of  experiments as $NN$ collisions in the few GeV energy region,
or di-electron production from pion beam reactions on nucleon targets $\pi^- p \to e^+ e^- n$, 
performed by the HADES (GSI) collaboration
as phase-0 experiments at FAIR~\cite{HADES17,Ramstein18,HADES22a}.
In these experiments one can access information on
the Dalitz decay of nucleon resonances ($N^\ast \to e^+ e^- N$),
which encode information on the multipole form factors
in terms of $q^2$ for $q^2 >  0$~\cite{Timelike2}.
Theoretical calculations of transition form factors in that region
are necessary for the analysis of the Dalitz decay rates
and for the experimental determination of the
Dalitz decay width of $N^\ast$ states~\cite{HADES17,Ramstein18}.
The first determination of a $N^\ast$ Dalitz decay, 
the  $\Delta (1232)$ at HADES~\cite{HADES17}, was performed with the assistance
of the covariant spectator quark model~\cite{Timelike2},
described in Sections~\ref{secCSQM} and \ref{sec-QM-MC}.
The final result for the $\Delta (1232) \to e^+ e^- N$ branching ratio,
$(4.2 \pm 0.7) \times 10^{-5}$, is published at PDG~\cite{PDG2022}.
Under study presently at HADES are the analysis
of the Dalitz decay rates of the $N(1520)$ and $N(1535)$ resonances~\cite{HADES22a},
guided also by calculations of transition form factors
from the covariant spectator quark model~\cite{N1535-TL,N1520TL,HADES22a}.

The knowledge of the transition magnetic moments of hyperons
discussed above (like $\Sigma^{0 \ast} \to \gamma \Lambda$,
$\Sigma^{+ \ast} \to \gamma \Sigma^+$, $\Sigma^{0 \ast} \to \gamma \Sigma^0$,
$\Xi^{- \ast}  \to  \gamma \Xi^-$, etc.) 
are also useful for the study of baryon Dalitz decays ($B^\prime \to e^+ e^- B$)
in order to determine how important the $q^2$-dependence is in the
region $0 < q^2 < (M_{B'}-M_B)^2$~\cite{DecupletDecays2,DecupletDecays3,HADES21a}.
The measurement of hyperon Dalitz decays is planned
for HADES in a near future~\cite{HADES21a,Ramstein18}.

\subsection{\it Electromagnetic form factors in the timelike region}

As proposed long time ago by Cabibbo and Gato~\cite{Cabibbo61a},
the electromagnetic structure of baryons for finite $q^2$, 
including baryons with heavy quarks, can be 
accessed by $e^+ e^-$ scattering or $p \bar p$ scattering.
The $e^+ e^- \to B \bar B$ reactions open a new window 
to study the rule of valence quark effects, cluster of two quarks 
and different quark 
compositions~\cite{Jakob93,Kroll93a,Jaffe03a,Wilczek04a,Selem06a,Dobbs17a}
which are not available in spacelike reactions,  
due to the short live of the baryons with heavy quarks. 
Data associated with the electromagnetic form factors  
of baryons with heavy quarks in the timelike region 
became available in facilities such as BaBar~\cite{BaBar}, 
BES-III~\cite{BESIII}, CLEO~\cite{Dobbs17a}  
and Belle~\cite{Belle}.
New data is also expected from PANDA at FAIR-GSI~\cite{PANDA}.

At the moment, the individual determination of the electric 
and magnetic form factors, complex functions of $q^2$ 
in the timelike region, is not possible, 
except for some particular baryons ($n$, $p$ and $\Lambda$).
However, the possibility exists of measuring
an effective function, which takes into account the 
shape of $G_E$ and $G_M$.

The  $e^+ e^- \to B \bar B$ experiments measure the integrated cross section 
in the $e^+ e^-$ center of mass ($\sigma_{\rm Born} (q^2)$),  
which can be related directly with the effective form factor 
$|G(q^2)|$~\cite{Pacetti14a,HyperonFF,Dobbs17a,Denig13}
\ba
|G(q^2)|^2 =
\frac{2 \tau_{\ms T} |G_M(q^2)|^2 + |G_E(q^2)|^2}{2 \tau_{\ms T} +1},
\label{eqGeff}
\ea
where $\tau_{\ms T} = \frac{q^2}{4 M_B^2}$,
and $G_E$, $G_M$ are the electric and magnetic form 
of the baryon for $J= \frac{1}{2}$. 
In the case $J > \frac{1}{2}$, $G_E$ and $G_M$ 
are replaced by combinations of the 
electric-type and 
the magnetic-type form factors~\cite{HyperonFF,Korner77a}.

Model calculations of  $G_E$ and $G_M$ form factors can be tested indirectly  
by experiments using the relation (\ref{eqGeff}).
In the recent years  $|G(q^2)|$ data associated to 
the baryon octet ($p$, $n$, $\Lambda$, $\Sigma^{0,\pm}$, $\Xi^{0,-}$) 
and $\Omega^-$ became available at BESIII~\cite{BESIII}
and CLEO~\cite{Dobbs17a} up to 18 GeV$^2$.

Timelike experiments are very promising tools 
to study baryon systems with heavy quarks 
for finite squared transfer momentum.
In the near future, we expect to access information 
on several baryons states, including baryons with $c$ quarks
such as the $\Lambda_c^+$~\cite{Belle,BESIII18}.


\section{Conclusions and Outlook \label{sec-conclusions}}

We have presented a review of the experimental and theoretical status 
of electromagnetic transition form factors of baryons
with a particular emphasis on the nucleon excitations $N^\ast$ 
($\gamma^\ast N \to N^\ast$ transitions) in the first,
second and third nucleon resonance regions.
The experimental and theoretical results collected here show
that a vast number of measurements  of
nucleon-resonance electrocouplings and helicity amplitudes has already been accomplished.

With these significant achievements, the resolution of the mass problem associated
with the $N(1440)\frac{1}{2}^+$ resonance, commonly known as the Roper,
serves as a notable example of progress made. However, a comprehensive
understanding of the $Q^2$-evolution of transition form factors,
particularly concerning constituent quarks transitioning to quarks with pointlike couplings,
remains elusive. The emergence of multiquark structures and the transformation
of constituent quarks with varying distance scales are then still
not fully comprehended in a unified manner.
Nevertheless, a promising approach based on Dyson-Schwinger equations
for non perturbative quark dynamics in baryons,
where constituent quark masses are dynamically generated, is gradually weaving an overall picture.

In addition to advancements in the baryon spectrum, calculations have consistently
confirmed that the empirical information gathered over the past two and a
half decades aligns with the conceptual dominance of the three valence quark model
in explaining the structure of nucleon excitation states $N^\ast$ at large momentum transfers.
In the experimentally scrutinized $Q^2$-region 
($Q^2=0$--4 GeV$^2$) these quarks do not refer  to
pointlike quarks as revealed by very large-$Q^2$ experiments,
but are extended quarks with some structure which emerge from 
gluon dressing and $q \bar q$ excitations.
At small $Q^2$, typically for $Q^2 < 2$ GeV$^2$, however, 
the limitations of the interpretation of the data exclusively 
in terms of these valence quark degrees of freedom is very explicit and universal.
In that region, analyses based on chiral symmetry 
which take into account excitations associated 
to light mesons (Goldstone bosons) and baryon meson dressing are crucial.

The $\Delta(1232)\frac{3}{2}^+$ provides a prime example 
that effects associated to the $q \bar q$  excitations
in the form of pion cloud extends the baryon quark core and 
can be as relevant as pure valence quark contributions.
The relevance of the meson cloud effects is also
manifest for instance in the $N(1440)\frac{1}{2}^+$ resonance which combines
meson-baryon degrees of freedom with
quark degrees of freedom.
In general, models which include the meson cloud dressing of the
three valence quarks baryon core describe better 
the transition form factor data, the mass of resonance states
and resonance decay modes.

We started by providing  a toolkit of definitions needed to express
both experimental and theoretical results on electroexcitation
of resonances with general
angular momentum and parity $J^P$.
We took the opportunity to review the
transformation relations between helicity amplitudes, favored by 
experimentalists in their data analysis, and the
kinematic-singularity-free form factors favored by theorists
due to the importance of gauge invariance symmetry.
This compilation is useful as a quick reference in many applications.
The conversion between helicity amplitudes and
multipole form factors was also presented and discussed.
The final formulas for the helicity transition amplitudes
include multiplicative factors that depend on $l=J-\frac{1}{2}$.
We have checked that the compiled
helicity amplitudes and multipole form factors formulas
are in agreement with the definitions
of the radiative decay widths used in data analyses.

We also discussed the limitations of some parametrizations 
of the data of the helicity amplitudes and multipole form factors,
when these representations do not take into account 
correlations between the different functions
(helicity amplitudes and multipole form factors)
that are imposed by their definitions from the independent
and kinematic-singularity-free form factors.
These correlations are revealed only at low $Q^2$, 
close to the pseudothreshold, and cannot be ignored 
in the parametrizations of the data.
The relevance of the pseudothreshold constraints 
is manifest very close to the photon point for the
$\gamma^\ast N \to \Delta(1232)$ amplitudes and amplitudes
of other low-lying nucleon excitations, and therefore in
the determination of the corresponding
nucleon-resonance electrocouplings.

From these analyses we conclude that accurate 
measurements of the helicity amplitudes below $Q^2=0.3$ GeV$^2$ which are
not yet presently available
are fundamental to determine the shape of the helicity amplitudes near $Q^2=0$.
The historically well studied $\gamma^\ast N \to \Delta(1232)$ transition
is the exception, with complete accurate data in the range $Q^2=0$--0.3 GeV$^2$.
New data in the range $Q^2 < 0.3$ GeV$^2$
can at present still be expected only from 
MAMI ($Q^2 < 0.2$ GeV$^2$) and JLab-12 GeV ($Q^2 > 0.05$ GeV$^2$).
In the opposite  extreme of large $Q^2$, it is also necessary to extend the measurements to larger
$Q^2$ values, such that one determines the onset
of perturbative QCD, and tests the evolution of quark masses and their electromagnetic couplings.
We reviewed and compiled the relations between the asymptotic form factors
that  allow the identification of that onset in a easy way,
as well as the expected power laws of their falloffs.

Promising results for the resonances 
$N(1710)\frac{1}{2}^+$, $N(1875)\frac{3}{2}^-$, 
$N(1880)\frac{1}{2}^+$, $\Delta(1900)\frac{3}{2}^-$,..., 
at low $Q^2$ and above 4 GeV$^2$ are expected 
with the emergence of the data analysis from JLab-12 GeV upgrade.
The extraction of properties of the baryons states has been possible
with the use of the partial wave analysis
of the amplitudes associated to the photo- and electro-excitation of the nucleon.
The inclusion of $\pi \pi N$ channels, as well as meson-hyperon channels
is crucial for the identification of resonances with masses above 1.6 GeV.

In the next decades the results from JLab-12 GeV-upgrade 
will be essential to confirm the quark counting rules for the helicity amplitudes, 
or to find particular deviations of those rules.
The possible extension of the JLab-12 GeV program to JLab-22 GeV 
presently under study will be fundamental to probe
the limits of validity of the descriptions based on
valence quarks and the dynamical dependence of the quark mass on $Q^2$.

Future experiments of electron scattering on neutron targets
and ongoing  experiments $e^+ e^-$ or $p \bar p$
scattering in the timelike kinematic regime can also provide complementary information
about the structure of baryons, in particular for hyperons.

Another main conclusion is the  necessity of
linking independent different theoretical approaches that are complementary, for
more model independence and uncertainty control. Excellent examples
are chiral effective field theory to guide extrapolations of lattice QCD to the physical region
and large $N_c$ limit relations that include $q\bar{q}$/meson degrees
of freedom to validate those extrapolations.
Also, the combination of lattice QCD with quark models leverages the strengths of both:
Lattice QCD can provide input parameters for quark models,
such as quark masses and hadron wave functions,
and can also test the validity of assumptions made in quark models.
Quark models, in turn, can provide insights into the interpretation of lattice QCD results.

For amusement of the reader we end this review with the answer that we obtained
by questioning ChatGPT about the importance of baryons:
"Baryons are paramount for our understanding of the universe. As the building blocks of matter,
they drive energy production in stars, shape cosmic structures through gravitational interactions,
and leave imprints on the cosmic microwave background. Studying exotic baryonic
states expands our knowledge of fundamental forces.
Baryons serve as crucial messengers, connecting the microcosm to the vastness of the cosmos,
illuminating its mysteries." Not a bad synthesis!


\subsection*{\bf Acknowledgments}

G.R.~was supported by the Basic Science Research Program 
funded by the Republic of Korea Ministry of Education 
(Grant No.~NRF--2021R1A6A1A03043957).
M.T.P.~was supported by the Portuguese Science Foundation
FCT under project CERN/FIS-PAR/0023/2021.
The present work was not possible without the collaboration 
and multiples discussions with the hadronic community.
We thank the following authors for sharing calculations, figures and discussions:
Inna Aznaryan, Volker Burkert, Lothar Tiator, Stefan Scherer,  Marius Hilt, 
Nikolas Sparveris, Sean Stave, Christian Fischer,
Gernot Eichmann, Diana Nicmorus,  Constantia Alexandrou,
Vladimir Pascalutsa, Mark Vanderhaeghen, Franziska Hagelstein, 
Toru Sato, Harry Lee, Satoshi Nakamura, Hiroyuki Kamano,
Thomas Gutsche, Igor Obukhvsky, Valery Lyubovitskij, 
Vladimir Braun, Igor Anikin, and Elena Santopinto.

\appendix

\renewcommand{\theequation}{A.\arabic{equation}}
\setcounter{equation}{0}

\section{Notation \label{appNotation}}

\subsection{\it Matrix tensor}

For the conversion between covariant $a^\mu$ and 
contravariant $a_\mu$ 4-vector we use the 
following convention for the metric tensor
\ba
g^{\mu \nu} = {\rm diag}(1,-1,-1,-1) =
\left(  
\begin{array}{c r r r} 1 &0 &0 &0 \cr 0 & -1 & 0 & 0\cr
0 & 0 & -1 & 0 \cr
0 & 0 & 0 & -1 \cr
\end{array}
\right).
\ea

\subsection{\it Dirac matrices}

We write the Dirac matrices $\gamma^\mu$ 
in the Dirac-Pauli representation~\cite{HalzenMartin,Griffiths}
\ba
\gamma^0 = 
\left(  
\begin{array}{r r} \Unit &0 \cr 
 0 & - \Unit \cr
\end{array}
\right),
\hspace{.5cm}
\gamma^i = 
\left(  
\begin{array}{r r} 0 & \sigma_i \cr 
 -\sigma_i & 0  \cr
\end{array}
\right),
\hspace{.5cm}
\gamma_5 = i 
\gamma^0 \gamma^1 \gamma^2  \gamma^3 =
\left(  
\begin{array}{r r} 0 & \Unit \cr 
 \Unit & 0  \cr
\end{array}
\right),
\ea
where $\Unit$ is the $2 \times 2$ unitary matrix, 
and $\sigma_i$ are the Pauli matrices
\ba
\sigma_1 = 
\left(  
\begin{array}{r r} 0 & 1 \cr 
 1 & 0  \cr
\end{array}
\right),
\hspace{.5cm}
\sigma_2 = 
\left(  
\begin{array}{r r} 0 & -i \cr 
 i & 0  \cr
\end{array}
\right),
\hspace{.5cm}
\sigma_3 = 
\left(  
\begin{array}{r r} 1 & 0 \cr 
 0 & -1  \cr
\end{array}
\right).
\ea

The Dirac matrices satisfay the commutation relation 
\ba
\left\{\gamma^\mu,\gamma^\nu \right\} = 2 g^{\mu \nu}.
\ea

An equivalent definition of the $\gamma_5$ matrix is given by 
\ba
\gamma_5 = \frac{i}{24!} 
\varepsilon_{\alpha \beta \sigma \rho} 
\gamma^\alpha \gamma^\beta \gamma^\sigma \gamma^\rho,
\ea
where $\varepsilon_{0123} =1$.

In the expressions of the transition currents we use also 
the definition
\ba
\sigma^{\mu \nu} = 
\frac{i}{2}(\gamma^\mu \gamma^\nu - \gamma^\nu \gamma^\mu).
\ea

\subsection{\it Normalization of spinors}

Spin $\frac{1}{2}$ and $\frac{3}{2}$ states are described by
 Dirac ($u$) and Rarita-Schwinger ($u_\alpha$)
spinors, respectively, with the normalizations~\cite{NDelta,Rarita41a}
\ba
\bar u(p,s) u(p,s) = 1, \hspace{1cm}
\bar u_\alpha (p,s) u_\alpha (p,s) = -1,
\ea
where $p$ is the momentum and 
$s$ is the spin projection along the $z$ direction.
States with higher  order non integer spins $J= l + \frac{1}{2}$ ($l=2,3,...$)
are represented by generalized Rarita-Schwinger spinors 
defined by  $l$ indexes, which are normalized 
according to~\cite{Krivoruchenko02,Rarita41a,Fierz39a}
\ba
\bar u_{\alpha_1 \alpha_2 ... \alpha_{\; \, {\sm l}} }(p,s)
u^{\alpha_1 \alpha_2 ... \alpha_{\; \, {\sm l}} } (p,s)= (-1)^l.
\label{eqSpinorJL}
\ea


\renewcommand{\theequation}{B.\arabic{equation}}
\setcounter{equation}{0}

\section{Multipole form factores for $J \ge \frac{3}{2}$ \label{appMultipole}}

We present here the explicit expressions 
for the multipole form factors $G_M$, $G_E$ and $G_C$
for $J \ge \frac{3}{2}$ for \mbox{$l=1,2,3, ...$}
in terms of the kinematic-singularity-free form factors $G_i$.
We use the notation from Section~\ref{sec-TransitionFF}.

\subsection{\it Form factors in terms of $h_i$}

In a first step we use the relation 
between the multipole form factors and 
the helicity form factors~\cite{Devenish76}.
For $J^P =\frac{3}{2}^+, \frac{5}{2}^-,  \frac{7}{2}^+, ...$, 
one has
\ba
G_M(Q^2) &=& - \left[ (l+2) h_2 + l h_3 \right] \frac{1}{l+1} Z_+ \\
G_E(Q^2) &=& - \left[ h_2 - h_3 \right] \frac{1}{l+1} Z_+ \\
G_C(Q^2) &=& h_1 Z_+,
\ea
where 
\ba
Z_+ = \frac{M}{3(M_R + M)} \sqrt{\frac{3}{2}}.
\label{eqZp}
\ea

For $J^P=\frac{3}{2}^-, \frac{5}{2}^+, \frac{7}{2}^-, ...$,  
one has
\ba
G_M(Q^2) &=&  \left[ h_2 + h_3 \right] \frac{1}{l+1} Z_- \\
G_E(Q^2) &=&  \left[ (l+2) h_2 - l h_3 \right] \frac{1}{l+1} Z_- \\
G_C(Q^2) &=& h_1 Z_-,
\ea
where 
\ba
Z_- = \frac{M}{3(M_R - M)} \sqrt{\frac{3}{2}}.
\label{eqZm}
\ea

The previous expressions are in the 
Aznauryan-Burkert representation~\cite{Aznauryan12a}.
The Devenish representation~\cite{Devenish76} exclude 
the factor $\sqrt{\frac{3}{2}}$ from $Z_\pm$ 
since the factor is canceled by the 
redefinition of the functions $G_i$ and $h_i$.

\subsection{\it Form factors in terms of $G_i$}

Using the explicit expressions for the 
$h_i$ from Eqs.~(\ref{eqH1})--(\ref{eqH3}),
we obtain the final expressions for 
the multipole form factors.

For $J^P =\frac{3}{2}^+, \frac{5}{2}^-,  \frac{7}{2}^+, ...$, one has
\ba
\hspace{-1.3cm}
G_M(Q^2) &=& \left\{
\left[ 2M_R (M_R + M) + l Q_+^2 \right]\frac{G_1(Q^2)}{M_R}
+ (M_R^2 - M^2 -Q^2) G_2(Q^2) - 2 Q^2 G_3(Q^2)  
\right\}\frac{2}{l+1}  Z_+ , \\
\hspace{-1.3cm}
G_E(Q^2) & =& \left\{
(M_R^2 - M^2 - Q^2) \frac{G_1(Q^2)}{M_R} + 
(M_R^2 - M^2 - Q^2) G_2(Q^2) - 2 Q^2 G_3(Q^2) 
\right\} \frac{2}{l+1}  Z_+, \\
\hspace{-1.3cm}
G_C(Q^2) &=& \left[ 2 M_R G_1 (Q^2)  + 
2 M_R^2 G_2 (Q^2) + (M_R^2 - M^2 - Q^2) G_3(Q^2)
\right] 2Z_+. 
\ea

For $J^P=\frac{3}{2}^-, \frac{5}{2}^+, \frac{7}{2}^-,...$, one has
\ba
G_M(Q^2)  &=& -Q_-^2\frac{G_1(Q^2)}{M_R} \frac{2}{l+1} Z_-, \\
G_E(Q^2)  &=& - \left\{
\left[ 2M_R (M_R -M) + l (M_R^2 -M^2 - Q^2)\right] 
\frac{G_1(Q^2)}{M_R}  \right. \nonumber \\
& & \left. +  (l+1) (M_R^2 - M^2 - Q^2) G_2(Q^2) - 2(l+1) Q^2 G_3(Q^2)   \frac{}{} \!
\right\} \frac{2}{l+1} Z_-, \\
G_C(Q^2) &=& \left[ 2 M_R G_1 (Q^2)  + 
2 M_R^2 G_2 (Q^2) + (M_R^2 - M^2 - Q^2) G_3(Q^2)
\right]  2Z_-.
\ea

The factors $Z_\pm$ are defined by Eqs.~(\ref{eqZp}) and (\ref{eqZm}).

The case $J= \frac{3}{2}$, for the $\Delta(1232)$ and $N(1520)$
are obtained setting $l=1$.

As mentioned earlier, the previous expressions differs from 
Devenish et al.~\cite{Devenish76} by the factor 
$\sqrt{\frac{3}{2}}$ in the functions $Z_\pm$.
Due to the normalization of $G_i$, the two representations 
provide the same numerical result for the multipole form factors, 
and helicity amplitudes.
To be more specific, the Aznauryan-Burkert and 
the Devenish representations provide the same result 
for the magnetic transition form factor 
of the $\gamma^\ast N \to \Delta(1232)$ transition:
$G_M(0) \simeq 3.0$, although they use different normalizations 
for $G_i$.

\renewcommand{\theequation}{C.\arabic{equation}}
\setcounter{equation}{0}

\section{Spin states of particles with spin $J$ \label{appSpinJ}}

We describe in this appendix how to derive spinors of particles 
with spin $J=\frac{3}{2}, \frac{5}{2},  \frac{7}{2}, ...$.
The starting point is the construction of states of spin with $J= \frac{3}{2}$.
Following the rules of angular momentum states, 
we can built a state of spin $\frac{3}{2}$ combining the 
states of angular momentum $j_1=1$ and $j_2= \frac{1}{2}$.
The states of spin $\frac{1}{2}$ are the Dirac spinors:
$u(P,s)$ where $P$ is the momentum and $s= \pm \frac{1}{2}$ 
the projection along the $z$-axis.
As for the states of spin $j_1=1$ they correspond to the 
polarization vectors $\varepsilon_{{\ms R} \lambda}^\alpha$, where $\lambda = 0,\pm$.
We use the label $R$ to indicate that the 
polarization vector respects the resonance, 
and not to the photon states ($\epsilon_\lambda$), 
as in Eqs.~(\ref{eqEpsilon}).

The consequence of this construction is that a state of spin $\frac{3}{2}$ 
(Rarita-Schwinger state) depend on a Lorentz index $\alpha$.
A state of spin $\frac{5}{2}$ can be derived from the combination 
of spin $j_1=1$ ($\varepsilon_{{\ms R} \lambda}^\alpha$) and $j_2= \frac{3}{2}$,
a Rarita-Schwinger state $u^\beta $, and include consequently two indexes ($u^{\alpha \beta}$).
We can keep going, step by step, creating states with 3 labels ($J=\frac{7}{2}$),
4 labels ($J=\frac{9}{2}$) etc.
Notice that, since we know how to define states with $J=\frac{3}{2}$, 
we can create states with arbitrary large $J$, 
adding the angular momenta states $j_1=1$ and $j_2 = J-1$, successively.

The polarization vectors for $\lambda= \pm$ take the form
\ba
\varepsilon_{{\ms R} \pm}  
= \mp \frac{1}{\sqrt{2}} (0,1,\pm i,0). 
\ea
The polarization vector for $\lambda=0$ 
take the form 
\ba
\varepsilon_{{\ms R} \pm}  = (0,0,0,1),
\ea
when the particle is at the rest frame.

The states associated to particles at rest,
when $p= \bar P =(M_R,0,0,0)$ are straight forward 
and are presented in Section~\ref{apRestFrame} 
for the cases $J=\frac{3}{2}$, $\frac{5}{2}$ and  $\frac{7}{2}$.
We discuss now how to determine a state for a particle 
with a momentum $p= (E_R,0,0, |{\bf p}|)$, 
where $E_R = \sqrt{M_R^2 + |{\bf p}|^2}$.
The results can be used to calculate amplitudes at the Breit frame 
(see Section~\ref{sec-largeQ2}).

The transformations of the $J$-spinors in a Lorentz 
transformation $\Lambda$ are similar to the transformation 
of the Dirac spinor (mass $M$) 
\ba
u(P,s) \to S(\Lambda) u(\bar P,s),
\ea
using the notation $p= (E,0,0,|{\bf p}|)$,
with $E= \sqrt{M^2 + |{\bf p}|^2}$.
and $\bar p= (M,0,0,0)$ for the initial state.

In the case of a $J$-spinor associated to the indexes 
$\alpha_1$, ...$ \alpha_2$ the transformation is 
\ba
u^{\alpha_1 \cdot \cdot \cdot \alpha_n} (p,s) =
S(\Lambda) \, \Lambda^{\alpha_1}_{\;\; \beta_1} \cdot \cdot \cdot  
\Lambda^{\alpha_n}_{\;\; \beta_n} \, u^{\beta_1 \cdot \cdot \cdot \beta_n} (\bar p,s), 
\ea 
where $\Lambda^{\alpha}_{\;\; \beta}$ 
is the Lorentz transformation  associated 
to the boost the frame where $\bar p= (M_R,0,0,0)$ 
to $p= (E_R,0,0, |{\bf p}|)$
\ba
\Lambda^{\alpha}_{\;\; \beta}
=
\left[
\begin{array}{cccc} 
 \frac{E_R}{M_R} & 0 & 0 & \frac{|{\bf p}|}{M_R} \cr
    0            & 1 & 0 & 0\cr
        0        & 0 & 1 & 0\cr
 \frac{|{\bf p}|}{M_R}  & 0 & 0 &  \frac{E_R}{M_R} \cr
\end{array}
\right]
\ea
Since  $\Lambda^{\alpha}_{\;\; \beta}$ acts on the polarization vectors, 
and $S(\Lambda)$ on the spin $\frac{1}{2}$ states, 
the boosted states can be determined using~\cite{HalzenMartin,Gross93}
\ba
S(\Lambda) = \sqrt{\frac{M_R + E_R}{2 M_R}} 
+ \sqrt{\frac{M_R + E_R}{2 M_R}} \frac{|{\bf p}|}{M_R + E_R} 
\left[
\begin{array}{cc}
 0 & \sigma_3 \cr 
\sigma_3 & 0 \cr
\end{array}
\right].
\ea 

In the calculations, we can use the results  
of the operation $\Lambda$ on the polarization states
\ba
& &
\Lambda^{\alpha}_{\;\; \beta} \; \varepsilon_{{\ms R} \pm}^\beta =   \varepsilon_{{\ms R} \pm}^\alpha \\
& &
\Lambda^{\alpha}_{\;\; \beta} \; \varepsilon_{{\ms R}  0}^\beta = 
\left[
\begin{array}{c}
    \frac{|{\bf p}|}{M_R} \cr
   0  \cr
   0 \cr
    \frac{E_R}{M_R}
\end{array}
\right].
\ea
Notice that at the l.h.s.~the polarization vectors are 
defined at the rest frame ($p \to \bar p$).
The boost preserves $\varepsilon_{{\ms R} \pm}^\alpha$ and 
modifies $\varepsilon_{{\ms R}  0}^\beta$.

The operator $S(\Lambda)$ transforms  
\ba
u_R (\bar p,s) = 
\left[
\begin{array}{c}
   1     \cr
   0     \cr
\end{array}
\right] \chi_s,
\ea 
into 
\ba
u_R (p,s) = 
 \sqrt{\sfrac{M_R + E_R}{2 M_R}} 
\left[
\begin{array}{c}
   1 \cr
   \frac{ |{\bf p}|  \; \sigma_3}{ M_R + E_R}     \cr
\end{array}
\right] \chi_s,
\ea 
where $\chi_s$ are Pauli spinors for $s= \pm \frac{1}{2}$.

\subsection{\it States of the particle at rest \label{apRestFrame}}

The states of the particles at rest follow 
the rules of the angular momentum,
according with the Clebsch-Gordan expansions.
The states $s= - \frac{1}{2}$ and $s= - \frac{3}{2}$ 
can be derived from $s= + \frac{1}{2}$ and $s= + \frac{3}{2}$ 
using the symmetry between Clebsch-Gordan coefficients
\ba
\left< j_1 j_2 ; - m_1 -m_2 \right| \left. J -M \right> 
= \left< j_1 j_2 ;  m_1 m_2 \right| \left. J  M \right>,
\ea 
when $J= j_1 + j_2$.

\subsubsection*{\it Spin $\frac{3}{2}$} 

\ba
& &
u^\alpha \left(\bar p, +  \sfrac{3}{2} \right)
=   \varepsilon_{{\ms R}+}^\alpha  \, u_R \!\left(\bar p, +\sfrac{1}{2}\right), \\
& &
u^\alpha \left(\bar p, +  \sfrac{1}{2} \right)
= 
\sqrt{\sfrac{2}{3}} \, \varepsilon_{{\ms R} 0}^\alpha \,  u_R \!\left(\bar p, 
+\sfrac{1}{2}\right) +
\sqrt{\sfrac{1}{3}} \, \varepsilon_{{\ms R} +}^\alpha \, u_R \!\left(\bar p, -\sfrac{1}{2}\right).
\ea

\subsubsection*{\it Spin $\frac{5}{2}$} 

\ba
& &
u^{\alpha \beta} \left(\bar p, +  \sfrac{3}{2} \right)
=  
\sqrt{\sfrac{2}{5}} \, \varepsilon_{{\ms R}0}^\alpha \, u^\beta\! \left(\bar p, +\sfrac{3}{2}\right) 
+ \sqrt{\sfrac{3}{5}} \, \varepsilon_{{\ms R}+}^\alpha \, u^\beta \!\left(\bar p, +\sfrac{1}{2}\right), \\
& &
u^{\alpha \beta} \left(\bar p, +  \sfrac{1}{2} \right)
=  
\sqrt{\sfrac{1}{10}} \, \varepsilon_{{\ms R}-}^\alpha \, u^\beta \! \left(\bar p, +\sfrac{3}{2}\right) 
+ \sqrt{\sfrac{3}{5}} \, \varepsilon_{{\ms R}0}^\alpha \, u^\beta \! \left(\bar p, +\sfrac{1}{2}\right) 
+ \sqrt{\sfrac{3}{10}} \,\varepsilon_{{\ms R}+}^\alpha \, u^\beta \! \left(\bar p, -\sfrac{1}{2}\right).
\ea

\subsubsection*{\it Spin $\frac{7}{2}$}

\ba
& &
u^{\alpha \beta \sigma} \left(\bar p, +  \sfrac{3}{2} \right)
=  
\sqrt{\sfrac{1}{21}} \, \varepsilon_{{\ms R}-}^\alpha \,
u^{\beta \sigma} \! \left(\bar p, +\sfrac{5}{2}\right) 
+ \sqrt{\sfrac{10}{21}} \, \varepsilon_{{\ms R}0}^\alpha \, u^{\beta \sigma} \! 
\left(\bar p, +\sfrac{3}{2}\right) + 
 \sqrt{\sfrac{10}{21}} \, \varepsilon_{{\ms R}+}^\alpha \, u^{\beta \sigma} \! 
\left(\bar p, +\sfrac{1}{2}\right), \\ 
& &
u^{\alpha \beta \sigma} \left(\bar p, +  \sfrac{1}{2} \right)
=  
\sqrt{\sfrac{1}{7}} \, \varepsilon_{{\ms R}-}^\alpha \, 
u^{\beta \sigma} \! \left(\bar p, +\sfrac{3}{2}\right) 
+ \sqrt{\sfrac{4}{7}} \, \varepsilon_{{\ms R}0}^\alpha \, u^{\beta \sigma} \! 
\left(\bar p, +\sfrac{1}{2}\right) + 
 \sqrt{\sfrac{2}{7}} \, \varepsilon_{{\ms R}+}^\alpha \, u^{\beta \sigma} \! 
\left(\bar p, +\sfrac{1}{2}\right).  
\ea

The coefficients for higher values of $J$ 
can be extracted from Clebsch-Gordan tables 
to obtain relations with the states with angular momentum $J-1$.

The previous states are normalized according to the relation 
(see Appendix~\ref{appNotation})
\ba
\bar u_{\alpha_1 \alpha_2 ... \alpha_{\; \, {\sm l}} }(p,s)
u^{\alpha_1 \alpha_2 ... \alpha_{\; \, {\sm l}} } (p,s)= (-1)^l.
\label{eqSpinorJL2}
\ea

\renewcommand{\theequation}{D.\arabic{equation}}
\setcounter{equation}{0}

\section{Asymptotic form of the transition form factors \label{appLargeQ2}}

\subsection{\it Cases $J^P= \frac{1}{2}^\pm$}

Using the $G_i$ representation 
and the relation with $F_i$
\ba
G_1 \; \propto \; \frac{1}{Q^6}, \hspace{1cm} 
G_2  \; \propto \; \frac{1}{Q^6}, 
F_1  \; \propto \; \frac{1}{Q^4}, \hspace{1cm} 
F_2  \; \propto \; \frac{1}{Q^6}. 
\ea

\subsection{\it Cases $J^P= \frac{3}{2}^\pm$, $\frac{5}{2}^\mp$, ...}

For convenience, we use the representation 
proposed by Jones and Scadron
\ba
\Gamma_{l}^{\alpha \mu}  (q)=
\left[q^\alpha \gamma^\mu - {\not \! q} g^{\alpha \mu} \right] G_1 (Q^2) + 
\left[q^\alpha P^\mu -  (P \cdot q)  g^{\alpha \mu} \right] G_2 (Q^2) +
\left[q^\alpha q^\mu -  q^2  g^{\alpha \mu} \right] G_3 (Q^2), 
\ea
where $G_i$ are free of kinematic singularities independent functions 
which differ from Eq.~(\ref{eqGamma1}) 
by the second term, where we replace $p^\prime$ by $P= \frac{1}{2}(p + p^\prime)$.
We obtain then
\ba
h_1 &=& 4M_R G_1 + (3 M_R^2 + M^2 + Q^2) G_2 + 2 (M_R^2  - M^2 - Q^2) G_3,
\label{eqH1p}\\
h_2 &=& - 2 (M_R\pm M) G_1 - (M_R^2 -M^2) G_2 + 2  Q^2 G_3, 
\label{eqH2p}\\
h_3 &=& - 2 [Q^2 + M (M \pm M_R)] \frac{G_1}{M_R} + (M_R^2 -M^2) G_2 - 2  Q^2 G_3.
 \label{eqH3p}
\ea
We recover the Devenish representation using
\ba
G_1 \to G_1, \hspace{1cm} 
G_2 \to G_2,  \hspace{1cm} 
G_3 \to G_3 + \frac{1}{2} G_2.
\ea

From the relations (\ref{eqH1p})--(\ref{eqH3p}), (\ref{eqAmps32}), 
(\ref{eqAmps52}) and (\ref{eqPQCD}), 
we conclude that
\ba
h_1 &=& 4M_R G_1 + (3 M_R^2 + M^2 + Q^2) G_2 + 2 (M_R^2  - M^2 - Q^2) G_3 
 \; \propto \; \frac{1}{Q^{2l + 4}}, 
\label{eqH1b}\\
h_2 &=& - 2 (M_R\pm M) G_1 - T_1  \; \propto \; \frac{1}{Q^{2 l+ 4}}, 
\label{eqH2b}\\
h_3 &=& - 2 [Q^2 + M (M \pm M_R)] \frac{G_1}{M_R}  
+ T_1   \; \propto \; \frac{1}{Q^{2 l+ 2}},
 \label{eqH3b}
\ea
where 
\ba
T_1 =  (M_R^2 -M^2) G_2 - 2  Q^2 G_3. 
\ea

Combining (\ref{eqH2b}) and (\ref{eqH3b}), 
we can write 
\ba
h_2 + h_3 &=& - 2 \frac{Q_\pm^2}{M_R} G_1  \; \propto \; \frac{1}{Q^{2 l+ 2}},
\label{eqH2H3}\\
T_2 &=& 
\frac{ M(M\pm M_R) + Q^2}{M_R  (M_R \pm M)} h_2 - h_3  
\nonumber \\
&=& - \left( 1 + \frac{ M(M\pm M_R) + Q^2}{M_R (M_R \pm M)}
 \right)  T_1  \; \propto \;  \frac{1}{Q^{2 l+ 2}}.
\label{eqT2}
\ea 

From Eqs.~(\ref{eqH2H3}), we conclude that 
\ba
G_1  \; \propto \; \frac{1}{Q^{2 l + 4}}.
\ea
From Eq.~(\ref{eqT2}), one conclude also that 
\ba
T_1 = (M_R^2 -M^2) G_2 - 2  Q^2 G_3  \; \propto \; \frac{1}{Q^{2 l + 4}}.
\label{eqT1b}
\ea

To obtain a new relation between $G_2$ and $G_3$, we use
\ba
T_3 &=& h_1 + \frac{2 M_R}{M_R \pm M} h_2  \nonumber \\
    &=& Q_\pm^2 G_2 + 2 
\left( M_R^2 - M^2  + Z Q^2
\right) G_3  \; \propto \;  \frac{1}{Q^{2 l + 4}},
\label{eqT3}
\ea
where $Z=  \frac{(M_R \pm M)^2}{M_R^2 -M^2}$.

We have now two equations for $G_2$ and $G_3$,
Eqs.~(\ref{eqT1b}) and (\ref{eqT3}).
Manipulating the two equations, one can derive 
the relations
\ba
T_4 &=& T_3 - \frac{Q_\pm^2}{M_R^2 -M^2} T_1 \nonumber \\
    &=& 2 
\left( M_R^2 - M^2 + 2 \frac{M_R^2 + M^2}{M_R^2 - M^2} Q^2
+ \frac{Q^4}{M_R^2 -M^2} \right) 
G_3  \; \propto \; \frac{1}{Q^{2 l + 2}},
\label{eqT4} \\
T_5 &=&  T_1 + \frac{M_R^2 - M^2 + Z Q^2}{Q^2} 
T_3 \nonumber \\
&=&  
\left(Q^2 + 2(M_R^2 + M^2) + \frac{(M_R^2-M^2)^2}{Q^2}   
\right) G_2  \; \propto \;  \frac{1}{Q^{2 l + 4}}.
\label{eqT5}
\ea  
From Eqs.~(\ref{eqT4}) and (\ref{eqT5}) 
one can conclude that $G_2  \; \propto \; 1/Q^{2l +6}$ 
and $G_3  \; \propto \; 1/Q^{2l +6}$.

To summarize, the form factors (\ref{eqH1b}), 
(\ref{eqH2b}) and   (\ref{eqH3b}) are the consequence 
of the relations 
\ba
G_1 \; \propto \; \frac{1}{Q^{2l + 4}}, 
\hspace{1cm}
G_2 \; \propto \; \frac{1}{Q^{2l + 6}}, 
\hspace{1cm}
G_3 \; \propto \; \frac{1}{Q^{2l + 6}}. 
\ea
The same proportions hold for the form factors associated to Eq.~(\ref{eqGamma1}).

The asymptotic forms for the multipole form factors 
can be obtained from Eqs.~(\ref{eqGM1p})--(\ref{eqGC1p}),
(\ref{eqGM1m})--(\ref{eqGC1m}), (\ref{eqGM2m})--(\ref{eqGC2m})
and   (\ref{eqGM2p})--(\ref{eqGC2p}):
\ba
G_E \; \propto \; \frac{1}{Q^{2l+2}}, \hspace{1cm}
G_M \; \propto \; \frac{1}{Q^{2l+2}}, \hspace{1cm}
G_C \; \propto \;\frac{1}{Q^{2l+4}}.
\label{eqGEMC-LQ2-A}
\ea




\bibliographystyle{phaip01}

\bibliography{biblo}

\end{document}